\documentclass[printer]{aa} % for a referee version
\usepackage{graphicx}
\usepackage{txfonts}
\usepackage{lscape}
\begin{document} 

   \title{A multi-wavelength view of magnetic flaring from PMS stars}

   \subtitle{}

\author{E. Flaccomio\inst{1} \and G. Micela\inst{1} \and S. Sciortino\inst{1} \and A. M. Cody\inst{2} \and M. G. Guarcello\inst{1} \and M. Morales-Calder\`on\inst{3} \and L. Rebull\inst{4}  \and J. R. Stauffer\inst{4}
}
\institute{
INAF - Osservatorio Astronomico di Palermo,  Piazza del Parlamento 1, I-90134 Palermo, Italy \\ \email{E. Flaccomio, ettoref@astropa.inaf.it}
\and
NASA Ames Research Center, Kepler Science Office, Mountain View, CA 94035, USA \and
Centro de Astrobiolog\'ia, Departamento de Astrof\'isica, INTA-CSIC, PO BOX 78, ESAC Campus, 28691 Villanueva de la Ca\~nada,
Madrid, Spain 
\and
Spitzer Science Center, California Institute of Technology, Pasadena, CA 91125, USA 
}

   \date{Received Month XX, YYYY; accepted Month ZZ, HHHH}

% \abstract{}{}{}{}{} 
% 5 {} token are mandatory
 
  \abstract
  % context heading (optional) {} leave it empty if necessary  
   { Flaring is an ubiquitous manifestation of magnetic activity in low
   mass stars including, of course, the Sun. Although flares, both from
   the Sun and from other stars, are most prominently observed in the
   soft X-ray band, most of the radiated energy is released at
   optical/UV wavelengths. In spite of decades of investigation, the
   physics of flares, even solar ones, is not fully understood. Even
   less is known about magnetic flaring in pre-main sequence (PMS)
   stars, at least in part because of the lack of suitable
   multi-wavelength data. This is unfortunate since the energetic
   radiation from stellar flares, which is routinely observed to be
   orders of magnitude greater than in solar flares, might have a
   significant impact on the evolution of circumstellar, planet-forming
   disks.   
   }
   { We aim at improving our understanding of flares from PMS stars. Our
   immediate objectives are constraining the relation between flare
   emission at X-ray, optical, and mid-infrared (mIR) bands, inferring properties of
   the optically emitting region, and looking for signatures of the
   interaction between flares and the circumstellar environment, i.e.
   disks and envelopes. This information might then serve as input for
   detailed models of the interaction between stellar atmospheres,
   circumstellar disks and proto-planets.
    }
   {  Observations of a large sample of PMS stars in the NGC~2264 star
   forming region were obtained in December 2011, simultaneously with
   three space-borne telescopes, {\em Chandra} (X-rays), {\em CoRoT}
   (optical), and {\em Spitzer} (mIR), as part of the “Coordinated
   Synoptic Investigation of NGC2264” (CSI-NGC2264). Shorter {\em
   Chandra} and {\em CoRoT} observations were also obtained in March
   2008. We analyzed the lightcurves obtained during the {\em Chandra}
   observations ($\sim$300\,ks and $\sim$60\,ks in 2011 and
   2008, respectively), to detect X-ray flares with an optical and/or
   mIR counterpart. From the three datasets we then estimated basic
   flare properties, such as emitted energies and peak luminosities.
   These were then compared to constrain the spectral energy
   distribution of the flaring emission and the physical conditions of
   the emitting regions. The properties of flares from stars with and
   without circumstellar disks were also compared to establish any
   difference that might be attributed to the presence of disks.
  } 
   { Seventy-eight X-ray flares (from 65 stars) with an optical and/or mIR counterpart
   were detected. The optical emission of flares (both emitted energy
   and peak flux) is found to correlate well with, and to be
   significantly larger than, the X-ray emission. The slopes of the
   correlations suggest that the difference becomes smaller for the most
   powerful flares. The mIR flare emission seems to be strongly affected
   by the presence of a circumstellar disk: flares from stars with disks
   have a stronger mIR emission with respect to stars without disks.
   This might be attributed to either a cooler temperature of the region
   emitting both the optical and mIR flux or, perhaps more likely, to the
   reprocessing of the optical (and X-ray) flare emission by the inner
   circumstellar disk, providing evidence for flare-induced disk heating.
   }
   {}


   \maketitle
%
%________________________________________________________________

\section{Introduction}

Low-mass stars are characterized by strong magnetic fields and an
associated diverse array of atmospheric phenomena, collectively
referred to as magnetic activity, such as chromospheres, coronae,
photospheric dark spots, and flares. While relevant at all ages,
magnetic activity is particularly strong during the first few
million years of pre-main-sequence (PMS) evolution, impacting the
evolution of both stars and their circumstellar environments. Indeed, the
X-ray/EUV/UV radiation from magnetic coronae and chromospheres is
believed to significantly heat and ionize circumstellar disks, driving
strong photo-evaporative disk outflows and affecting disk viscosity
\citep[e.g.,][]{gla97,pas09,bai11,erc16}. Therefore, the formation and
early evolution of proto-planets within circumstellar disks is also most
likely affected \citep[e.g.][]{mor16a}. Furthermore, in addition to
driving the aforementioned array of classical activity phenomena, the
magnetic fields of PMS stars also play a central role in mass-accretion
from circumstellar disks and in the launching and collimation of
protostellar jets.

Observationally, non-accreting PMS stars (weak line T Tauri stars,
WTTSs) resemble, in several respects, the most active main sequence (MS)
stars, for example when comparing {\em normalized} coronal and
chromospheric luminosities such as $L_X/L_{bol}$ or
$L_{H\alpha}/L_{bol}$. By extension, Solar-like magnetic activity is
generally inferred, although at much enhanced levels \citep[e.g. by
3-4\,dex in $L_X/L_{bol}$,][]{pre05a}. The X-ray activity levels of
stars that are still accreting from their circumstellar disk (Classical 
T Tauri stars, CTTSs), however, are observed to be significantly lower
on average, and with a larger scatter at any given mass or spectral
type \citep{dam95a,fla03b}. Moreover, \citet{fla12} demonstrated
that CTTSs are also significantly more variable in X-rays with respect
to WTTSs. Whether these differences between activity on CTTS and WTTS
are intrinsic or not, e.g. due to unaccounted for absorption by
circumstellar structures, is still an open question.

In this work we will focus on magnetic flaring, outbursts ubiquitously
observed on all coronal sources, including, of course, the Sun, and
whose origin can be traced to the release of magnetic energy in the
higher corona. Flares are most prominently observed in the soft X-ray
band, where the contrast with the out-of-flare emission is the highest.
However, at least for solar flares, most of the radiated energy is known
to be emitted at longer wavelengths, in the optical and UV bands.
Moreover, about as much of the total flare energy may be transformed into
kinetic energy, for example in coronal mass ejections (CMEs), as is
radiated away \citep{ems05}.

According to the standard model \citep[e.g.,][]{fle11a}, flares are the result of magnetic reconnection events high
in the corona, a sudden rearrangement of the magnetic field
configuration. The release of previously accumulated magnetic energy
results in streams of energetic particles flowing downward (as well as
upward). In the prevailing ``thick target'' models, the downward
electrons ``hit'' the dense chromosphere, heat the plasma locally,
evaporating it to fill the overlaying magnetic loops, and producing
non-thermal hard X-ray (HRX) emission at the loop feet. The
plasma-filled loops are then responsible for the gradual phase of the
flare, which is best observed in soft X-rays, characterized by the slow
cooling of the plasma. Along with hard X-rays, the impulsive heating
phase is also characterized by closely associated optical/UV emission
from the vicinity of the loop feet and well correlated in space and time
with the HRX emission. The emitting regions are very compact and bright,
but their precise location, whether in the lower chromosphere or in the
photosphere, and the physical mechanism responsible for them, is not
fully understood. This is unfortunate since most ($>$90\%) of the {\em
radiated} energy in flares is actually in this component, while both the
soft and the hard X-ray components make up a much smaller
fraction of the total radiative output\footnote{As indicated, comparable
energy is released in kinetic form both in CMEs and in the accelerated
electron streams}. The spectral energy distribution of this optical/UV
emission from Solar and stellar flares is also not fully characterized:
a black body component at $\sim$10$^4$\,K is generally inferred from
observations and several other components seem to be present, e.g Balmer
continuum and lines in the UV, and a cooler $\sim$5000\,K black body
component, each evolving on different timescales \citep{kow16a}.
Moreover, realistic models of flare heating so far fail to explain these
characteristics \citep{kow15a}.  

In spite of decades of observations of Solar and stellar flares
\citep[see e.g.,][]{ben10}, the physics involved is thus still not well
understood. This is surely even more true for flares from the youngest
PMS stars which, in the soft X-ray band, appear to be up to several
orders of magnitudes more powerful than Solar ones \citep[][this
work]{fav05}. Total irradiance measurements available for some of the
largest Solar flares indicate radiated energies
$\sim$10$^{31}$-10$^{32}$\,ergs. This is to be compared to
$\sim$10$^{34}$\,ergs {\em in the soft X-ray band alone}, for some of
the smallest and frequently detected X-ray flares on PMS stars and to up
to $>$10$^{36}$\,ergs for some of brightest ones (again in soft X-rays).
There is thus no guarantee that the physics of PMS flares is a scaled up
version of the solar events and that the latter may be used as a
reasonable template. In addition to the widely different energies, the
presence of circumstellar disks and accretion streams on PMS stars might
also complicate the picture, both by modifying the properties of the
involved magnetic field structures, and by affecting the transport of
emitted radiation, through absorption and re-emission. For example,
modeling of the flaring soft X-ray lightcurves indicates that at least
some PMS flares likely originate in extended magnetic loops, several
stellar radii long \citep{fav05,lop16a} that might even connect the star
with the inner circumstellar disks.

% \citet{mit05b} - UV vs. X-ray flares in M stars, complete with correlations

Although soft X-ray flares from PMS stars are routinely observed
\citep[e.g.,][]{car07}, observations in other bands or, even better,
simultaneous multi-band observations, e.g., in the optical/UV and X-ray
bands, have been conspicuously missing. Therefore, even the basic energy
budget of PMS flares is poorly constrained. As a result, we have no
constraint on bolometric energies, on the ratio between X-ray and
optical/UV emitted energy, or on the spectra of the optical/UV emission.
This not only precludes a better understanding of the flare physics, but
also, very importantly, hinders any assessment of the impact of flaring
activity on circumstellar disks and planet formation. Indeed, although
the energetic soft X-ray emission, which is well understood, will surely
contribute to the heating and ionization of disk material \citep{erc16},
other manifestations of the same energy release process might be even
more relevant. In particular the optical/UV counterparts to the soft
X-ray flares are expected to deposit more energy onto circumstellar
disks (energetic particles in CMEs will not be discussed in this paper
but may also play an important role). Moreover, no direct observational
signature has been thus far observed of the interaction between flare
emission and disks.   

We have obtained valuable simultaneous optical/mIR and X-ray lightcurves
of a sample of young PMS stars as part of the {\em Coordinated Synoptic
Investigation of NGC\, 2264} (CSI NGC2264). The project \citep{cod14a} involved
a number of space and ground based observations of the young stars in
the well known, $\sim$3\,Myr old NGC\,2264 star forming region \citep{dah08}.
Several studies have been published based on the CSI NGC2264 data
focusing, for the most part, on accretion and circumstellar disk
structures \citep{cod14a,sta14a,sta15b,mcg15a,sta16a,gua17a}. We will here exploit the same dataset for an unprecedented
multi-band exploration of flaring. In particular we will use data
obtained with three satellites: {\em CoRoT} in the optical band, {\em Spitzer}
in the mIR, and {\em Chandra} in soft X-rays, to try to constrain the
optical/UV component of PMS flaring emission in terms of flux and SED,
and to look for signatures of the reaction of circumstellar disks to
flaring. 

The paper will be organized as follows: section 2 presents the data, and in
\S\,3 we discuss the detection of flares and the ensemble
characteristics of the stars from which flares were detected. In \S\,4
we show how we characterize flares in the three bands. Section \S\,5
compares and correlates the characteristics of flares in the different
bands. The results are then discussed in \S\,6 and, in \S\,7, we finally
summarize our conclusions.

%---
%
%- Questions that can be addressed with our own data
%
%- open questions, on: 
%%	- Coronal structures (e.g. sizes, geometry in general) 
%	
%	- High energy emission from PMS stars and effects on
%	circumstellar disks (heating, ionization, evaporation) 
%	
%	- Flares, how they are started, the total energy involved,
%	compared to the photospheric emission. Contribution of flares to
%	the emitted flux at different wavelengths. DO flare dominate at
%	short wavelengths? How about with respect to accretion spots for
%	CTTSs?

%1e31 to 1e32 ergs... i.e. ~5 dex lower than in NGC2264 (comparing with large
%flares, but cumulative observing time should also be considered. Since we
%observed <200 members with CoRoT and Chandra, simultaneously, for 300ks, the
%accumulated observing time is <2yr)

%Fletcher et al. (2011) \cite{fle11a} on solar flares - Flares stem from the conversion of
%magnetic energy - Most of the emission is in the optical/UV - emitted energy
%comparable to both the energy in CMEs and in accelerated electrons (inferred
%from HRXs) 
%HXRs a small contribution but easy to interpret

%- timing - Neupert effect. Cross correlation analysis to find delays between HXR
%and SXR (~10s), interpreted as the hydrodinamic time-scale for the
%redistribution of energy deposited by non-thermal electrons. Can the same be
%attempted with CoRoT and SXR?

%---

%--

%Hudson (2007) seem to be an interesting review (from the abstract)..

\section{The data} 
\label{sect:data}

We base our analysis of flare properties on data collected by the CSI
NG2264 project, described by \citet{cod14a}. In particular, we make use
of three of the acquired datasets: the {\em Chandra}-ACIS X-ray data,
and two photometric time series from {\em CoRoT} (broadband optical) and {\em
Spitzer} (mIR, 3.6$\mu m$ and 4.5$\mu m$), the latter limited to the
high-cadence, small fields of view (FoV), staring-mode observations.
We also analyze simultaneous {\em CoRoT} and {\em Chandra}
data from a previous campaign conducted in 2008 \citep{fla10}.

%The {\em Chandra} and {\em CoRoT} lightcurves are also studied by Guarcello et al.
%(2016) to investigate the correlation between X-ray and optical {\em
%non-flaring} variability.

We restrict our analysis to NGC\,2264 stars observed by {\em Chandra}
and at least one of the other two telescopes. Both in 2008 and 2011, {\em
Chandra} observed the central-southern part of NGC\,2264 with ACIS-I
(FoV $\sim17^\prime\times17^\prime$), with overlapping pointings. Two
exposures were taken during the 2008 {\em CoRoT} observations, 28 and 30\,ks
long, with the first starting on 12 March at 17:56 UT with a gap of 15.5
days in between. The observations were co-pointed at R.A. 06:41:12, Dec.
+09:30:00, with similar but not identical roll-angles, 270$^\circ$.4 and
266$^\circ$.6. The 2011 `CSI' {\em Chandra} campaign consisted of four
exposures for a total of 297\,ks (3.4\,d), spanning 5.9 days. The
observations were co-pointed at R.A. 06:40:58.700, Dec. +09:34:14.00,
with almost identical roll-angle, 63$^\circ$.95. The first exposure
started on 3 Dec. 2011 at 1:22 UT and lasted 75\,ks, followed by three
more exposures lasting 94, 61, and 67\,ks, with intervening gaps of 99,
111, and 4.5\,ks. The last two exposures are therefore almost adjacent
in time. A full description of the X-ray data reduction and analysis
will be presented in Flaccomio et al. (2018, in preparation). Briefly, we
treated the six exposures separately, preparing them for scientific
analysis using standard {\sc ciao} tools and procedures. Source
detection was then performed with {\sc pwdetect} \citep{dam97} on each
of the exposures {\em and} on co-added datasets. The resulting source
lists were then merged. Source and background photons, and relative
time-averaged X-ray spectra, were finally obtained using {\sc
acis-extract} \citep{bro10}.

The {\em CoRoT} and {\em Spitzer} observations from the CSI project, and
relevant data reduction, are fully descried by \citet{cod14a}. The 2008
{\em CoRoT} observation was discussed by \citet{fla10}. In both cases the full
NGC\,2264 region was included in one of the two {\em CoRoT} CCDs in the
exoplanet field, with a FoV of $1^\circ.3\times1^\circ.3$ \citep[see
Fig.\,1 in][]{cod14a}. The resulting optical broad-band lightcurves were
$\sim$23 and $\sim$40 days long in 2008 and 2011, respectively, with a
cadence of 32 or 512 seconds, depending on target. {\em CoRoT} performs source
photometry on-board, from pre-selected windows. There were 3642 and 4235
such windows in the 2008 and 2011 runs, respectively. Of these, 332 and
379 were centered on likely NGC\,2264 members, in 2008 and 2011,
respectively, for a total of 498 observed members. The {\em
Spitzer}-IRAC staring-mode observations in 2011 covered two much smaller
{\em central} fields, $5^\prime.2\times5^\prime.2$, in the 3.6$\mu m$
and 4.5$\mu m$ channels, respectively. These imaging observations were
timed to be simultaneous with {\em Chandra} pointings (see below), and
had a cadence of $\sim$15\,s. We will not discuss the longer {\em
Spitzer} mapping-mode observations because, by design, they are not
simultaneous with the {\em Chandra} data.

We construct the parent sample for our multi-band study of flares
starting from the 744 X-ray sources in the {\em Chandra} FoVs. Of these,
587 are uniquely identified with a optical/IR object in the field,
almost all of which are likely NGC\,2264 members according to the
classification presented by \citet{cod14a}. We will further focus our
search on two subsamples: X-ray sources with simultaneous optical ({\em
CoRoT}) coverage and with mIR ({\em Spitzer}) coverage.  The {\em
Chandra}/{\em CoRoT} sample comprises 179 {\em CoRoT} targets associated
with at least one of the above X-ray sources, which reduces to 154 stars
when considering unambiguously identified stars only. The {\em
Chandra}/{\em Spitzer} sample comprises 176 stars, almost all of which
are likely NGC2264 members (173). The intersection of the two samples,
that is likely members observed simultaneously with {\em Chandra}, {\em
CoRoT}, and {\em Spitzer}, and with likely unique cross-identifications,
comprises 44 stars. Alternatively, the union of likely members with data
in X-rays and either the optical or mIR, and with likely unique
cross-identifications, comprises 289 stars.

\begin{figure*}[!t!]
\centering
\includegraphics[width=9.0cm]{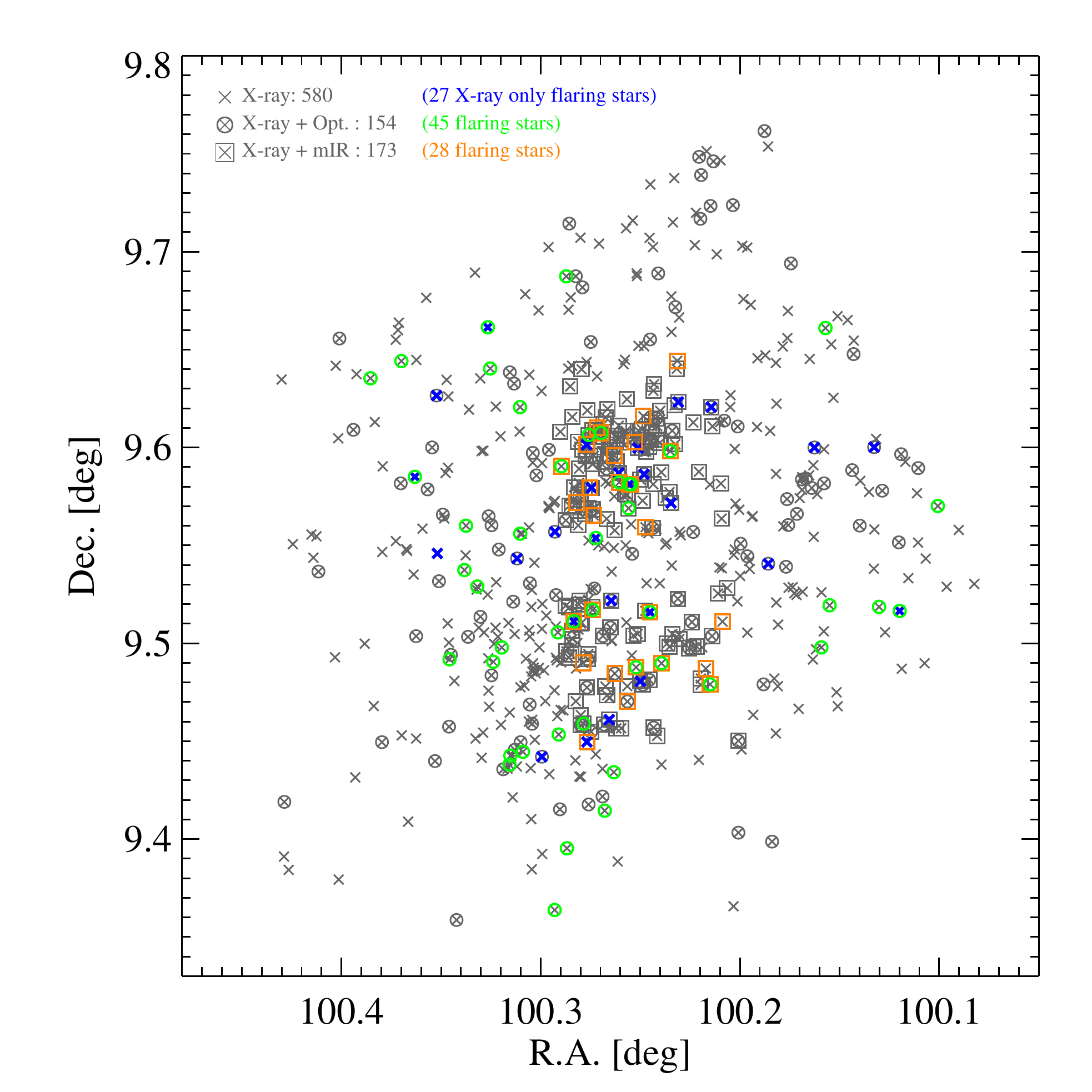}
\includegraphics[width=9.0cm]{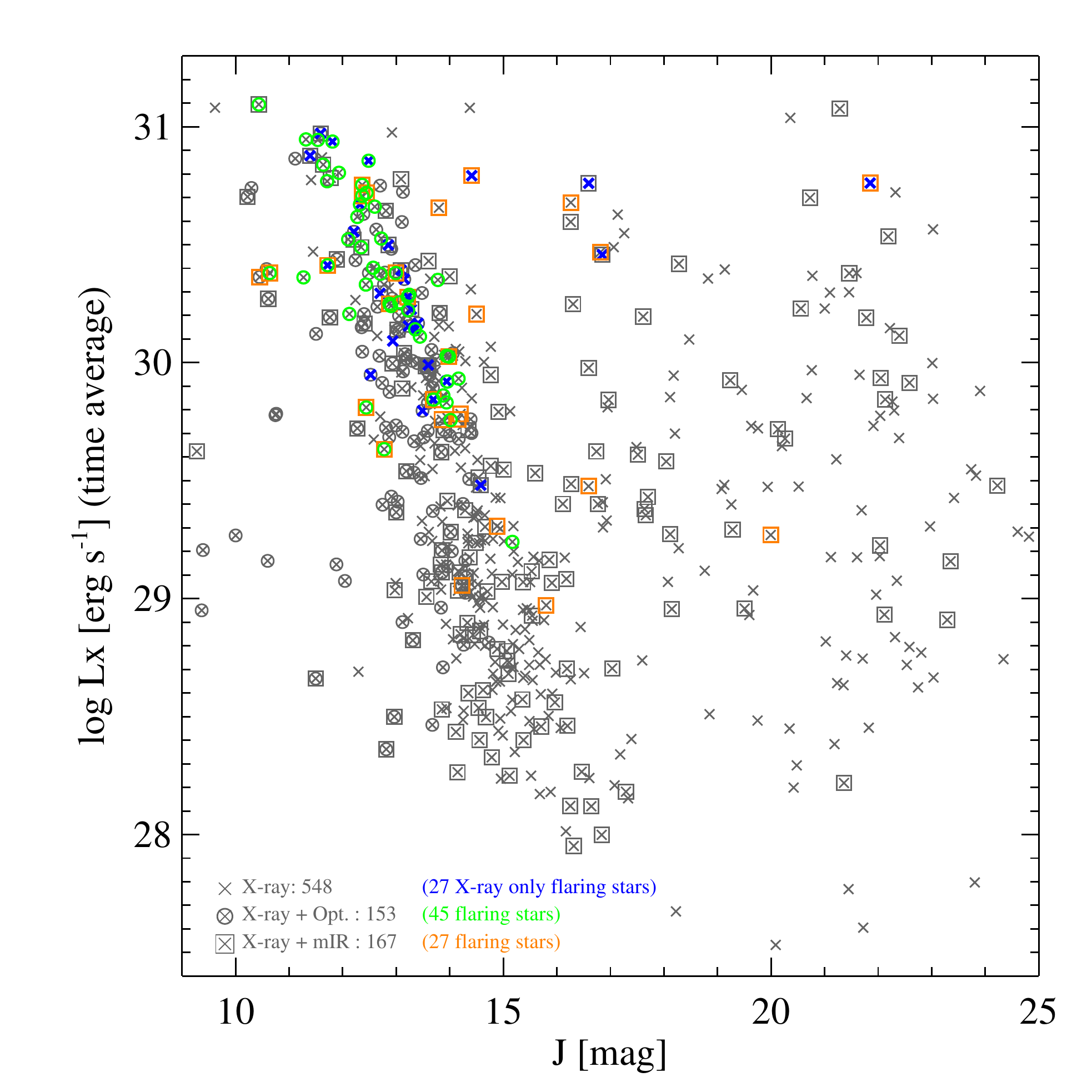}
\caption{[Left]: spatial distribution of likely NGC\,2264 members
with simultaneous lightcurves, highlighting those undergoing
simultaneous X-ray+optical and/or X-ray+mIR flaring. X-symbols: all {\em
Chandra} sources in the {\em simultaneous} FOVs. Circles: {\em
Chandra}+{\em CoRoT} sources (with unique identifications). Squares: {\em
Chandra}+{\em Spitzer} sources. Green and orange symbols indicate
sources with simultaneous X-ray+optical and X-ray+mIR flaring, respectively. Blue crosses indicate stars with a well observed X-ray flare having no counterpart in other available bands (\S\,\ref{sect:xray_only} and Appendix\, \ref{app:LC_Xrayonly_Oallin}). Note the smaller spatial coverage, but better completeness with respect to the X-ray sources, of the mIR sample with respect to the optical one. [Right]: same as above in the $\log L_X$ vs. J-magnitude plane, showing the deeper coverage of mIR data and the tendency of flares to be detected on the  stars with higher average X-ray luminosity.}
\label{fig:RaDec_Lx_R}
\end{figure*}

These two main subsamples, and their intersection, are clearly subject
to severe selection biases. Figure\,\ref{fig:RaDec_Lx_R} shows the
spatial distribution and the $L_X$ vs. J-band magnitudes scatter plot
for the two samples, allowing a comparison with the full list of X-ray
detected likely members. The {\em Chandra}/{\em CoRoT} sample covers the
full {\em Chandra} FoV. A stellar luminosity bias is produced by the
{\em CoRoT} photometric limits: the predefined target list was limited
to R$\lesssim$17 (V$\lesssim$18, I$\lesssim$16, corresponding to a broad
range of minimum mass, 0.8-0.2 M$_\odot$). Moreover, the target selection
introduced a preference toward Classical T Tauri stars (CTTSs), and
toward known members, based on pre-existing X-ray data, spectroscopic
and photometric H$_\alpha$ data, and mIR excesses data. The {\em
Chandra} data is more simply flux-limited, but with a spatially varying
sensitivity limit. This may translates into a selection in terms of mass
and accretion/circumstellar disk properties \citep{fla03,pre05a}.
Figure\,\ref{fig:RaDec_Lx_R}, however, indicates that the X-ray flux
limit is probably less severe than the {\em CoRoT} flux limit. The {\em
Chandra}/{\em Spitzer} sample is instead severely biased in spatial
terms, being limited to the two $5^\prime.2\times5^\prime.2$ IRAC FOVs.
However, it reaches to fainter, lower mass, and/or more absorbed stars
with respect to the {\em Chandra}/{\em CoRoT} sample.

%\begin{figure}[!t!]
%\centering
%\includegraphics[width=9.0cm]{FoV_draft.pdf}
%\caption{[figures with Chandra and/or optical image showing the position of
%{\em Chandra}/{\em CoRoT} and {\em Chandra}/{\em Spitzer} samples and that of flaring sources?]}
%\label{fig:FoV}
%\end{figure}

\section{Flare detection and sample definition}
\label{sect:detection}

Magnetic flares are most easily detected in the soft X-ray band, where
the gradual phase, corresponding to the cooling of hot thermal plasma
confined in coronal structures, is commonly and easily observed with a
large contrast with respect to the `quiescent' coronal emission. We thus
started our search for flares from the {\em Chandra} X-ray lightcurves
and then inspected the simultaneous {\em CoRoT} and {\em Spitzer}
lightcurves in order to identify optical and mIR counterparts. To
simplify the detection of the X-ray flares, and their ensuing spectral
characterization, we made use of a lightcurve segmentation algorithm
developed by ourselves, and already used on several occasions
\citep{wol05,pre05a,car07,alb07a}. Briefly, our method provides a
representation of X-ray lightcurves by using temporal segments during
which the photon count-rate is constant or, more precisely, compatible
with being constant at a given confidence level. The algorithm is based
heavily on the Bayesian Blocks algorithm introduced by \citet{sca98} in
that it proceeds by dividing lightcurves at their most likely
`break-point' and recursively repeating the process on the resulting
segments until the significance of the break-points fall below a set
threshold, $P^{ML}_{min}$. The differences with respect to
\citet{sca98} are: $i$), we use a maximum likelihood approach in place
of a Bayesian one, $ii$) significance thresholds for the segmentation
process are established via Monte Carlo simulations of constant
count-rate sources, $iii$) at each of the recursive steps we attempt to
fragment the lightcurve by testing two-segment models (with one
break-point), {\em as well as} three-segment models (with two
break-points), $iv$) we are able to set the minimum number of photons
that the resulting time-segments will include, $N^{ML}_{MinPh}$. The last two
characteristics are particularly useful for our goals. A three-segment
model is indeed more sensitive for detecting impulsive variability (such
as flares) than a two-segment model. Moreover, in order to follow the
plasma evolution during flares, we need to perform a spectral analysis
of the X-ray emission, which requires a minimum number of photons in
each segment.

We first computed the segments with a 95\% confidence threshold and
setting the minimum number of photons per segment to 20. These segments
were then used to define a criterion for automated detection of flares,
as described in \citet{car07}, and, once detected, to determine their
basic properties in as much of an unbiased way as possible. Start and
end times, for example, as well as peak X-ray emission fluxes, were
defined based on these segments. It should be noted that our flare
detection algorithm, tuned to detect events that follow our preconceived
idea of flares (i.e. both elevated flux and time-derivative of the
flux), produces mostly reasonable results, but is clearly not
perfect. All the {\em Chandra} X-ray light curves were thus inspected to
determine whether other flare-like events had escaped automatic
detection and/or had been improperly defined by our algorithm. Several
ad-hoc choices were made:

\begin{itemize}

\item Eleven flares were detected with our automated procedures adopting
segments with $N^{ML}_{MinPh}=1$ (instead of 20) and the default
$P^{ML}_{min}=95$\%. These are for the most part faint flares, defined
by a very small number of photons.

\item We ``forced'' the detection of nine X-ray flares with an obvious
counterpart in the other bands. In these cases the segmented X-ray
lightcurves showed a corresponding elevated X-ray flux which, however,
did not qualify as a flare according to our automated
criterium\footnote{We, however, discarded, some tentative X-ray flares
only found by inspection but whose counterpart at longer wavelengths was
either questionable or too difficult to define.} (for the 2$^{nd}$
flare on ACIS\,\#\,677, we also lowered the significance threshold for
the segmentation). The X-ray properties of three of these where actually
considered uncertain and the flares are not included in our main study
sample (see below). We also verified that the exclusion of these nine
flares from our sample does not change any of the results discussed
below. 

\item For 19 automatically detected flares, we adopted a different set
of segments to more accurately describe the shape of flares with respect
to those used to detect them. Most often (14 cases) the flare was
detected using default segments, i.e. ($N^{ML}_{MinPh},
P^{ML}_{min}$)=(20, 95\%), but we define it using segments with as few
as one photon/segment $N^{ML}_{MinPh}=1$. The opposite choice was
adopted in one case (ACIS\,\#\,1018), and altogether different sets were
preferred in the four following cases: ACIS\,\#\,297
($N^{ML}_{MinPh},P^{ML}_{min}$)=(1, 99\%), 405 (1, 93\%), 677
(first flare) and 789 (20, 60\%)

\item In eleven more cases the automatic definition of ``flaring''
intervals was adjusted by hand. Four flares were defined separating two
pairs of events, each of which had been detected as single event (on
ACIS\,\#\,871 and 924). In five other cases we included/excluded
segments that appeared related/unrelated to the events (ACIS\,\#\,110,
600, 664, 693, and 1000) . Finally, in two cases (flares on ACIS\,\#\,630 and 713), a flare was found having the rise phase at the end of
the 2nd-last {\em Chandra} observing intervals and continuing in the
last interval, thus also spanning the short 1.24\,h gap between the two:
we then modified the default flare definition, which by design is
limited to one interval, and included all the relevant segments. 

\item In some cases, only the tail of an X-ray flare was detected at the
beginning of one of the {\em Chandra} observing segments. We looked at
the {\em CoRoT} and {\em Spitzer} data to search for optical/mIR counterparts
that might show the onset of the flare before the beginning of the X-ray
observation. Eleven such cases were found: in five of the cases (flares
on ACIS\,\#\,58, 585, 591, 690, and 1018) the optical/mIR counterpart was
reasonably well defined and a large fraction of the X-ray flare, assumed
to start at or after the onset of the optical/mIR one, was inferred to
have actually been observed. We thus decided that the X-ray flares could
actually be defined, although with some uncertainty, and we added them
to our sample. We discarded the remaining six cases, however, since
either too large a fraction of the X-ray flare may not have been
observed or the optical/mIR counterpart could not be uniquely defined.

\item In one case, two X-ray sources with flares were associated with a
single {\em CoRoT} source, due to the large {\em CoRoT} photometric windows:
ACIS\,\#\,536 had one X-ray flare with a corresponding event in the
{\em CoRoT} lightcurve, while ACIS\,\#\,541 showed three X-ray flares,
two of which with a counterpart in the {\em CoRoT} lightcurve. Even though the
non-flaring {\em CoRoT} emission has ambiguous origin, since we are
exclusively interested in the flaring emission,  we exploited the time
coincidence to confidently associate the {\em CoRoT} and X-ray events.

\end{itemize}

In Table\,\ref{tbl:flares} we list the 78 X-ray flares in our sample
(from 65 stars), all with a simultaneous optical and/or mIR
counterpart. The first three columns list the {\em Chandra} ACIS source
number (from Flaccomio et al. 2018), the corresponding {\em Mon}
identifier from \citet{cod14a}, and the {\em Chandra} observation id (or
ids) during which the flare occurred. In Figure \ref{fig:lc_examples} we
show six representative examples of ``good quality'' X-ray detected
flares, while in Appendix\,\ref{app:lightcurves} we show the full sample
of lightcurves\footnote{In addition to the maximum likelihood segmentation of the X-ray lightcurves discussed above, the figures also show, for comparison, a more traditional representation with fixed bins, with duration varying on a source-by-source basis. This latter is, however, not used in the following.}.

\begin{table*}
\caption{List of flares, deduced physical quantities, and host-star properties}             
\label{tbl:flares}      
\centering          
\resizebox{\textwidth}{!}{
 	\begin{tabular}{r@{}lrrr@{}lr@{}lr@{}lr@{}lr@{}lr@{}lrrrrrrrrrr}
\hline\hline
           Src \# &                &      Mon &          Obs &             E$_X$ &    &        L$_{X,pk}$ &    &         E$_{Opt}$ &    &      L$_{Opt,pk}$ &    &          E$_{IR}$ &    &       L$_{IR,pk}$ &    &  Class &      H$\alpha$ EW &        A$_V$ &                N$_H$ &    Sp.T. &       V &       R &       I &        CoRoT &         Prot \\
\footnotesize
                  &                &          &              &    [10$^{35}$erg] &    &  [10$^{32}$erg/s] &    &    [10$^{35}$erg] &    &  [10$^{32}$erg/s] &    &    [10$^{35}$erg] &    &  [10$^{32}$erg/s] &    &        &           [$\AA$] &        [mag] & [10$^{22}$cm$^{-2}$] &          &   [mag] &   [mag] &   [mag] &         type &       [days] \\
\hline
              25 &               &    1061 &       14368 &             1.19 &   &             0.07 &   &             2.61 &   &             0.40 &   &               -- &   &               -- &   &    II &             20.1 &        2.39 &                0.19 &      M0 &  16.71 &  16.39 &  14.96 &           S &             \\
              25 &               &    1061 &       13610 &             1.27 &   &             0.15 &   &             6.05 &   &             0.80 &   &               -- &   &               -- &   &    II &             20.1 &        2.39 &                0.19 &      M0 &  16.71 &  16.39 &  14.96 &           S &             \\
              42 &               &    1275 &       14368 &             8.10 &   &             2.61 &   &            46.56 &   &            23.15 &   &               -- &   &               -- &   &   III &              1.8 &        1.63 &                0.00 &      K4 &  14.93 &  14.14 &  13.27 &           S &             \\
             104 &               &    1027 &       13611 &             5.70 &   &             0.82 &   &             4.78 &   &             1.68 &   &               -- &   &               -- &   &   III &              4.4 &          -- &                0.00 &         &  17.33 &  16.09 &  14.84 &           P &        1.15 \\
             110 &               &     990 &       14368 &             0.33 &   &             0.02 &   &             2.98 &   &             0.41 &   &               -- &   &               -- &   &   III &              2.0 &        1.04 &                0.00 &      K7 &  16.85 &  16.07 &  15.08 &           N &             \\
             121 &               &    1076 &       13610 &             0.06 &   &             0.04 &   &             1.44 &   &             0.46 &   &               -- &   &               -- &   &    II &              2.6 &        0.07 &                0.00 &      M1 &  17.43 &  16.30 &  15.26 &           U &             \\
             297 &               &     249 &       14368 &             0.57 &   &             0.16 &   &               -- &   &               -- &   &            26.51 &   &             4.40 &   &   III &              4.5 &        0.00 &                0.20 &      M5 &  19.92 &  18.17 &  16.42 &             &        2.16 \\
             336 &               &     663 &       14368 &             0.32 &   &             0.03 &   &               -- &   &               -- &   &            56.22 &   &             4.35 &   &    II &             36.4 &        0.29 &                0.81 &      M4 &  19.40 &  17.97 &  16.27 &             &        7.22 \\
             331 &               &     808 &       13610 &             5.26 &   &             0.20 &   &            11.41 &   &             1.33 &   &           233.91 &   &            78.56 &   &    II &             50.2 &        1.09 &                0.01 &      K4 &  15.79 &  15.02 &  14.28 &           B &             \\
             384 &               &     643 &       14369 &             0.44 &   &             0.01 &   &               -- &   &               -- &   &            14.70 & : &             0.88 & : &   III &               -- &          -- &                0.49 &         &  20.99 &  19.47 &  17.66 &             &             \\
             405 &          $^*$ &     567 &       14368 &             0.63 & : &             0.03 & : &            24.02 & : &             2.25 & : &          3741.73 & : &           289.30 & : &    II &             84.1 &        1.57 &                1.78 &      K3 &  16.73 &  15.46 &  14.61 &           B &             \\
             424 &               &     869 &        9769 &             0.63 &   &             0.03 &   &             2.25 &   &             0.15 &   &               -- &   &               -- &   &   III &              4.0 &          -- &                0.00 &         &  17.62 &  16.58 &  15.41 &           P &        8.64 \\
             424 &               &     869 &       14369 &             0.40 &   &             0.07 &   &             0.47 & : &             0.20 & : &             6.55 & : &             1.37 & : &   III &              4.0 &          -- &                0.00 &         &  17.62 &  16.58 &  15.41 &           P &        8.64 \\
             433 &               &     502 &       13611 &             9.97 &   &             0.31 &   &               -- &   &               -- &   &           156.61 & : &             8.59 & : &    II &              7.3 &        1.08 &                0.16 &      K7 &  16.65 &  15.57 &  14.57 &             &        3.65 \\
             468 &               &     774 &       13610 &             0.74 &   &             0.12 &   &            14.48 &   &             1.76 &   &           223.10 &   &            44.55 &   &    II &             14.3 &        0.50 &                0.02 &    K2.5 &  13.97 &  13.32 &  12.72 &           S &        3.49 \\
             488 &               &      94 &       13610 &            18.90 &   &             4.72 &   &               -- &   &               -- &   &          1561.59 &   &          1163.51 &   &    II &               -- &          -- &                0.78 &         &  16.89 &     -- &  14.82 &             &             \\
             496 &               &     712 &       13610 &             0.28 &   &             0.01 &   &               -- &   &               -- &   &            18.07 &   &             1.31 &   &    II &             12.3 &        0.00 &                0.53 &      M6 &  21.99 &  19.84 &  17.65 &             &        5.43 \\
             502 &    $^\dagger$ &     649 &       13610 &             0.76 & : &               -- &   &               -- &   &               -- &   &            39.82 &   &             8.05 &   &    II &               -- &          -- &                  -- &         &     -- &     -- &     -- &             &             \\
             523 &               &     660 &       13611 &             2.04 &   &             0.36 &   &             4.04 &   &             0.82 &   &           102.67 &   &            14.20 &   &    II &             16.6 &        0.36 &                0.27 &      K4 &  14.31 &  13.68 &  13.10 &         QPD &        5.12 \\
             536 &   $^\ddagger$ &     433 &       14368 &             1.64 &   &             0.08 &   &             3.92 &   &             1.14 &   &            18.41 &   &             3.65 &   &   III &              7.0 &          -- &                0.00 &         &  16.74 &  15.85 &  14.83 &         QPD &        9.79 \\
             528 &               &     953 &       13610 &             0.99 &   &             0.11 &   &               -- &   &               -- &   &            22.67 &   &             5.81 &   &    II &               -- &          -- &                1.19 &         &  23.13 &  20.02 &  18.37 &             &        2.82 \\
             541 &               &     218 &       13610 &             0.60 &   &             0.07 &   &               -- &   &               -- &   &            32.21 &   &             7.17 &   &    II &             27.9 &          -- &                0.00 &         &  17.09 &  16.65 &  15.42 &             &             \\
             541 &               &     218 &       13611 &             0.13 &   &             0.03 &   &             2.42 &   &             0.53 &   &            18.81 &   &             7.54 &   &    II &             27.9 &          -- &                0.00 &         &  17.09 &  16.65 &  15.42 &             &             \\
             541 &               &     218 &       14369 &             0.18 &   &             0.01 &   &             6.17 &   &             1.25 &   &               -- &   &               -- &   &    II &             27.9 &          -- &                0.00 &         &  17.09 &  16.65 &  15.42 &             &             \\
             542 &               &     784 &       14368 &             1.61 &   &             0.12 &   &             4.47 &   &             0.63 &   &               -- &   &               -- &   &   III &              1.8 &        0.43 &                0.00 &      K5 &  14.75 &  14.06 &  13.33 &           P &        10.0 \\
             546 &               &     892 &       13610 &             5.40 &   &             1.69 &   &               -- &   &               -- &   &            42.92 &   &            21.71 &   &   III &               -- &        0.04 &                0.00 &      A0 &  10.73 &  10.70 &  10.69 &         QPS &        2.41 \\
             571 &               &     236 &       13610 &             5.89 &   &             1.01 &   &            34.24 &   &             7.61 &   &           142.55 &   &            17.21 &   &   III &              0.6 &        0.51 &                0.01 &      K0 &  14.40 &  13.80 &  13.30 &           P &        1.97 \\
             588 &               &     361 &       13611 &             0.08 &   &             0.02 &   &               -- &   &               -- &   &             9.53 &   &             1.63 &   &    II &             14.6 &          -- &                0.30 &       M &  18.54 &  17.06 &  15.83 &             &        2.30 \\
             592 &               &     777 &        9768 &             1.38 &   &             0.11 &   &             3.61 &   &             0.90 &   &               -- &   &               -- &   &   III &              3.9 &        0.17 &                0.00 &      M0 &  17.73 &  16.53 &  15.60 &           U &        3.63 \\
             600 &               &     879 &       13611 &             0.36 &   &             0.18 &   &             1.87 &   &             1.15 &   &               -- &   &               -- &   &    II &             16.5 &        0.00 &                0.03 &      M1 &  16.03 &  15.20 &  14.37 &           S &       0.910 \\
             630 &               &     510 & 13611+14369 &            10.46 &   &             0.19 &   &            17.59 &   &             1.75 &   &               -- &   &               -- &   &    II &            101.8 &        0.01 &                0.08 &      M0 &  15.82 &  14.95 &  14.06 &           B &             \\
             649 &               &     517 &       13610 &             1.19 &   &             0.30 &   &             6.48 &   &             2.71 &   &            19.71 &   &             5.88 &   &   III &               -- &        0.24 &                0.00 &      A0 &  10.99 &  10.98 &  10.93 &             &             \\
             662 &               &  143538 &       13610 &             0.93 &   &             0.04 &   &               -- &   &               -- &   &            45.85 &   &             5.14 &   &    II &               -- &          -- &                1.92 &         &     -- &     -- &     -- &             &             \\
             664 &               &     226 &       14368 &             0.21 &   &             0.06 &   &             1.48 &   &             0.54 &   &               -- &   &               -- &   &   III &              2.7 &        0.05 &                0.01 &      K5 &  15.39 &  14.40 &  13.76 &           P &        1.20 \\
             664 &               &     226 &       13610 &             1.32 &   &             0.09 &   &             7.99 &   &             1.32 &   &               -- &   &               -- &   &   III &              2.7 &        0.05 &                0.01 &      K5 &  15.39 &  14.40 &  13.76 &           P &        1.20 \\
             671 &               &     834 &       14368 &             1.32 & : &             0.04 & : &               -- &   &               -- &   &            29.99 & : &             2.44 & : &    II &               -- &          -- &                2.69 &         &     -- &     -- &  21.28 &             &             \\
             677 &               &     357 &       14368 &             3.25 &   &             0.33 &   &             9.61 &   &             1.03 &   &           285.51 &   &            30.57 &   &    II &              8.0 &        0.99 &                0.05 &      K5 &  15.25 &  14.32 &  13.47 &           N &             \\
             677 &               &     357 &       13610 &             0.32 &   &             0.05 &   &             2.13 &   &             1.39 &   &            19.20 & : &             5.16 & : &    II &              8.0 &        0.99 &                0.05 &      K5 &  15.25 &  14.32 &  13.47 &           N &             \\
             693 &               &     177 &       14369 &             0.65 &   &             0.17 &   &             6.28 &   &             1.98 &   &               -- &   &               -- &   &    II &             10.0 &        0.49 &                0.05 &      G5 &  13.52 &  13.00 &  12.52 &         QPS &        3.02 \\
             704 &               &     855 &       13610 &             0.41 &   &             0.02 &   &               -- &   &               -- &   &           443.52 &   &           114.63 &   &     I &               -- &          -- &                1.08 &         &  23.81 &  20.21 &  18.51 &             &             \\
             706 &               &     648 &       14368 &             0.64 &   &             0.08 &   &               -- &   &               -- &   &            17.21 &   &             3.33 &   &    II &               -- &          -- &                5.17 &         &     -- &     -- &     -- &             &             \\
             713 &         $^\S$ &     474 & 13611+14369 &             4.74 &   &             1.27 &   &            53.89 &   &            13.93 &   &               -- &   &               -- &   &    II &            104.7 &          -- &                0.00 &       G &  12.99 &  12.30 &  11.69 &           B &             \\
             714 &               &     736 &       13610 &             0.40 & : &             0.02 & : &               -- &   &               -- &   &            46.69 & : &             4.73 & : &    II &             27.4 &        0.00 &                0.27 &      M0 &  17.69 &  15.71 &  15.52 &             &        3.14 \\
             739 &               &    1363 &       14368 &             4.94 &   &             0.17 &   &               -- &   &               -- &   &            15.73 & : &             1.69 & : &     I &               -- &          -- &               18.64 &         &     -- &     -- &     -- &             &             \\
             747 &               &     200 &       13610 &             4.58 &   &             0.49 &   &             2.02 & : &             0.45 & : &            29.43 &   &             3.58 &   &   III &              0.5 &        0.60 &                0.00 &      K4 &  15.69 &  14.81 &  14.18 &         QPS &             \\
             769 &               &     354 &        9769 &             0.78 &   &             0.33 &   &             7.26 &   &             3.77 &   &               -- &   &               -- &   &   III &              2.4 &        0.00 &                0.00 &      M0 &  14.73 &  13.86 &  13.00 &         QPS &        1.73 \\
             771 &               &     881 &       13611 &             0.22 & : &             0.04 & : &             1.16 & : &             0.17 & : &               -- &   &               -- &   &   III &              1.0 &        0.20 &                0.00 &      K5 &  15.10 &  14.52 &  13.84 &           P &        3.92 \\
             781 &               &     344 &       13610 &             0.43 &   &             0.05 &   &             1.96 & : &             0.30 & : &               -- &   &               -- &   &   III &              3.9 &        0.00 &                0.00 &      M2 &  15.98 &  14.67 &  13.82 &           P &       0.856 \\
             781 &               &     344 &       13610 &             0.47 &   &             0.03 &   &             1.80 & : &             0.48 & : &            15.61 & : &             2.49 & : &   III &              3.9 &        0.00 &                0.00 &      M2 &  15.98 &  14.67 &  13.82 &           P &       0.856 \\
             789 &               &     810 &       13611 &             2.34 &   &             0.25 &   &            20.30 &   &             2.89 &   &               -- &   &               -- &   &   III &              3.1 &        0.55 &                0.00 &      K5 &  14.15 &  13.37 &  12.61 &           P &        2.92 \\
             789 &               &     810 &       14369 &            10.31 &   &             0.46 &   &            29.16 &   &            12.73 &   &               -- &   &               -- &   &   III &              3.1 &        0.55 &                0.00 &      K5 &  14.15 &  13.37 &  12.61 &           P &        2.92 \\
             791 &               &     519 &       14368 &             0.79 &   &             0.31 &   &             5.10 &   &             3.22 &   &               -- &   &               -- &   &   III &               -- &          -- &                0.02 &         &  17.49 &  16.31 &  15.27 &         QPS &        6.00 \\
             791 &               &     519 &       14368 &             0.17 &   &             0.05 &   &             0.79 &   &             0.42 &   &               -- &   &               -- &   &   III &               -- &          -- &                0.02 &         &  17.49 &  16.31 &  15.27 &         QPS &        6.00 \\
             804 &               &     172 &        9769 &             0.40 &   &             0.04 &   &             3.05 &   &             1.02 &   &               -- &   &               -- &   &   III &              2.7 &          -- &                0.00 &         &  16.87 &  15.78 &  14.43 &           U &             \\
             874 &               &     749 &       14369 &             0.13 &   &             0.01 &   &             2.57 &   &             0.31 &   &               -- &   &               -- &   &   III &              1.9 &        0.00 &                0.00 &      M1 &  16.61 &  15.33 &  14.59 &           P &        1.44 \\
             879 &               &     606 &        9768 &             0.16 &   &             0.04 &   &             1.77 &   &             0.33 &   &               -- &   &               -- &   &   III &              1.9 &        0.00 &                0.00 &      K5 &  15.40 &  14.36 &  13.78 &           P &        10.7 \\
             879 &               &     606 &       14368 &             1.23 & : &             0.05 & : &             1.91 & : &             0.15 & : &               -- &   &               -- &   &   III &              1.9 &        0.00 &                0.00 &      K5 &  15.40 &  14.36 &  13.78 &           P &        10.7 \\
             879 &               &     606 &       13610 &             2.49 &   &             0.34 &   &             1.88 &   &             0.38 &   &               -- &   &               -- &   &   III &              1.9 &        0.00 &                0.00 &      K5 &  15.40 &  14.36 &  13.78 &           P &        10.7 \\
             880 &               &     770 &       13610 &             1.39 &   &             0.05 &   &            12.93 & : &             0.72 & : &               -- &   &               -- &   &   III &              4.1 &          -- &                0.00 &         &  17.20 &  16.07 &  15.17 &         QPS &        5.44 \\
             893 &               &     607 &        9769 &             0.41 &   &             0.03 &   &             0.63 & : &             0.67 & : &               -- &   &               -- &   &   III &              5.5 &        0.00 &                0.06 &    M3.5 &  16.79 &  15.45 &  14.50 &             &       0.610 \\
             896 &               &     477 &       13611 &             0.33 &   &             0.07 &   &             3.35 &   &             1.54 &   &               -- &   &               -- &   &   III &              1.9 &        0.33 &                0.00 &      K5 &  13.82 &  13.11 &  12.41 &           P &        6.22 \\
             905 &               &     425 &       13611 &             0.37 &   &             0.05 &   &             2.57 &   &             0.39 &   &               -- &   &               -- &   &    II &              6.3 &        0.53 &                0.06 &      K5 &  14.29 &  13.46 &  12.71 &           S &        7.51 \\
             920 &               &     657 &       13611 &             0.34 &   &             0.04 &   &             3.38 &   &             0.90 &   &               -- &   &               -- &   &   III &              2.4 &        0.00 &                0.00 &      M3 &  15.78 &  14.43 &  13.55 &         QPS &        2.43 \\
             920 &               &     657 &       14369 &             0.42 &   &             0.01 &   &             4.20 &   &             0.79 &   &               -- &   &               -- &   &   III &              2.4 &        0.00 &                0.00 &      M3 &  15.78 &  14.43 &  13.55 &         QPS &        2.43 \\
             924 &               &     931 &       13611 &             0.47 &   &             0.18 &   &             1.14 &   &             0.66 &   &               -- &   &               -- &   &   III &             32.5 &        1.17 &                0.00 &      M3 &  18.75 &  17.31 &  15.68 &           N &             \\
             924 &               &     931 &       13611 &             0.06 &   &             0.04 &   &             0.47 &   &             0.39 &   &               -- &   &               -- &   &   III &             32.5 &        1.17 &                0.00 &      M3 &  18.75 &  17.31 &  15.68 &           N &             \\
             931 &               &     273 &       14368 &             0.24 &   &             0.01 &   &             1.63 &   &             0.38 &   &               -- &   &               -- &   &    II &            123.5 &        0.25 &                0.00 &      M1 &  16.59 &  15.73 &  14.65 &           N &             \\
             943 &               &     198 &       14368 &             0.35 &   &             0.20 &   &             5.38 &   &             1.71 &   &               -- &   &               -- &   &   III &              1.5 &        0.25 &                0.00 &      K5 &  14.40 &  13.65 &  12.97 &         QPS &        4.99 \\
             955 &               &     119 &       14369 &             0.48 &   &             0.02 &   &             5.27 &   &             0.98 &   &               -- &   &               -- &   &    II &             10.6 &        0.00 &                0.00 &      K6 &  15.18 &  14.12 &  13.55 &          QP &             \\
             957 &               &     279 &       13610 &             1.22 &   &             0.02 &   &             2.39 &   &             1.39 &   &               -- &   &               -- &   &    II &              5.8 &        0.00 &                0.00 &    M2.5 &  17.74 &  16.41 &  15.42 &         QPS &        8.46 \\
             967 &               &     878 &       13610 &             0.14 &   &             0.05 &   &             0.71 &   &             0.33 &   &               -- &   &               -- &   &   III &               -- &          -- &                0.12 &         &  18.75 &  17.56 &  16.48 &           N &        5.08 \\
             992 &               &     948 &       13610 &             0.28 & : &             0.02 & : &             0.60 & : &             0.47 & : &               -- &   &               -- &   &   III &              1.1 &        0.25 &                0.00 &      K2 &  14.25 &  13.68 &  13.16 &         QPS &        1.54 \\
            1000 &               &     667 &       13610 &             0.79 &   &             0.16 &   &             2.59 &   &             1.10 &   &               -- &   &               -- &   &    II &               -- &          -- &                0.04 &         &  14.37 &  13.76 &  13.12 &           D &        5.92 \\
              58 &               &    1279 &       13611 &             1.29 & : &             0.03 & : &             4.04 & : &             0.53 & : &               -- &   &               -- &   &   III &              1.7 &        0.63 &                0.00 &      K6 &  15.40 &  14.64 &  13.80 &           P &        1.97 \\
             585 &               &     553 &       13611 &             0.12 & : &             0.02 & : &               -- &   &               -- &   &           324.27 & : &            60.98 & : &    II &             34.1 &          -- &                0.55 &         &     -- &  17.44 &     -- &             &             \\
             591 &               &     422 &       14368 &             1.12 & : &             0.05 & : &               -- &   &               -- &   &           120.94 & : &            19.45 & : &    II &               -- &          -- &                3.40 &         &     -- &     -- &     -- &             &             \\
             690 &               &     287 &       13610 &            13.66 & : &             0.44 & : &               -- &   &               -- &   &            69.03 & : &            11.73 & : &   III &               -- &          -- &                2.65 &         &  21.30 &  19.32 &  20.41 &             &             \\
            1018 &               &     695 &       13610 &             3.91 & : &             0.08 & : &             3.34 & : &             0.61 & : &               -- &   &               -- &   &   III &              1.3 &        0.45 &                0.00 &      K6 &  14.75 &  14.11 &  13.31 &           P &        3.23 \\
\hline\hline
\end{tabular}

}
\tablefoot{
\scriptsize
%\\ 
$^*$Src \#\,405: All energies and peak luminosities are considered highly uncertain because of the large discrepancy between optical and X-ray extinction estimates (\S\,\ref{sect:results}). 
$^\dagger$Src \#\,502: X-ray flare detected with 4.5 counts. No estimate of extinction was possible since  $A_V$ is not available and $N_H$ could not be derived from the fit of the X-ray spectrum because of its low statistics.
%\\ 
$^\ddagger$Src \#\,536: Classified as class III on the basis of the \citet{sun09} {\em Spitzer} photometry. \citet{cod14a} indicate, however, that the star has a mIR excess. This is based on the MIPS 240$\mu m$ flux \citep[not reported by][]{sun09}, which is, however, strongly contaminated by a nearby YSO associated with ACIS source \#\,541.  
%\\
$^\S$Src \#\,713: The {\em Chandra} flare falls in the gap between the last two observations -  $E_X$ was multiplied by 1.497 to approximatively correct for the missing exposure time.  
}
\end{table*}

Out of our 78 X-ray flares, 58 (from 46 stars) have a {\em CoRoT}
counterpart, 32 (from 30 stars) a {\em Spitzer} counterpart, and 13
(from 12 stars) have both. These flare samples are obviously affected by
selection biases, which might be even more severe than those affecting
the samples of stars simultaneously observed in two or three bands (see
\S\,\ref{sect:data}), and which ought to be taken into account when
interpreting our results. Figure\,\ref{fig:RaDec_Lx_R} shows the
distribution of the flaring stars in our samples in sky coordinates and
in the time-averaged $L_X$ vs. J-band magnitudes scatter plot.
Stars on which we detected simultaneous X-ray/optical flares are
systematically brighter in X-rays than the stars with just simultaneous
X-ray and optical lightcurves (median $\log L_X$=30.4\,erg/s vs.
30.0\,erg/s), which are in turn systematically brighter than the general
population of X-ray sources (median $\log L_X$=29.6\,erg/s). The same is
true for stars with simultaneous X-ray and mIR flares with respect to
the stars with just simultaneous X-ray and mIR lightcurves, (median
$\log L_X$=30.3\,erg/s vs. 29.5\,erg/s). This latter sample, even if
limited to two small regions in the sky seems, however, to be quite
representative of the full X-ray population {\em in those regions} in
terms of X-ray luminosities. However, because of the chosen {\em
Spitzer} pointing toward active star forming regions, all mIR data are
biased toward stars in early evolutionary stages (class\,II and
Class\,I) and with larger than average extinction.

We inspected the {\em CoRoT} and {\em Spitzer} lightcurves to try to
identify flares independently from the X-ray lightcurves. This is not
straightforward as faint flares are harder to identify against the
strong quiescent and, most often, highly variable optical or mIR
emission. Searching the {\em CoRoT} lightcurves, we only found one
convincing flare-like feature with no corresponding X-ray counterpart.
In another 5 stars, we detected suggestive low-significance features
which might actually be associated with a low-significance X-ray
feature. We found only one convincing {\em Spitzer}-only flare which,
however, occurred while the relative X-ray source was extremely faint
(just one detected photon in the last 67\,ks long observing segment, and
moreover coincident in time with the mIR flare). We conclude that our
data indicate that while X-ray only flares are rather common (see
\S\,\ref{sect:xray_only}), optical- and/or mIR-only flares are rare, if
at all present.

\section{Flare characterization}

In this section we describe, separately for the three spectral bands,
how we characterized our flares. For the scope of the present paper, we
will focus solely on time-integrated emitted energy and on peak flare
luminosity. The estimates of these two quantities in the three bands
will be discussed and compared in the following section.

\subsection{X-ray data}

As described in the previous section, each X-ray flare is defined as a
group of one or more consecutive ``maximum likelihood'' (ML) time
intervals. In order to estimate the luminosity at the flare peak,
$L_{X,pk}$, and the total emitted energy, $E_X$, both in the
0.5-8.0\,keV band, we first evaluated absorption-corrected X-ray
luminosities for each of the individual intervals, $L_{X,i}$. The
maximum of these values was then taken as $L_{X,pk}$, while $E_X$ was
taken as the sum of $L_{X,i}\times\Delta T_i$, where the $\Delta T_i$
indicates the duration of the segments in seconds.

In most cases, the $L_{X,i}$ were estimated through spectral fitting of
the X-ray spectra extracted for each segment. The spectral fits were
performed using the XSPEC package, modeling the flaring emission with
the {\sc apec} isothermal plasma emission model, subject to absorption 
from intervening interstellar and circumstellar material ({\sc tbabs}).
The non-flaring emission, also contributing to the flux in each of the
intervals but of no interest for our purposes, was taken into account
by adding a suitable spectral model (see below) with no free parameters.
The absorption-corrected fluxes of the flaring components were then
converted to luminosities adopting a distance of 760\,pc.

In many cases the spectra of individual segments are defined by a small
number of photons, resulting in large uncertainties on the best-fit
fluxes, mostly due to uncertainties on the absorbing column density,
$N_H$. Since the $N_H$ does not generally vary during flares\footnote{In
a small number of cases increases in the $N_H$ during flares have been
reported in the literature, possibly related to the associated Coronal
Mass Ejection. We do no find evidence of this phenomenon during the
flares analyzed here.}, we significantly reduced the uncertainties by
fixing $N_H$ to the best-fit, time-averaged, value obtained by Flaccomio
et al. (2018, in prep.) fitting the combined spectrum from all the
available {\em Chandra} data with a physically meaningful model. 

The spectral model for the quiescent or characteristic X-ray emission
\citep{wol05} during each flare was determined, for each
source, from a distinct set of ML time intervals defining the source
{\em characteristic} emission, as defined in \citet{wol05} and
\citet{car07}. By excluding all flares or other times of elevated flux,
the X-ray emission in these sets of ML intervals well approximates the
pre- and post-flare emission. A spectral model representative of the
characteristic X-ray emission was obtained by simultaneously fitting the
spectra extracted from these segments with a model with a large number
of free parameters\footnote{Specifically an absorbed- three-temperature
plasma model, {\sc tbabs(vapec+vapec+vapec)} in XSPEC, with variable
abundances for all elements}, with little regard for their physical
meaning. Although parameters were most often not constrained, these
models served our sole intention of obtaining an accurate representation
of the observed out-of-flare spectrum.

In a minority of cases, the spectra of some, or all, of the ML intervals
defining a flare contained too few photons to perform a meaningful
spectral fit. In these cases we estimated the energy flux from the
absorbed photon flux for the segment (in ph\,s$^{-1}$\,cm$^{-2}$),
subtracted by the characteristic photon flux, the time average for all
time intervals defining the characteristic emission level (see above).
The thus corrected photon flux is finally converted to
absorption-corrected energy flux multiplying it by a conversion factor,
taken as the ratio between the time-averaged absorption corrected
energy-flux of the source (from spectral fittings in Flaccomio et al.
2018) to its observed time-averaged photon-flux. This approach is
equivalent to approximating the X-ray spectrum during a given ML segment
with the time-averaged one. Thus neglecting the increase in plasma
temperature that usually characterizes flaring emission, we generally
obtain slightly smaller fluxes with respect to those obtained by
spectral fitting. The difference is, however, small and negligible with
respect to all other sources of uncertainties\footnote{For 57 flares
spectral fits were available for all of the defining segments. For this
sample, the emitted soft X-ray energies estimated from the conversion
factors are lower than those from spectral fits by 0.08\,dex in the
median ($\pm1\sigma$ quantiles of the ratio distribution: 0.03 and
0.14\,dex). An indistinguishable distribution (same median and
quantiles) is found for the ratio of 64 peak X-ray luminosities
estimated with both methods.}.

Finally, for the flare on ACIS\,\#\,713, one of the two including the
observing gap between the two last observing segments, discussed in
\S\,\ref{sect:detection}, we approximately accounted for the missing
observing time, by multiplying $E_X$, as determined above, by the ratio
between the duration of the flare and the observed time (1.50). No
correction was considered for the other similar case since the observed
fraction of the flare is, in this case, significantly longer than the
gap. 

\begin{figure*}[!t!]
\centering
\includegraphics[width=6.0cm]{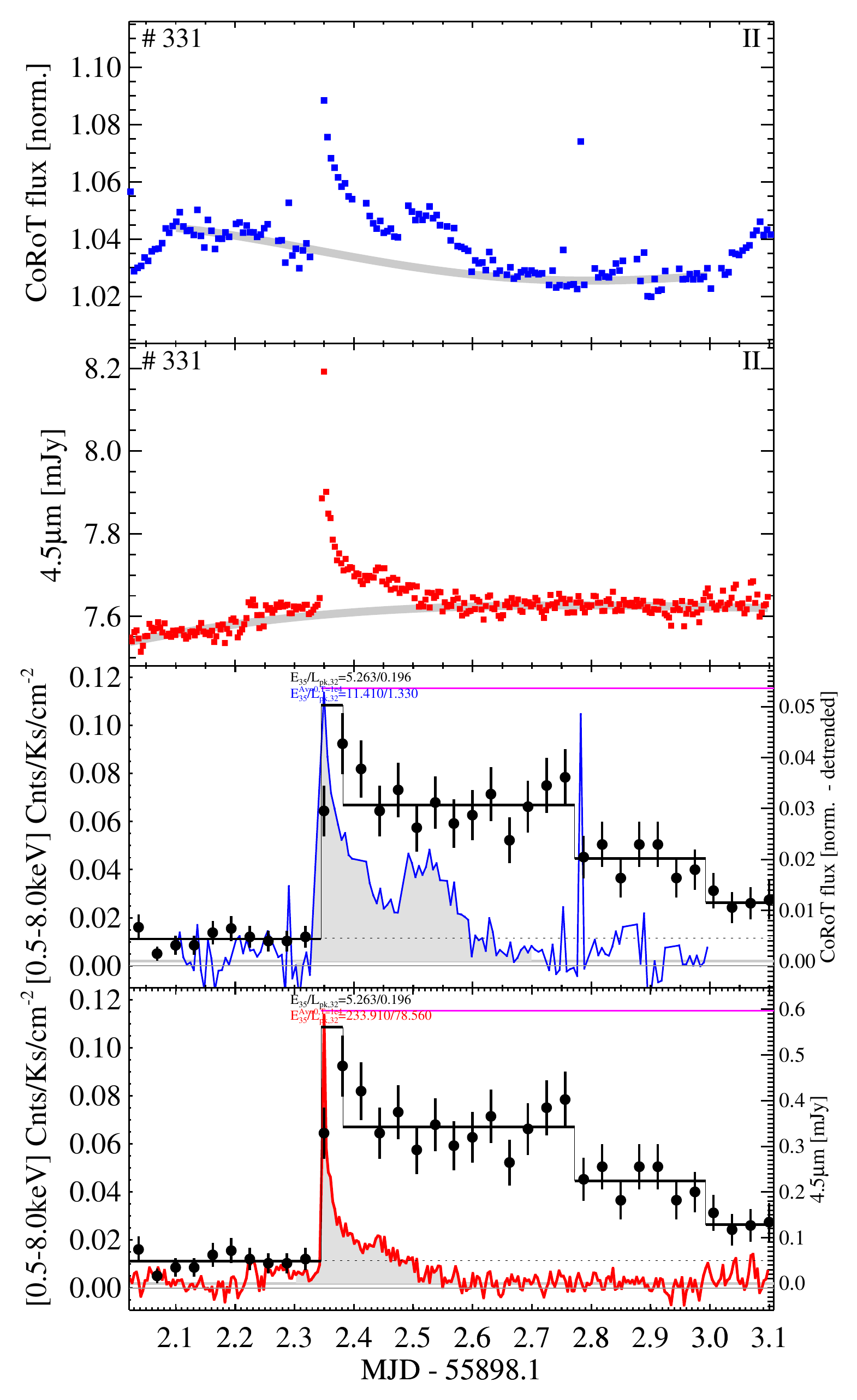}
\includegraphics[width=6.0cm]{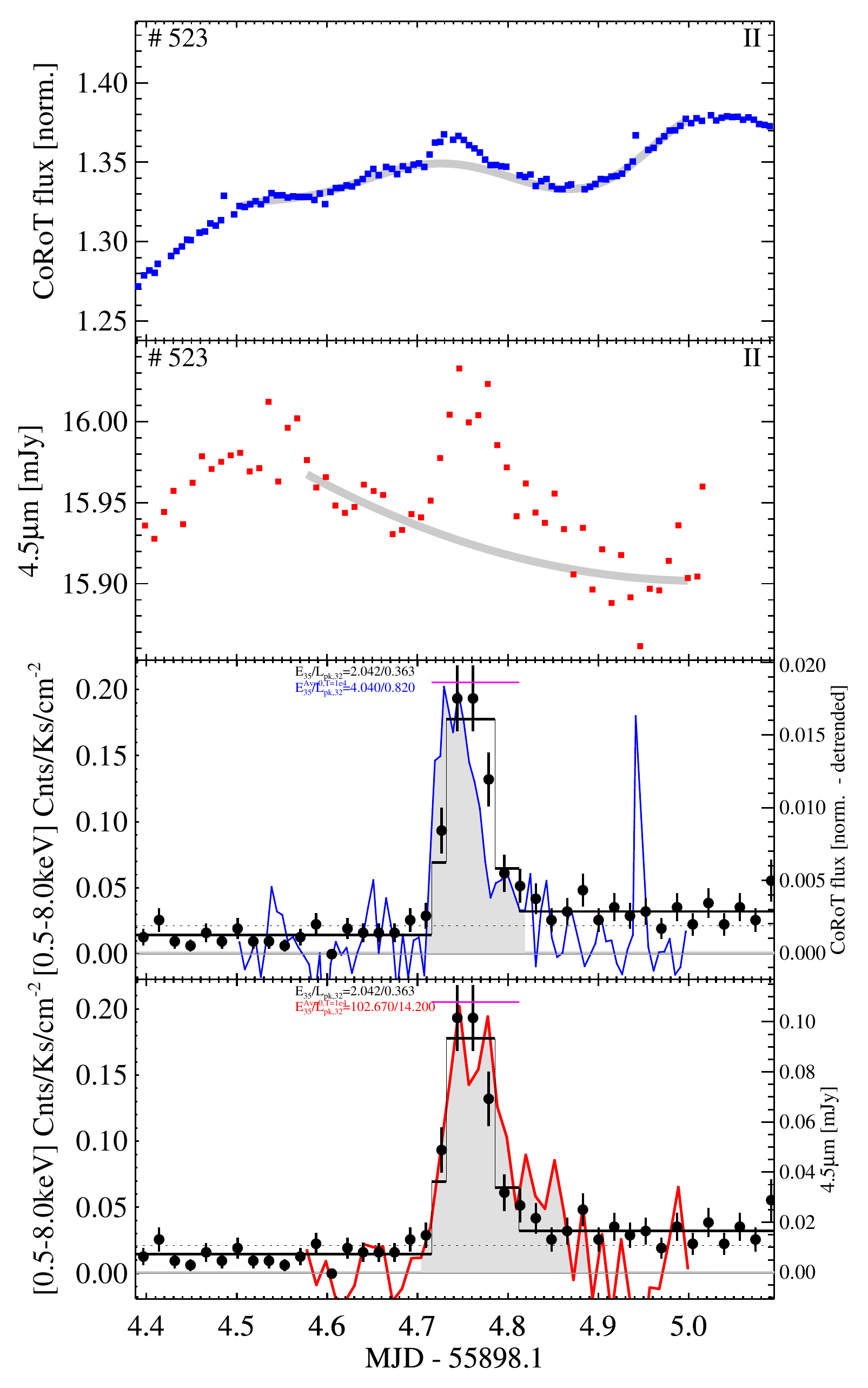}
\includegraphics[width=6.0cm]{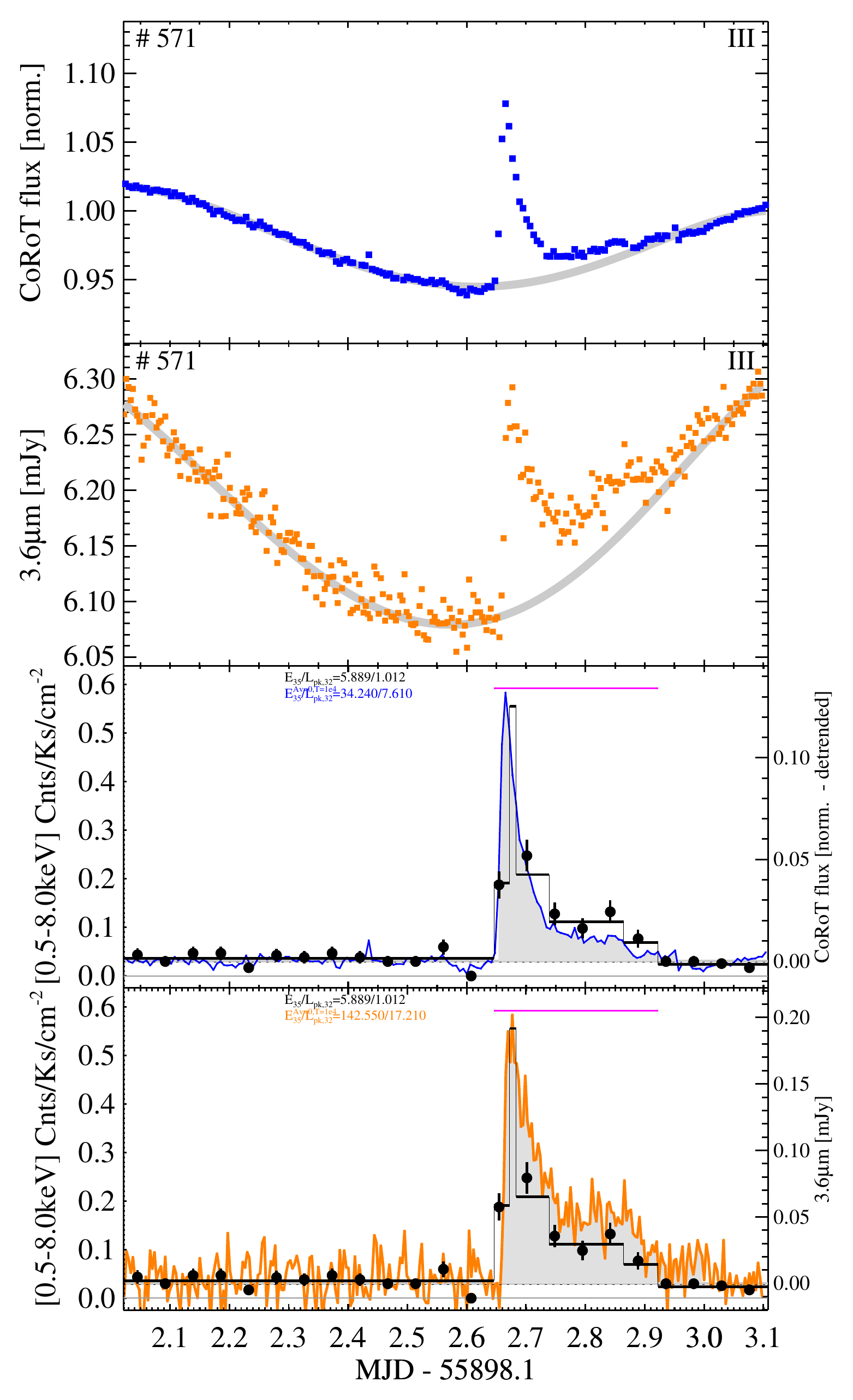}
\includegraphics[width=6.0cm]{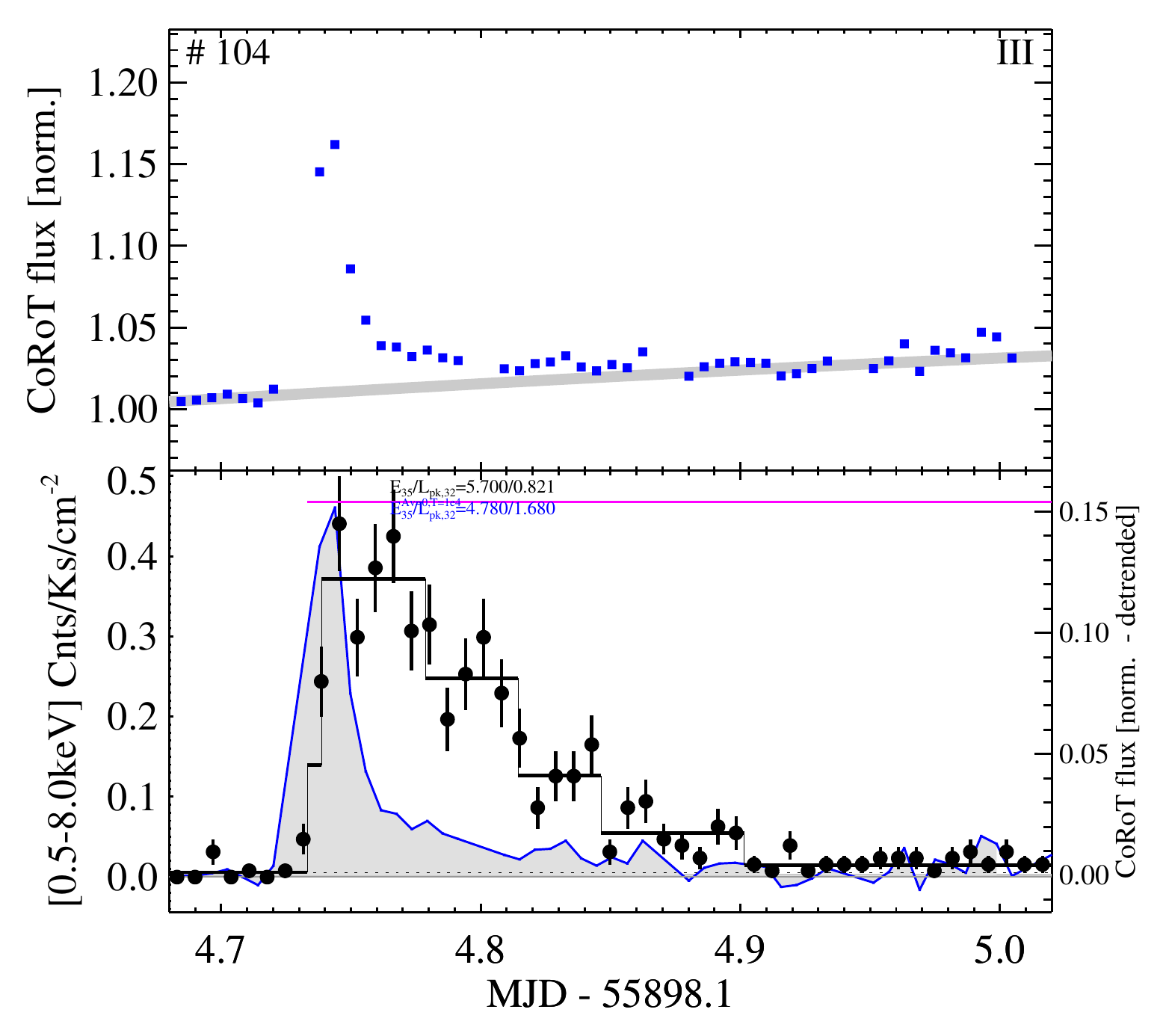}
\includegraphics[width=6.0cm]{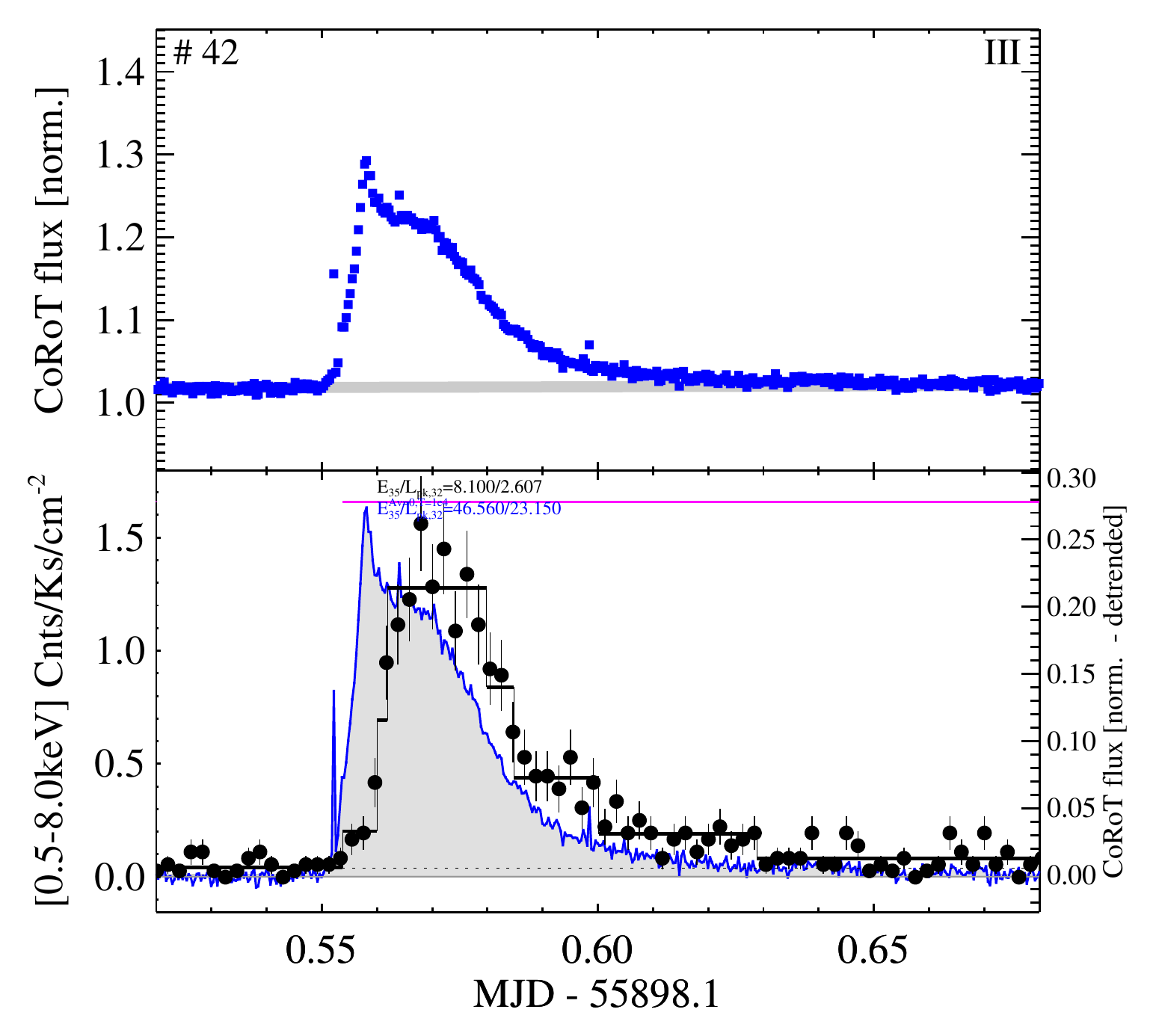}
\includegraphics[width=6.0cm]{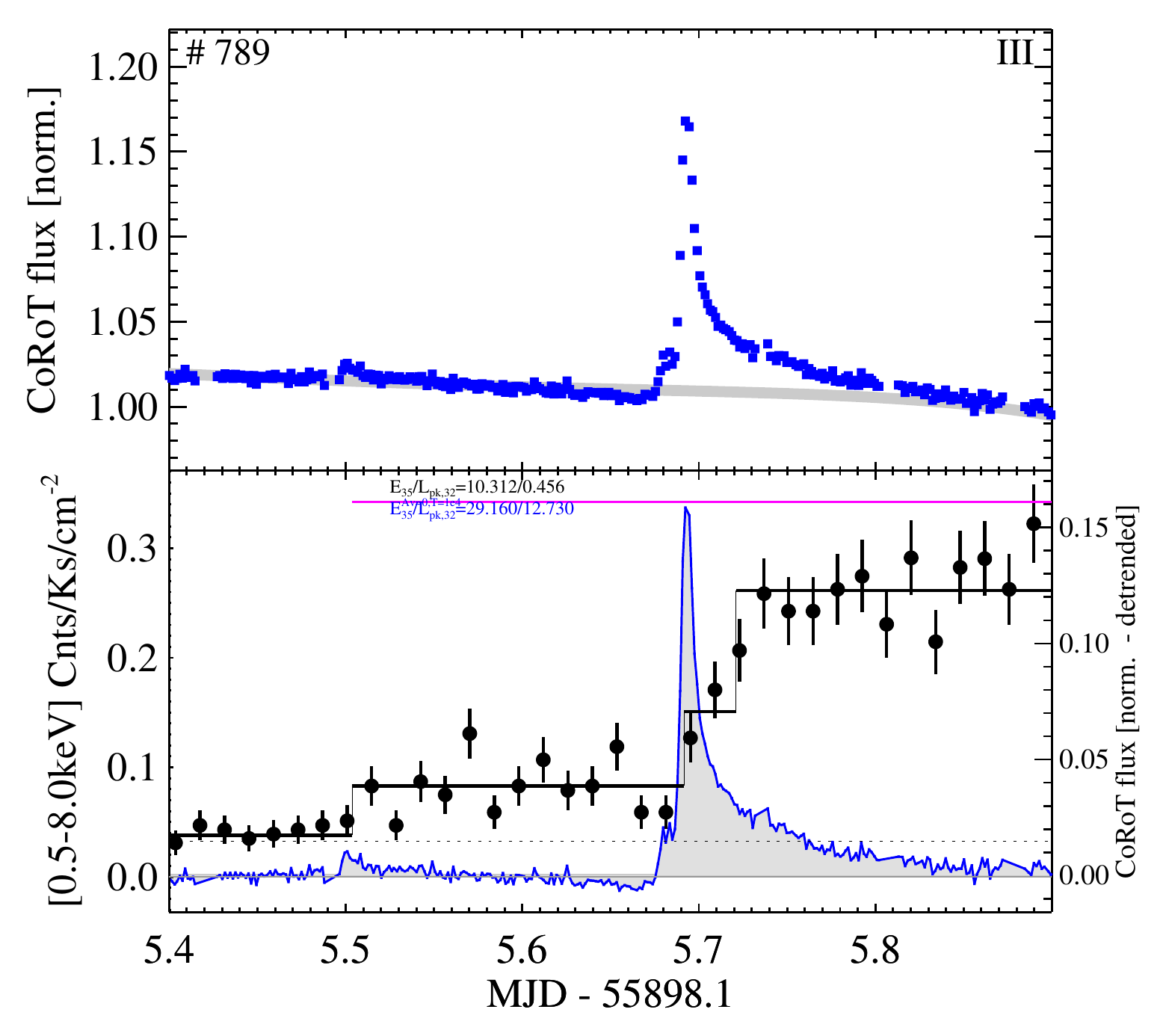}
\caption{Lightcurves of six of the X-ray flares with optical and/or mIR
counterpart discussed in the text. Lightcurves for the full sample can
be found in Appendix\,\ref{app:lightcurves}, while lightcurves for well defined X-ray
flares for which no optical/mIR counterpart was detected are shown in
Appendix\,\ref{app:LC_Xrayonly_det}. The three panels at the top refer to
flares with simultaneous data in all three bands. Within each panel the
first two sub-panels, from top to bottom, show the {\em CoRoT} lightcurve,
normalized to the median of the observed {\em CoRoT} flux in the whole
exposure (as opposed to the short segment shown), and the {\em Spitzer}
lightcurve in the IRAC 3.6$\mu m$ or 4.5$\mu m$ bands (as indicated by
the y-axis label and in orange and red, respectively). For both panels a
gray line indicate the polynomial intended to represent the non-flaring
emission (see text). The ACIS source number of the star and its mIR
class are given at the top-left and top-right corners, respectively. The
third and fourth panels show the same {\em CoRoT} and {\em Spitzer} lightcurves,
this time subtracted by the non-flaring emission (units given in the
right-hand y-axes). The areas shaded in gray indicate our choice for the
definition of the optical and mIR flares. The same two panels show the
simultaneous {\em Chandra} X-ray lightcurve, both binned (black dots
with error bars), and using the piece-wise representation discussed in
the text (black broken line). Units are shown on the left-hand y-axis.
The temporal extension of the X-ray flare, according to our definition, is
indicated by a magenta horizontal bar at the top of each panel. The
three panels at the bottom show three more X-ray flares with
simultaneous {\em CoRoT} data, but lacking {\em Spitzer} data.}
\label{fig:lc_examples}
\end{figure*}

\subsection{{\em CoRoT} data}

The  optical counterparts to the X-ray flares, in the {\em CoRoT}
lightcurves, were most often harder to define with respect to the X-ray
events. We followed an iterative approach. For each flare, we started by
examining the {\em CoRoT} lightcurve in the same time interval spanned
by the {\em Chandra} observation segment during which the flare was
detected\footnote{In some cases, for X-ray flares for which only the
decay phase was detected, the rising phase presumably falling before the
beginning of the {\em Chandra} observing segment, we considered a longer
portion of the {\em CoRoT} lightcurve, starting a few hours before the
beginning of the {\em Chandra} lightcurve}. In order to account for the
large non-flaring variability of our stars we then subtracted the {\em
quiescent} emission, determined through a polynomial fit to the {\em
CoRoT} lightcurve in the considered time interval. The order of the
polynomial was initially chosen as 3 or 5 for time segments shorter and
longer than 0.4 days, respectively. In order to reduce the influence of
positive deviations, such as flares, we settled, after some
experimentation, on a robust asymmetric sigma-clipping
procedure\footnote{We made use of the {\sc robust\_poly\_fit} IDL
routine contained in the {\sc astrolib} library. After an initial fit
with {\sc robust\_poly\_fit} we determined the standard deviation of
residuals, $\sigma$, and excluded outlying points, those with offsets
from the best fit curve smaller than -3$\sigma$ and larger than
1.5$\sigma$, and repeated the fit. This simple sigma-clipping procedure
was repeated three times.}. We then inspected both the original {\em
CoRoT} lightcurve and the continuum-subtracted one to visually search
for the optical counterparts of the X-ray flare. If unsuccessful, we
also tried to rebin the {\em CoRoT} lightcurve (in cases of low signal)
and to vary the standard filtering for removal of bad
data-points\footnote{Each {\em CoRoT} datapoint is flagged by the
standard pipeline for a number of potential issues (see
\url{http://idoc-corotn2-public.ias.u-psud.fr/jsp/doc/DescriptionN2v1.3.
pdf}) and we have initially adopted the {\em status=0} condition to
filter out all possibly affected data-points. We, however, realized that
accepting data flagged for certain conditions can, in several cases,
result in more flare-like lightcurves and in a better matches with X-ray
data (and/or mIR data, see below). This is possibly because $i)$ some
data-points are flagged for potential issues that do not always affect
the quality of the data, at least for our science, and $ii)$ the
impulsive phase of flares appears to spuriously trigger some of the
conditions used to set some specific flags. We thus quite often accepted
data-points for which the status flags 1, 2, 3, 4, 5, and 8 were set. In
one single case, one highly discrepant point in the {\em CoRoT}
lightcurve of ACIS\,\#\,331 was removed, in spite of its status
flag being zero, to exclude an obviously spurious peak.}. Once the
presence of an optical flare was determined, we refined the fit of the
out-of-flare lightcurve by excluding the time interval during which the
flare was observed, adjusting the degree of the polynomial, and, in a
small number of cases in which the fit was unsatisfactory, limiting the
fitted temporal interval. We then refined the choice of binning and
light-curve filtering (see above) that resulted in a {\em better
looking} flare in the continuum-subtracted lightcurve. We finally
defined ad-hoc start and end times and preceded to estimate the
time-integrated and peak fluxes, both in instrumental units.

The conversion of instrumental fluxes to intrinsic source luminosities
(and time-integrated fluxes to total emitted energies) is not
straightforward, especially for a wide-band telescope such as {\em
CoRoT}, moreover not optimized for absolute photometry. Appendix
\ref{app:corot_fluxes} illustrates how we proceeded in order to derive
the extinction law for the {\em CoRoT} band and conversion factors from
instrumental to bolometric fluxes\footnote{Since for our flares we are
only interested in variations of the flux, as opposed to total source
fluxes, our conversion from the observed {\em CoRoT} units to physical
units is not strongly affected by the significant background correction
issues discussed by \citet{cod14a}, as long as the background is
approximatively constant during the duration of each flare.}. Both of
these derivations depend critically on the source spectrum. Lacking
precise information, we considered two alternative shapes for the
optical spectrum of our flares: a stellar photospheric spectrum and a
black body. In both cases the extinction law and conversion factor
depend on the source temperature, while, for stellar-like spectra, the
dependence on surface gravity turns out to be negligible. In the
following we will mainly assume, for our flares, a black body spectrum
with $T=10^4$\,K. As we will indicate, however, our main results will
not depend significantly on this assumption. All absorption-corrected
fluxes were finally converted to luminosities multiplying by
$4\pi$d$^2$, with d=760\,pc.

\subsection{{\em Spitzer} data}

The analysis of the {\em Spitzer} lightcurves proceeded in much the same way
as that of the {\em CoRoT} ones. Unlike the {\em CoRoT} lightcurves, the {\em Spitzer}
staring-mode data does not extend beyond the {\em Chandra} observing
intervals, and we therefore could not search for the rise phase of mIR
flares before the beginning of the {\em Chandra} observations. Like for
the {\em CoRoT} case, we convert the observed fluxes, in this case provided in
physical units (mJy), to bolometric luminosities (in
erg\,s$^{-1}$cm$^{-2}$). The conversion factor for the 3.6$\mu$m and
4.5$\mu$m IRAC bands, function of the source spectrum and of the
intervening extinction, were derived for photospheric models and black
body spectra, in the same way described in
Appendix\,\ref{app:corot_fluxes} for the {\em CoRoT} band. We will initially
assume a  black body spectrum with $T=10^4$\,K, as we did for the
optical quantities. In doing so we are assuming that the mIR emission
originates from the same spectrum as the optical emission detected by
{\em CoRoT}. In the following we will, however, test this initial assumption.

%\section{Properties of X-ray flares}
%
%- Estimation of lenghts using various approaches (decay and rise phases)
%- comparison of various estimates l 
%- loop lengths on CTTSs vs. WTTSs
%
%- discussion on large uncertainties on the definition of flares, in many cases. And of integration of light-curves
%

\section{Results}
\label{sect:results}

In Table\,\ref{tbl:flares} we present, for our sample of X-ray detected
flares, estimates for the X-ray energy and peak flux in the
0.5-8\,keV band, and for the {\em bolometric} energies and peak fluxes
from the optical and mIR lightcurves, assuming a $T=10^4$\,K black body
spectrum.

The X-ray quantities are corrected for absorption. The listed
optical and IR quantities are instead not corrected for extinction, which is small in {\em most} cases. In the following we will, however, use
extinction-corrected energies and luminosities. In
Table\,\ref{tbl:flares} we report two independent estimates of
intervening material: $A_V$, based on spectral types and source
photometry from the literature, also listed in Table\,\ref{tbl:flares},
and the column density of neutral hydrogen $N_H$, estimated by Flaccomio
et al. (2018) fitting the time-averaged X-ray source spectra with
absorbed thermal plasma emission models (with either one or two
components). Both estimates suffer from considerable uncertainties. In
adopting an extinction correction for our flare peak luminosities and
total energies, we considered both estimates. We find that the X-ray
values, converted to an optical extinction through the relation
$A_V$=$N_H/2.1\times10^{21}$ \citep{zhu17} produces slightly more
significant correlations between the optical and X-ray measurements (see
below). This may be attributed to two facts: $i)$ the estimates of $N_H$
are available for more sources/flares, and $ii)$ the $N_H$ is estimated
from data which is, on average, much more simultaneous with the flares
with respect to the $A_V$ values\footnote{$N_H$ comes from the X-ray
spectrum averaged over all existing {\em Chandra} observations. For the
vast majority of flares, observed during the 300\,ks {\em Chandra}
observations within the CSI campaign, data from additional exposures, up
to 160\,ks long, were then included. For $A_V$, however, spectral types
and photometry were in all cases obtained years before the events.}. In
the following we will thus correct our optical and mIR quantities using
the X-ray derived extinction values, but will note whether our results
depend on this choice. We made an exception for the optical+mIR flare on
source ACIS\,\#\,405, for which a large discrepancy is found between
optical and X-ray absorption estimates: we adopt the optical estimate
and cosider all energies and peak luminosities as highly
uncertain\footnote{The $N_H$ would convert into an unphysical $A_V$
implying, e.g., $(V-I)_0=-1.15$ and $(R-I)_0=-1.04$ for the K3 host
star, and a peck optical flare luminosity  $\sim$100 times larger than
the stellar bolometric luminosity, in spite of the fact that the CoRoT
flux increases by only $\sim$15\%. We do not understand the reason of
these discrepancies, but note that the star is significantly accreting,
so that the disk material and accretion streams might well play a role
in the X-ray absorption.}, thus effectively excluding the flare from our
main sample (see below). 

In addition to the quantities described above, Table\,\ref{tbl:flares}
also list data from the literature for the flaring stars: mIR class
\citep{sun09,cod14a}\footnote{Four YSOs listed by \citet{sun09} as
possessing transition disks are here treated as Class II sources}, 
H$\alpha$ equivalent widths \citep{dah05,reb02}, spectral types
\citep{wal56,mak04,dah05}, V,R,I magnitudes \citep{lam04,sun08}, CoRoT
light curve type \citep{cod14a,ven17a} and rotational periods
\citep{lam04,ven17a}.

\subsection{Optical and X-ray emission}

\begin{figure*}[!t!]
\centering
\includegraphics[width=17.0cm]{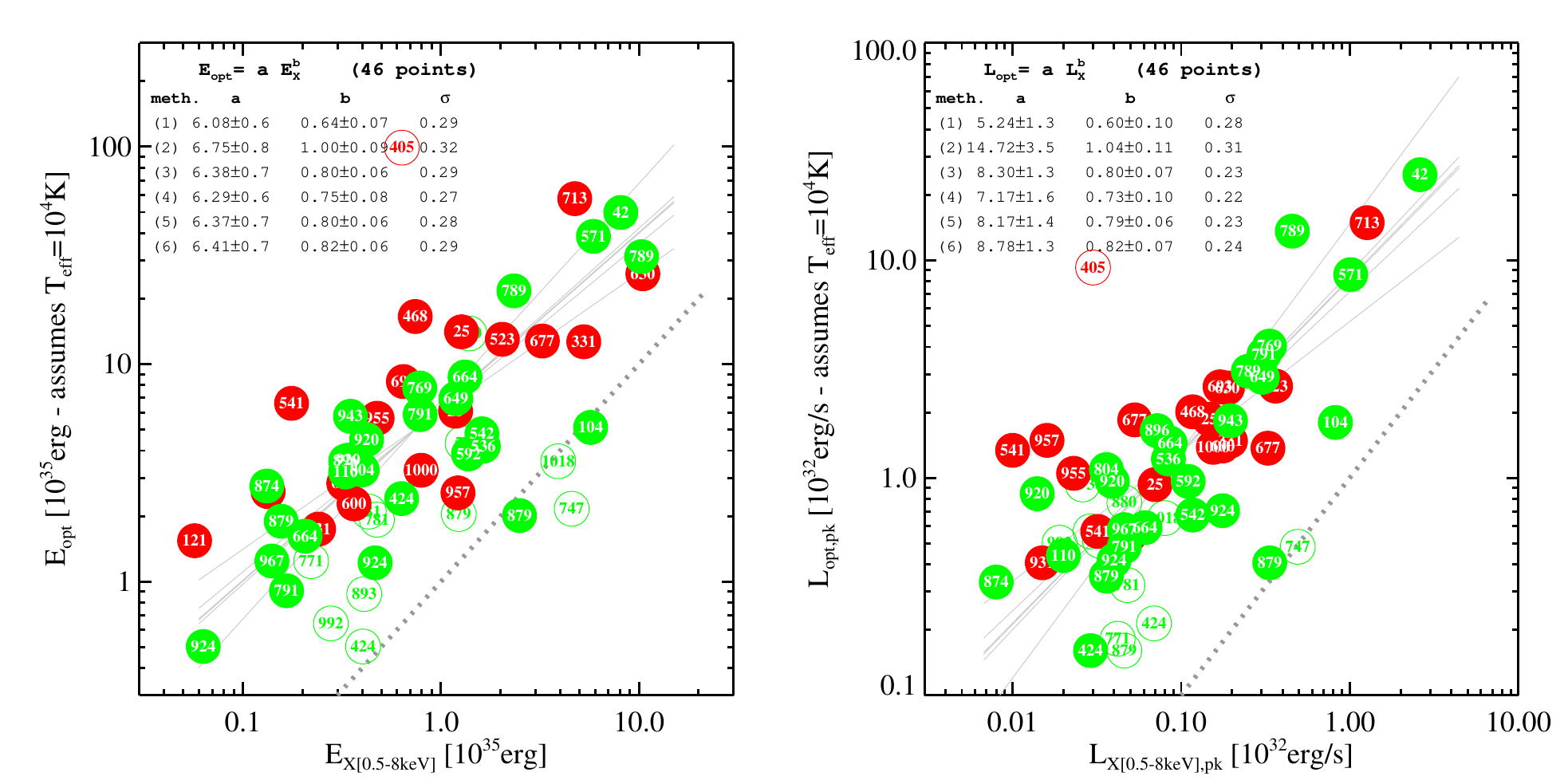}
\caption{[left]: Total {\em bolometric} emitted energy, $E_{opt}$, estimated from the {\em CoRoT} lightcurves (see text) vs. energy emitted in the 0.5-8.0\,keV band, $E_X$. Green and red circles indicate flares from class\,III and class\,II stars, respectively. Filled circles refer to the flares in our main samples, empty ones to flares with particularly uncertain estimates. The dashed lines shows the unit relation, while the solid lines show the results of six different linear fits (in the log-log plane) performed with different methods, and whose parameters are shown in the upper part of the panel along with the 1$\sigma$ dispersion of residuals. The fitting methods are: (1) Ordinary Least Squares Y vs. X, (2) Ordinary Least Squares  X vs. Y, (3) Ordinary Least Squares Bisector, (4) Orthogonal Reduced Major Axis, (5) Reduced Major-Axis, (6) Mean ordinary Least Squares. [right]. Same as the plot on the left, but for peak luminosities instead of energies.  
}
\label{fig:EL_XOpt}
\end{figure*}

In Figure\,\ref{fig:EL_XOpt} we show the relation between optical {\em
bolometric} and X-ray emitted energies and peak fluxes. We indicate with
different colors flares from stars with and without indication of
circumstellar disks. Empty symbols indicate flares for which the
estimates of either of the plotted quantities were deemed highly
uncertain.

A clear correlation is observed for both the emitted energies and for
peak luminosities. The optical values are almost always significantly
larger than the soft X-ray quantities. We fit the logarithms of the
plotted quantities (filled symbols only) with straight lines using six
different methods as provided by the {\sc sixlin} routine in the {\sc
astrolib} IDL library. The results for the 46 flares depicted as solid
symbols are shown within each panel, along with the 1$\sigma$
dispersion computed from the corresponding quantiles of the distribution
of y-axis residuals. For emitted energies we obtain $E_{opt}=a_E\times
E_X^{b_E}$ with $a_E\sim6.3$ and $b_E$ ranging between 2/3 and 1. For
peak luminosities we obtain $L_{opt}=a_L\times L_X^{b_L}$ with $a_L$
ranging from 5.2 to 14.7 and $b_L$ between 0.6 and 1.0. In spite of a
couple of discrepant points, lying in both panels close to the unity
relation (the gray dotted lines), the 1$\sigma$ scatter about the
best-fit relations are as low as 0.27\,dex and 0.22\,dex for energies
and luminosities, respectively.

We note that a different assumption for the optical flare spectrum, such
as assuming a photospheric spectrum instead of a black body, or a
different temperature, would imply an almost rigid shift in the y-axis
(since the effect of source-dependent extinction is small). The amount
of this shift can be read from Fig.\,\ref{fig:Fbol_Fcorot} and is
$<$0.3\,dex, toward lower values, for $4000<T$(or $T_{eff}$)$<10000$\,K.
An unaccounted-for source-dependent $T$ (or $T_{eff}$), surely a likely
occurrence, would contribute to the observed scatter. As for the choice
of extinction, had we adopted the $A_V$ values from
Table\,\ref{tbl:flares} to correct the optical energies and
luminosities, the correlations would be similar to the ones shown in
Fig.\,\ref{fig:EL_XOpt}, but would include only 35 flares (instead of
46) and would have slightly larger scatters, both for flare energies
(average for the six regressions: 0.33\,dex vs. 0.28\,dex) and, even
more, for luminosities (average $\sim$0.38\,dex vs. $\sim$0.25\,dex).
The coefficients of the correlations would also vary, with  $a_E\sim9$,
$b_E$ between 0.7 and 1.2, $a_L$ between 9 and 34, and $b_L$ between 0.7
and 1.3.

\subsection{mIR emission}

\begin{figure*}[!th!]
\centering
\includegraphics[width=17.0cm]{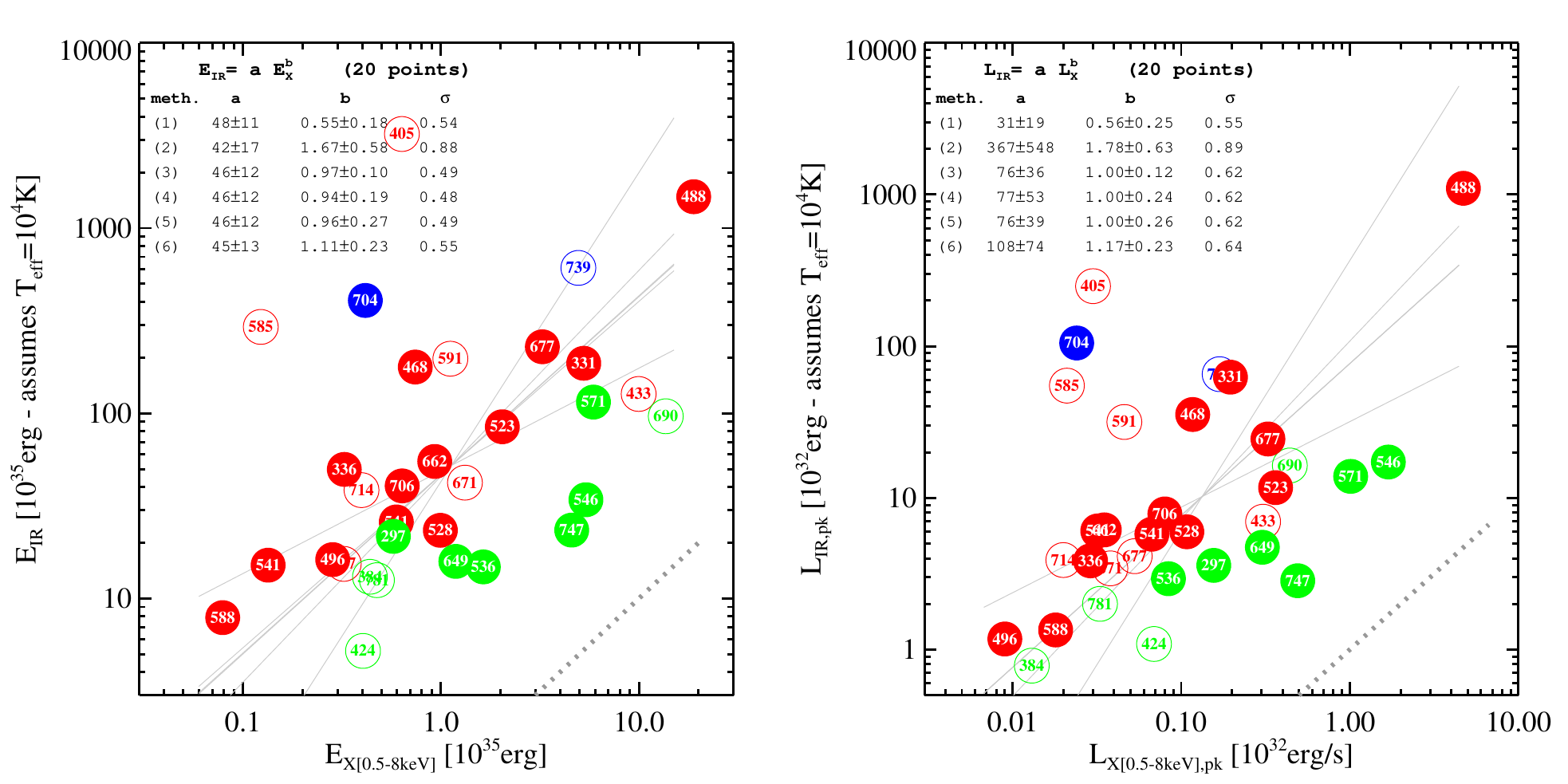}
\caption{[left] {\em Bolometric} flare emitted energy, estimated from the {\em Spitzer} lightcurves assuming a 10$^4$\,K black body spectrum, E$_{IR}$, vs. $E_X$ the energy emitted in the 0.5-8.0\,keV band. Symbols and lines as in Fig.\,\ref{fig:EL_XOpt}, with the addition of blue circles, which indicate flares from class\,I sources. [right] Same as the plot on the left, but for peak luminosities instead of energies.}
\label{fig:EL_XmIR}
\end{figure*}

In Figure \ref{fig:EL_XmIR} we show the run of the bolometric energies
and peak luminosities, as computed from the mIR {\em Spitzer}
lightcurves ($E_{IR}$ and $L_{IR,pk}$, either from the 3.6$\mu$m or
$4.5\mu$m data) vs. the corresponding X-ray quantities ($E_X$ and
$L_{X,pk}$). A correlation between energies may be observed, but it is
surely less significant than the optical vs. X-ray correlation. We
applied the Spearman's $\rho$ and Kendall's $\tau$ correlation tests
obtaining null probabilities of 3.2 and 2.3\%, respectively. Limiting
the sample to class\,II surces, the correlation becomes more significant
($P_{null}$=0.01/0.04\%). The most striking feature of the plot is,
however, the fact that flares from class\,II stars appear to have a
higher $E_{IR}/E_X$ ratio with respect to those from stars without
disks. This in confirmed by a Kolmogorov-Smirnov (KS) test which
indicates that the likelihood that two samples are drawn from the same
population is 0.19\% (0.01\% if we include the uncertain points plotted
as empty symbols). Similar conclusions can be drawn for peak
luminosities: the correlation between the two quantities is slightly
less significant for the whole sample ($P_{null}$=5.5/4.4\%), but still
significant for flares from class\,II stars ($P_{null}$=0.005/0.04\%). For
$L_{IR}/L_X$ the KS test again indicate a significant difference between
flares from stars with and without disks, with $P_{null}$=0.05\% (0.004\%
including uncertain points). Finally, the two flares from class\,I
sources may have even larger emission in our mIR bands with respect to the
class\,II (and III) samples.

In contrast to the {\em CoRoT} instrumental-to-bolometric flux
conversion, which is rather insensitive to the incoming spectrum thanks
to the {\em CoRoT} broad wavelength response, the estimation of
bolometric fluxes from the observed mIR fluxes is highly dependent on
the assumed spectrum. Assuming a $10^4$\,K photospheric spectrum would
increase $E_{IR}$ and $L_{IR,pk}$ by $\sim$25\%, while choosing a cooler
black body for the optical/mIR flare emission would significantly
decrease the estimated bolometric flux, for example by a factor of
$\sim$4 for $T= 6000$\,K. Significant systematic uncertainties and
scatter might therefore be introduced by our assumption of similar
flaring optical/mIR spectra. If the assumption holds, however, the
slopes of the regressions would not be affected. Also substantially
unaffected would be the results of the KS test on the difference of
$E_{IR}/E_X$ and $L_{IR}/L_X$ between flares on stars with and without
disks. Had we corrected the mIR flare fluxes using the optically derived
$A_V$ instead of $N_H$ the number of available points would have been
reduced from 20 to 11, since $A_V$ is not available for some of the most
absorbed stars, especially those with disks. The correlation tests are
in this case inconclusive. The difference in the distributions of
$E_{IR}/E_X$ and $L_{IR}/L_X$, between stars with and without disks,
remains somewhat significant: $P_{null}$=1.83\% and 0.23\%, for energies
and luminosities, respectively (0.66\% and 0.40\% including uncertain
data-points).

We will now investigate the ratio between mIR and optical flare
emission. Unfortunately the number of flares with a good
characterization of flares in the two bands is small: eight flares, five
and three from stars with and without disks, respectively. However, all
IR flares in our sample have an X-ray counterpart, and their optical
energy and peak luminosity may be approximatively estimated from the
correlations shown in Fig.\,\ref{fig:EL_XOpt}. Adopting the ``Ordinary
Least Squares`` relations, we thus estimate $E_{opt}$ and $L_{opt,pk}$
also for flares with no optical counterpart.

\begin{figure*}[!t!]
\centering
\includegraphics[width=17.0cm]{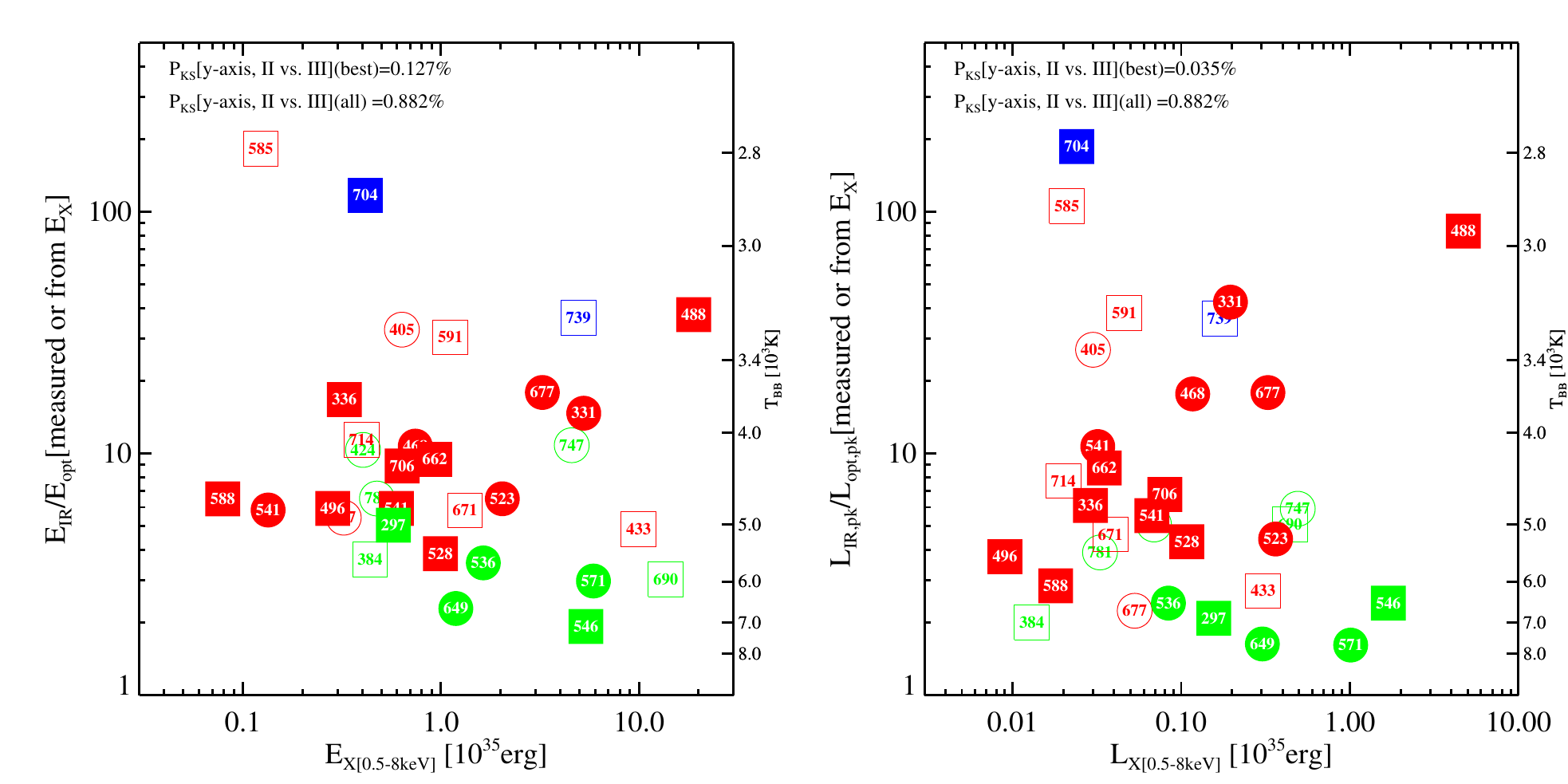}
\caption{E$_{IR}$/E$_{opt}$ vs. $E_X$ [left] and L$_{IR,pk}$/E$_{opt,pk}$ vs. $E_{X,pk}$ [right]. $E_{opt}$ and $L_{opt}$ are derived either from the analysis of the {\em CoRoT} lightcurves (shown as circles) or estimated from the X-ray quantities and the correlation with the optical ones (squares). The remaining symbols are as in Fig.\,\ref{fig:EL_XmIR}. The vertical scale on the right-hand axes indicate the temperature of the black body spectrum, assumed to be responsible for the emission in the optical and IR bands. The results of KS  tests comparing the distributions of the optical/IR ratios for flares from class II and class III sources are shown in the upper-left corner, both for the higher quality flares (filled symbols) and for all flares. Note that, at least for the higher quality flares, the null probabilities reported indicate that the distributions are significantly different. }
\label{fig:EL_XmIROpt}
\end{figure*}

In Fig.\,\ref{fig:EL_XmIROpt} we use these estimates to plot the
$E_{IR}/E_{opt}$ ratio vs. $E_X$, and $L_{IR,pk}/L_{opt,pk}$ vs.
$L_{X,pk}$. Circles indicate measured values while squares the ones
estimated from the X-ray quantities. No clear correlation is observed.
However, it is quite clear that flares on class\,II (and, even more,
class\,I) sources have a significantly larger IR/optical ratio than
those on class\,III stars.

If the flaring optical and mIR emission, detected with {\em CoRoT} and
{\em Spitzer}, came from the same emitting regions, and thus probed different
parts of the same spectrum, and if this spectrum were the same for all
flares, the ratio of the $E_{bol}$ ($L_{bol,pk}$) values estimated from
the two bands would be a constant. If the spectra were precisely those
we have assumed (10$^4$\,K black bodies), the ratios would be equal to
one. The fact that all ratios are larger than one then indicates that
the spectra depart from our assumption. The large scatter, moreover,
tells us that the spectra are not all the same. If we assume that, for a
given flare, the optical and mIR emission comes from the same spectrum
and that these are black bodies (or photospheric-like spectra), we can
easily relate the optical/IR ratios to the (effective) temperature of
the optically/mIR emitting region. The relation depends only very
slightly on the IR band adopted to derive $E_{IR}$ and $L_{IR}$, i.e.
3.6$\mu$m or 4.5$\mu$m. The y-axis on the right-hand side of the two
panels in Fig.\,\ref{fig:EL_XmIROpt} shows the mean
correspondence\footnote{the difference between the temperatures
corresponding to given IR/opt ratio in the two bands is always
$<$2.2\%.}.

A KS test shows that well characterized flares from class\,II and
class\,III sources have different distributions of $E_{IR}/E_{opt}$, (or
of the derived temperatures, $P_{null}$=0.13\%), as well as
$L_{IR,pk}/L_{opt,pk}$ (or of the derived temperatures,
$P_{null}$=0.04\%). The significance of these conclusions are not
affected by the assumption of a black body vs. photospheric spectra.
Flares on class\,III stars seem to originate from hotter material with
respect to those from class\,II (and class\,I) sources, and to span a
much narrower range of temperatures, especially when considering the
temperatures derived from peak luminosities (T=7000-8000\,K).
Alternatively, the optical and mIR flares we observe in stars with
circumstellar material (class\,II and I) might originate from different
regions, so that the temperatures we estimated would be meaningless: one
can easily envisage a scenario in which the optical flares originate at
the feet of the flaring loops (and trace the plasma heating phase) while
the mIR emission is dominated by the emission from inner disk (or
envelope), heated or otherwise affected by the optical and X-ray
emission of the flare.

\subsection{Duration and start times}

Measuring the duration of flares from the lightcurves is not always
straightforward, especially for faint/low signal events. We here define
the flare duration $\tau$, as the ratio between the emitted energy and
the peak luminosity. This definition, which corresponds to the decay time
for a pure exponential decay, can be easily applied to the three bands.
It is, however, subject to biases, most notably from the underestimation
of the flare peak luminosities, which is unavoidable with our direct
measurement procedure and limited statistics and/or temporal resolution.
Durations may thus be systematically overestimated. This bias may be
more severe in the X-ray band where the signal-to-noise ratio is the
lowest.

\begin{figure*}[!t!]
\centering
\includegraphics[height=6.1cm]{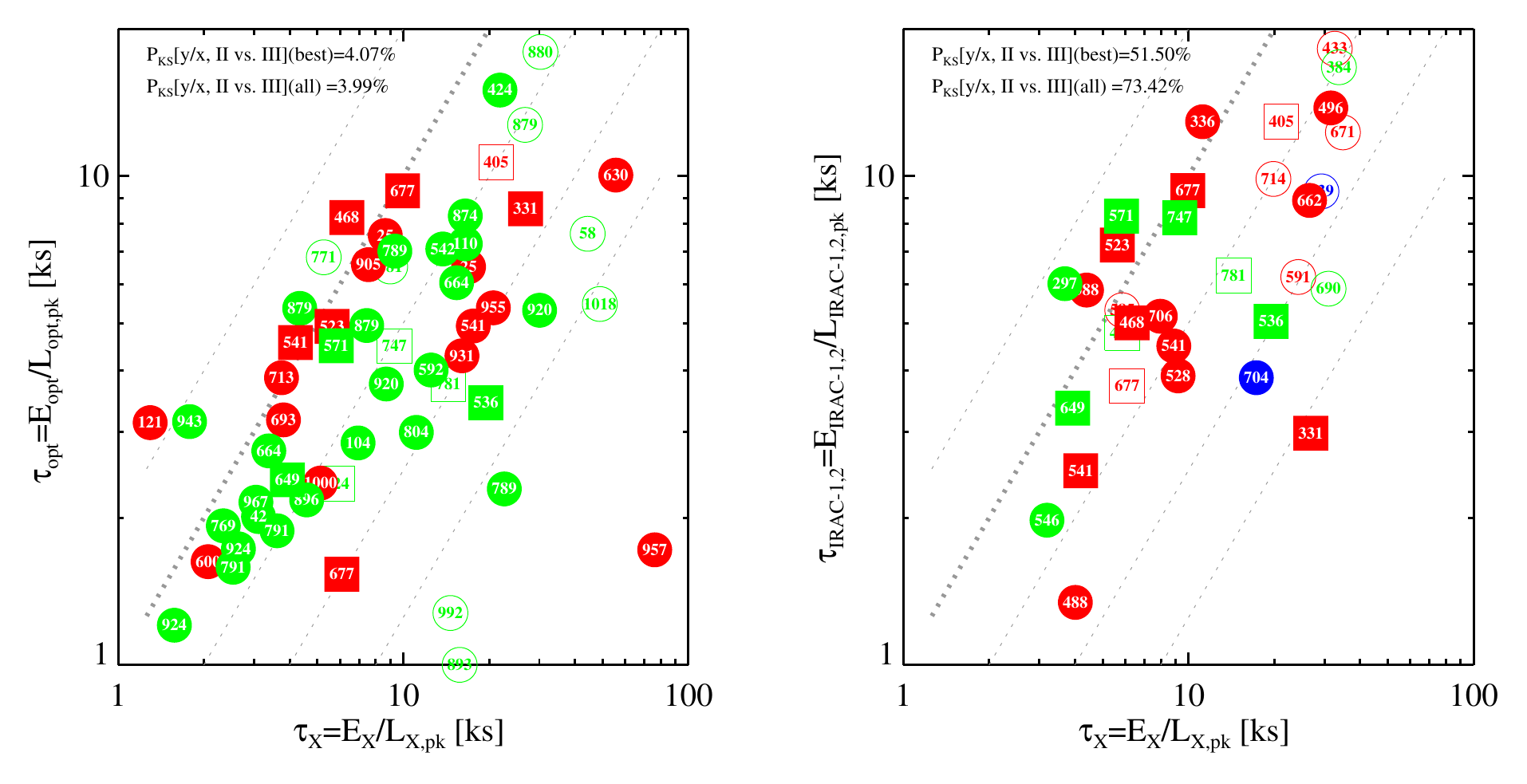}
\includegraphics[height=6.1cm]{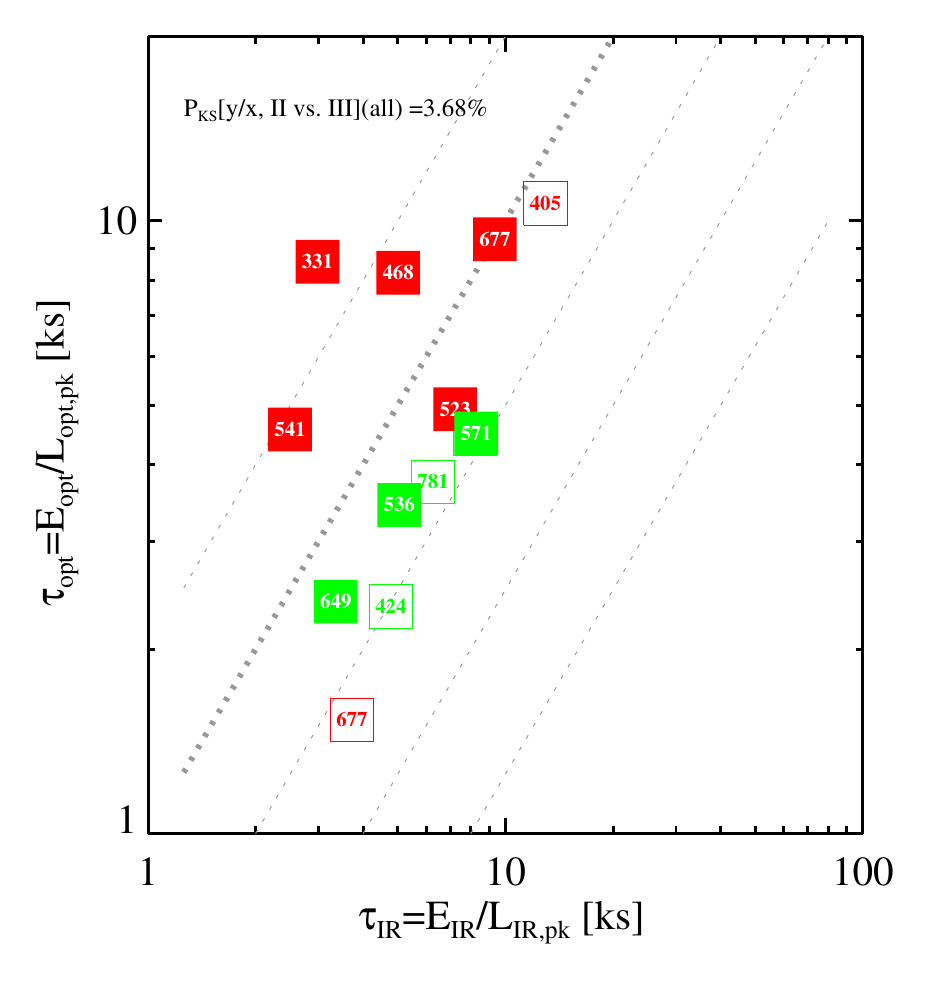}
\caption{Comparisons between the duration of flares in different bands.
The three panels show all three possible comparison between the X-ray,
optical, and mIR bands. Duration are defined as described in the text,
as the ratio between integrated energy and peak luminosity. Flares plotted as squares are those observed in all three bands. The remaining symbols and colors are as in Figures\,\ref{fig:EL_XOpt} and \ref{fig:EL_XmIR}. The thick gray diagonal line indicates the unit relation. Thinner lines deviations by factors of 0.5, 2.0, 4.0, and 8.0. 
The results of KS  tests comparing the distributions of the ratios of the two plotted quantities for flares from class II and class III sources are shown in the upper-left corner, both for the higher quality flares (filled symbols) and for all flares. The null probabilities reported do not evidence any significant difference. }
\label{fig:tau_vs_tau}
\end{figure*}

In Figure\,\ref{fig:tau_vs_tau} we compare the duration of flares with
simultaneous coverage in the three bands. We observe that X-ray flares
are on average longer than both their optical and mIR counterparts,
while the optical and mIR flares have similar durations (within a factor
of $\sim$2). No clear difference is observed between flares on stars of
different classes.

\begin{figure*}[!t!]
\centering
\includegraphics[height=6.1cm]{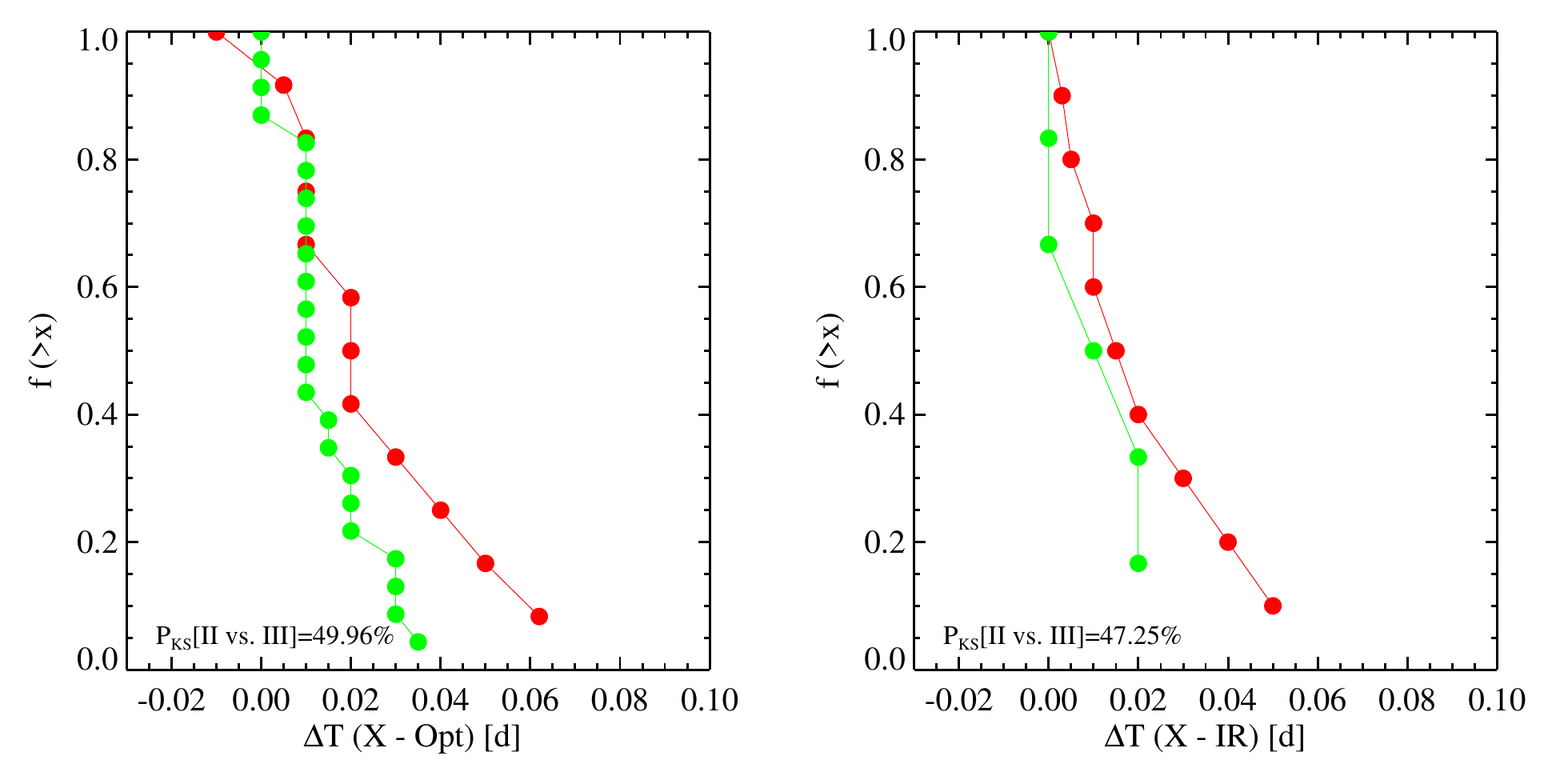}
\includegraphics[height=6.1cm]{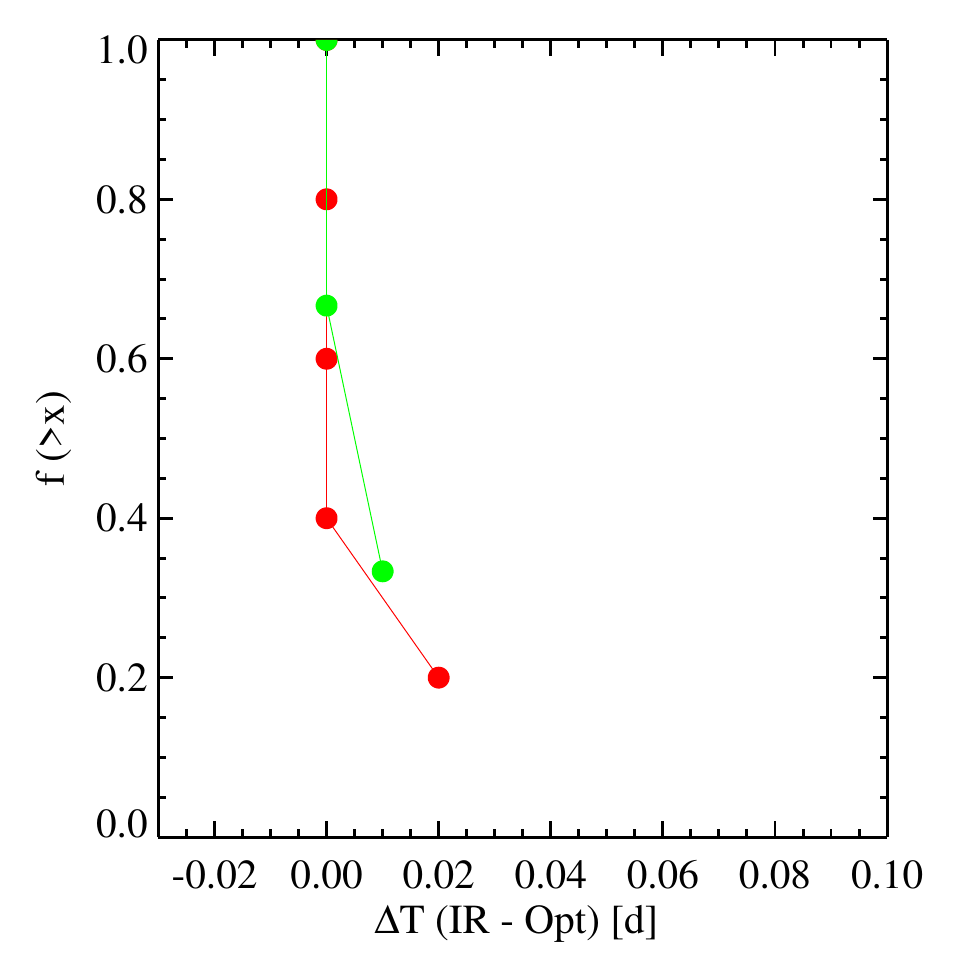}
\caption{Distributions of time delays between the flare start times in the optical band vs. X-rays [left panel], IR vs. X-rays [center], and optical vs. IR bands [right]. Distributions for flares on class\,II and class\,III stars are plotted separately in each panel in red and green, respectively. 
The results of KS  tests comparing the two distributions are shown in the lower-left corner, indicating in all cases that the distributions are not significantly different.
}
\label{fig:deltaT}
\end{figure*}

We also attempted to estimate the start times of flares in the three
bands: however, while in X-rays we can profitably make use of the
maximum-likelihood segmentation described in \S\,\ref{sect:detection},
determining the start time in the {\em CoRoT} or {\em Spitzer}
lightcurves is not straightforward. Whenever reasonable, we have
estimated the delay between the onset of an X-ray event and that of
the {\em CoRoT}/{\em Spitzer} one by simple inspection of the
lightcurves shown in Appendix\,\ref{app:lightcurves}. These estimates
are probably good within 0.01 days ($\sim$14 minutes). Some prominent
examples of such delays can be observed for the flares from ACIS sources
\#\,42, \#\,104, \#\,677 (first flare), and \#\,713.
Figure\,\ref{fig:deltaT} shows the distributions of these delays,
separately for Class\,II and Class\,III stars. The X-ray flares almost
invariably trail both the optical and the IR flares, by $\sim$0.01 days.
Observational biases might well affect this result: for example, the
X-ray events might be detected with a delay simply because of the
limited photon statistics. At face value, however, stars with disks
appear to have slightly longer delays, with respect to both the optical
and the IR counterpart. 
%[Is this true? or due to an observational
%effect? E.g. are flares on class\,II are X-ray fainter (does not seems
%so)? If true why is that?] 
In no case, however, are the distributions statistically incompatible
with each other. A comparison between the two delays (optical vs. X-ray
and mIR vs. X-ray), shows that the two are equal or within the
(significant) uncertainties. This, together with the similarity in
duration, may point toward a common origin of the optical and mIR
flares.

%\subsection{X-ray only flares}

%\begin{figure*}[!t!]
%\centering
%\includegraphics[width=17.0cm]{EL_X_vs_Tflare.pdf}
%\caption{}
%\label{fig:EL_Tflare_X}
%\end{figure*}

\section{Discussion}

Our characterization of flares in the X-ray, optical and mIR bands is
plagued by a significant number of uncertainties. Some are mostly
stochastic, e.g. those related to the uncertain definition of the shape
of the underlying emission (from the corona, photosphere, or the inner
disk), the choice of the flare start and end times, and the absorption
correction. Some may be systematic, such as those related to the
uncertain nature of the flare optical/IR spectrum and the conversion
between observed and physical quantities. In spite of these large and
hard to quantify uncertainties, we are able to draw several conclusion.

\subsection{Energetics} 
\label{sect:disc_ene}

Much more energy is emitted in the optical band with respect to the soft
X-ray band. This is consistent with previous finding for solar and
stellar flares \citep{fle11a}. Here we are able to derive a correlation
spanning about two orders of magnitude, for flares that are several
orders of magnitude more energetic than the ones observed on the Sun.
The correlation is rather tight and consistent with the idea that the
optical emission traces the plasma heating process, and that a fraction
of this energy is then radiated away by the plasma-filled loops in the
soft X-ray band. The fact that optical flares are almost always shorter
than their X-ray counterparts, and that they also usually start $\sim$15
minutes earlier, agrees with this picture. Moreover, although a
detailed analysis of the lightcurves of the brightest of our flares is
beyond the scope of the present work, we notice that some flares seem to
show the Neupert effect \citep{neu68} in that the time integral of the
optical emission appears to track the soft X-ray lightcurve (e.g. flare on source \#\,789 in Fig.\,\ref{fig:lc_examples}). 

Since the presence of a circumstellar disk seems irrelevant for the
optical/X-ray correlation, either the coronal loops involved are
unaffected by disks (and accretion) or the modifications are not
relevant for the heating of the plasma in the flaring structures and for
its radiative cooling. The slope of the log$E_{opt}$ vs. log$E_X$
correlation is such that, as flares become more powerful, either more of
the total energy is converted to X-ray radiation, or a smaller fraction
is emitted in the optical band, or both.

The correlation between peak X-ray and optical luminosities is even
tighter than for total energies. Since optical flares are shorter, their
peak luminosity, $L_{opt,pk}$ is even larger with respect to the X-ray
peak luminosity $L_{X,pk}$ than $E_{opt}$ is with respect to $E_X$. The
slope of the correlation, however, appears similar. Again the presence
of a circumstellar disk does not appear to make a difference.

%--- COMPARISON WITH THE SUN ----
%
%Conclusions of \citet{woo06a}:
%
%From a comparison of total solar irradiance (TSI), GOES, and other X/UV instruments on a very small sample of large X-class flares (mainly 4, two at the disk center, two at the limb):
%- Total (TSI) radiated flare energy >1e32 ergs) 
%- Total energy ~105 times the GOES energy (0.1-0.8 nm) 
%- VUV (<200 nm) energy ~1/2 of the total, the rest in NUV, optical, IR
%- Spatial variations observed: limb flares have "fully" suppressed >200um emission, while the X-ray one is unchanged.
%- This support the thick target scenario: "... particles ... absorbed in the transition region and upper chromosphere, which in turn respond by significantly increasing the emissions during the impulsive phase to further heat the chromosphere and photosphere".

%E[GOES band (0.1-0.8um = 1.5498-12.3984 keV)] =E[0.5-8.0 keV] * [$\sim$0.5 (kT$\sim$2keV), $\sim$0.8 (kT$\sim$5keV), $\sim$1.0 (kT >10keV) ]
%=> 2 * E(NUV+Optical+IR) (~total) = 105*(0.5-0.8) E[0.5-0.8keV] 
%=>  E(NUV+Optical+IR)= (25-40) E[0.5-0.8keV]  for flares with E(NUV+Optical+IR)$\sim$1e32 ergs

%According to our correlations
%Eopt/Ex=a$^{1/b}$ Eopt[35]$^{1-1/b}$. At E(NUV+Optical+IR)$sim$1e32 ergs, depending on the adopted fit:
%Eopt/Ex = 5-550, with 4 values out of 6 in the 30-65 range, fully compatible with the above for solar flares. 

How does the correlation we find compare with what is observed for Solar
flares? \citet{woo06a} find that the total irradiance of four bright
solar flares ($\gtrsim 10^{32}$\,ergs) is $\sim$105 times the energy in
the GOES band  (0.1-0.8\,nm), which translates to 50-80 times the soft
X-ray energy in our 0.5-8.0\,keV band, assuming a reasonable range of
average flaring plasma temperatures between 2 and 5\,keV.  Since about
1/2 of the total energy is found to be in the near UV+optical+IR bands,
we infer that, for flares with E$_{opt}\sim10^{32}$\,ergs,
E$_{opt}=(25-40) \times E_{X(0.5-8keV)}$. Excluding the two linear fits
with the most extreme slopes in our Fig.\,\ref{fig:EL_XOpt}, the
E$_{opt}$/E$_X$ values we extrapolate for E$_{opt}\sim10^{32}$\,ergs
are, for the four remaining linear fits, 55, 110, 59, and 45.  We
consider these values compatible with what inferred for the bright solar
flares of \citet{woo06a}, given the significant uncertainties of
both estimates. 

\subsection{Origin of the optical/mIR flares}

A couple of flares in Fig.\,\ref{fig:EL_XOpt} appear to lie below the
general correlations so that their optical and X-ray energies (and peak
luminosities) are similar. This may, again, be consistent with the idea
that the optical emission originates at the feet of the flaring loops,
on the chromosphere or photosphere, while X-rays are emitted by extended
coronal loops. Low optical-emission flares might occur close to the
stellar limb so that, while the X-ray loops are fully in view, the feet
of the loops are either only partly visible, or obscured by a large
amount of intervening material, or the viewing angle reduces the
fraction of the optical emission that reaches the observer, for example
because of a projection effect or limb darkening. Among these optically
faint flares, two, albeit with poor-quality estimates, have mIR
counterparts (ACIS \#\,424 and \#\,747) and both are Class\,III stars.
Since the optical and mIR emission in Class\,III stars are likely to
share the same physical origin (\S\,\ref{sect:mIR_flares}) and the mIR
bands are much less affected by extinction with respect to the {\em
CoRoT} band, these flares allow us to test the extinction hypothesis. Of
the flares from Class\,III sources, they are the ones with highest ratio
between mIR and optical emission, thus providing support for this
picture.

\begin{figure}[!t!]
\centering
\includegraphics[width=9cm]{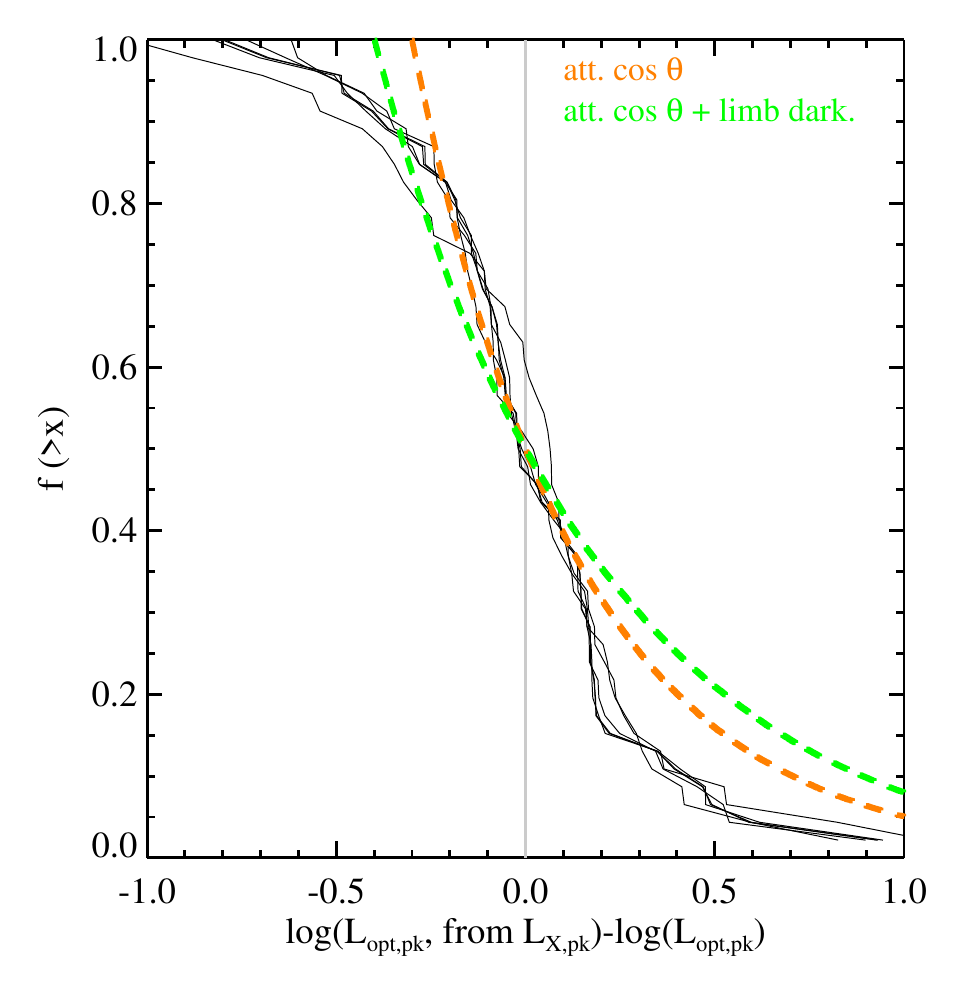}
\caption{Solid black lines: cumulative distributions of residuals in the
linear fits to the $\log L_{opt,pk}$ vs. $\log L_{X,pk}$ scatter plot.
Each line refers to one of the six linear fits shown in
Fig.\,\ref{fig:EL_XOpt}. Dashed orange line: predicted distributions
assuming the optical peak luminosity is perfectly correlated to the
X-ray peak luminosity and that the observed optical flux, from an optically thick slab-like region, is attenuated
by projection effects (see text). Dashed green line: same as above with
an additional attenuation due to limb darkening. The fact that the observed scatter is smaller than predicted by these assumption, even ignoring measurement uncertainties and flare-to-flare variations, indicates that the assumptions on the optical emitting regions are not correct.}
\label{fig:L_XvsOpt_disp}
\end{figure}

Further considerations on the observed scatter in the optical vs. X-ray
relations (Fig.\,\ref{fig:EL_XOpt}) may constrain the nature of the
optically emitting regions. Are they optically thick, as we have
implicitly assumed approximating their spectrum with a black body or a
photospheric emission model? Or are they optically thin? In both cases
we might expect to see a signature in the residuals of optical vs. X-ray
correlations. If the emission from the loop feet is optically thin, we
should expect no effect in the residuals due to the viewing geometry of
the flaring loop: when we see the feet of the loop we should presumably
also see the X-ray emitting loop and both emissions should be
unattenuated.  If, on the other hand, the emission from the feet of the
loop is optically thick, and assuming a slab-like geometry,  we would
expect that the observed emission is attenuated due to the reduced
projected area of the emitting region, by a factor $\cos \theta$,
where $\theta$ is the angle between the line of sight and the normal to
the emitting surface. Moreover, making the rough assumption that the
thermal structure of the emitting region is similar to that of an
unperturbed  stellar photosphere, we would expect a further attenuation
due to limb darkening. Figure\,\ref{fig:L_XvsOpt_disp} shows the
cumulative distribution of residuals from the $L_{opt,pk}$ vs.
$L_{X,pk}$ relation in the right-hand panel of Fig.\,\ref{fig:EL_XOpt}.
Solid black distributions refer to the residuals according to each of
the six linear regressions performed in the log-log plane  (a very
similar plot is obtained for the $E_{opt}$ vs. $E_X$ relation). If we
assume that the angle $\theta$ is uniformly distributed between 0 and
$\pi/2$ we can easily derive the distribution of the expected
attenuations due to both projection effects and limb
darkening\footnote{We have adopted the limb darkening law for the {\em CoRoT}
band derived by \citet{cla11a}, adopting $T_{eff}$=8000\,K, $\log
g=4.0$, Z=0.0, $\xi$=2.0: $I(\theta)/I_0= 0.437+
0.872\cos\theta-0.309(\cos\theta)^2]$}. These are plotted as thick
dashed lines, one for the projection $\cos\theta$ effect only, and the
other also taking into account limb darkening. Both are shifted along
the x-axis so to have zero median value. We see that the observed
scatter around the best fit regression line is actually {\em smaller}
than that predicted by these attenuation models. This is quite striking,
since we can identify several sources of stochastic uncertainties in our
estimates of $L_{opt,pk}$ and $L_{X,pk}$ (or $E_{opt}$ and $E_X$), which
surely contribute significantly to the observed scatter (an accurate
analysis of uncertainties is, however, not straightforward). The
intrinsic scatter in the optical vs. X-ray relations is thus likely much
smaller than we predict assuming optically thick emission and an uniform
spatial distribution of the emitting spots on the stellar disk. We take
this as a suggestion that the flare emission in the {\em CoRoT} band is
optically thin.

%Scatter in the Ex-Eopt correlations. Uncertainties:
%
%- Issue in the definition of non-flaring optical (and X-ray) emission.
%This is probably worse for peak luminosities than for total energies.
%However the definition of the {\em CoRoT} quiescent level is often problematic
%
%- Uncertainties on the optical spectrum, and subsequent uncertainty on
%the {\em CoRoT} flux to blolometric luminosity conversion. Especially relevant
%for hot ~1e4K temperatures. For cooler temperatures the conversion
%factor is more stable. While a different temperature affects the
%optical/X-ray flux/energy ratio, flare-to-flare variability should
%induce a scatter.
%
%- Uncertainty in the absorption correction, also dependent on the
%footpoint spectrum. In this case it seems that the dependence is steeper
%for T<1e4 than for hotter values. (this is all assuming that
%photospheric spectra are adequate, which is far from obvious)

%limb darkening should be multiplied by the geometrical projection factor
%cos(psi) (confirmed by the Landza et al. 2014 paper, for dark spots). Of
%course, whether to use the projection factor depends on whether the
%emission is optically thick or not. The limb darkening correction
%depends on the vertical stratification of temperature, and the above
%formulae only work for the photosphere. I can immagine that the heating
%might be different for the flaring foot-points. One might as well have a
%limb brightening if the photosphere is heated from above. In this case
%the correction would be in the opposite direction w.r.t. the projection
%factor.

\subsection{Optically undetected flares}
\label{sect:xray_only} 

We now discuss the X-ray flares for which no optical/mIR counterpart
could be detected. One obvious physical scenario in which this could
happen is when we observe the extended X-ray emitting loops while the
optically bright foot-points fall behind the stellar limb. The
occurrence rate of such a geometry depends on the hight of the X-ray
emitting loops. We will thus try to relate the statistics of optically
detected/undetected flares with the average extension of the flaring
magnetic loops. Although this is not straightforward for a number of
reasons (e.g., we must try to account for band-dependent sensitivity in
the detection of flares, as well as for false positives and negatives)
we feel that the effort is justified since this could well be one of the
very few available handles on the extension of coronae on PMS stars.

In addition to our main sample of 78 X-ray flares with reasonably
defined optical and/or mIR counterparts, the main focus of this paper,
our automatic detection algorithm for the detection of X-ray flares,
with default parameters, also singles out 97 more X-ray
events\footnote{Three more X-ray flares from two sources were discarded
since the two sources are close-by and basically unresolved stars,
severely hindering the attribution of features in both the X-ray and
optical lightcurves.}. For a fraction of these, a likely optical/mIR
counterpart is actually found but we assessed that the optical and/or
mIR and/or X-ray flare could not be satisfactorily defined, and thus
included in our main sample, for one of the following reasons: $i)$ the
X-ray event is detected at the beginning of the {\em Chandra} observing
segments and the likely optical/mIR counterpart significantly precedes
the beginning of X-ray observation, implying that we are not observing a
significant fraction of the X-ray flare; $ii)$ the X-ray event is at the
very end of the {\em Chandra} observing segments, we see no hint of a
decay phase, and the X-ray event could not be defined; $iii)$ the X-ray
event is contained within the {\em Chandra} observing segment and some
likely associated optical/mIR feature is observed, but cannot be easily
isolated. Appendix\,\ref{app:LC_Xrayonly_det} shows the 24 flares (from
21 stars) that fall into one of the above categories and which, in the
following, we will consider as detected in the optical/mIR band,
alongside the flares in our main sample. For consistency with the
present analysis, however, we will only consider the subset of flares in
our main sample that were detected with our default procedure
(\S\,\ref{sect:detection}): focusing, from now on, on flares with
optical (CoRoT) counterparts, this reduces our main sample to 49 flares,
to which we must add 18 of the 24 flares from the above selection of
events with likely optical counterparts. We thus have a total of 67
X-ray flares with optical counterparts. These must be compared to the
total of 62 detected X-ray events with CoRoT data and no significant
hint of an optical counterpart. Our starting estimate for the fraction
of X-ray flares with no optical counterpart, $f_{X,noOpt}$, is thus
62/(62+67)$=$48.1\%.

As already indicated, assuming that optically undetected X-ray flares are
due to flaring loops with feet behind the stellar limb, $f_{X,noOpt}$
can constrain the average hight of flaring loops, $h_f$, relative to the
stellar radius. Indeed, if $h_f \gg R_\star$ the fraction would approach
1/2. If, on the other hand, $h_f \ll R_\star$ the fraction would be
close to zero. We derive a relation between $f_{X,noOpt}$ and $h_f$
through a simple geometrical zero-order approximation of flaring loops,
i.e. taking them as 1D straight segments, of height $h_f$, extending
radially from the stellar surface. We assume that optical flares
originate from the feet of these segments and are detected anytime these
latter are in view, i.e. not behind the stellar limb. X-ray flares are
instead assumed to be detected whenever any part of the segment is in
view. With these assumptions, and assuming flares are uniformly
distributed on the stellar surfaces, we can estimate for a given value
of $h_f$, the fraction of flares for which we would detect the X-ray
emission but not the optical counterpart, i.e. $f_{X,noOpt}$. We perform
this estimate adopting straightforward Monte Carlo methods for a range
of $h_f$ values, thus deriving the relation between average
$f_{X,noOpt}$ and $h_f$. 

We finally obtain that the observed $f_{X,noOpt}$=62/(62+67),
corresponds to a nominal $h_f = 1.65R_\star$, while the $1\sigma$
uncertainty range, assuming a binomial distribution for the number of
optically undetected X-ray flares is [0.56,$\infty$]$R_\star$,
unconstrained in the upper limit. Two significant issues, however, are
likely to artificially increase our estimate of $f_{X,noOpt}$ (and thus
$h_f$). First, while close to 100\% of the events we identify as X-ray
flares with optical counterparts will indeed be coronal flares (because
of the temporal coincidence in the two bands), some of the X-ray-only
events, might actually not be bona-fide flares. This particularly
applies to faint X-ray events detected at the beginning or at the end of
the {\em Chandra} observing segments and whose duration cannot be
determined. For faint X-ray events contained within one of the observing
segments, the short duration provides some confirmation of the
flare-like nature of the event. Secondly, our X-ray flares with no
optical counterpart appear to be significantly fainter in X-rays than
those with optical counterparts, with median $E_X$ and $L_{X,pk}$ lower
by a factor of 1.9 and 2.5, respectively. The correlations between
optical and  X-ray flare properties imply that, if the optical
counterparts to these X-ray flares were observed, they would be fainter
and some might fall below our detection sensitivity. 

In order to reduce the two aforementioned biases, both leading to an
overestimation of the typical loop hight, we take two measures: $i)$ we
only consider X-ray flares whose peak is fully contained within its {\em
Chandra} observing segments (i.e. for which we observe both the rise and
at least the beginning of the decay phase) and, $ii)$ we consider
subsamples of bright X-ray flares, with $E_X$ and/or $L_{X,pk}$ above
set thresholds. Taking the first measure, our sample is reduced to 45
optically detected and 25 optically undetected flares (lightcurves show
in Appendix \S\,\ref{app:LC_Xrayonly_Oallin} along with those for seven
more similar flares with missing mIR counterpart). With
$f_{X,noOpt}=$35.7\%, our best estimate for $h_f$ is 0.20$R_\star$, with
a 90\% confidence upper limit of 0.51$R_\star$. We then also applied our
second bias-mitigation measure by considering flare subsamples with the
following conditions: (1) $L_{X,pk}>4\times10^{30}$, (2)
$E_X>3\times10^{34}$, (3) $L_{X,pk}>3\times10^{30}$ and
$E_X>3\times10^{34}$ and, (4) $L_{X,pk}>6\times10^{30}$ and
$E_X>4\times10^{34}$. This results in lower estimates for $f_{X,noOpt}$
(19-31\%) and stronger constrains on $h_f$, with best-guess estimates
ranging from 0.03 to 0.11$R_\star$ and 90\% upper confidence intervals
always lower than 0.29$R_\star$ (lower than 0.09$R_\star$ for the most
constraining sub-sample, \#4). We note, however, that in our physical
scenario, the X-ray emitting loops of flares with no optical
counterpart might be partly hidden by the stellar limb, so that the
observable X-ray emission would be reduced. This implies that, in order
to properly determine the optical detection frequency of a complete
sample of flares, we should not apply the same cut on $L_{X,pk}$ (and/or
$E_X$) for optically detected and undetected flares. Using our
simplified physical model and our Monte Carlo simulations, we estimate
that that the average fraction of the X-ray emitting loop that is
visible and contributes to the observed emission depends on $h_f$,
ranging from a minimum of 2/3 for the shortest loops to 1.0 for
infinitely long ones. Making the reasonable assumption that the
(time-averaged) 0.5-8.0\,keV emission from the flaring loop is
distributed quite uniformly along the hight of the loop (cf. Fig.\,8  of
\citealt{rea18}, and associated on-line animation), we thus repeat the
above analysis reducing the thresholds on $L_{X,pk}$ and $E_X$ for the
optically undetected flares to 2/3 those adopted for the optically
detected ones. By thus increasing the number of optically undetected
flares in our sample, our estimate for the average loop length
increases: our most-likely values for $h_f$ range between 0.10 and
0.21$R_\star$ and the 90\% confidence upper estimate remains always
lower than 0.64$R_\star$ ($<$0.37$R_\star$ for subset \#4). 

We conclude that, although uncertainties are large, the average flaring
magnetic loops area rather compact with respect to the stellar
dimensions. This is consistent with the results of \citet{fla05} on the
rotational modulation of coronal emission in the COUP dataset but, of
course, does not preclude the existence of rare extremely long flaring
loops \citep{fav05,rea18}.

\subsection{mIR flare emission and physical scenarios}
\label{sect:mIR_flares}

The most striking result of our investigation is possibly the very large
mIR emission we observe from our flares and, more specifically, from
those occuring in stars with circumstellar disks and envelopes. Stars
with no evidence of circumstellar disks, on the other hand, have
significantly fainter mIR flares: their mIR and optical emission levels
are, moreover, compatible with a single physical origin, most likely
emission at the feet of the flaring loops. If this is the case we
estimate the temperatures of the emitting region to be in the
7000-8000\,K range (assuming black body emission spectra, 6000-7000\,K
in case of photospheric spectra). We have, however, obtained indications
that the emission is optically thin (\S,\,\ref{sect:disc_ene}), making
our assumptions for the optical/IR spectra unlikely to be fully accurate.

It is tempting to assume that the optical to IR flux and energy ratios
that we observe for Class\,III stars are actually representative of the
loop feet for all flares. In this interpretation, the {\em excess} IR
emission observed for flares in Class\,II stars may be attributed to the
heating of circumstellar disks, possibly the inner regions, due to the
illumination from optical and X-ray flare emission. 
This effect may be particularly prominent for the two flares from
Class\,I YSOs, for which heating of the envelope or the different
properties of the disks might explain the large IR excesses.

This scenario is plausible since, i) the dust grains in the inner
disk, largely responsible for the mIR emission, are known to be heated
by the stellar radiation and, ii) the cooling time of these dust grains,
following the absorption of optical or X-ray photons, should be short
when compared to the duration of our flares \citep{boc13}.

\subsection{Temperature of the optically emitting regions}

As discussed in the previous section, the ratio between optical and mIR
quantities should give us, at least for Class\,III stars, an indication
of the temperature of the optically emitting region. As can be read from
Fig.\,\ref{fig:EL_XmIROpt} we almost invariably obtain somewhat lower
temperatures from the ratio of integrated energies, which might be
interpreted as a time-averaged value, than from those of peak
luminosities (with the single exception of a flare for which the optical
values are indirectly obtained from the X-ray data). The difference
between ``average'' temperatures vs. temperatures at the flare peak
might be interpreted as indication that the emitting region cools down
during the decay phase. Also, this appears consistent with what inferred
from the spectral analysis of moderate/large flares on M dwarfs
\citep{kow16a} for which the flux longword of $\lambda$>4000$\AA$
shows two black-body-like components, one at 1.0-1.2$\times 10^4$\,K and
the other at $\sim$5000\,K, with this latter decaying on a longer time
scale with respect to the hot component. In the two-ribbon flare
scenario, the hot black body might originate in newly heated kernels,
while the cooler component might be associated with the previously
heated ribbons.

\subsection{Effect of disks vs. accretion}

We have so far discussed the difference between flares in stars with and without circumstellar disks (Class\,IIs and Class\,IIIs). We now briefly discuss flares from accreting and non-accreting stars, or classical and weak-line T Tauri stars (CTTS and WTTS), as traced by the H$\alpha$ equivalent width (EW). However, while we have mIR classifications for all flaring sources, we have H$\alpha$ data for only a fraction, especially for the more embedded sources. Specifically, we have EW(H$\alpha$) values for 61 flares, 78\%, of our sample, which reduces to 13/20 (65\%) for the flares with good-quality X-ray+mIR lightcurves (20/32, 62.5\%, including lower quality X-ray+mIR flares).

The indications of accretion largely overlap with those of disks: taking
EW(H$\alpha$)=10$\AA$ as the threshold between CTTS and WTTS, the two
classifications ``agree'' for 87\% of our flares (i.e. 19 and 33 flares
from CTTS/Class\,II and WTTS/Class\,III stars, respectively). For only 8
flares, the two classifications differ: 6 flares (from 5 stars) are from
Class\,II WTTSs and 2 flares (from 1 star) are from a Class\,III CTTS.

We have checked that, had we separated our sample in CTTS and WTTS instead of Class\,II and Class\,III sources, the results discussed above would not change. Most of the significances of correlations and of two-population tests would, however, be reduced. For example, the probability that good quality flares from CTTSs and WTTSs have the same distribution of $E_{IR}$/$E_{opt}$ (cf. Fig.\,\ref{fig:EL_XmIROpt}) is 4.8\% vs. 0.13\% for Class\,IIs and Class\,IIIs. It would be tempting to infer that disks, rather than accretion, are responsible for the systematic difference in $E_{IR}/E_{opt}$, but the smaller sample size is probably to blame for most of the reduction in significance.

\section{Summary and Conclusions}

As part of the NGC2264 CSI project, we have observed a significant
sample of young stars in the $\sim$3\,Myr old NGC~2264 star forming
region, obtaining an unprecedented set of simultaneous lightcurves in
the soft X-rays, optical, and mIR bands. We have here focused on the
study of magnetic flares, known to be extremely powerful in PMS stars,
with the goals of gaining insights on the physics of these
strong events and to assess their impact on the evolution of
circumstellar disks and protoplanets. We have here conducted a
statistical investigation, mainly constraining and correlating the
energetics of a sizable sample of flares in the three bands. A detailed
analysis of individual events is left for a later study. We are able to
draw a number of novel conclusions, among which:

- A clear correlation between the soft X-ray and optical emission is
observed. The correlation is such that the ratio between emitted
energies in the soft X-ray band and in the optical bands range between
$\sim$1/10 to $\sim$1/4 for $E_X$ between $10^{34}$ and $10^{36}$\,ergs.
These ratios are significantly lower than what inferred for the most
powerful solar flares, with bolometric energies $\sim$3\,dex lower than
those of our least powerful flares. The slope of our correlation,
however, is roughly consistent with these solar events, pointing toward
a common physical mechanism.

- The durations of flares in the three bands are generally consistent
with the accepted picture for solar-like flares in that X-ray flares are
almost invariably longer than optical ones, perhaps indicating that the
X-ray emission from the cooling coronal loops always follows the heating
and subsequent evaporation of the plasma into these loops, as traced by
the optical flares. In many cases, however, the X-ray and optical
durations are comparable, indicating either compact loops with short
cooling times, or prolonged heating. mIR and optical flares, on the
other hand, have comparable durations, suggesting that the two have the
same physical origin, possibly the feet of the loops, or (see
below) that the mIR emission is due to reprocessing of the optical one. 

- The mIR flares on stars with disks (and circumstellar envelopes) are
significantly more intense with respect to their optical counterparts
than they are on stars without disks. At least two possible
interpretations are possible: $i)$ both optical and mIR emission come
from the  feet of the flaring loops and the spectrum of the emission is
much redder for stars with disks, indicating cooler emitting regions;
$ii)$ the feet of flaring loop actually have the same spectra and we are
observing mIR excesses due to the response of the inner disks to the
optical and X-ray flares. In this latter hypothesis, which we tend to
favor, the optical emission from the loop feet could come from a region
at 7-8$\times10^3$\,K, as inferred from the peak luminosities of flares
on diskless stars, while most of the observed mIR flux might be
reprocessed emission by the circumstellar material. Interestingly the
mIR excesses of flares from Class\,I stars, with both circumstellar
disks and envelopes, are among the strongest.

%	\citet{kow16a} Introduction: optical SEDs of (moderate/large)
%	flares on M dwarfs consistent with BB@90000K (or greater),
%	implying heating at high densities. However, SEDs using
%	broad-band filters are ambiguous as an optically thin emission
%	(hydrogen recombination continuum+ emission lines) produce a
%	similar SED. Spectra show several continuum components: Balmer
%	continuum and lines at l<4000AA, for l>4000AA (i.e. MOST of the
%	{\em CoRoT} band, but there is a tail down to 3000AA...) the flux is
%	dominated by a 1.0-1.2e4K BB-like component and a 0.5e4
%	component for l>5000AA. The 1e4K components evolves most
%	rapidly, apparently cooling to 0.8e4K, while the cooler BB (and
%	Balmer) continuum decays on longer time scales (the Balmer lines
%	are even slower). In the two-ribbon flare scenario, the hot BB
%	might originate in newly heated kernels, while the cooler BB
%	(and Balmer) continuum in the gradual phase in spatially
%	extended, previously heated ribbons. [This might explain why
%	many flares in the {\em CoRoT} band, mostly at 5000K, are long. Also
%	the two BBs can explain the 7-8e3K temperatures I derive for the
%	peak fluxes, AND the slighly cooler average temperatures, down
%	to 5e3K, that I get from energies]. The refined analysis in this
%	work confirms the two BB components, although not all flares
%	produce an hot component. 

A more involved line of reasoning also allows us to speculate on the
physical nature of the optical source in flares, based on the
surprisingly small dispersion in the relation between optical and X-ray
emitted energies and peak luminosities ($E_{opt}$ vs. $E_X$ and
$L_{opt,pk}$ vs. $E_{X,pk}$). Although a rigorous analysis of
uncertainties on the two pairs of quantities is not straightforward,
given the numerous approximations and assumptions made in the process,
it is reasonable to assume that a large fraction of the observed scatter
may be attributed to uncertainties. This leaves little room for physical
flare-to-flare variability. In particular, assuming as reasonable that
the X-ray emission comes from optically thin plasma, we can exclude that
the observed optical flares are strongly affected by the location of the
flare on the stellar surface with respect to the observer. This probably
indicates that the optical source is not too deeply set in the stellar
atmosphere as to be strongly obscured when viewed close to the stellar
limb, and that it is probably optical thin, so not to be subject to
projection or limb-darkening effects. In principle, other scenarios may
also be possible, however, such as a spherically symmetric optically
thick optical source located high up in the atmosphere. 

Finally, we constrain the typical hight of coronal flaring loops from
the frequency of detected X-ray flares with no optical counterparts.
Under the hypothesis that these events are produced by X-ray bright
loops whose optically emitting feet are hidden behind the stellar limb,
we estimate that the loops most likely extend up to a small fraction of
the stellar radius.

\begin{acknowledgements}
We than	Robert A. Stern for careful reading of the manuscript and insightful comments that helped improve this work, and Fabio Reale and Cesare Cecchi-Pestellini for useful discussions on the properties of flaring loops and on the effects of flares on circumstellar dusts. E. F., S.S., G.M., and M.G.G. acknowledge financial
support from PRIN-INAF 2012, as well as, in a modest measure, from the
ASI-INAF agreement n.2017-14.H.O. 
\end{acknowledgements}

\bibliographystyle{aa} %aa.bst\
\bibliography{bibtex.bib}

\appendix

\section{{\em CoRoT} fluxes}
\label{app:corot_fluxes}

The spectral response of the {\em CoRoT} Planer Finder channel \citep{auv09a}
spans a broad wavelength range, from $\sim 300$ to $11000\mu m$. The
conversion from instrumental flux to physical units (erg s$^{-1}$
cm$^{-2}$) thus depends significantly on the incident source spectrum,
which, in turn, depends on the intrinsic spectrum and the intervening
extinction. We have derived conversion factors from instrumental fluxes
to absorption-corrected source fluxes, adopting the spectral response
shown in Fig.\,14 of \citep{auv09a}, the extinction law of
\citet{wei03a} (R=3.1), and model source spectra. These latter were
either Black Bodies with varying temperatures, or the ATLAS9 stellar
atmospheric models provided by \citet{kur93a} as a function of effective
temperature, surface gravity, and metallicity. For the flares from our
PMS stars, we considered models computed for Solar abundances and four
values of $\log g$, between 3.0 and 4.5.

\begin{figure}[!t!]
\centering
\includegraphics[width=8.0cm]{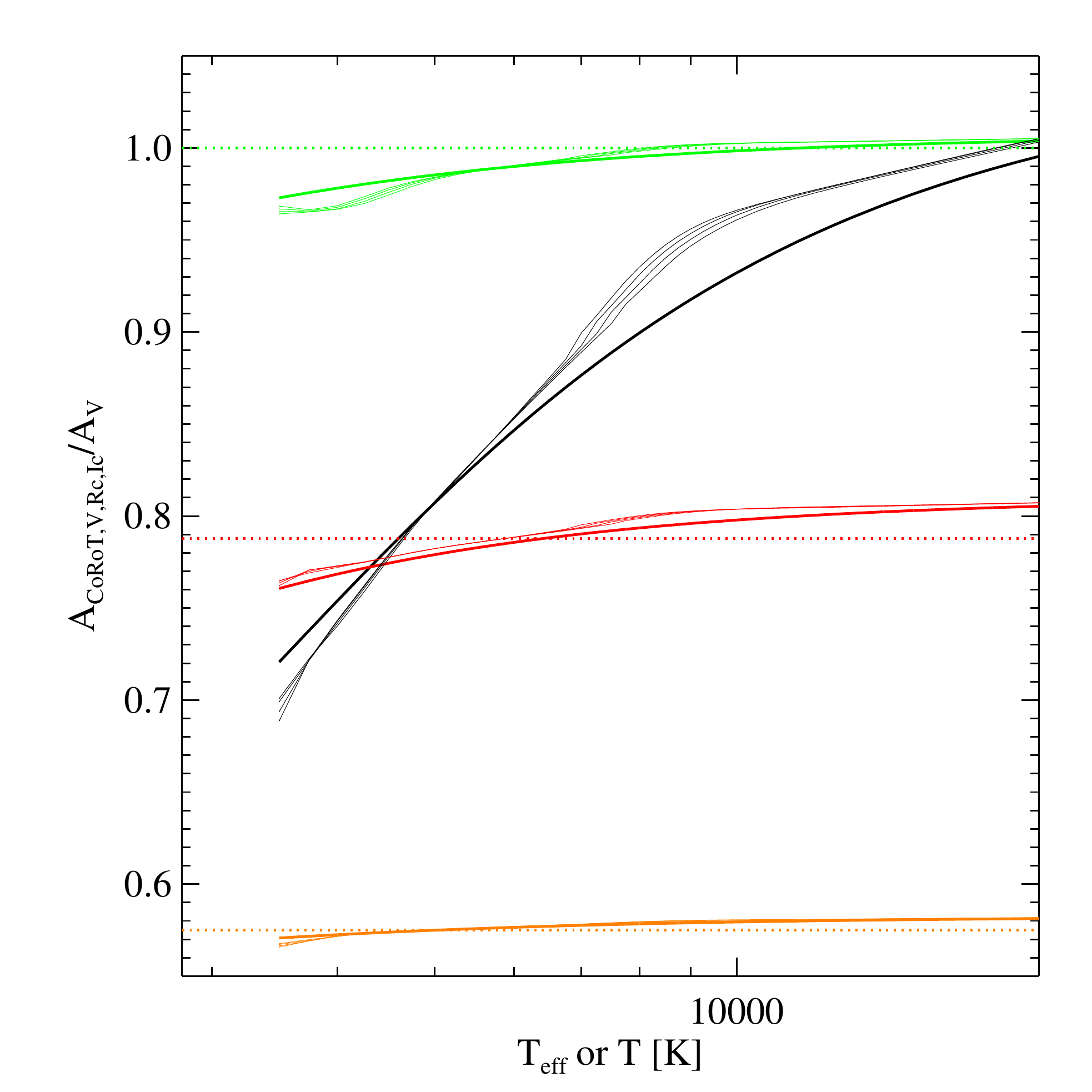}
\caption{Extinction laws for the {\em CoRoT} band (in black) and for the  V, R$_c$,
and I$_c$ optical filters (green, red, and orange, respectively). The thin lines show the variation of A$_{(band)}$/A$_V$ with effective temperature for the stellar-like spectra with $\log g$ between 3.0 and 4.5. Thicker lines show the trend of A$_{(band)}$/A$_V$ with temperature for black body spectra.}
\label{fig:AcorotAv}
\end{figure}

We started by deriving the extinction law for the {\em CoRoT} band. This is a
function of the spectral parameters ($T_{eff}$ and $g$ for the stellar
spectra and $T$ for the black bodies), which we will generically
indicate with $p$ in the following. 

\begin{equation}
\label{eq:Acorot_Av}
\frac{A_{\rm CoRoT}(p)}{A_V}=-2.5\log
	\frac{\int_0^{\infty}F_\star(\lambda,p)10^{-0.4A(\lambda)/A_V}Q_{\rm CoRoT}(\lambda)\cdot d\lambda}
	     {\int_0^{\infty}F_\star(\lambda,p)Q_{\rm CoRoT}(\lambda)\cdot d\lambda}
\end{equation}

where $F_\star(\lambda,p)$ is the intrinsic source spectrum,
$A(\lambda)/A_V$ is the extinction law of \citet{wei03a} assuming
$R=A_V/E(B-V)=3.1$, and $Q_{\rm CoRoT}(\lambda)$ is the normilized
quantum efficiency of the {\em CoRoT} Planer Finder camera.
Figure\,\ref{fig:AcorotAv} shows the thus derived extinction law for the
{\em CoRoT} band as a function of spectral parameters, alongside those derived
in the same way for the V, R$_c$, and I$_c$ optical filters. For
comparison, the dotted horizontal lines show spectrum-independent
approximations, as most commonly adopted, derived from the \citet{mat90}
extinction law. Note the small dependence of the extinction laws on
gravity for stellar-like spectra, and, at least for the {\em CoRoT} extinction
law, the non-negligible difference between stellar-like spectra and
black bodies.

%in the same way, we also obtained  the  extinction law also for the optical and mIR bands whisenuthat we will use (V,R,I,IRAC1, andIRAC2), as function of the source spectrum. We did so both for consistency and to take took into account, also for the standard photometric bands, the small and usually neglected dependence on Teff and g (this latter only for stellar models). 

Next, we estimated $k_{\rm CoRoT}$, the conversion
factor between {\em CoRoT} incident flux, in erg/s/cm$^{-2}$, and the flux in
instrumental units, $F_{\rm CoRoT}^{obs}$:

\begin{equation}
F_{\rm CoRoT}^{obs}(p)=k_{\rm CoRoT} \int_0^{\infty} F_\star(\lambda,p)10^{-0.4A(\lambda)/A_V\cdot A_V}Q_{\rm CoRoT}(\lambda)\cdot d\lambda
\end{equation}

where, $A_V$ is the extinction suffered by the source. We estimate
$k_{\rm CoRoT}$ by comparing, for a suitable sample of stars with known
spectral types and extinction, the observed flux in instrumental units
with that predicted for our {\em CoRoT} sources on the basis of the known
stellar spectra (from spectral types), extinction ($A_V$), and
photometry (in the R-band). More specifically, we take
$F_\star(\lambda,p)$ as the \citet{kur93a} model with the $T_{eff}$
indicated by the spectral type, $A_V$ as derived from the spectral type
and observed optical colors, and we set the normalization of the model
spectrum so to reproduce the flux measured in the R-band. For each star
in the sample we thus estimate one value of $k_{\rm CoRoT}$:

\begin{equation}
\label{eq:k_corot}
k_{\rm CoRoT}=\frac{F_{\rm CoRoT}^{obs}(p)}{F_{\rm R}^{obs}}
	\frac{\int_0^{\infty}F_\star(\lambda,p)10^{-0.4A(\lambda)/A_V\cdot A_V}Q_{\rm R}(\lambda)\cdot d\lambda} {\int_0^{\infty}F_\star(\lambda,p)10^{-0.4A(\lambda)/A_V\cdot A_V}Q_{\rm CoRoT}(\lambda)\cdot d\lambda}
\end{equation}

%\begin{equation}
%\begin{split}
%k_{\rm CoRoT}=
%\frac{F_{\rm CoRoT}^{obs}(p) }{ \int_0^{\infty}F_\star(\lambda,p)10^{-0.4A(\lambda)/A_V\cdot A_V}Q_{\rm CoRoT}(\lambda)\cdot d\lambda} \times \\
%\frac{\int_0^{\infty}F_\star(\lambda,p)10^{-0.4A(\lambda)/A_V\cdot A_V}Q_{\rm R}(\lambda)\cdot d\lambda}{F_{\rm R}^{obs}}
%\end{split}
%\end{equation}

where $Q_{\rm R}(\lambda)$ is the R-band filter response function
\citep[from][]{bes90a} and $F_{\rm R}^{obs}$ is the measured flux in the
R-band, obtained from the R magnitudes as:

%\begin{equation}
%F_{\rm R}(p)= 10^{-0.4({\rm R}-ZP_{\rm R})}=10^{-0.4(R-0.03)}\int_0^{\infty}F_{Vega}(\lambda)Q_R(\lambda)\cdot d\lambda
%\end{equation}

\begin{equation}
F_{\rm R}^{obs}=10^{-0.4(R-0.03)}\int_0^{\infty}F_{Vega}(\lambda)Q_R(\lambda)\cdot d\lambda
\end{equation}

\begin{figure}[!t!]
\centering
\includegraphics[width=8.0cm]{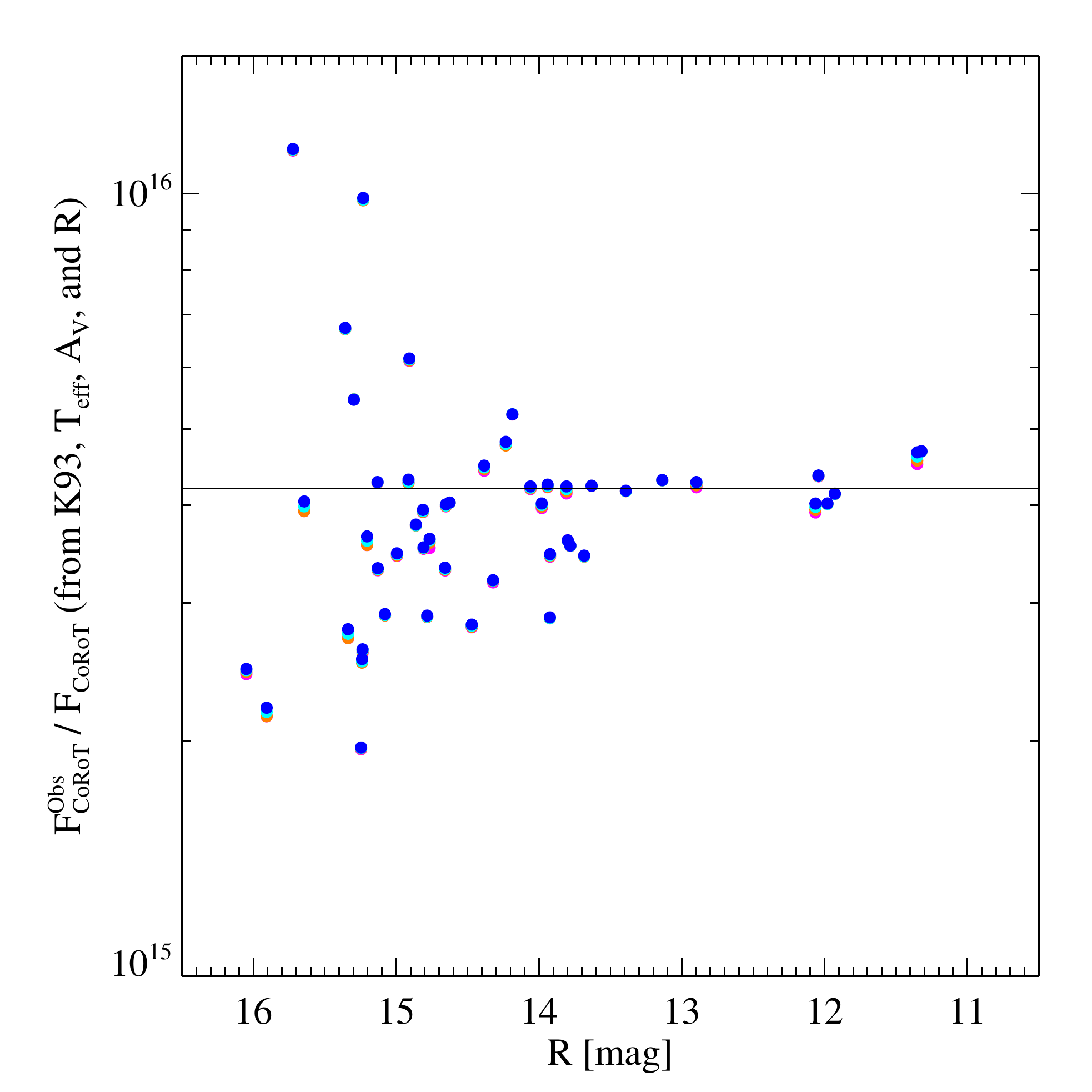}
\caption{Ratio between observed {\em CoRoT} flux, in instrumental units, and the flux predicted from model spectra vs. observed R magnitudes. The plotted points refer to a subsample of suitable and well characterized NGC~2264 members (see text), and symbols of different colors, largely overlapping with each other, refer to estimates obtained assuming four different values of $\log g$ between 3.0 and 4.5. The horizontal line refers to the adopted value, 4.2$\cdot 10^{15}$ erg$^{-1}$s\,cm$^{-2}$}
\label{fig:CoRoT_ZP}
\end{figure}

$F_{Vega}(\lambda)$ is here the flux-calibrated model spectrum of Vega
provided by \citet{kur93a}, whose R magnitude is assumed to be 0.03. If
the model spectra, relative parameters, and extinction values were
perfectly known, $k_{\rm CoRoT}$ would be the same for all stars in the
sample. In order to reduce uncertainties we select a sample of NGC~2264
members observed by {\em CoRoT}, with well identified counterpart in the
R-band, and with no evidence of disks or accretion (in order to avoid
accretion/disk-induced spectral excesses). Figure\,\ref{fig:CoRoT_ZP}
shows, as a function of the stellar R magnitude, $k_{\rm CoRoT}$ as
estimated for the above sample from equation \ref{eq:k_corot}. Given the
uncertainties on the gravity of our stars, we plot, with different
colors, values of $k_{\rm CoRoT}$ obtained assuming four different
values of $\log g$ between 3.0 and 4.5, showing that the effect of
surface gravity on our estimates is negligible.

A significant scatter can be noticed at R$\gtrsim$14, up to a factor of
$\sim$2 at the faint end. This can probably be attributed to already
identified issues with background subtraction of the {\em CoRoT} photometry
\citep[see e.g.][]{cod14a}. Indeed the brighter stars all have low
estimated background contributions, and therefore presumably small
errors on the background correction, explaining the small scatter in the
$k_{\rm CoRoT}$ values. We take $k_{\rm CoRoT}$ as the median of values
for R<14: 4.2$\cdot 10^{15}$ erg$^{-1}$s\,cm$^{-2}$. We note that
uncertain background corrections are irrelevant for the derivation of
flare fluxes, the focus of this paper. 

Finally, we can derive the relation between observed {\em CoRoT} flux and
bolometric flux for a given source spectrum and absorption:

\begin{equation}
\label{eq:CF}
\frac{F_{bol}(p,A_{\rm CoRoT})}{F_{\rm CoRoT}^{obs}}=\frac{1}{k_{\rm CoRoT}}
\frac{\int_0^{\infty}F_\star(\lambda,p)\cdot d\lambda} {\int_0^{\infty}F_\star(\lambda,p)Q_{\rm CoRoT}(\lambda)\cdot d\lambda}
10^{-0.4A_{\rm CoRoT}(p)}
\end{equation}

\begin{figure}[!t!]
\centering
\includegraphics[width=8.0cm]{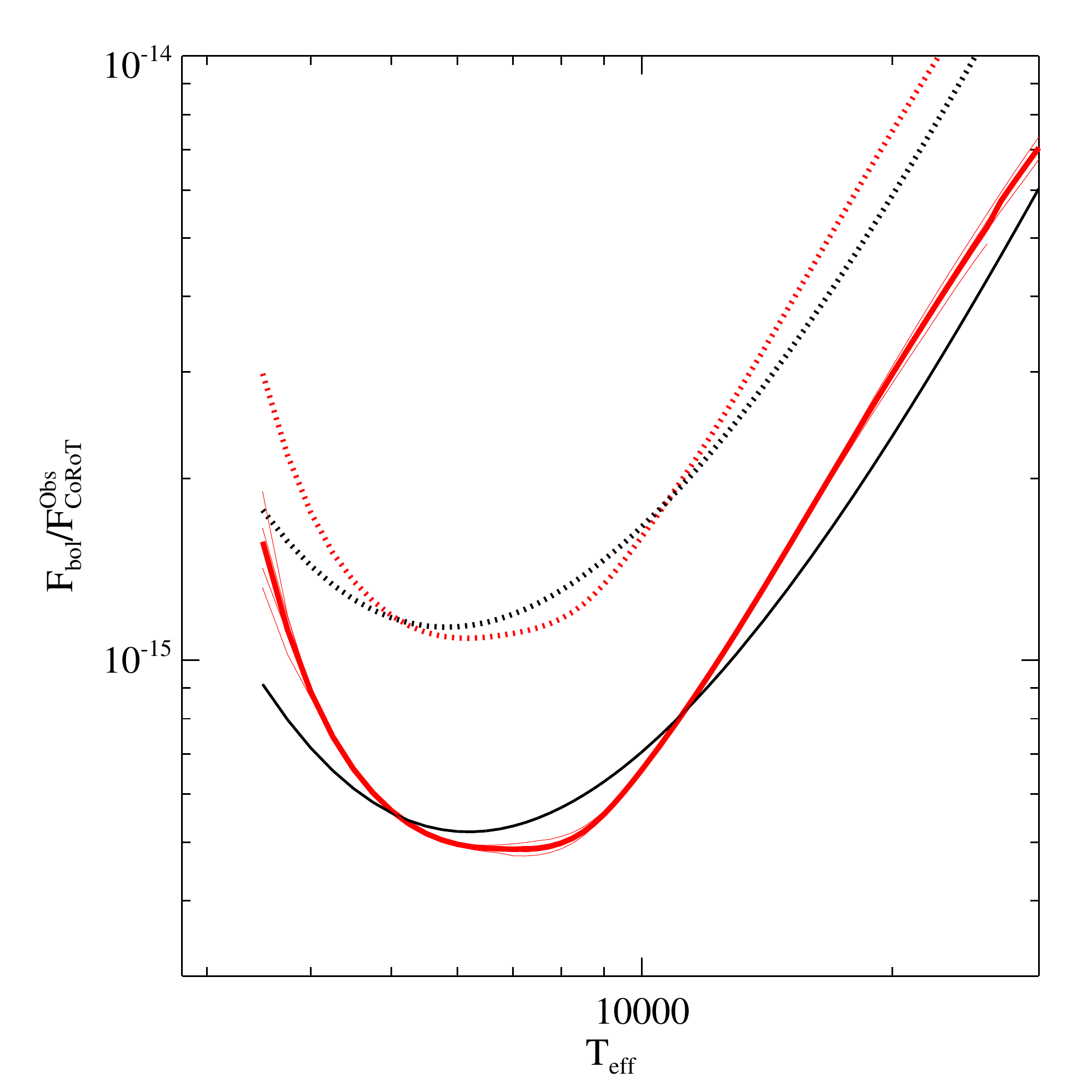}
\caption{Conversion factor between observed {\em CoRoT} flux and bolometric
flux as a function of spectral parameters/shape and extinction. The
$F_{bol}/F_{\rm CoRoT}^{obs}$ ratio is plotted as a function of
temperature for different source spectra. All solid lines refer to
unabsorbed source spectra. The four thin red lines refer the stellar
case with four values of $\log g$ between 3.0 and 4.5. The thicker red
line shows the mean of the four values at each $T_{eff}$. The black line
refers instead to black body emission spectra. The dotted red and black
lines show the $F_{bol}/F_{\rm CoRoT}^{obs}$ ratio for the stellar and
black-body cases, respectively, for an $A_V$=1.0 mag interstellar
extinction.}
\label{fig:Fbol_Fcorot}
\end{figure}

Where $A_{\rm CoRoT}$ can be derived from $A_V$ and
Eq.\,\ref{eq:Acorot_Av}. Figure\,\ref{fig:Fbol_Fcorot} shows the ratio
in Eq.\,\ref{eq:CF} as a function of spectral parameters, $T_{eff}$ and
$\log g$ for stellar spectra, or temperature for black body spectra, and
of interstellar extinction, $A_V$. All solid lines refer to the case of
$A_V$=0.0. The four thin red lines refer to the stellar case with the
four values of $\log g$ that we have explored. The thicker red line
shows the mean of the four values at each $T_{eff}$, and is the curve we
have adopted throughout this paper when converting observed flare fluxes to
bolometric fluxes, when adopting stellar-like spectra. The thick black line
refers to black body emission spectra. Finally, the dotted red and black
lines show the $F_{bol}/F_{\rm CoRoT}^{obs}$ ratio for the stellar and
black-body cases, respectively, for an $A_V$=1.0 mag interstellar
extinction.

\section{Lightcurves of X-ray flares with CoRoT and/or {\em Spitzer} counterparts}
\label{app:lightcurves}

\begin{figure*}[!t!]
\centering
\includegraphics[width=6.0cm]{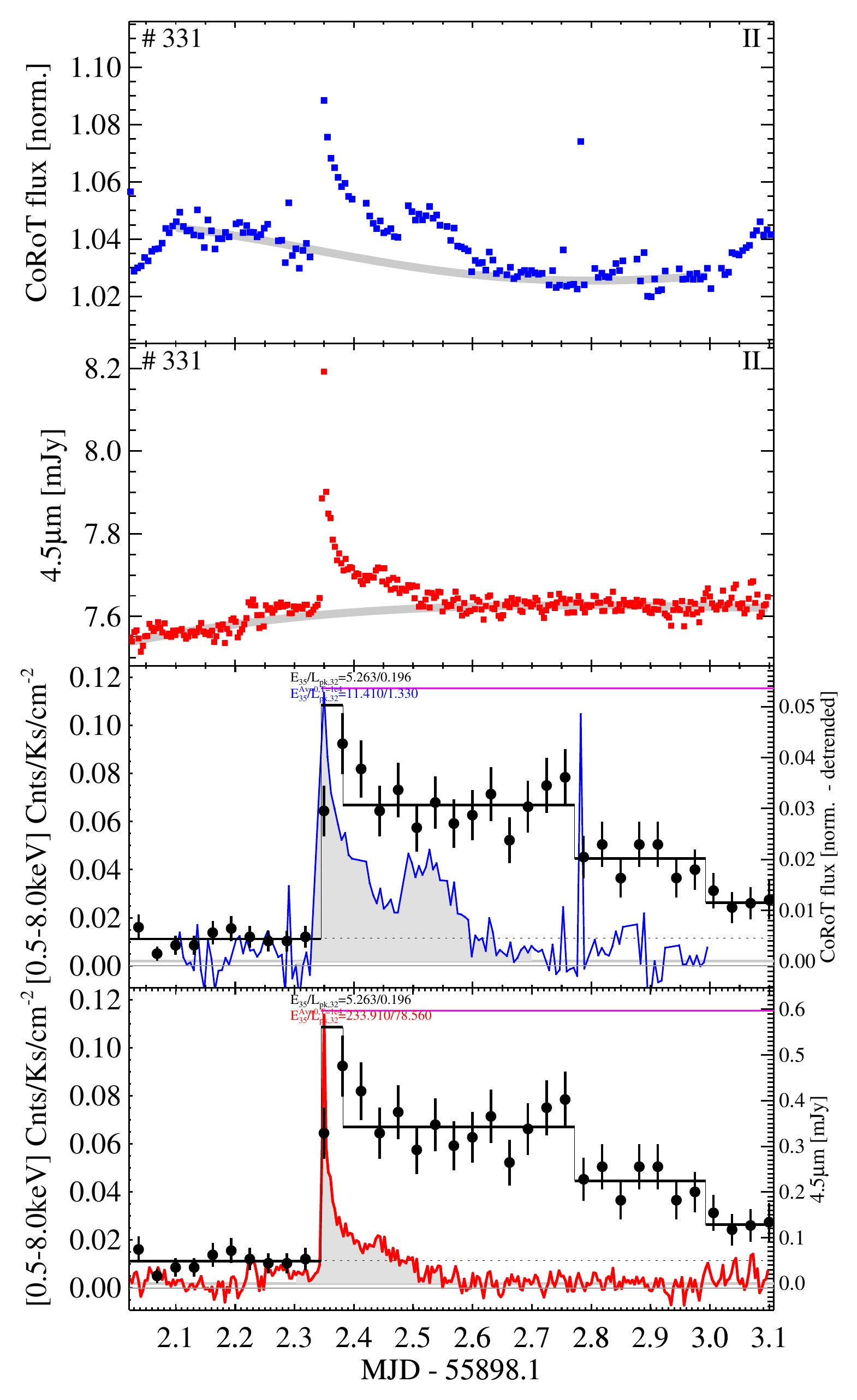}
\includegraphics[width=6.0cm]{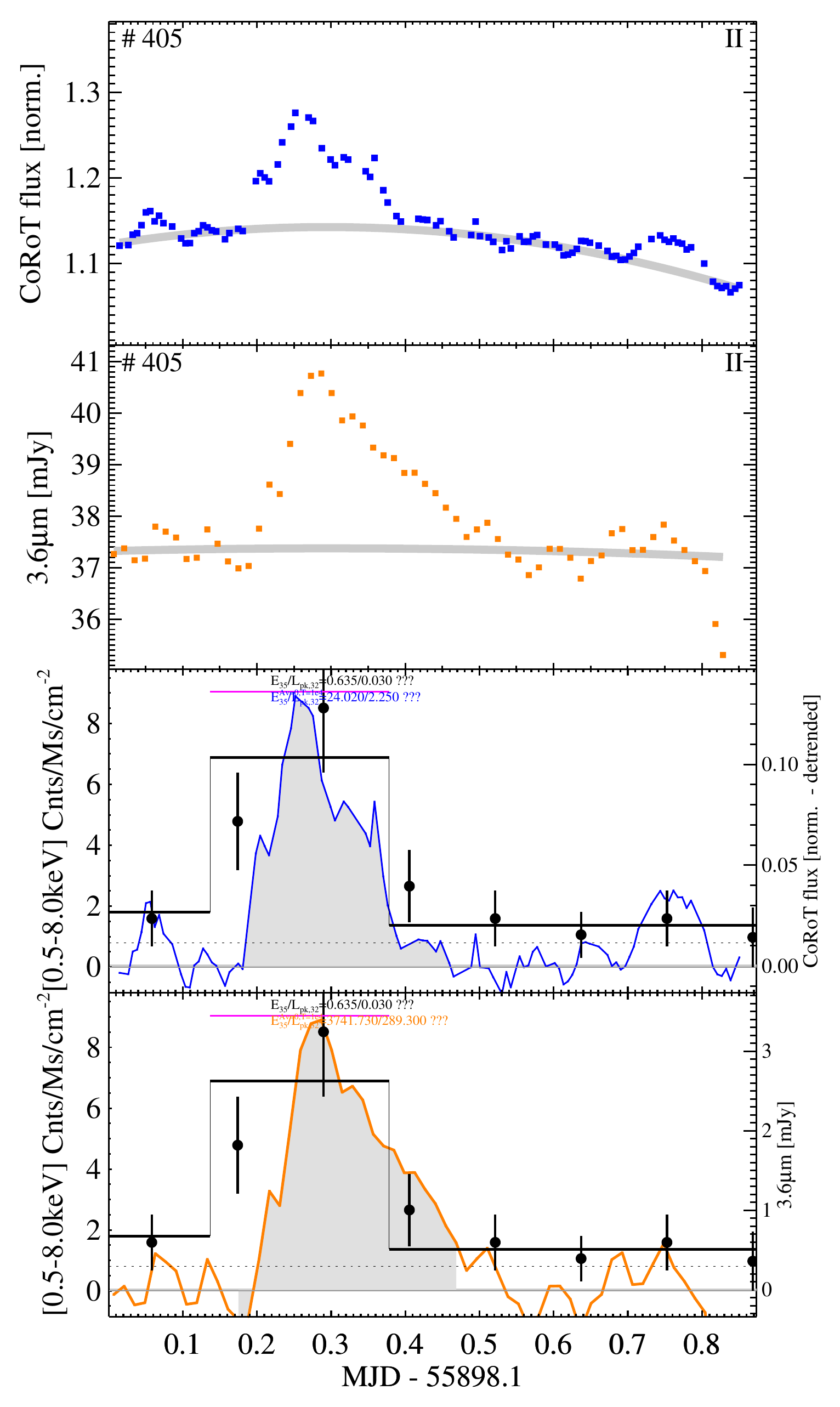}
\includegraphics[width=6.0cm]{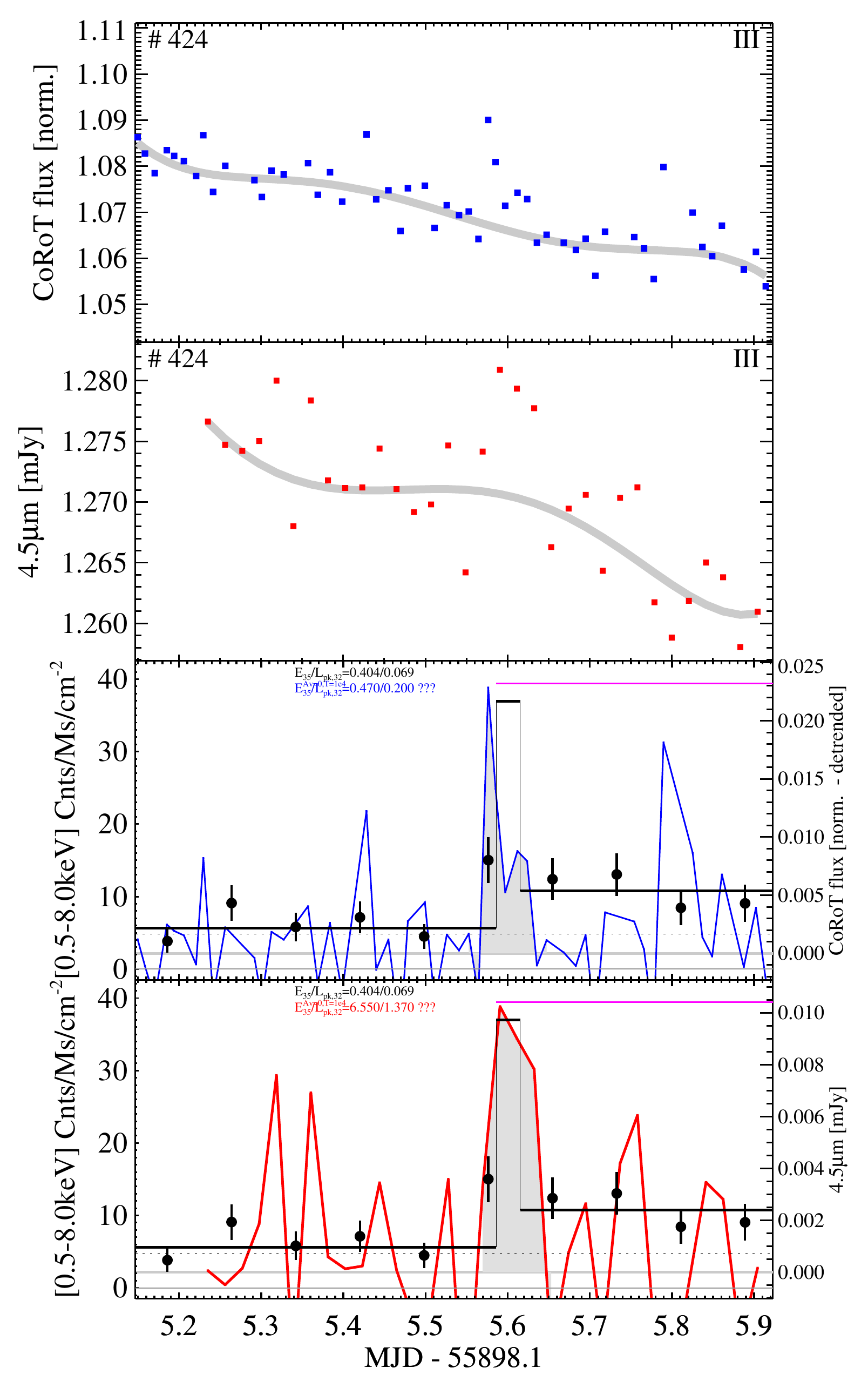}
\includegraphics[width=6.0cm]{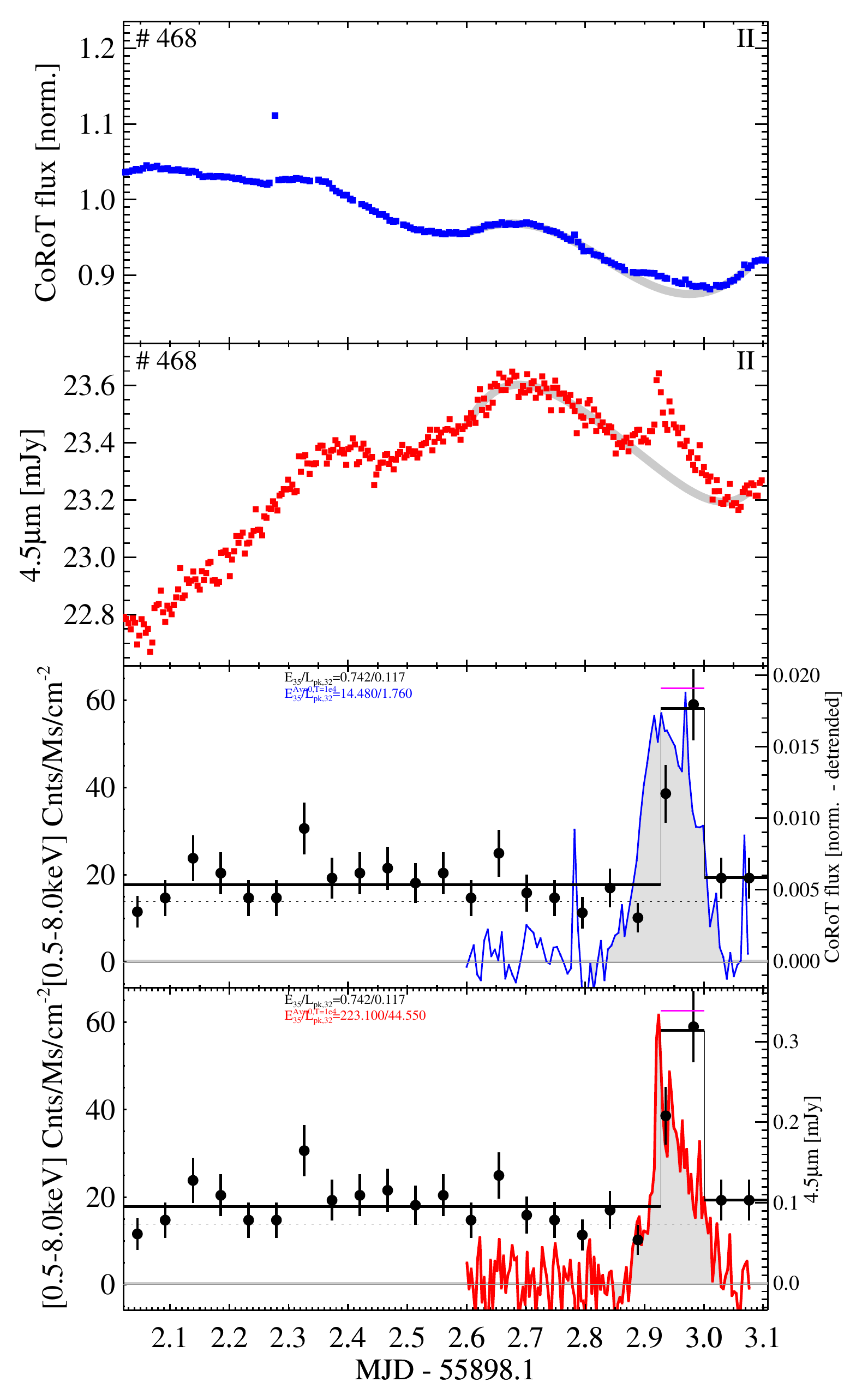}
\includegraphics[width=6.0cm]{lc_s565.pdf}
\includegraphics[width=6.0cm]{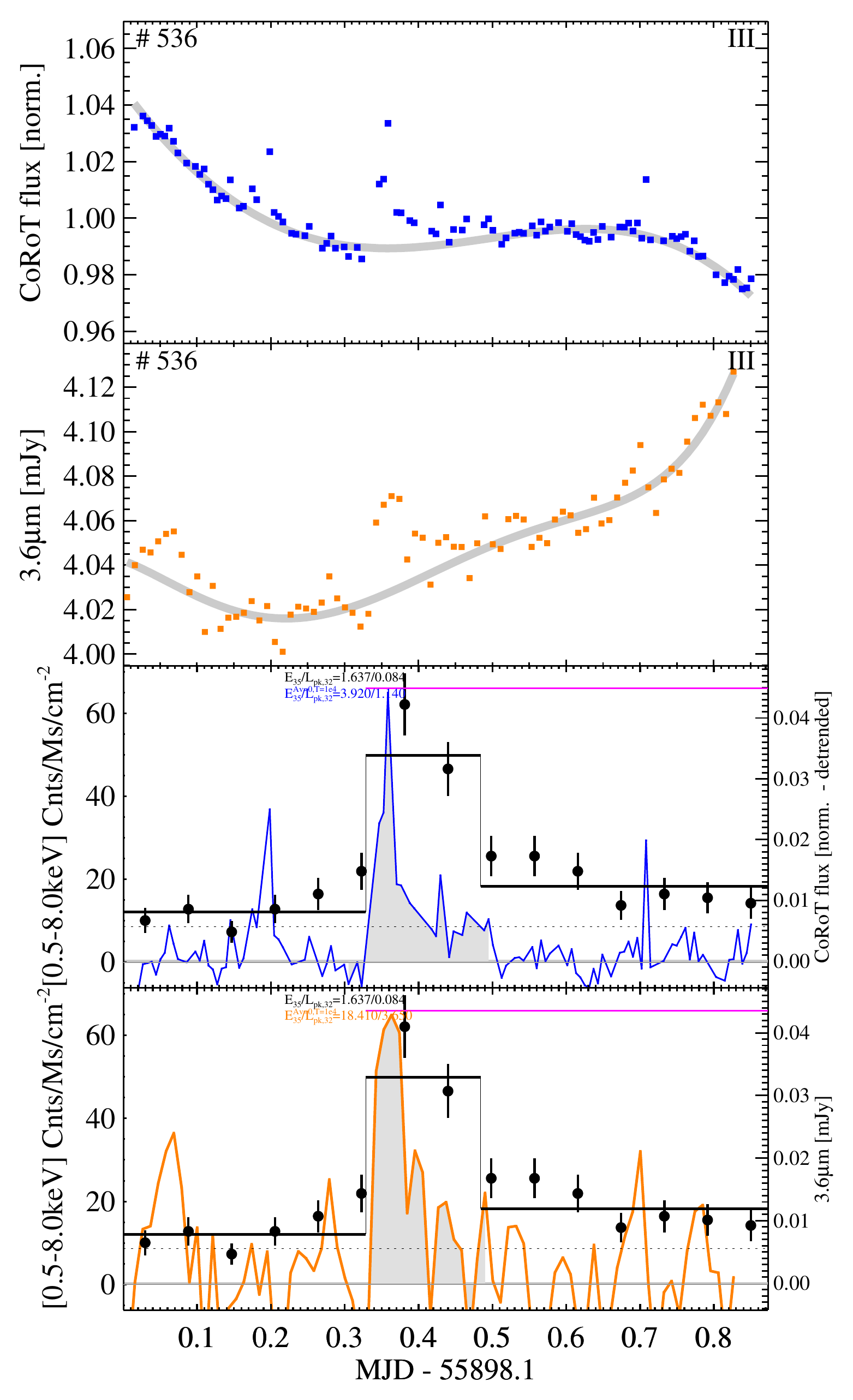}
\caption{Lightcurves of all X-ray flares with counterparts in the optical and/or mIR bands. See caption of Fig.\,\ref{fig:lc_examples} for a full description of the content of each panel.}
\label{fig:}
\end{figure*}

\addtocounter{figure}{-1}

\begin{figure*}[!t!]
\centering
\includegraphics[width=6.0cm]{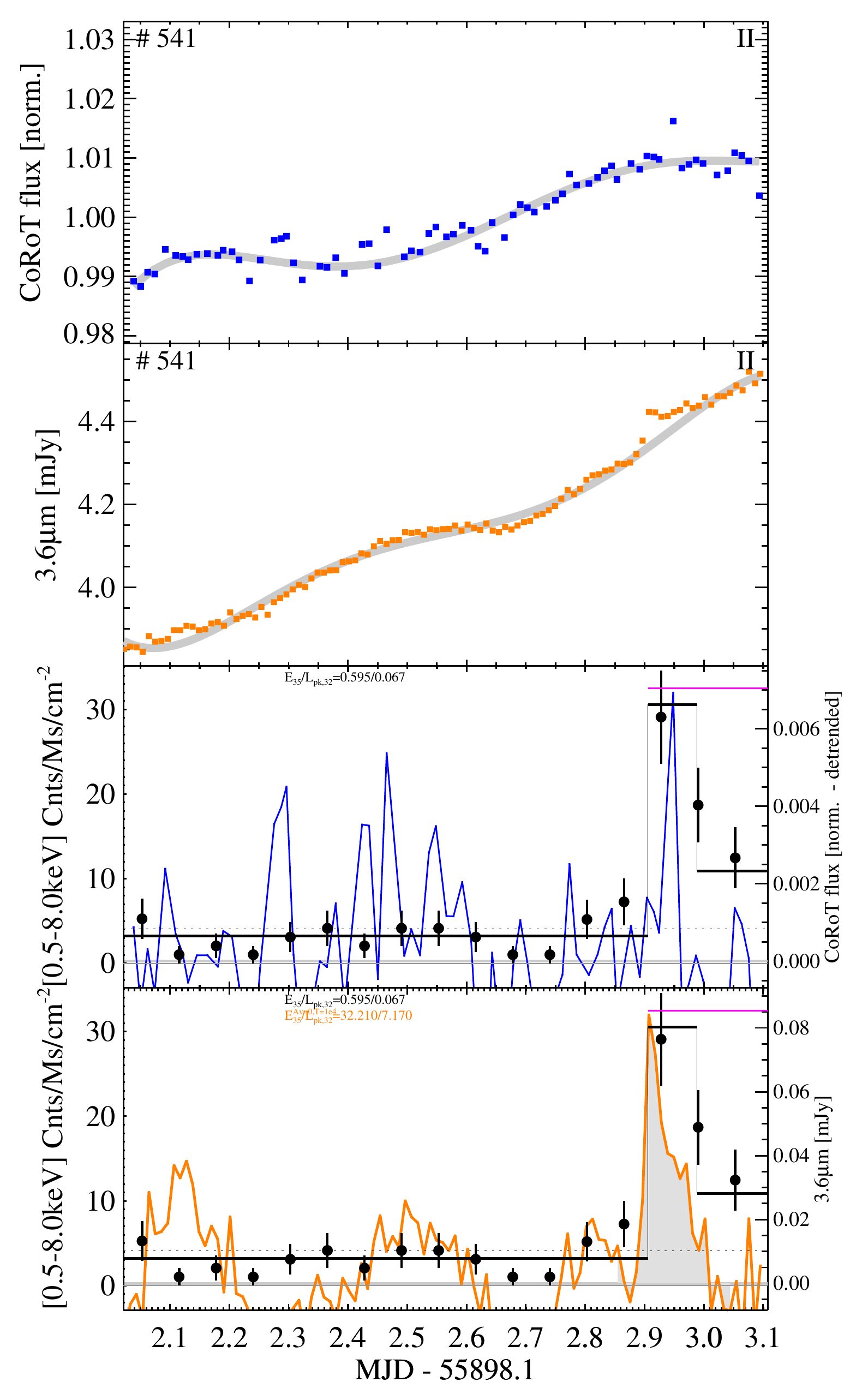}
\includegraphics[width=6.0cm]{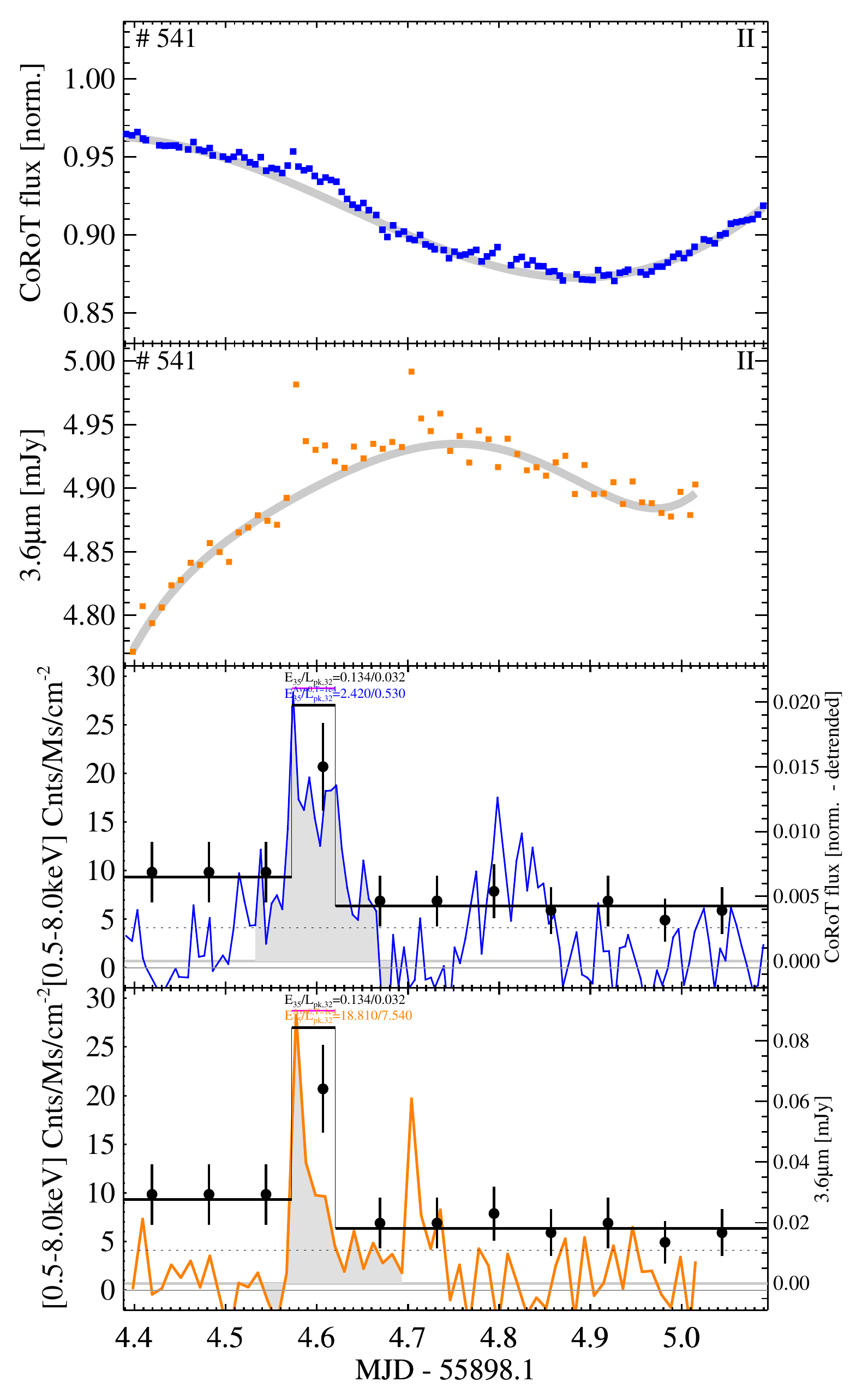}
\includegraphics[width=6.0cm]{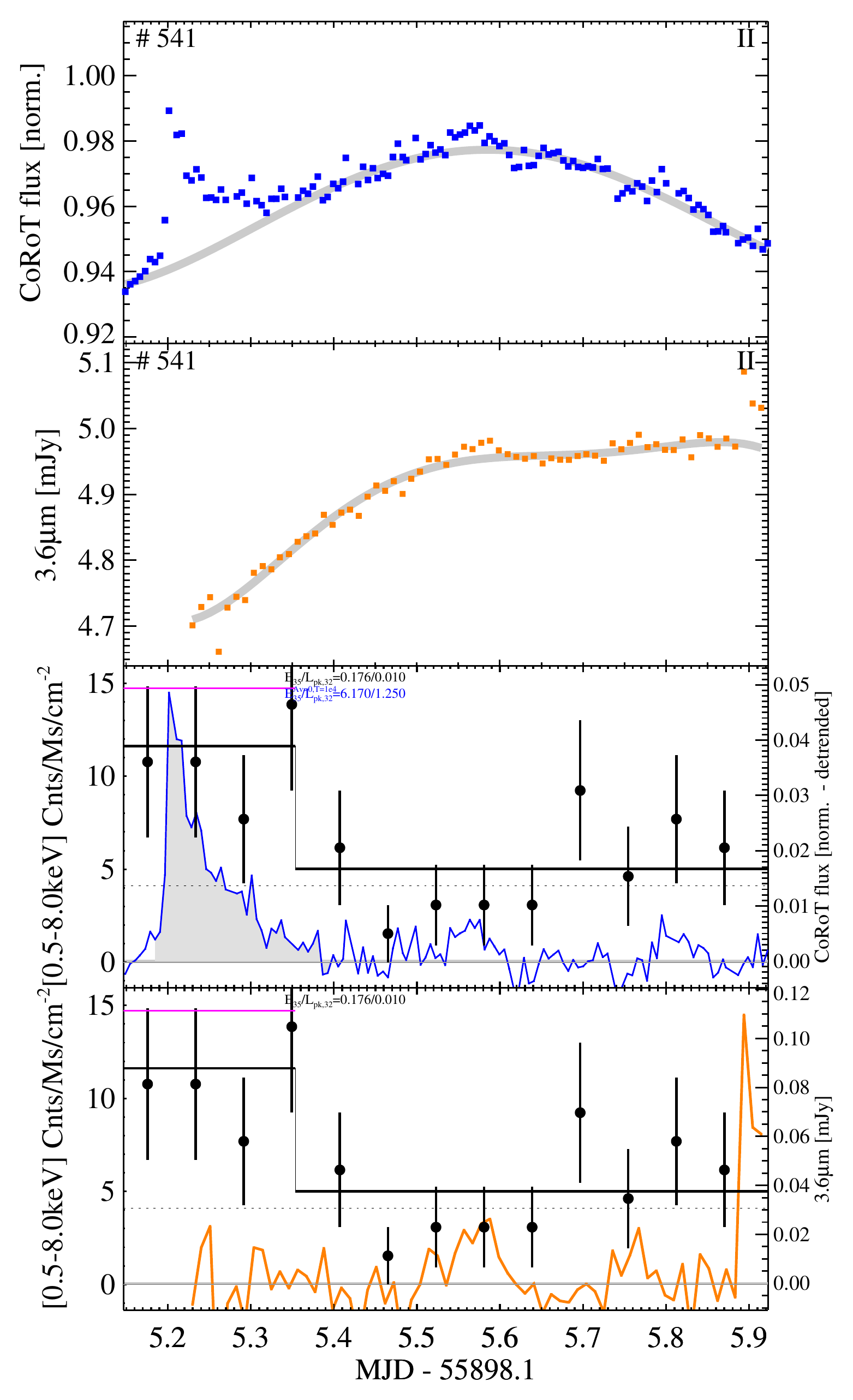}
\includegraphics[width=6.0cm]{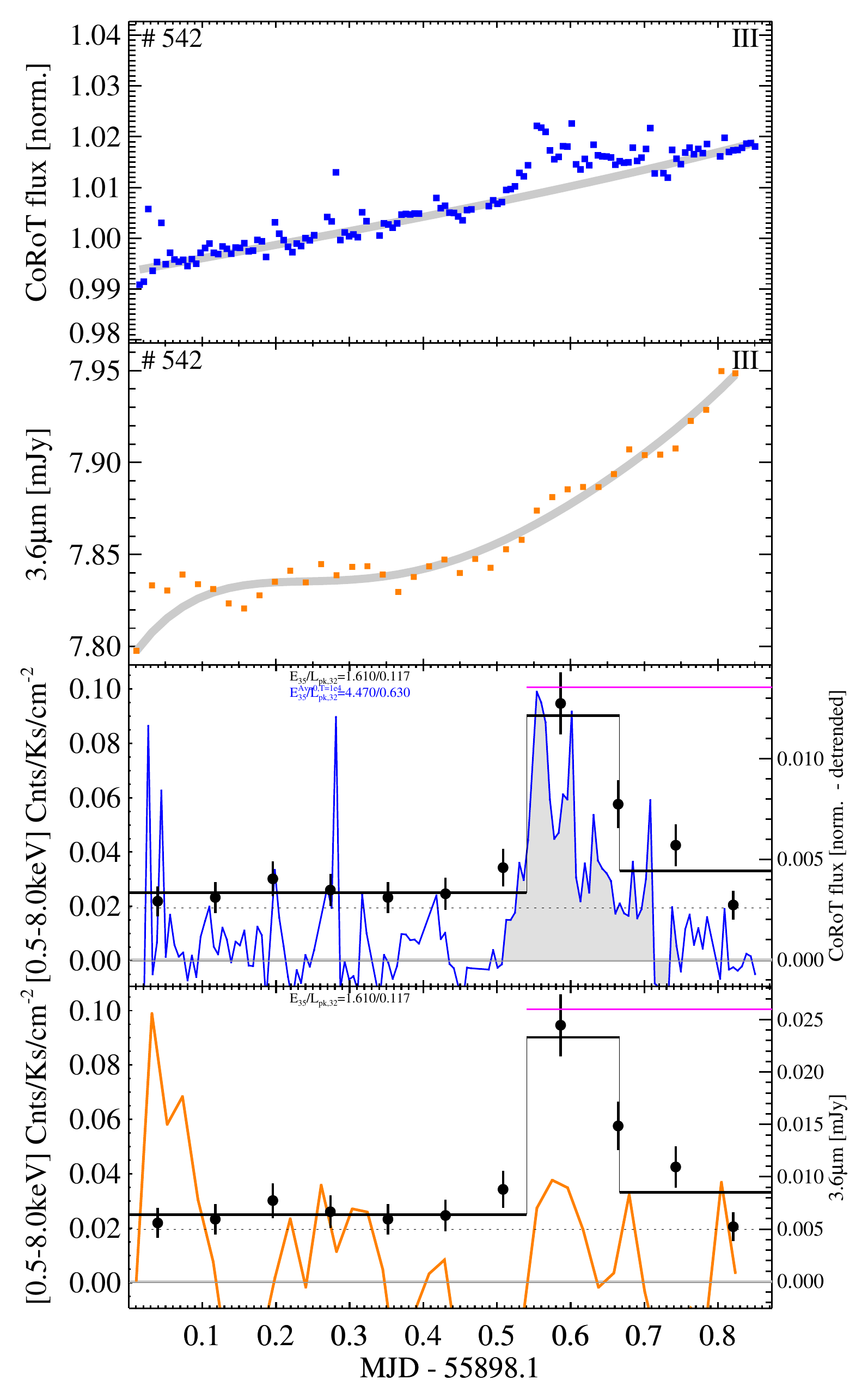}
\includegraphics[width=6.0cm]{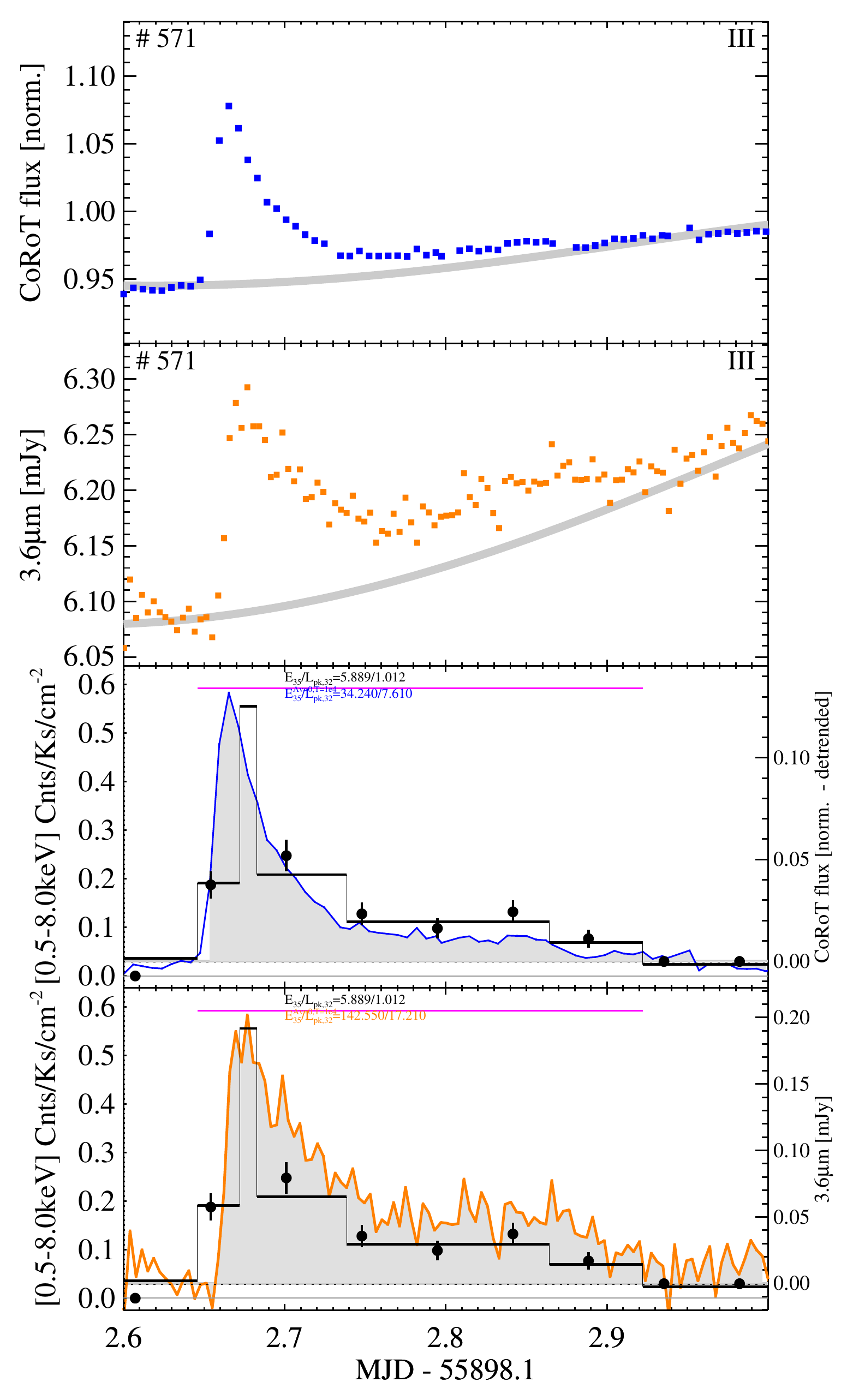}
\includegraphics[width=6.0cm]{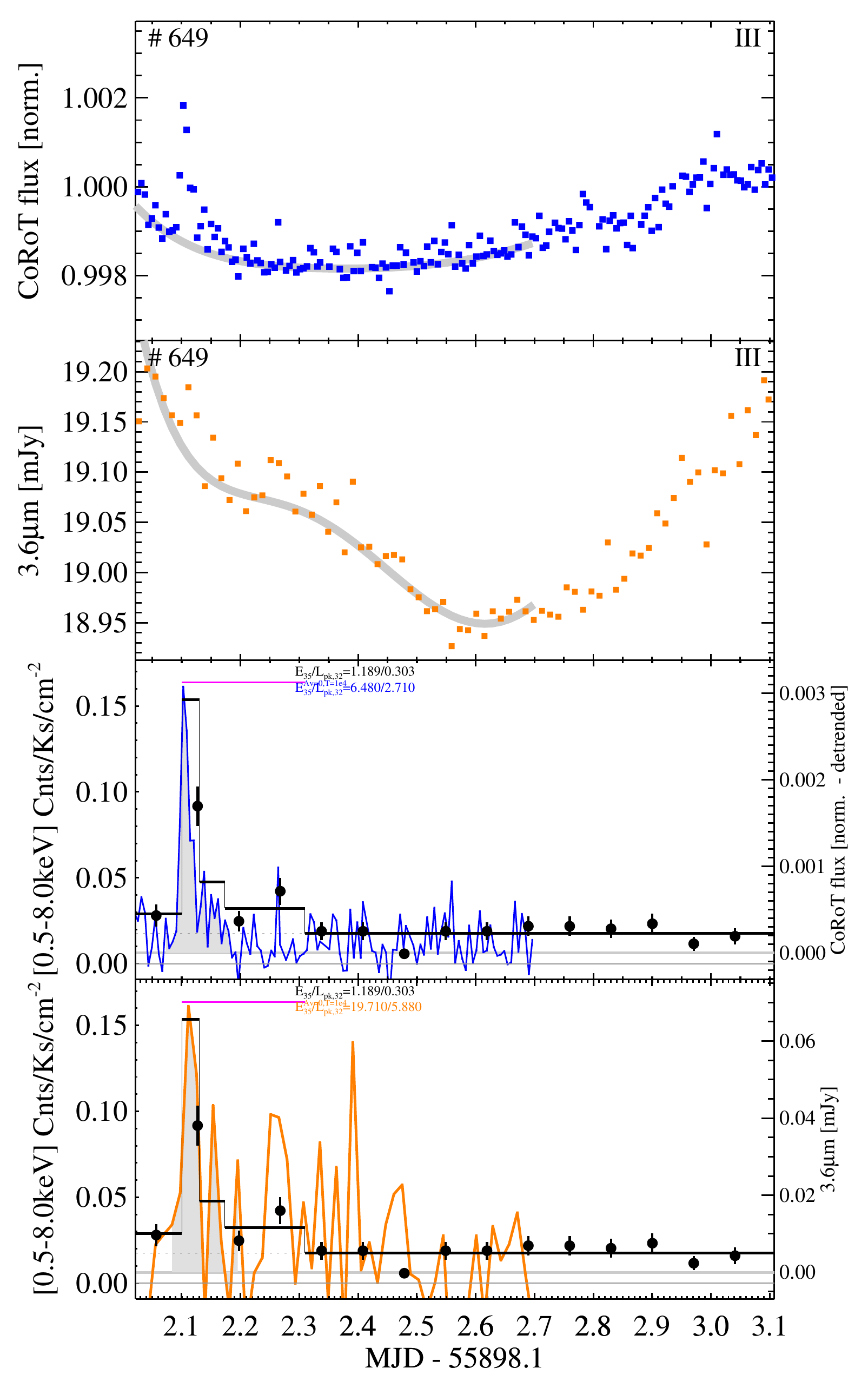}
\caption{(continued)}
\label{fig:}
\end{figure*}

\addtocounter{figure}{-1}

\begin{figure*}[!t!]
\centering
\includegraphics[width=6.0cm]{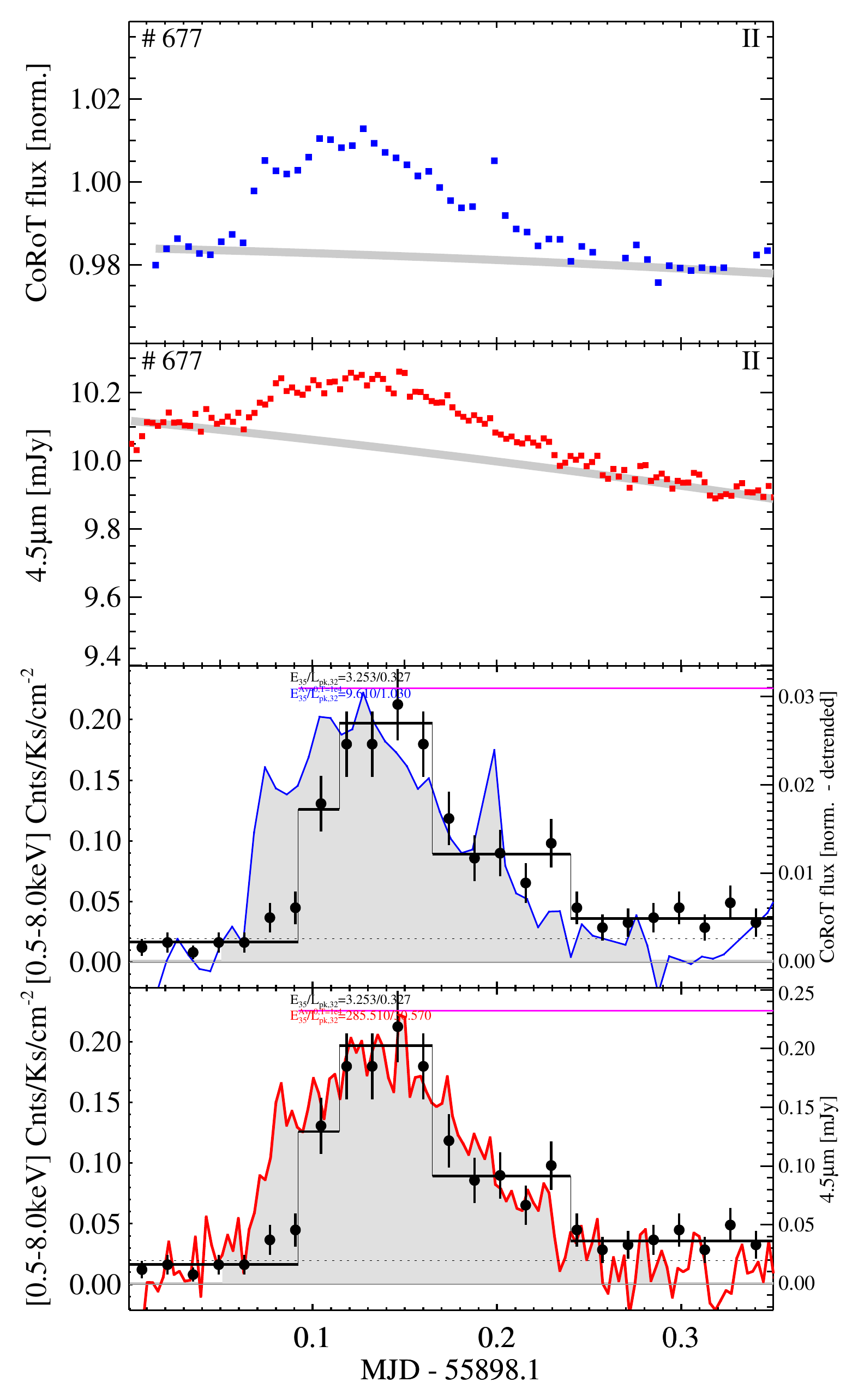}
\includegraphics[width=6.0cm]{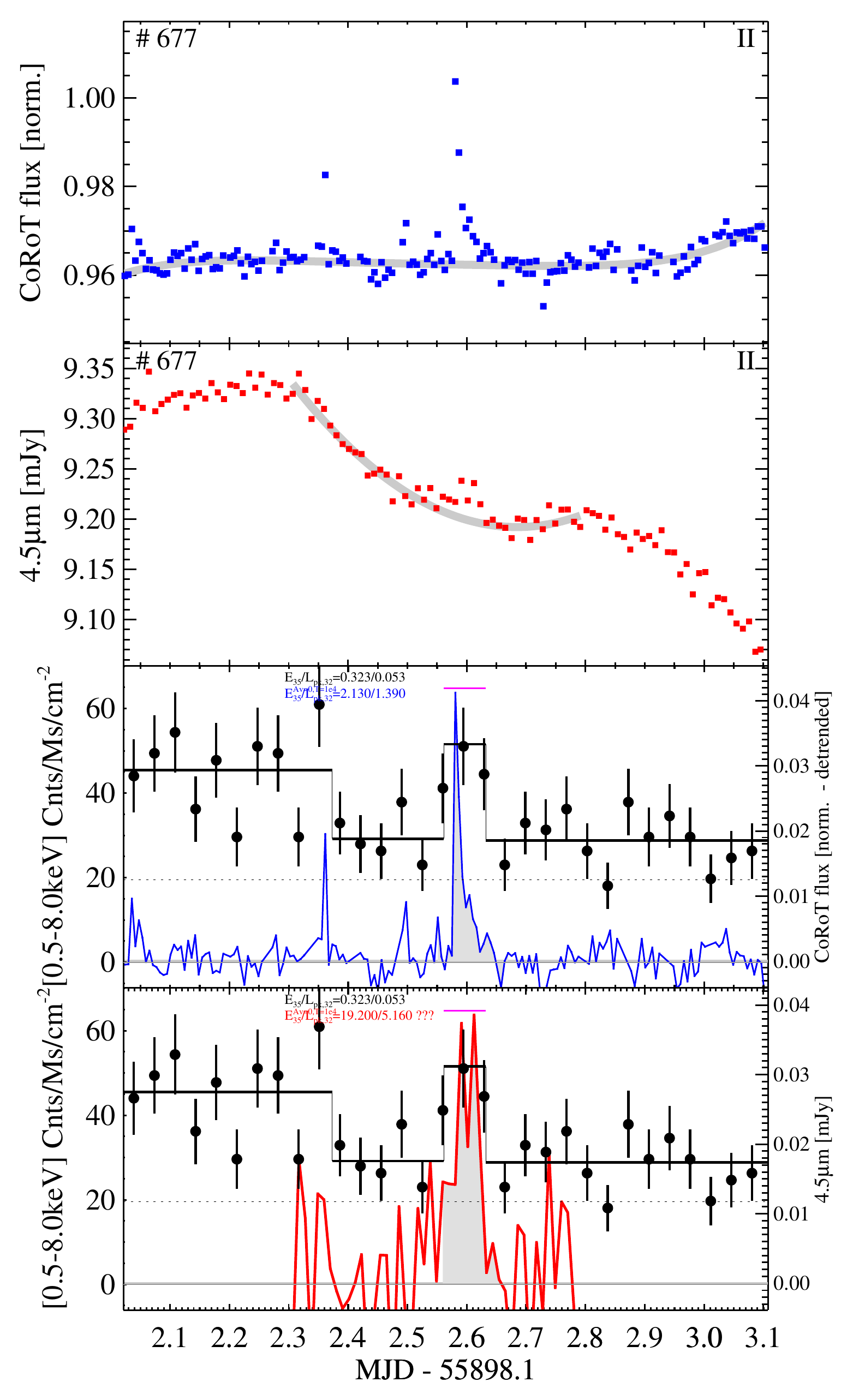}
\includegraphics[width=6.0cm]{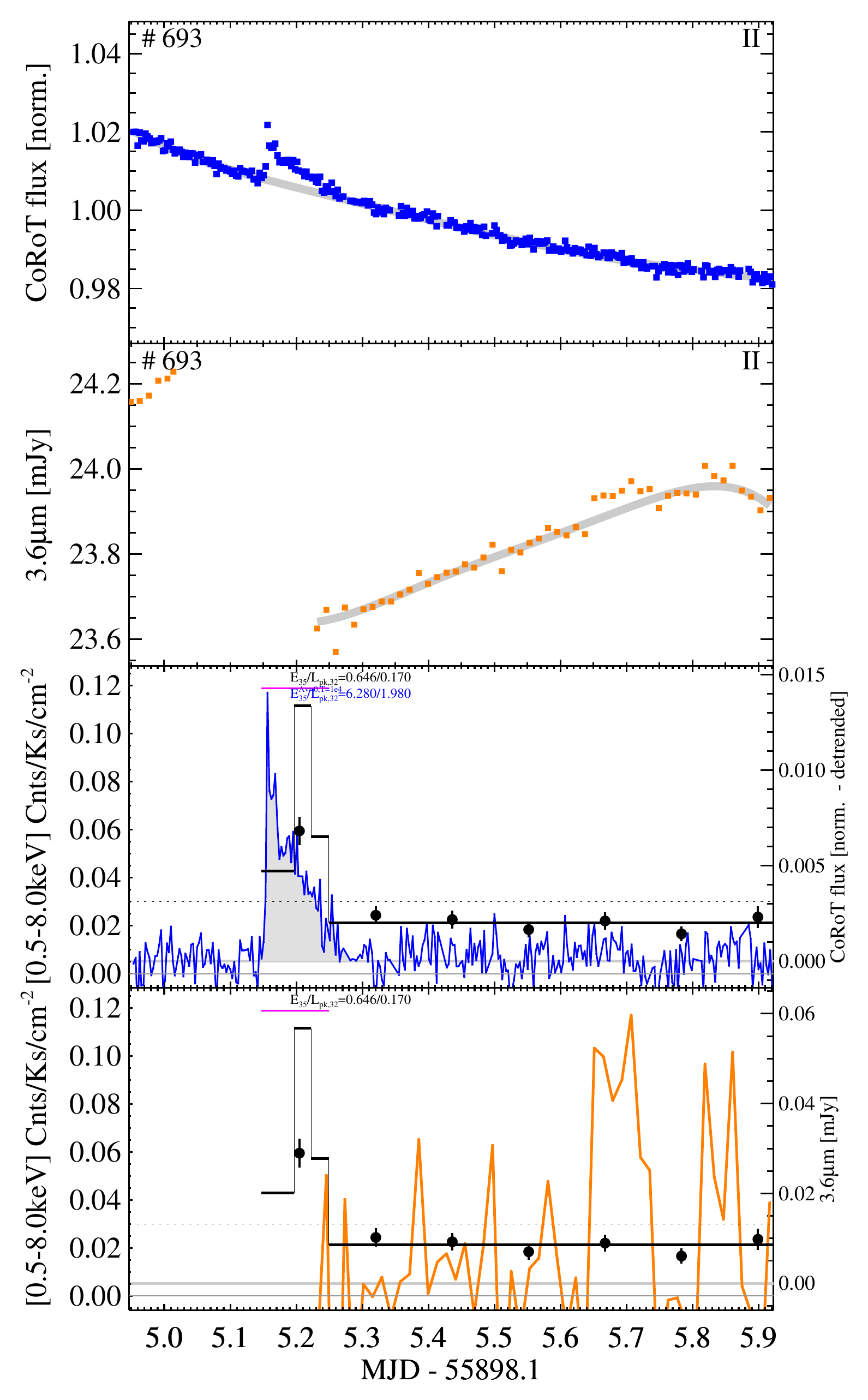}
\includegraphics[width=6.0cm]{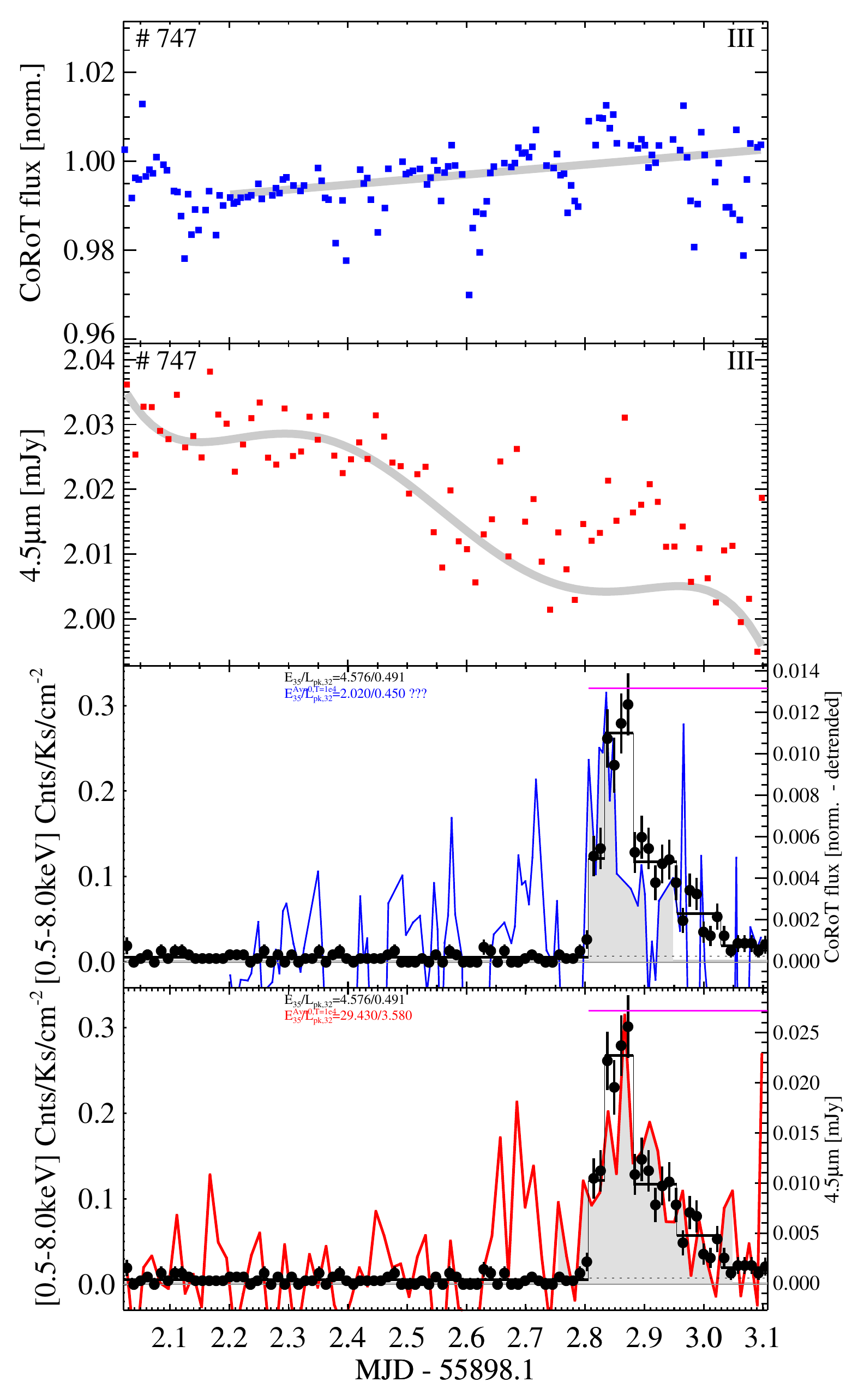}
\includegraphics[width=6.0cm]{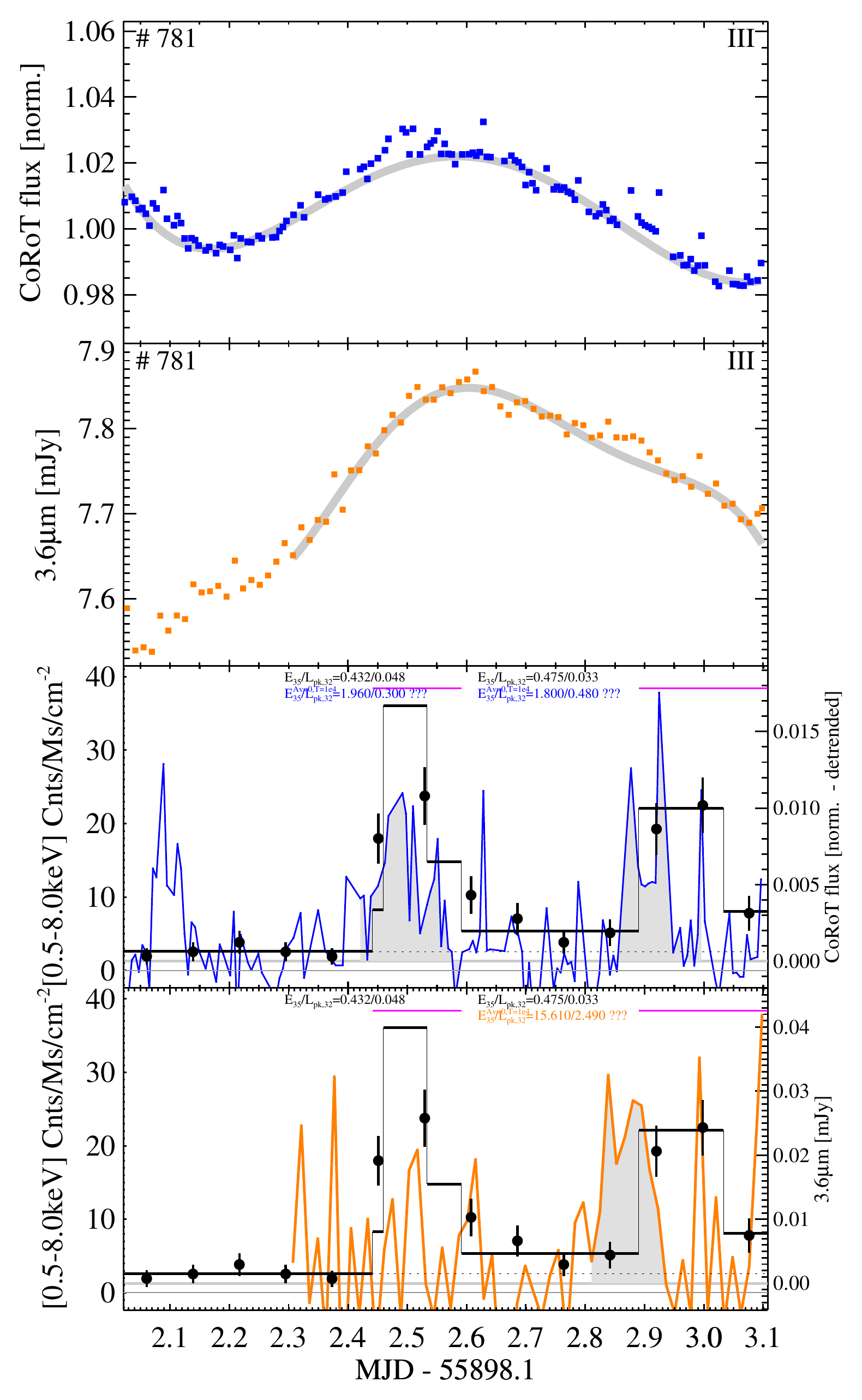}
\caption{(continued)}
\label{fig:}
\end{figure*}

\addtocounter{figure}{-1}

\begin{figure*}[!t!]
\centering
\includegraphics[width=6.0cm]{lc_zoom_s45.pdf}
\includegraphics[width=6.0cm]{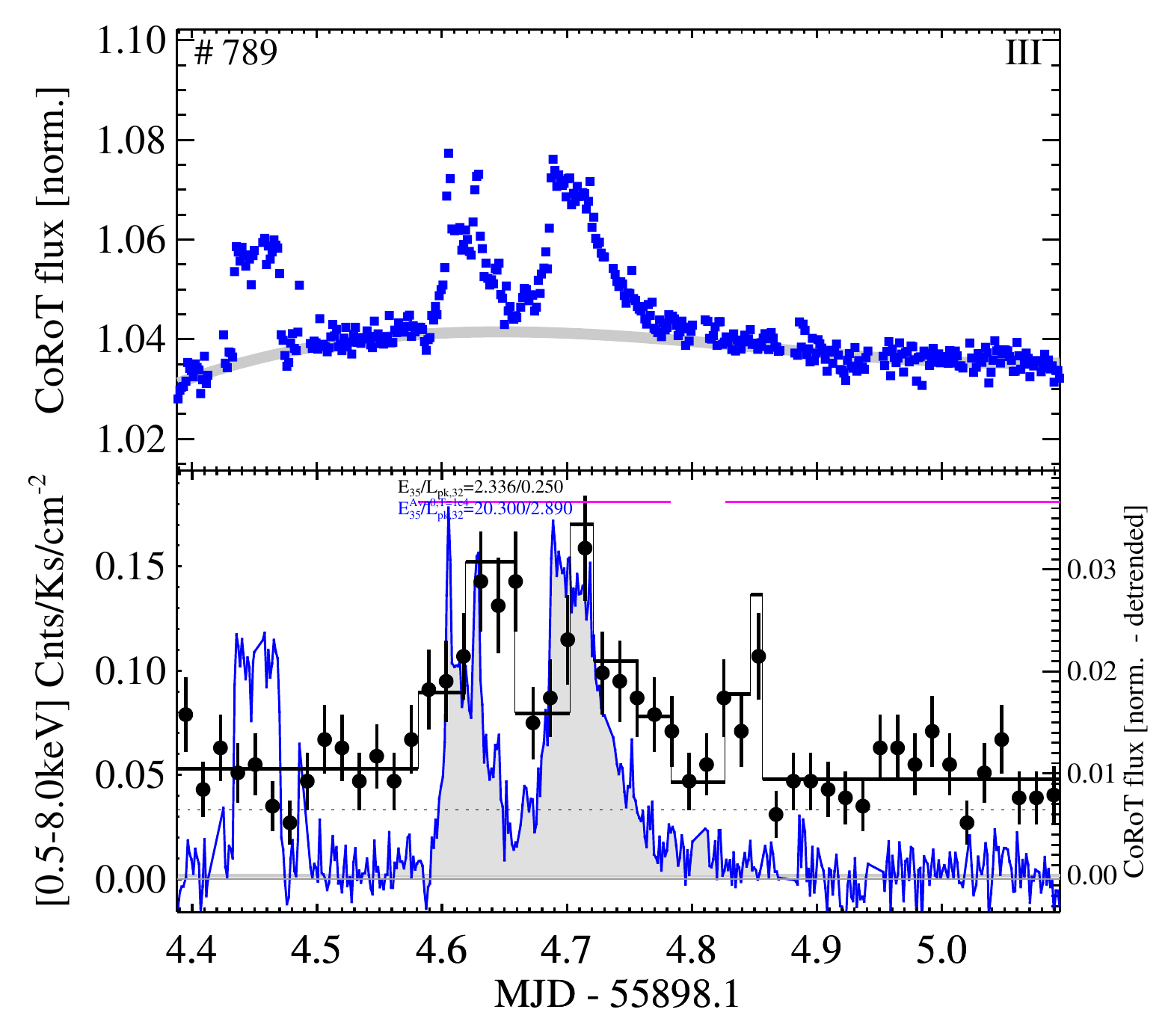}
\includegraphics[width=6.0cm]{lc_zoom_s863-2.pdf}
\includegraphics[width=6.0cm]{lc_zoom_s107.pdf}
\includegraphics[width=6.0cm]{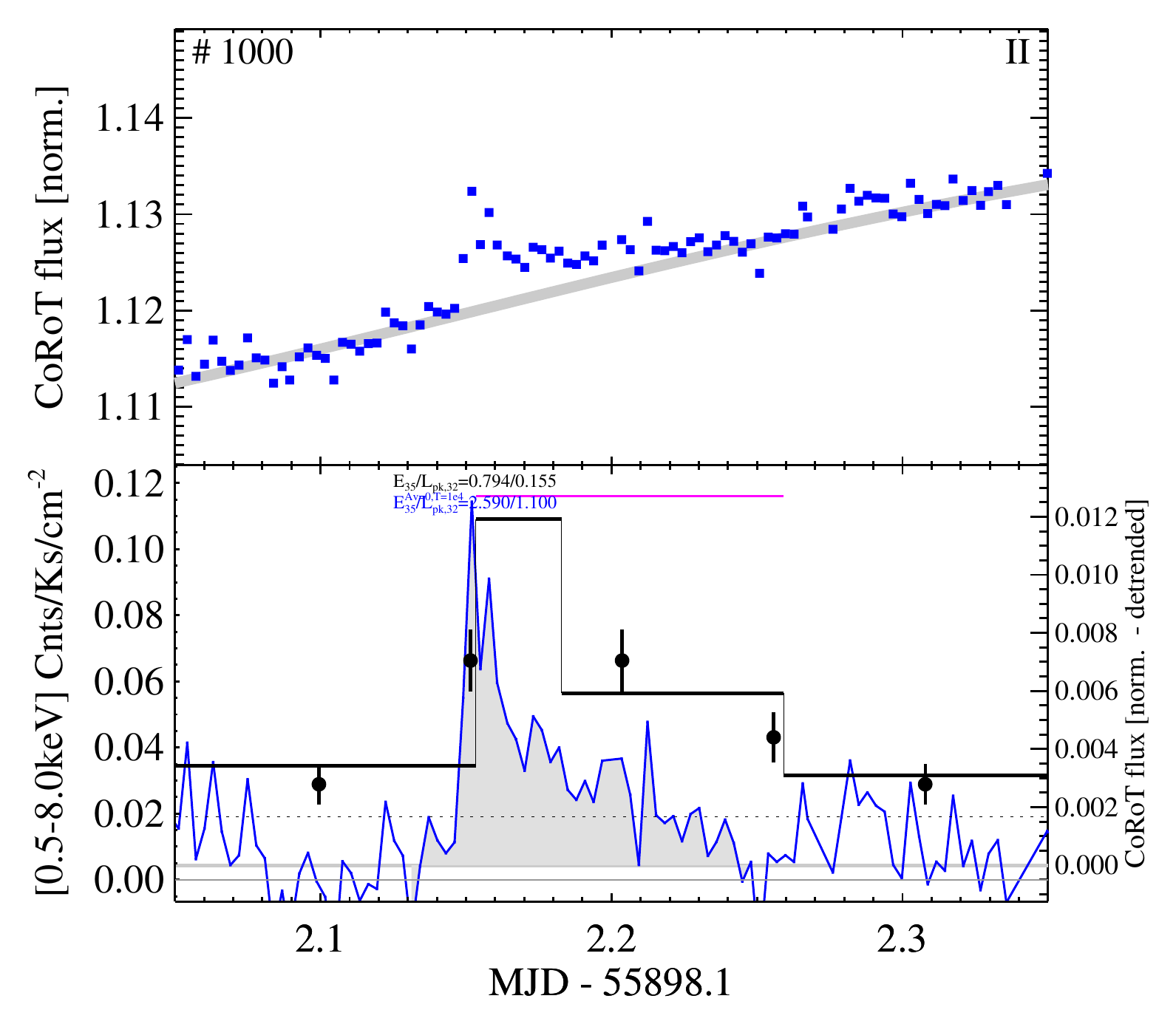}
\includegraphics[width=6.0cm]{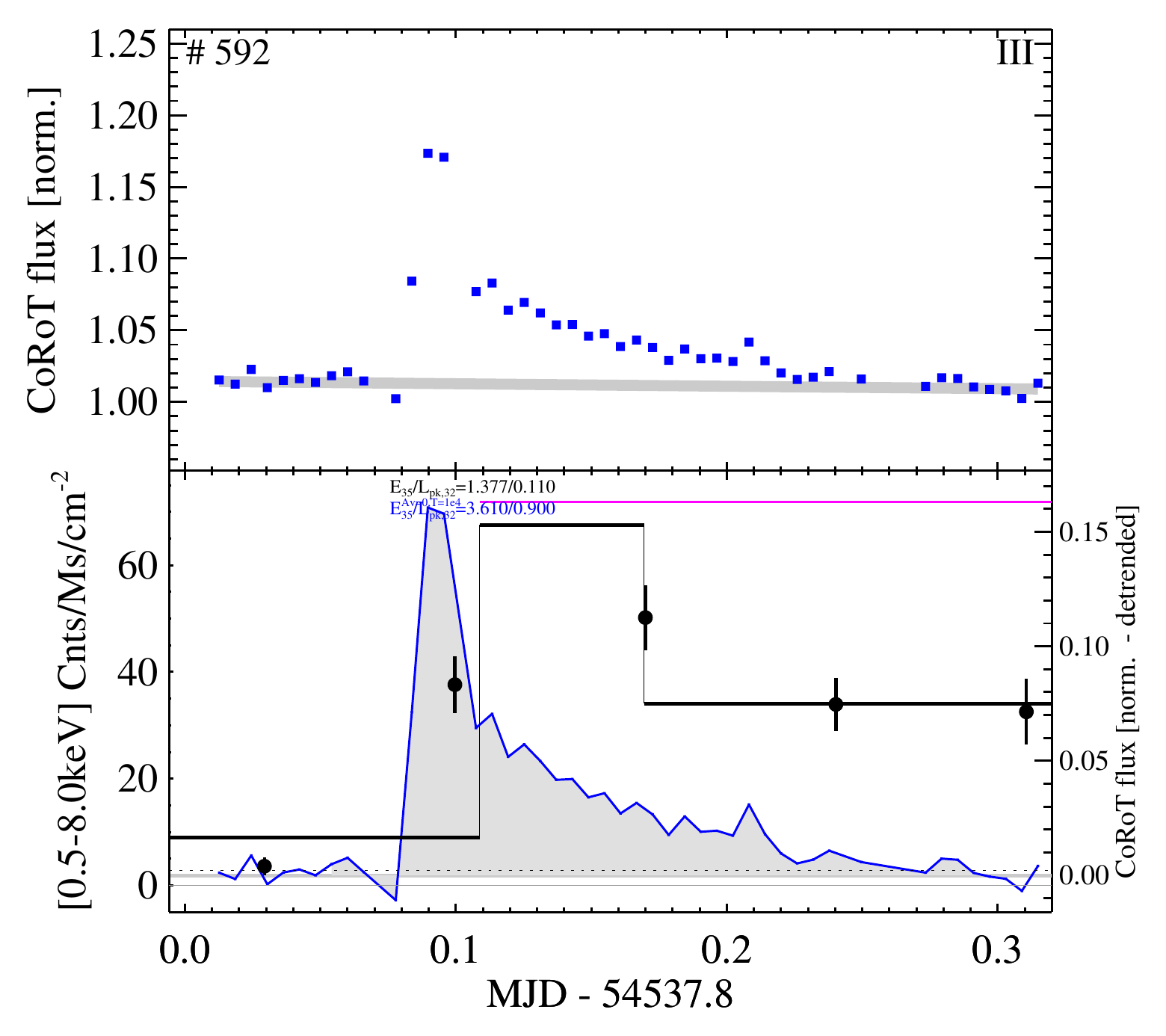}
\includegraphics[width=6.0cm]{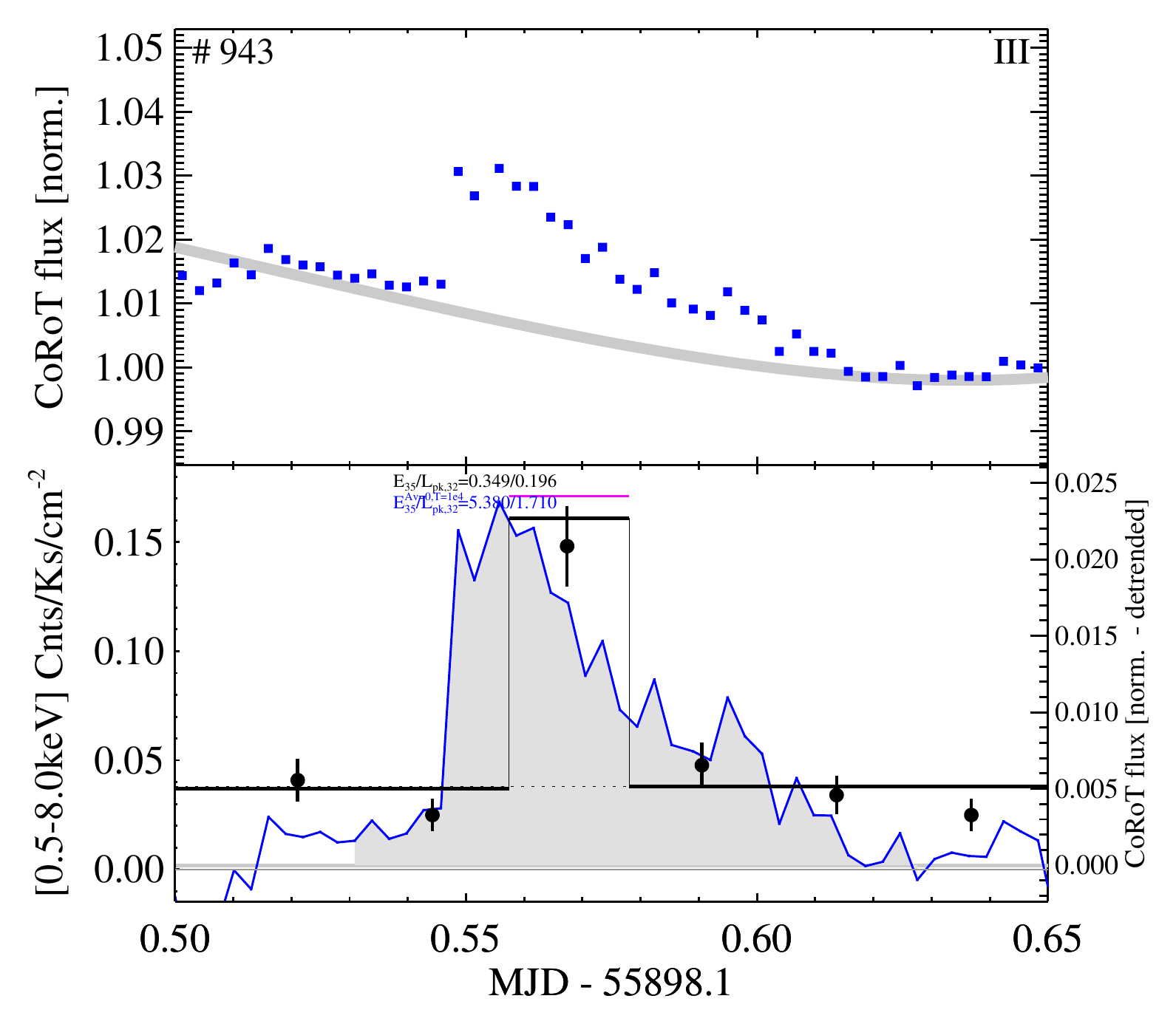}
\includegraphics[width=6.0cm]{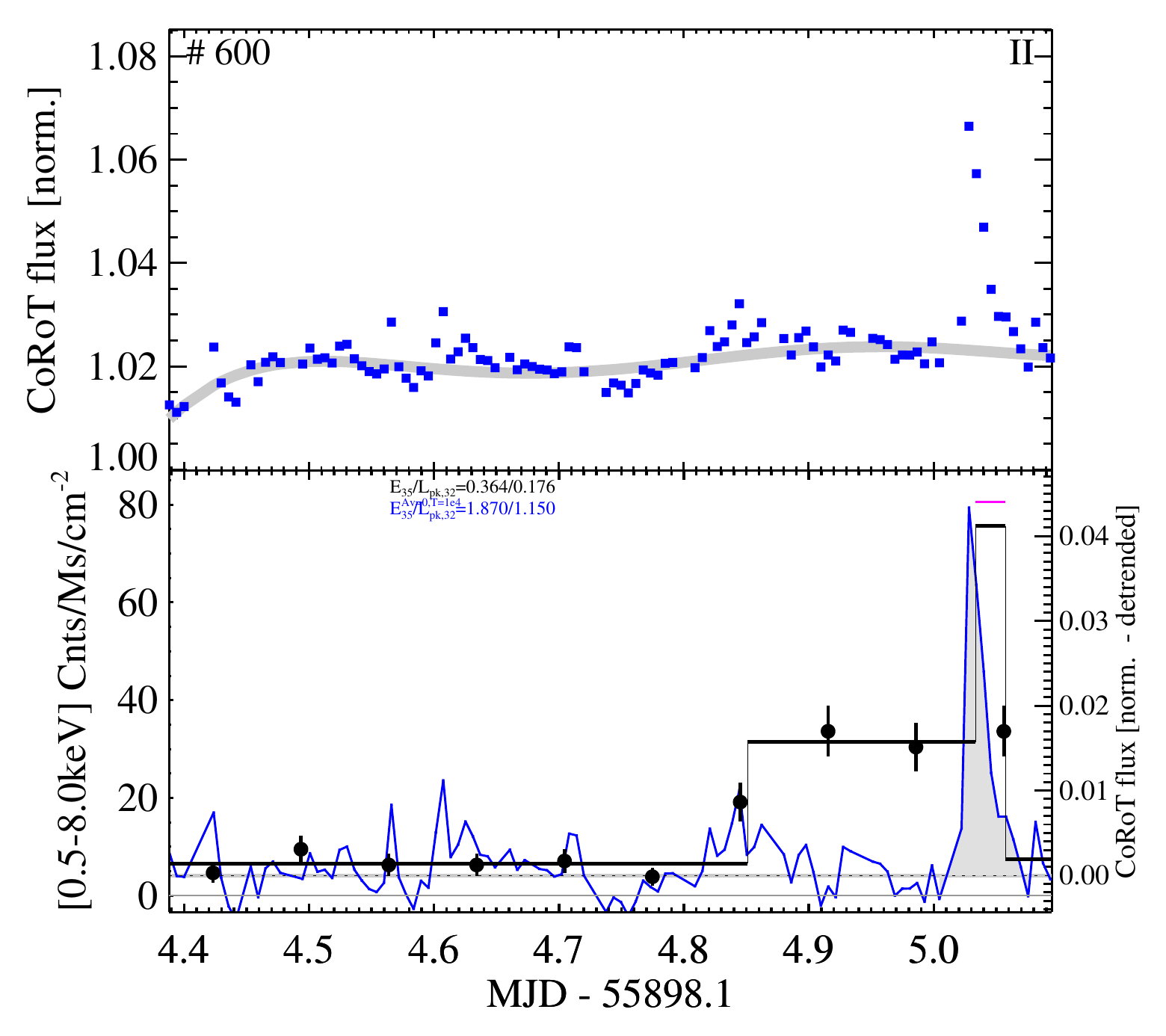}
\includegraphics[width=6.0cm]{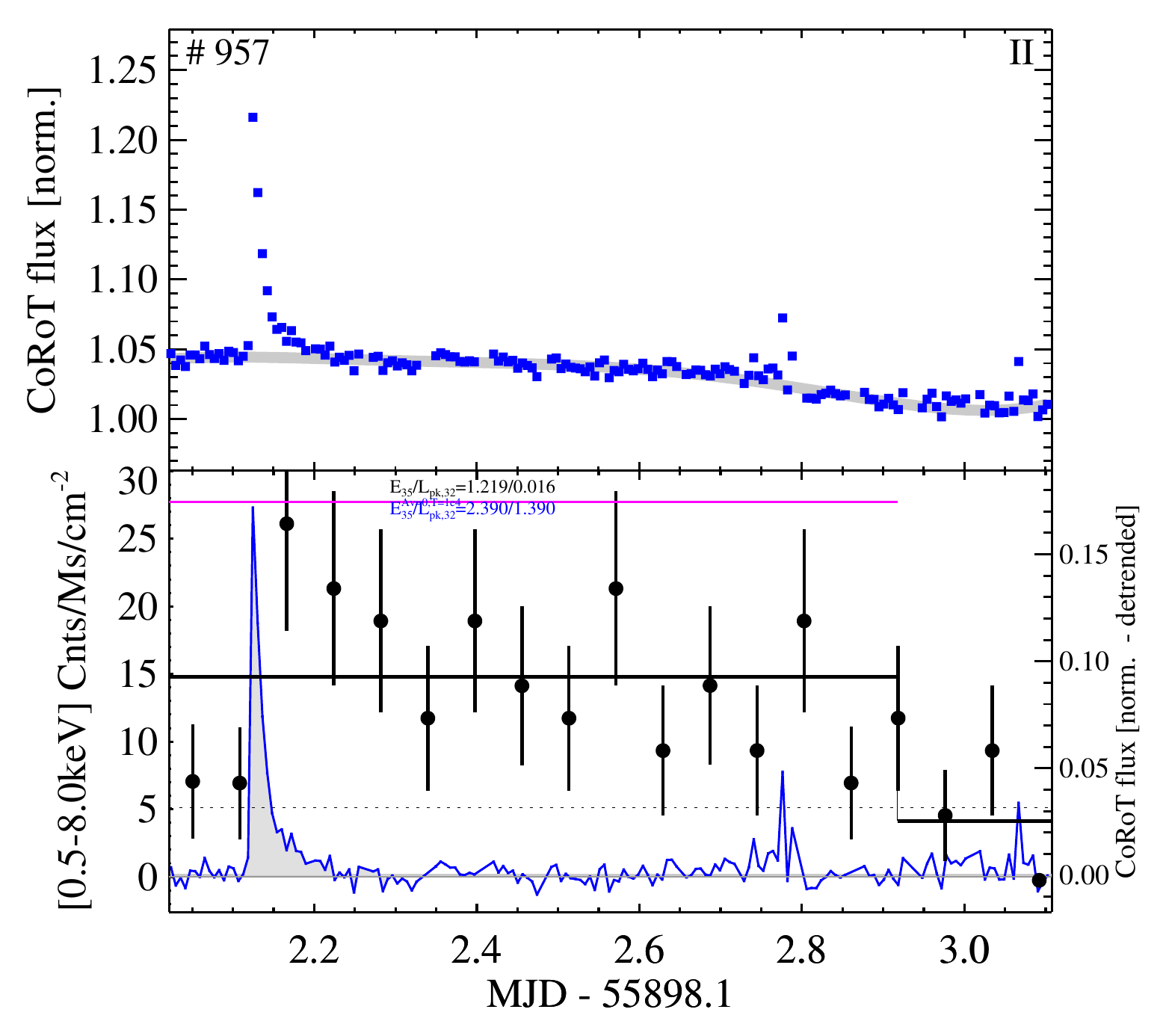}
\includegraphics[width=6.0cm]{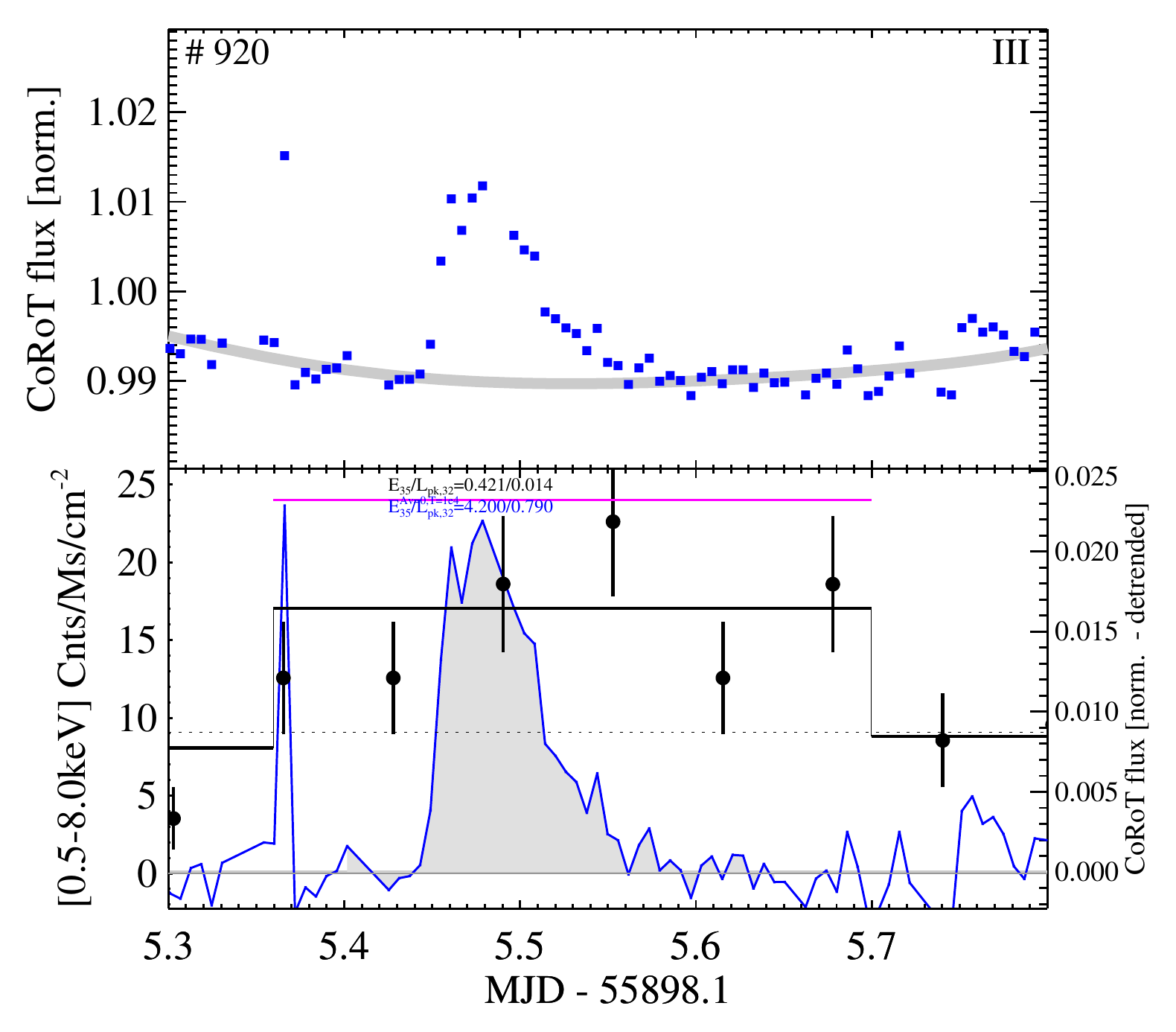}
\includegraphics[width=6.0cm]{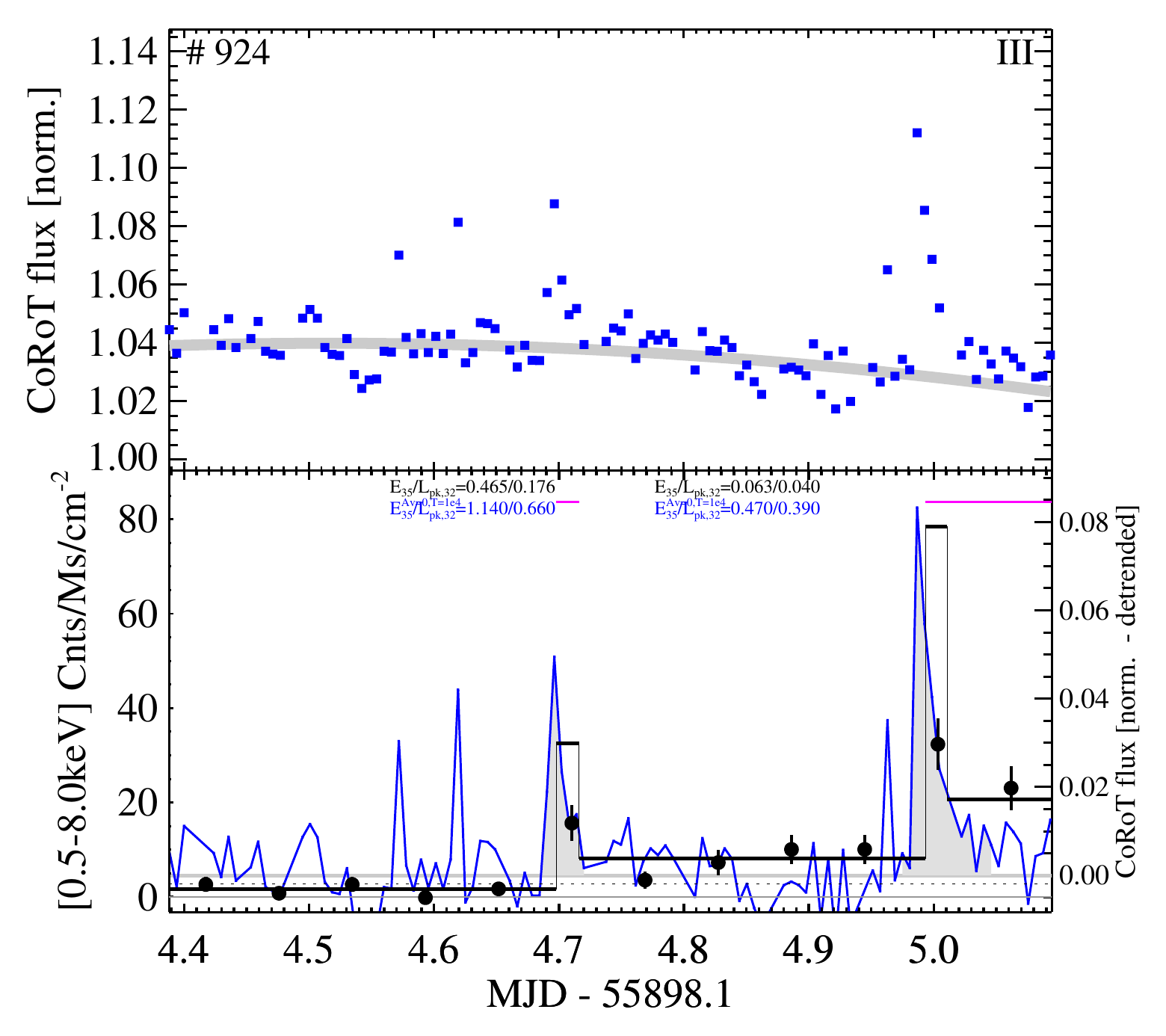}
\includegraphics[width=6.0cm]{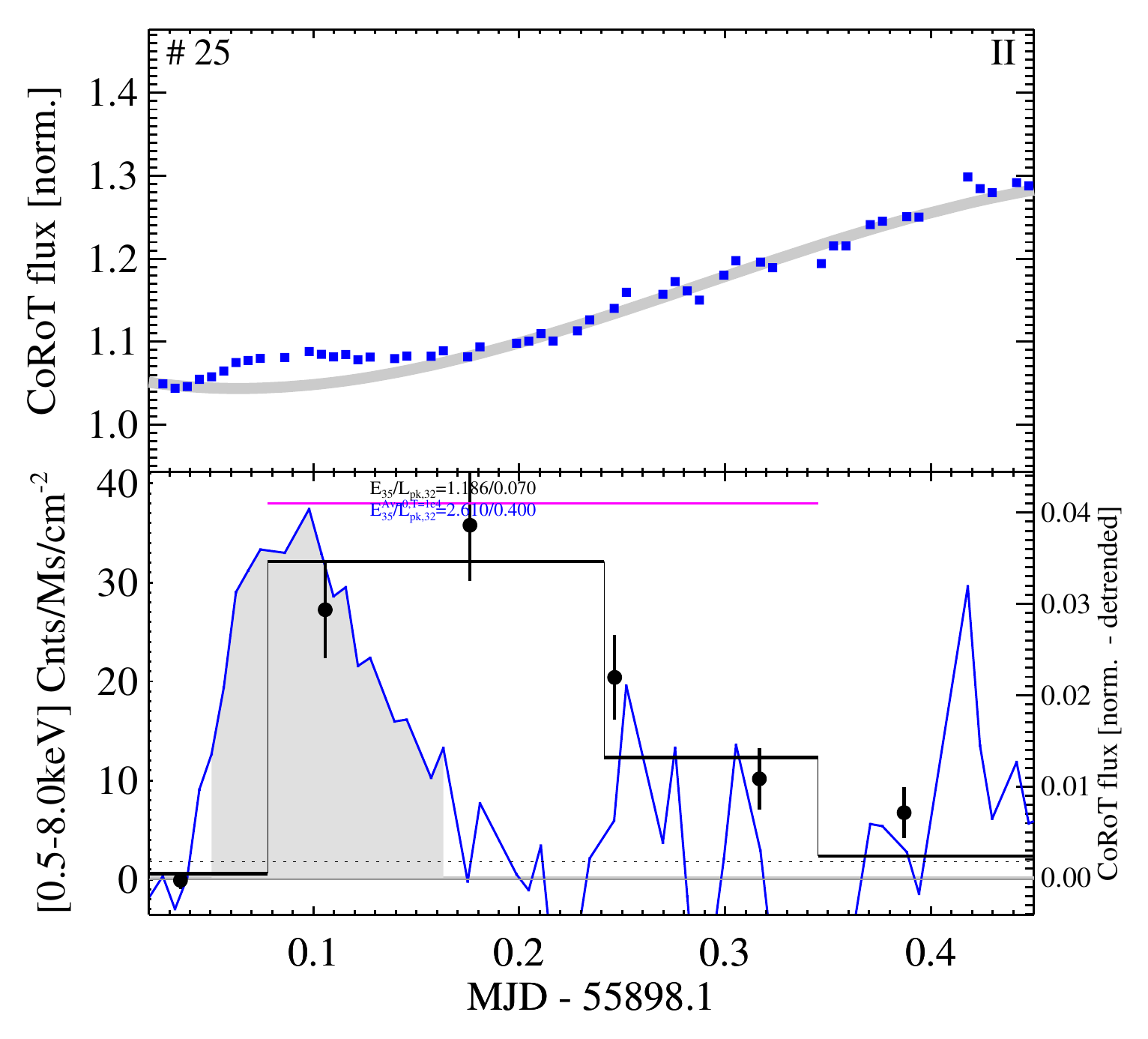}
\caption{(continued)}
\label{fig:}
\end{figure*}

\addtocounter{figure}{-1}

\begin{figure*}[!t!]
\centering
\includegraphics[width=6.0cm]{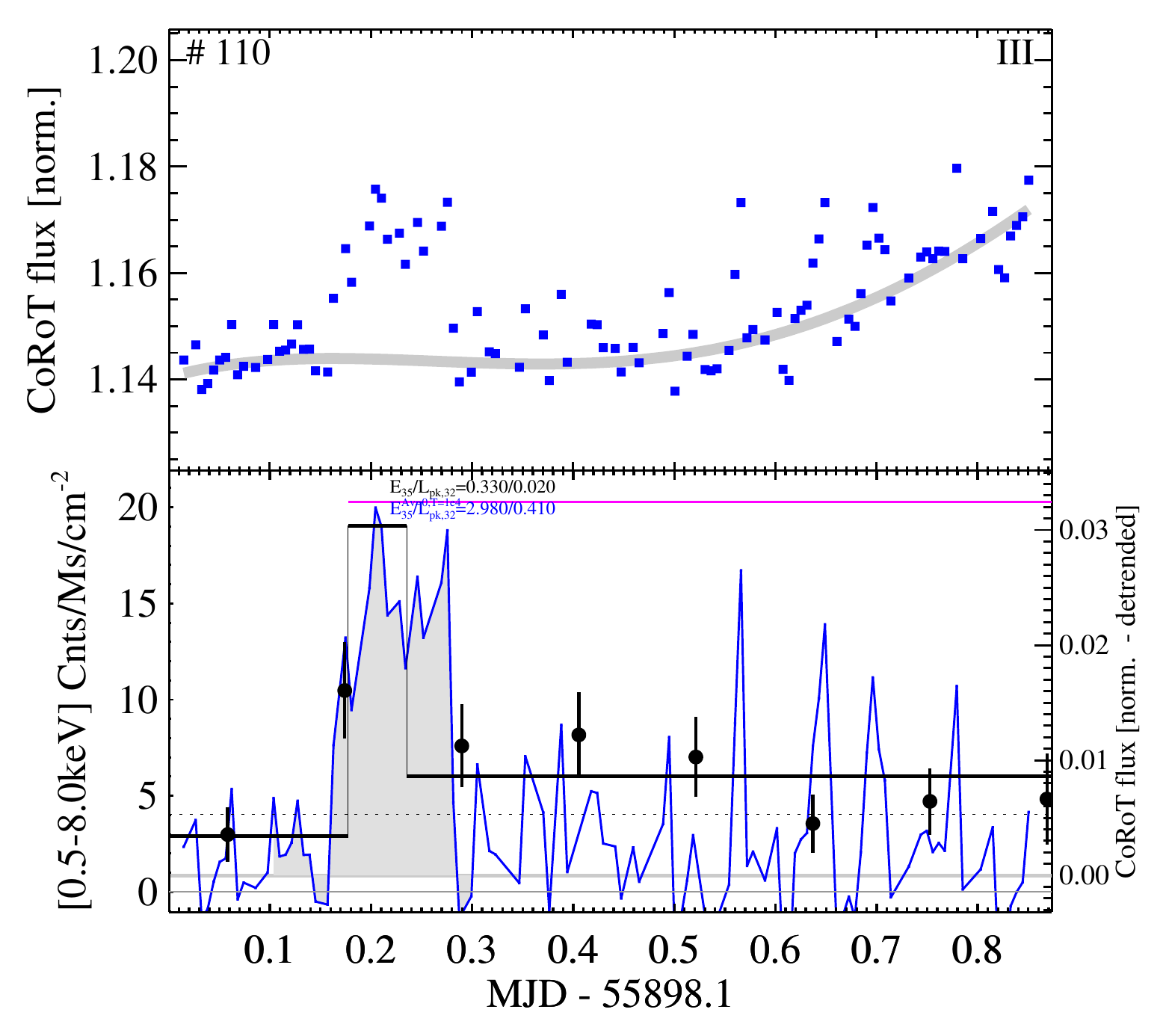}
\includegraphics[width=6.0cm]{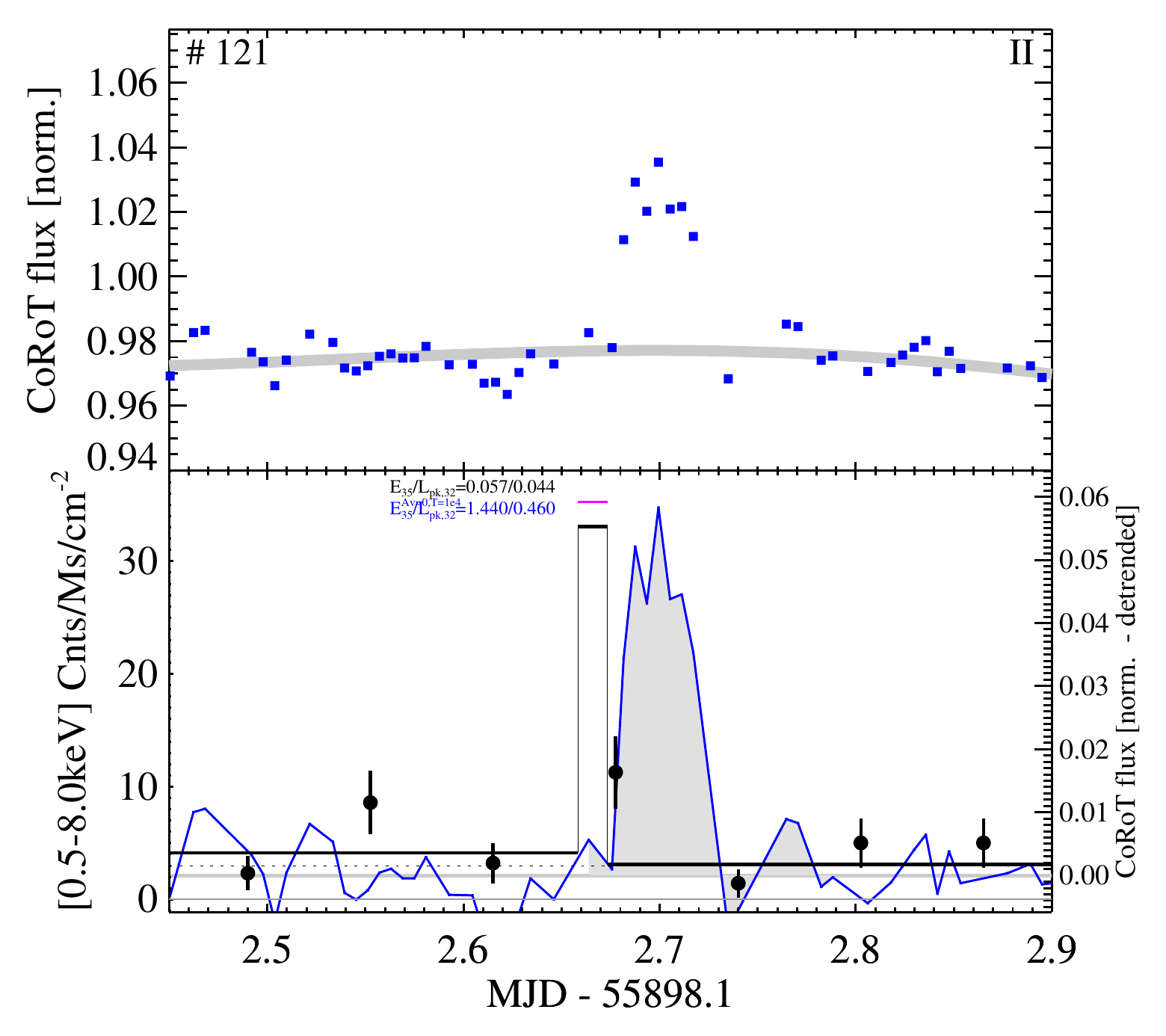}
\includegraphics[width=6.0cm]{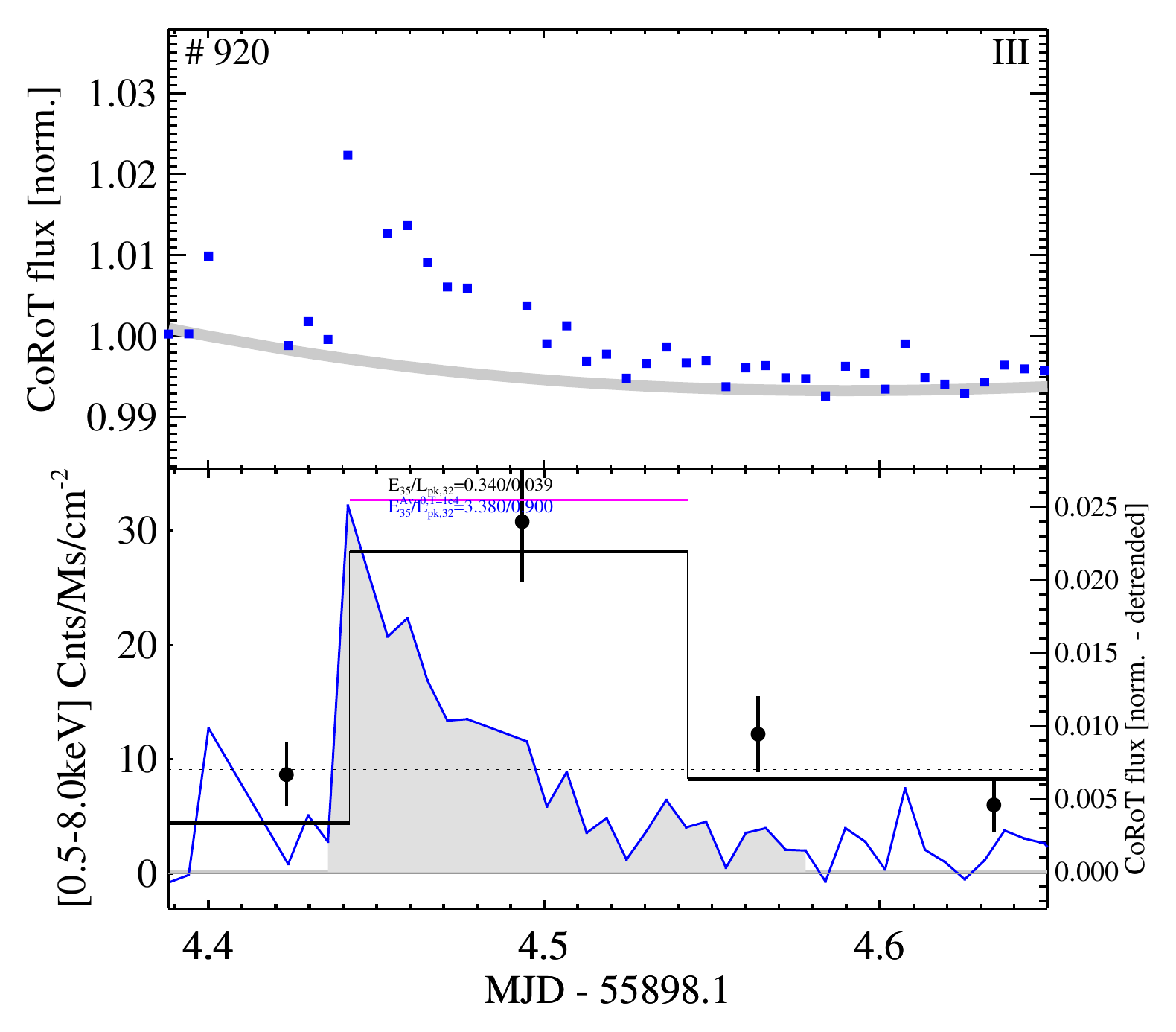}
\includegraphics[width=6.0cm]{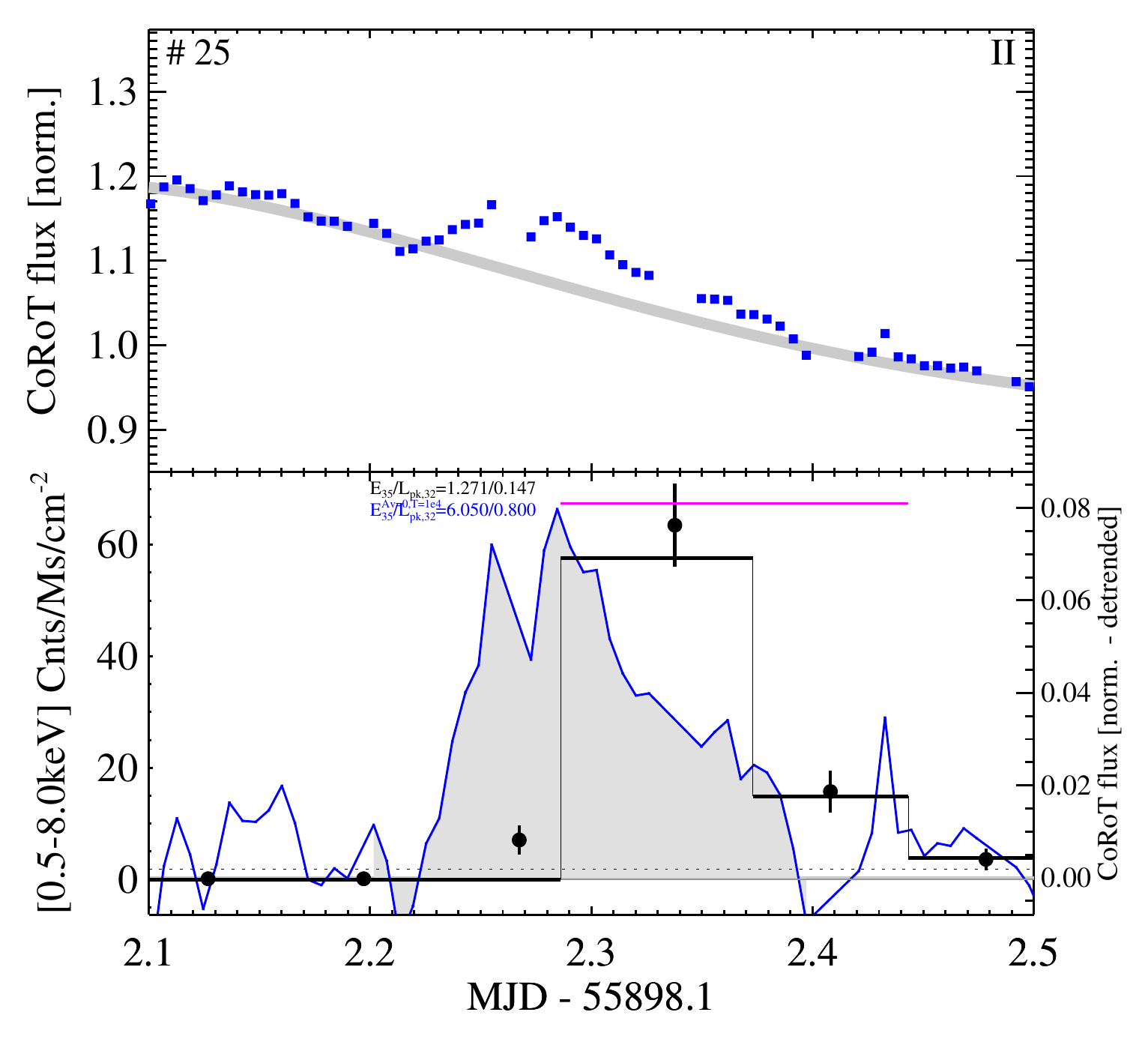}
\includegraphics[width=6.0cm]{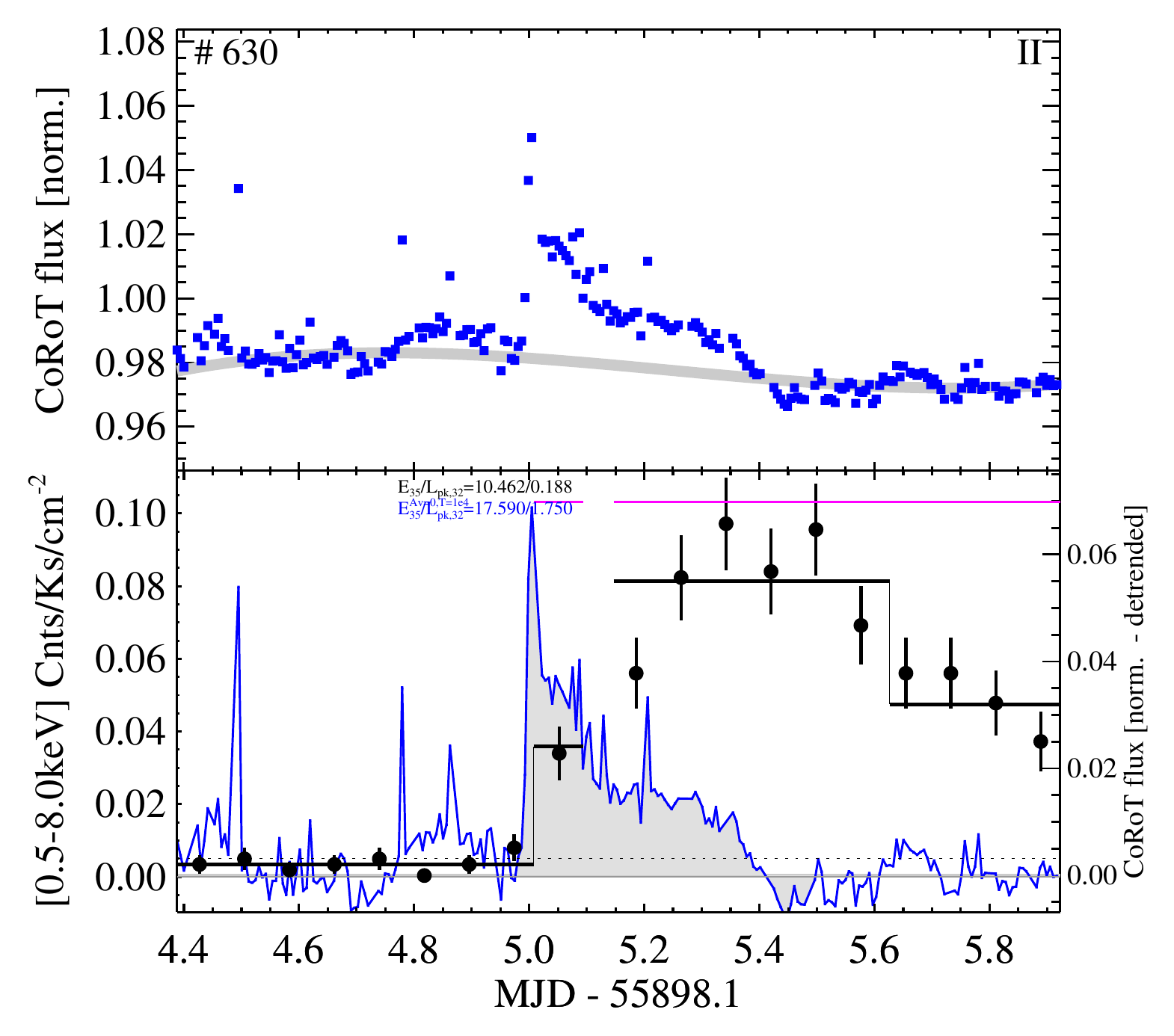}
\includegraphics[width=6.0cm]{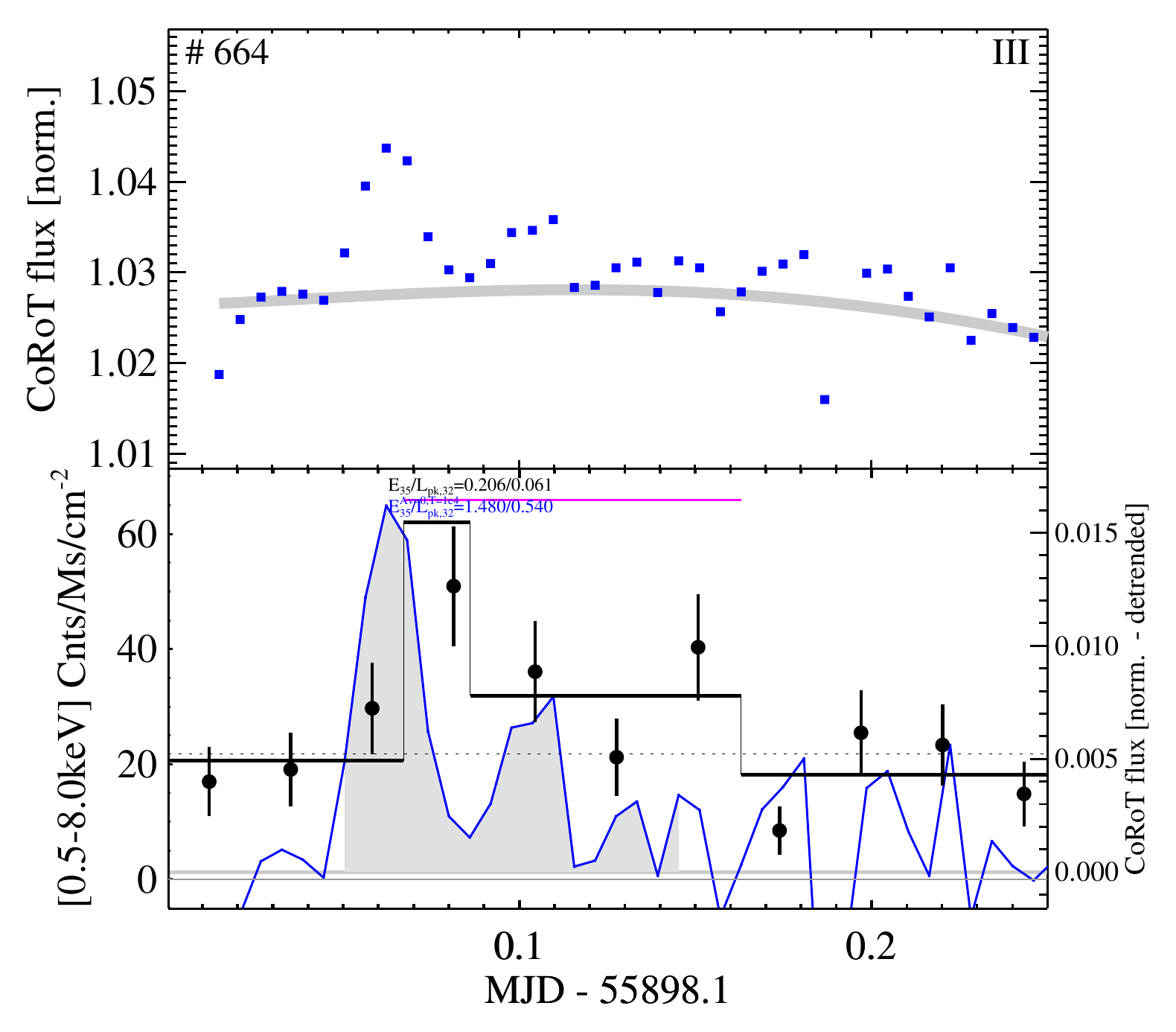}
\includegraphics[width=6.0cm]{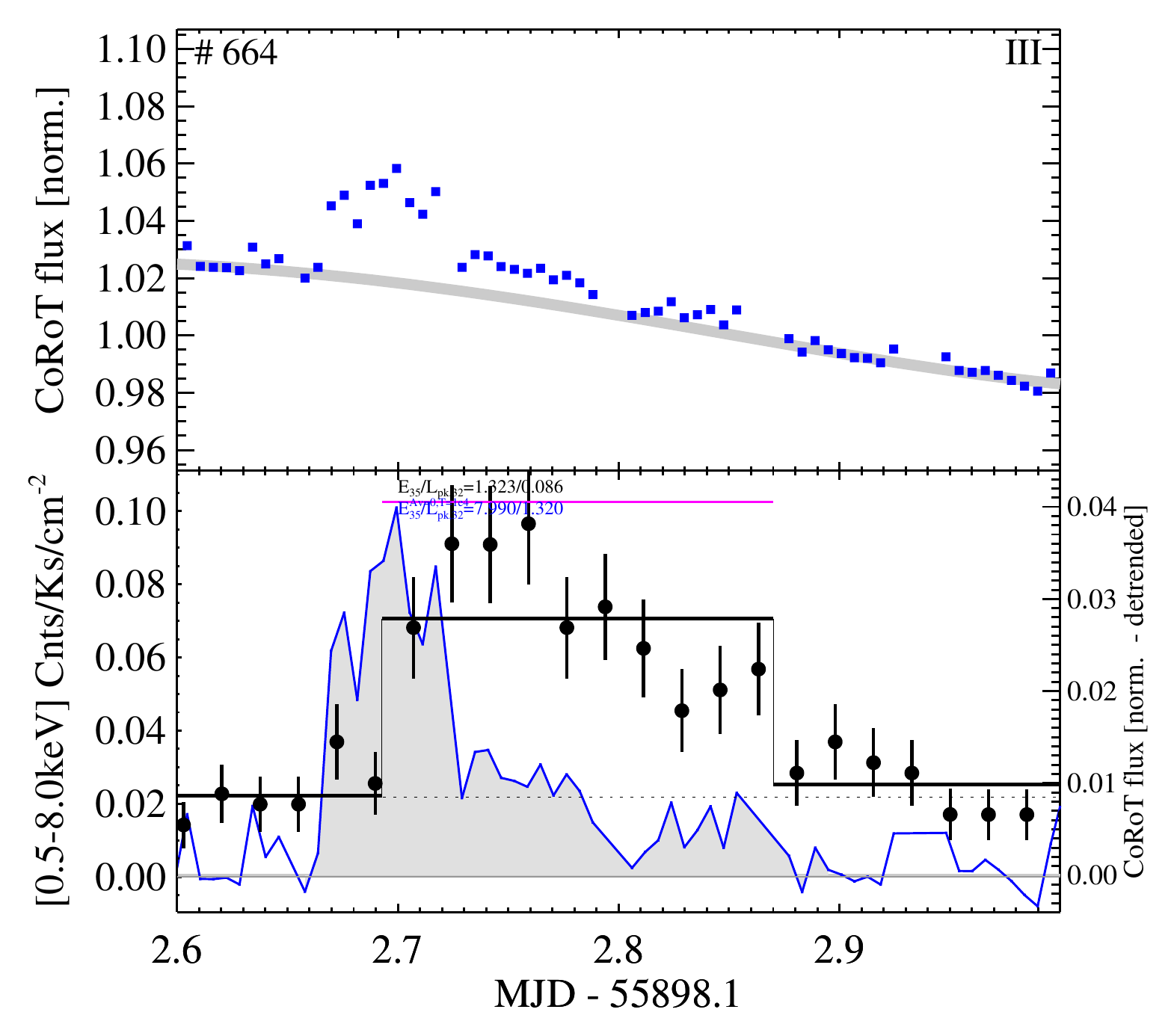}
\includegraphics[width=6.0cm]{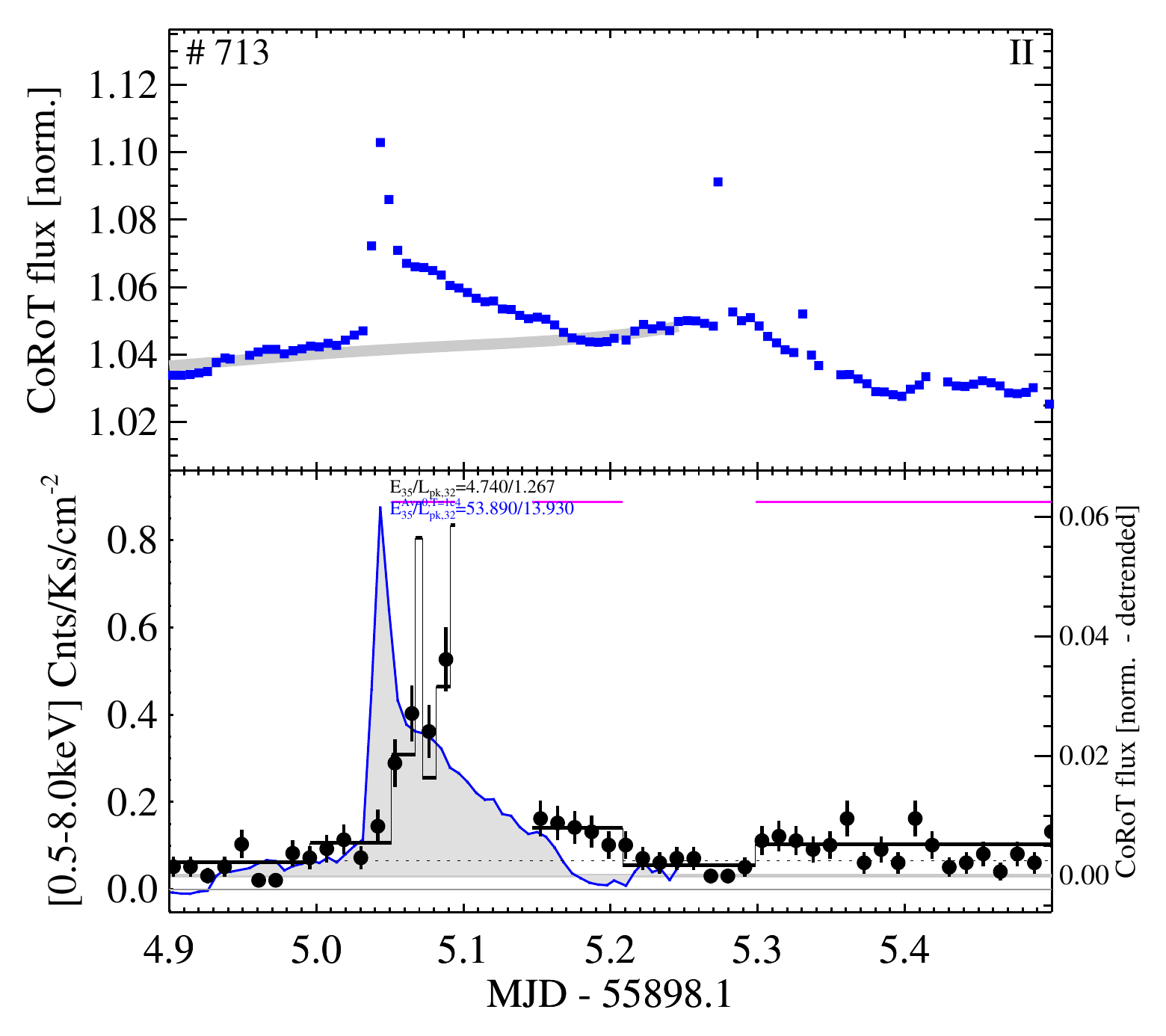}
\includegraphics[width=6.0cm]{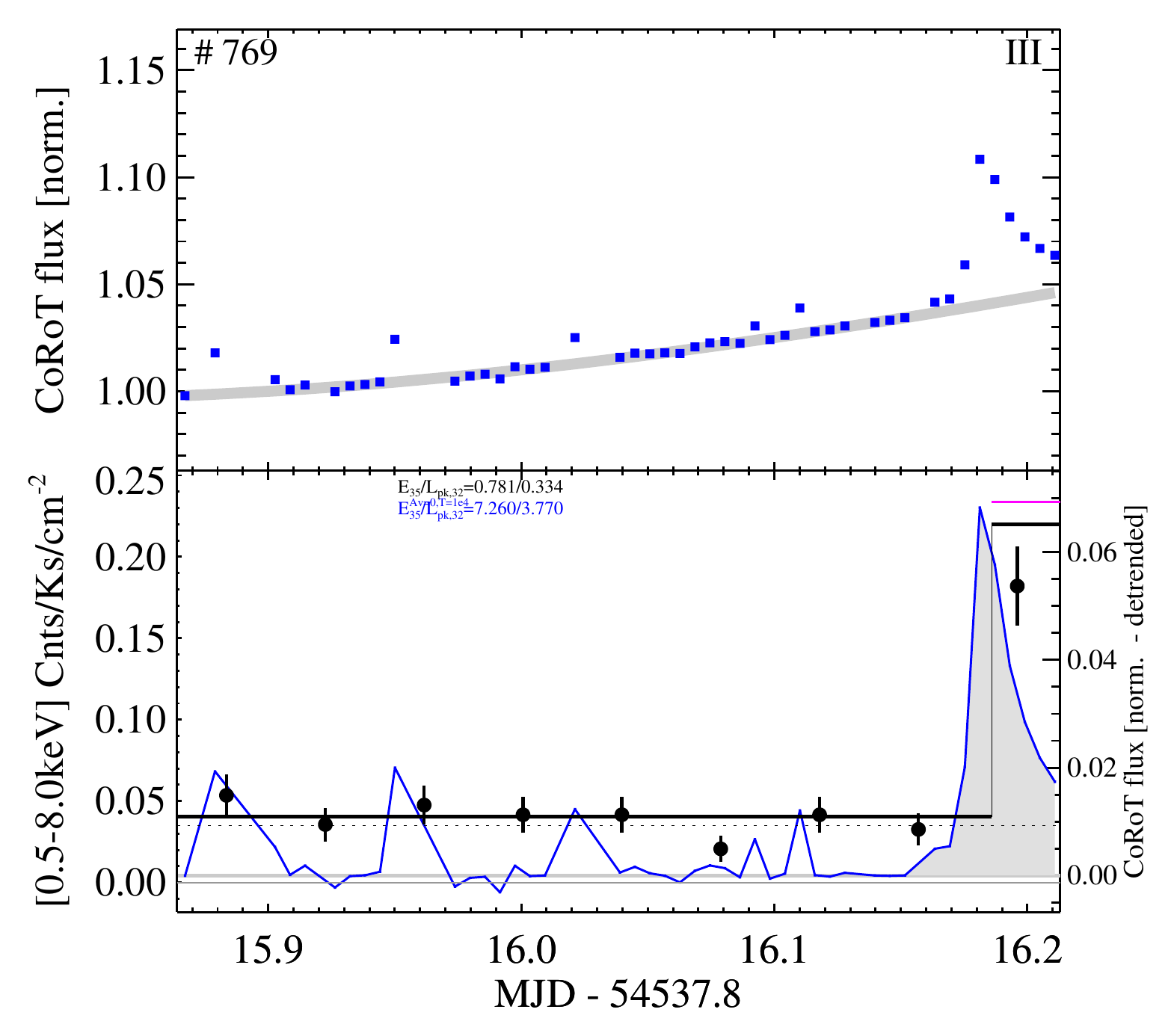}
\includegraphics[width=6.0cm]{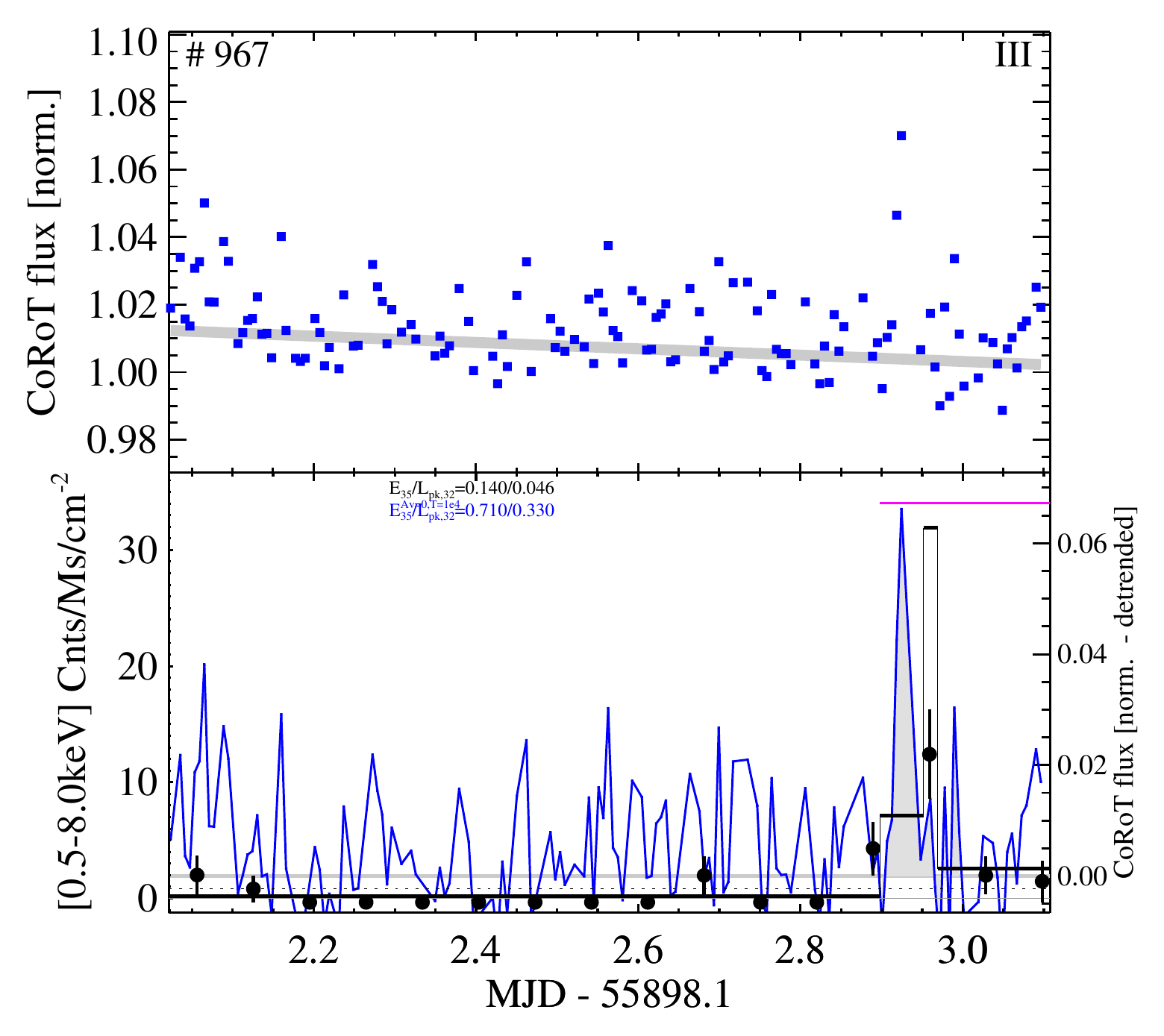}
\includegraphics[width=6.0cm]{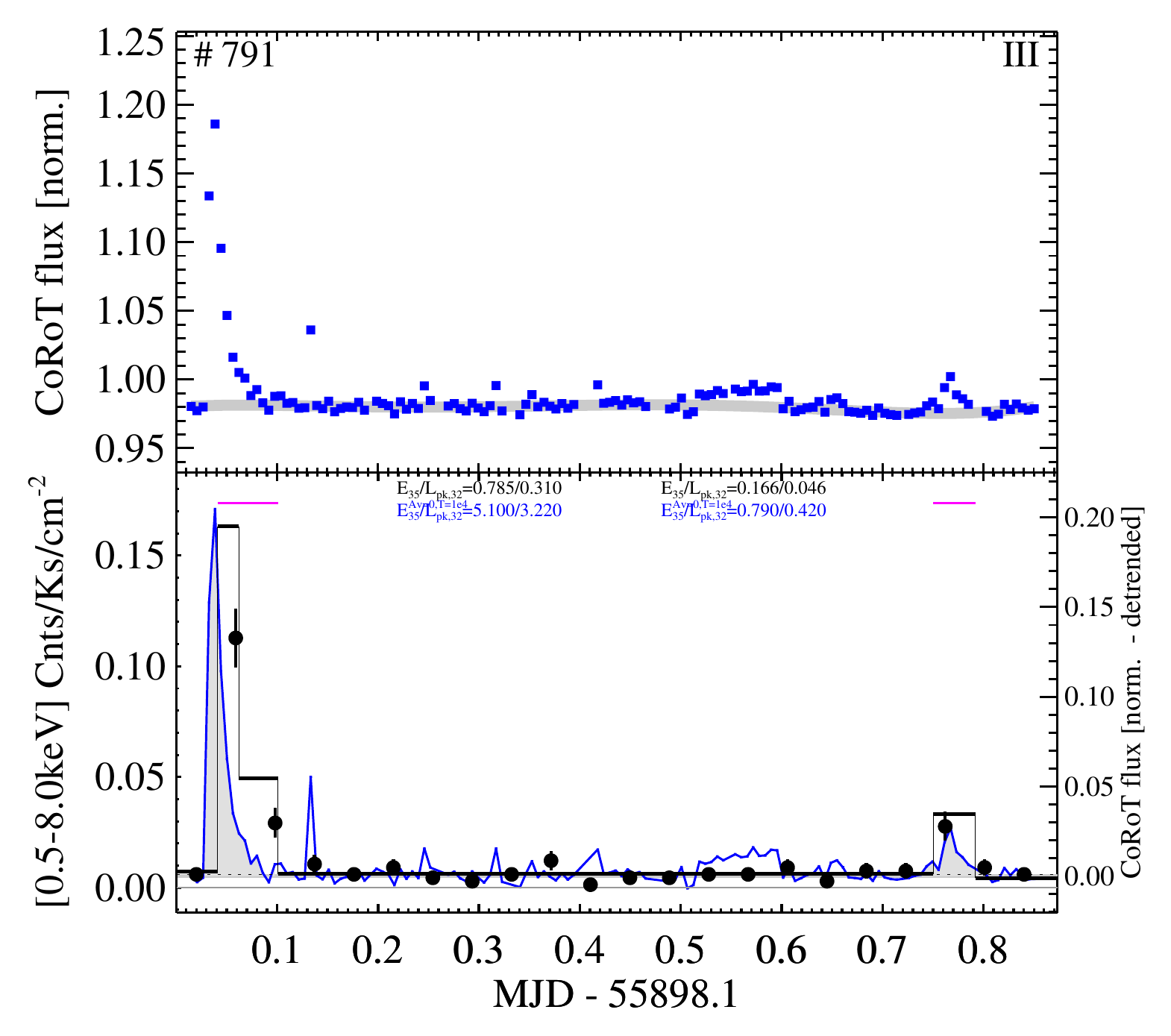}
\caption{(continued)}
\label{fig:}
\end{figure*}

\addtocounter{figure}{-1}

\begin{figure*}[!t!]
\centering
\includegraphics[width=6.0cm]{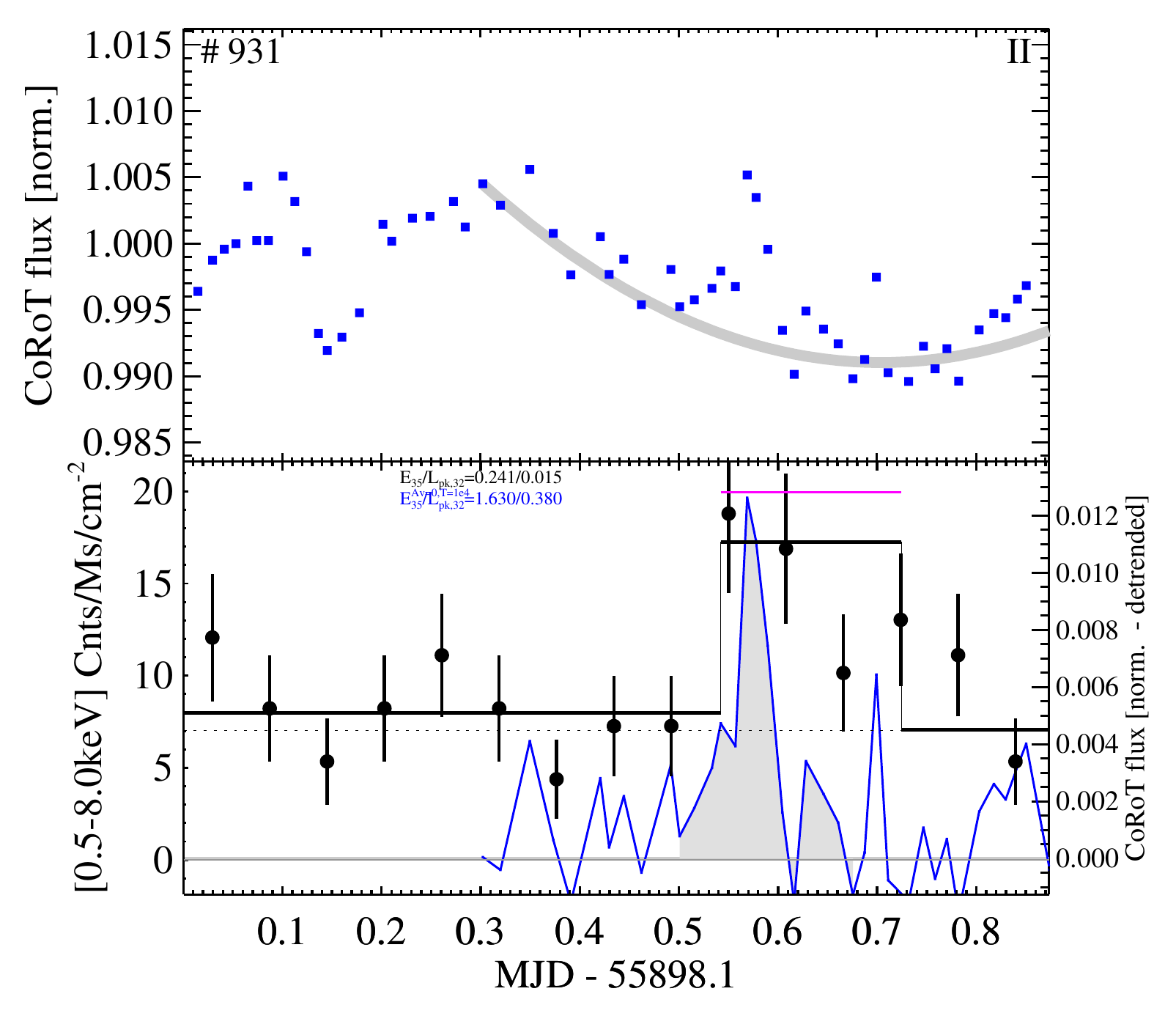}
\includegraphics[width=6.0cm]{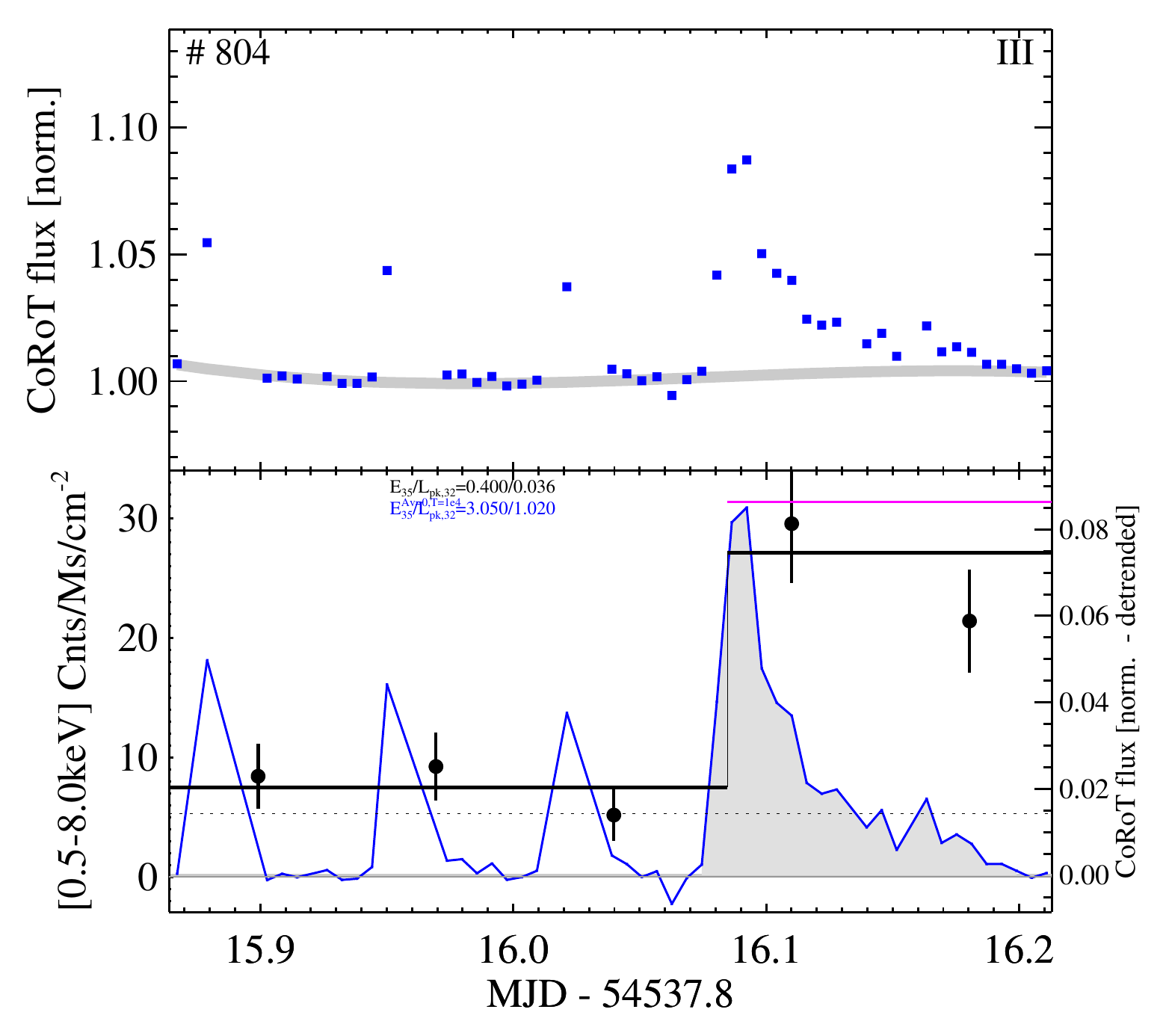}
\includegraphics[width=6.0cm]{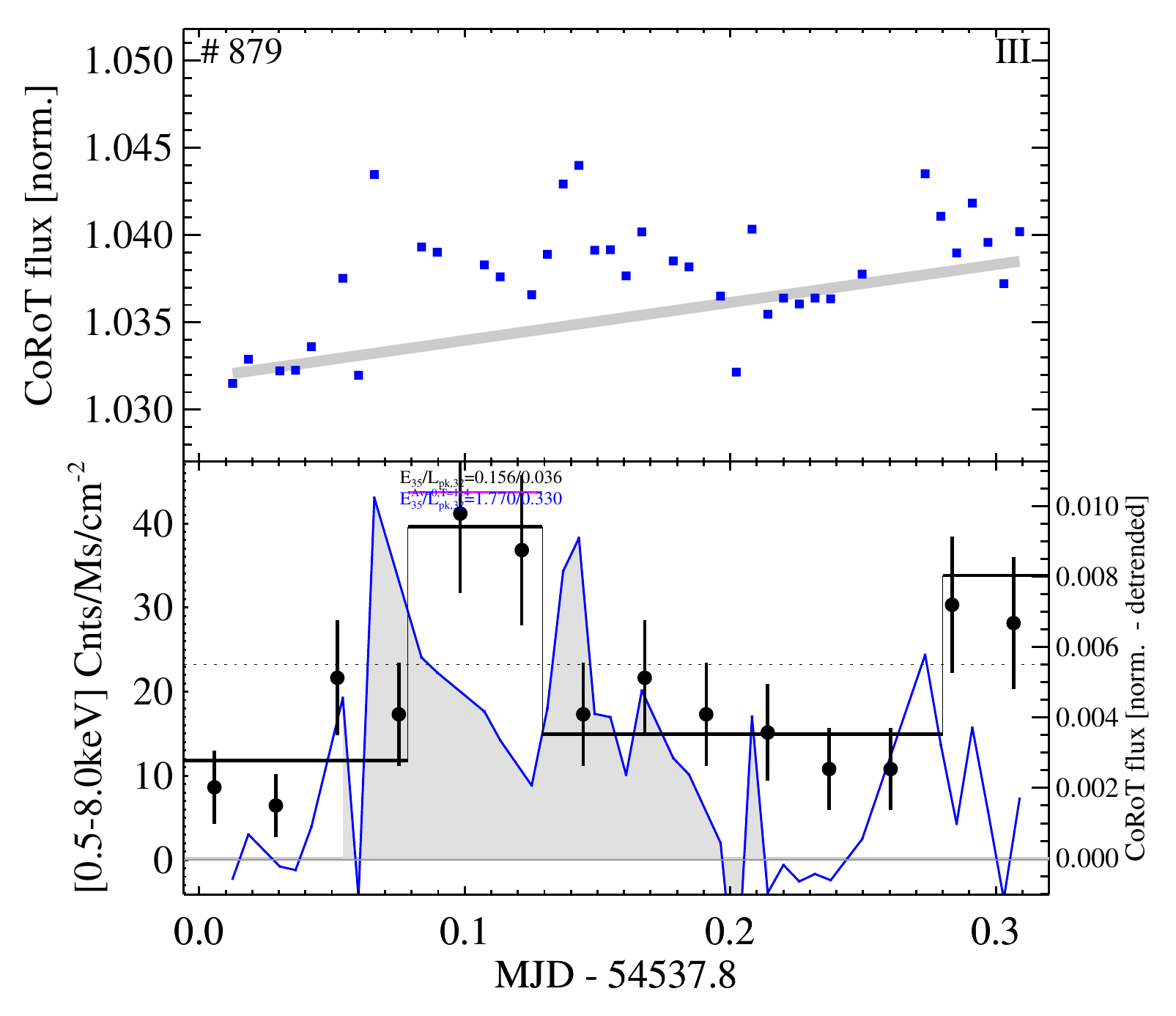}
\includegraphics[width=6.0cm]{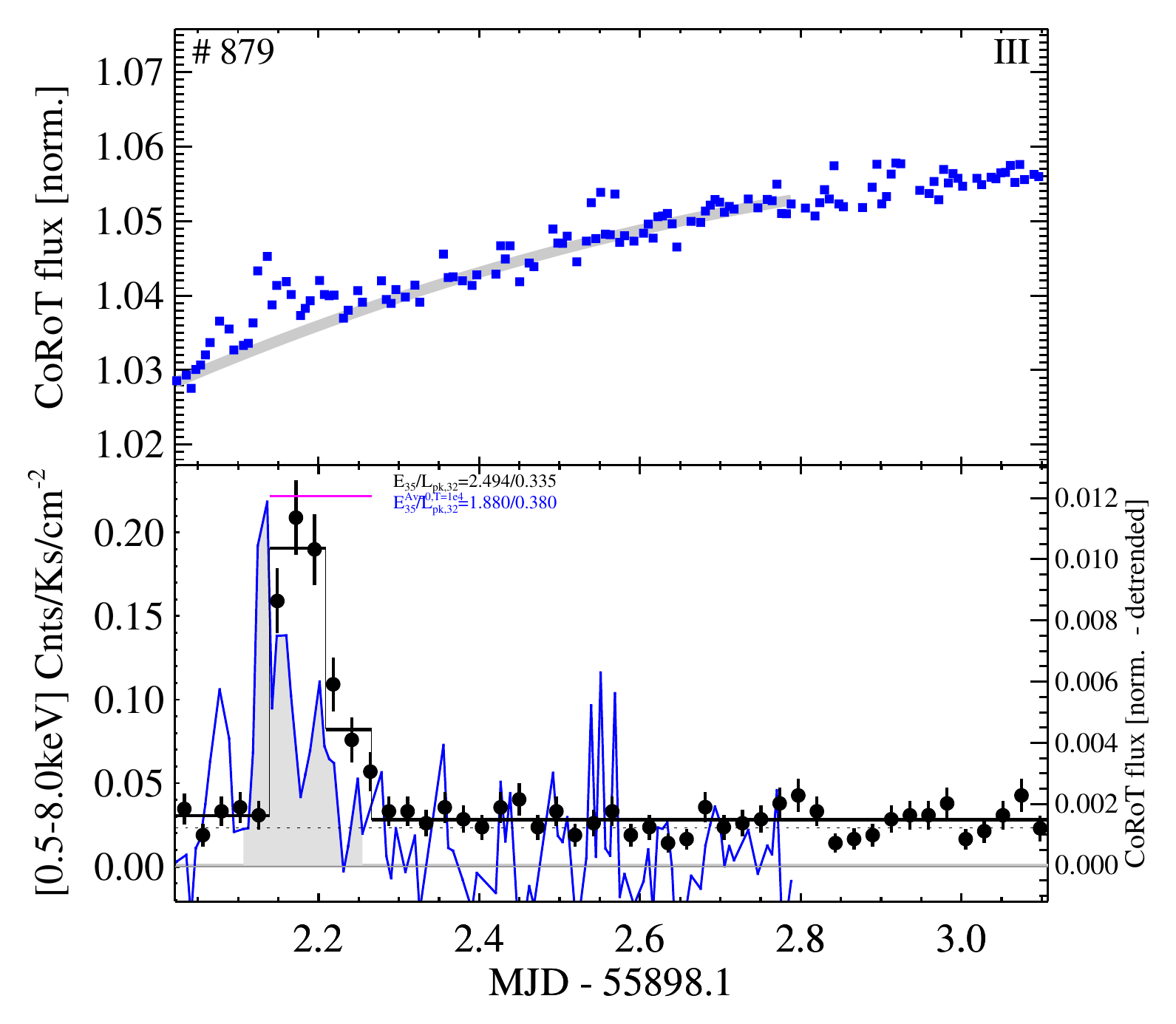}
\includegraphics[width=6.0cm]{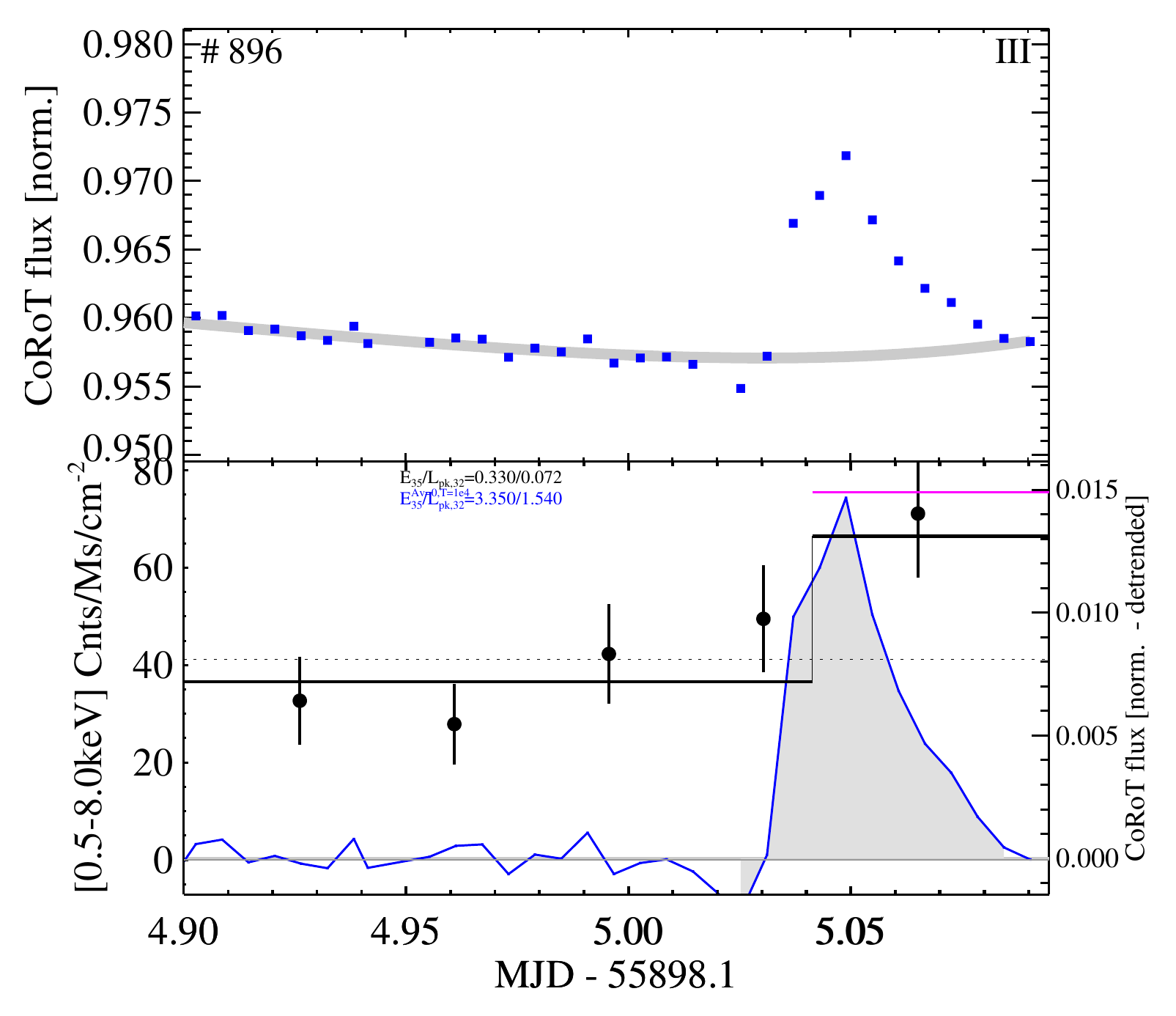}
\includegraphics[width=6.0cm]{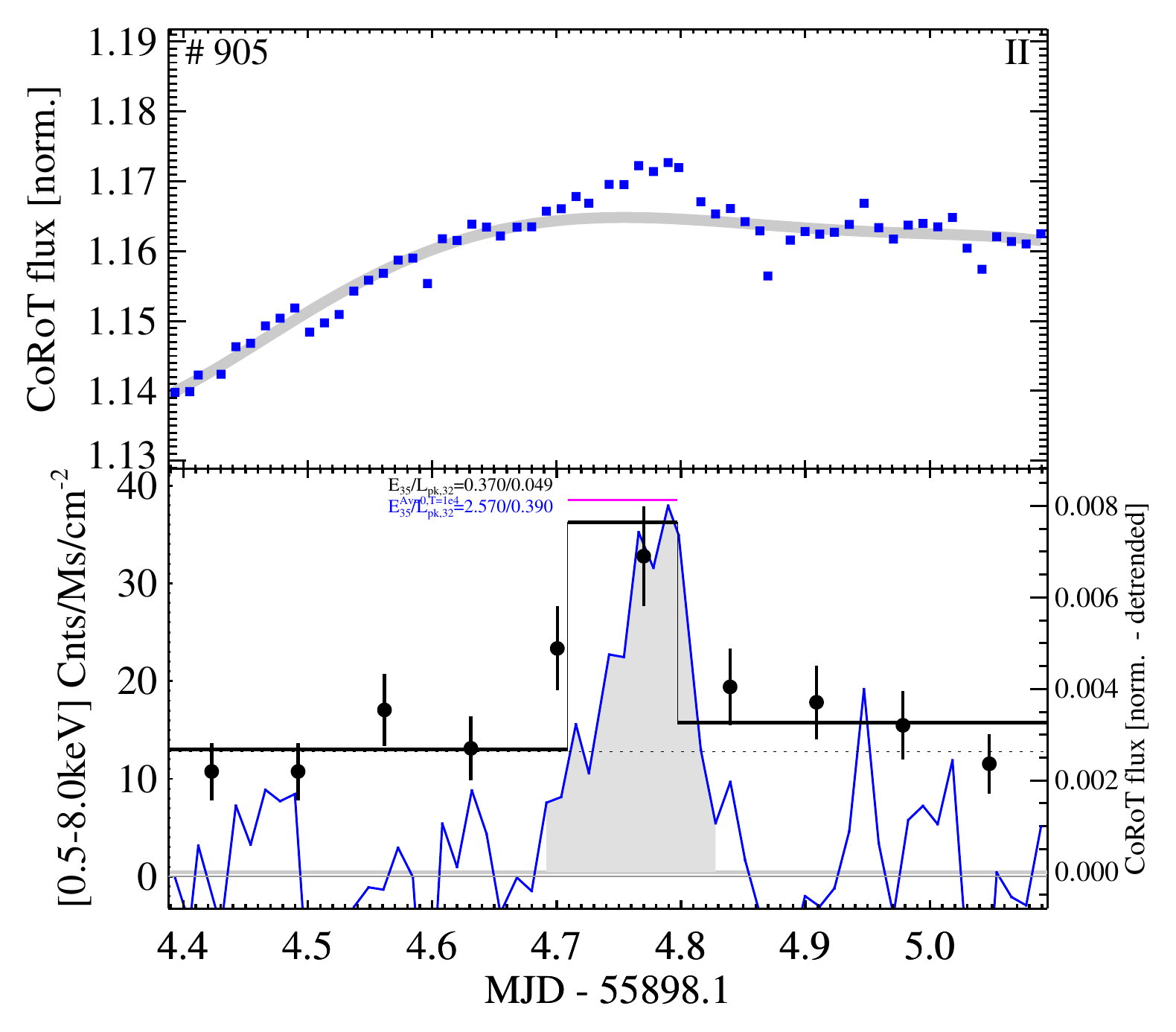}
\includegraphics[width=6.0cm]{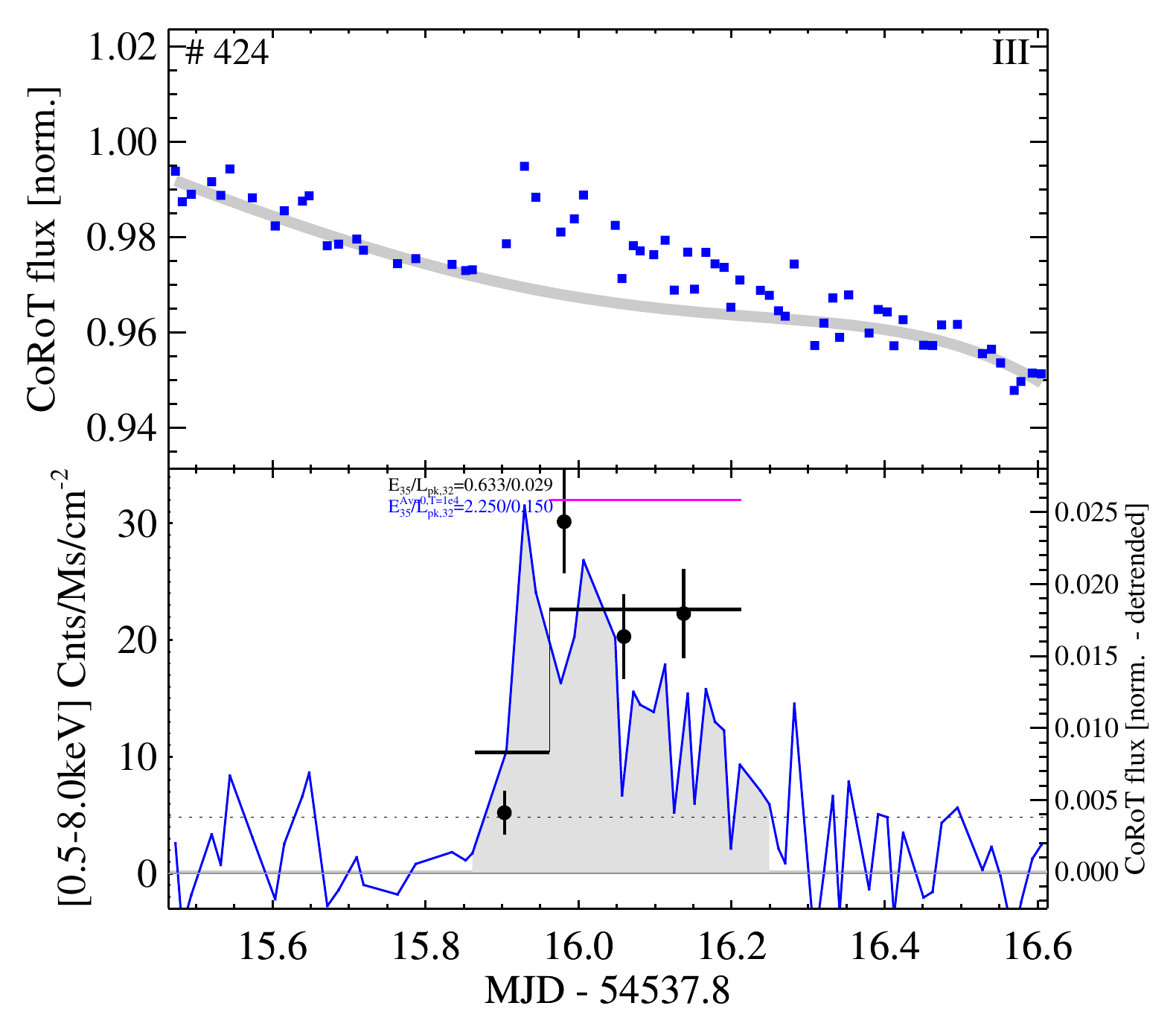}
\includegraphics[width=6.0cm]{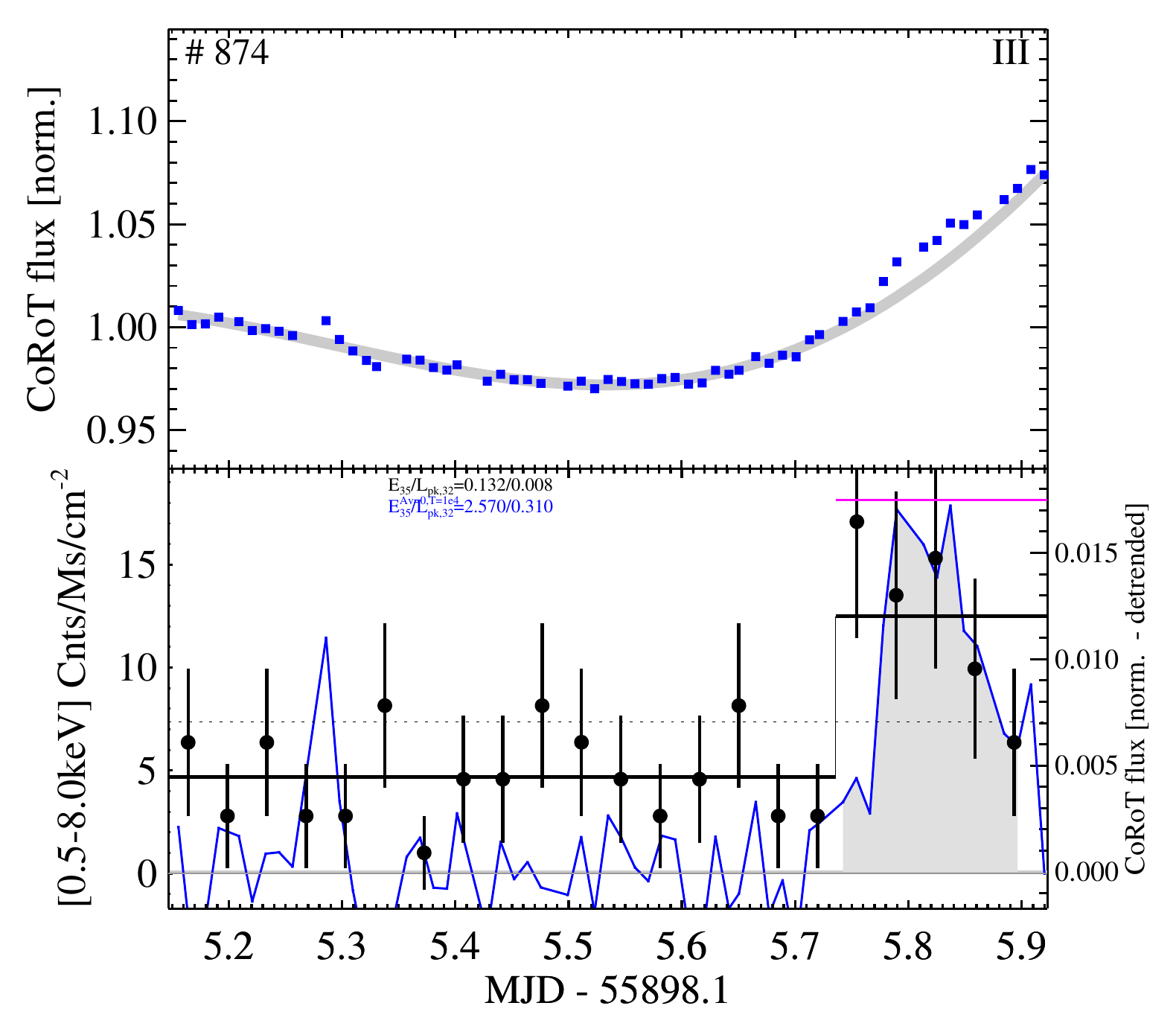}
\includegraphics[width=6.0cm]{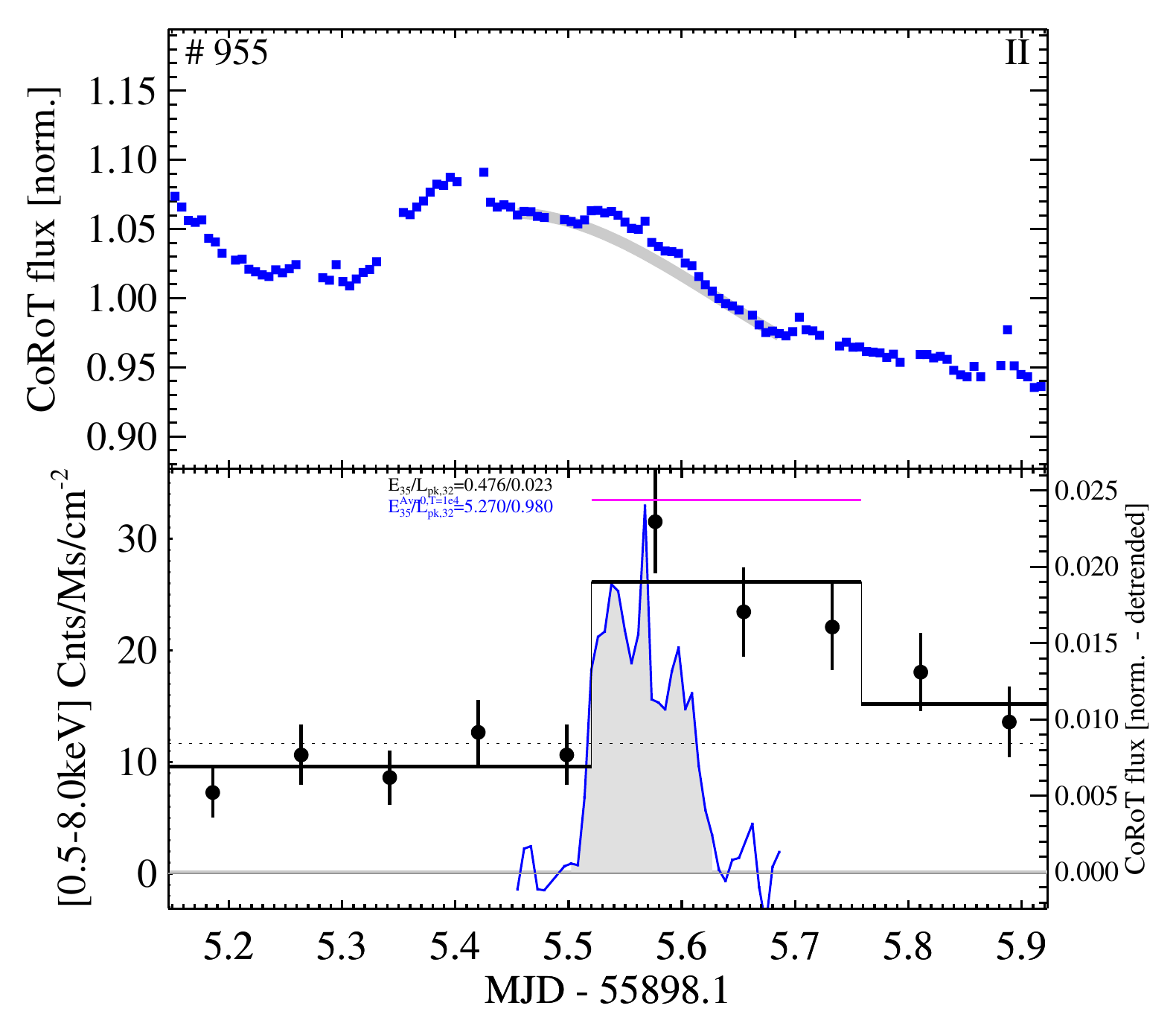}
\caption{(continued)}
\label{fig:}
\end{figure*}

\addtocounter{figure}{-1}

\begin{figure*}[!t!]
\centering
\includegraphics[width=6.0cm]{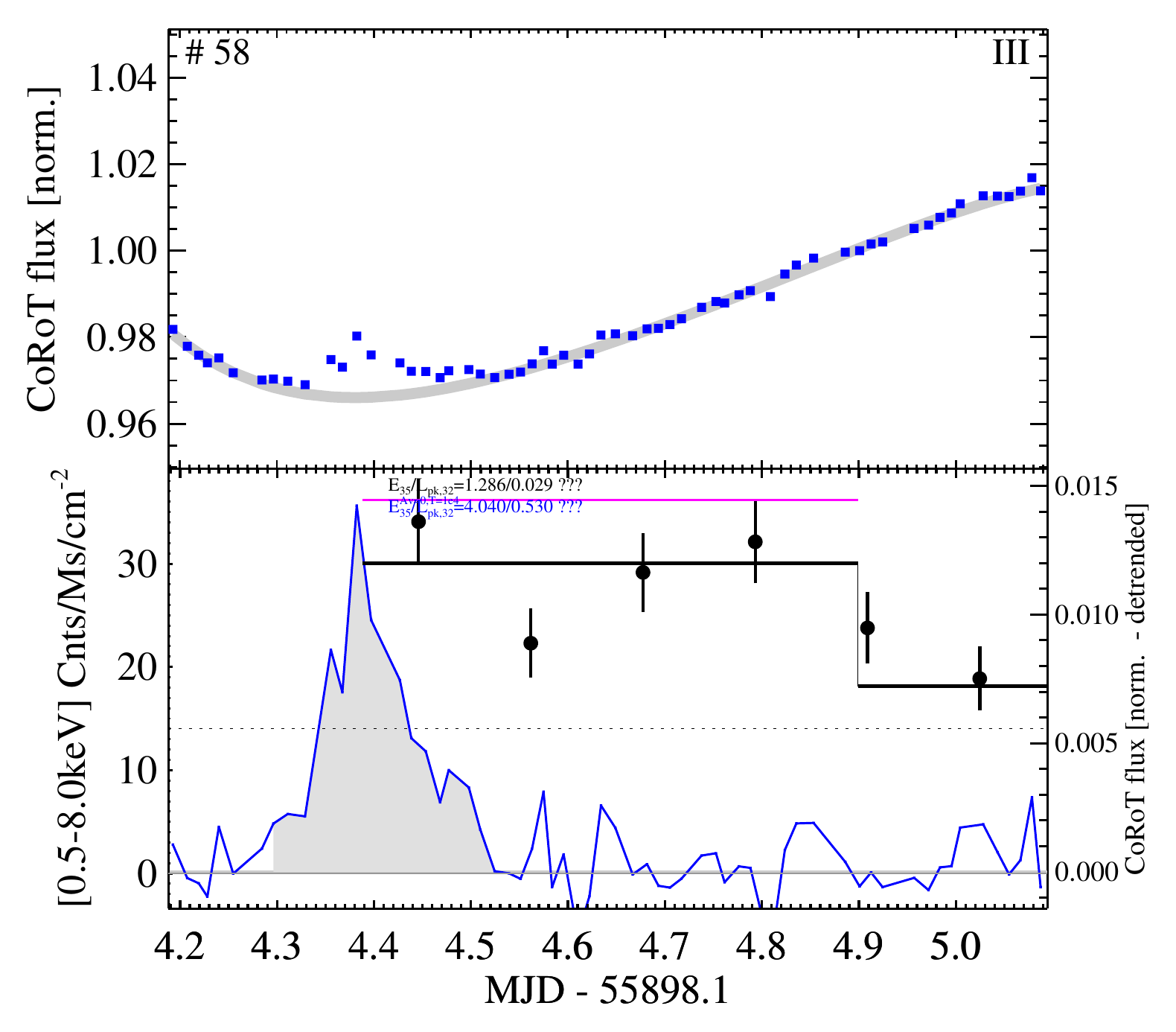}
\includegraphics[width=6.0cm]{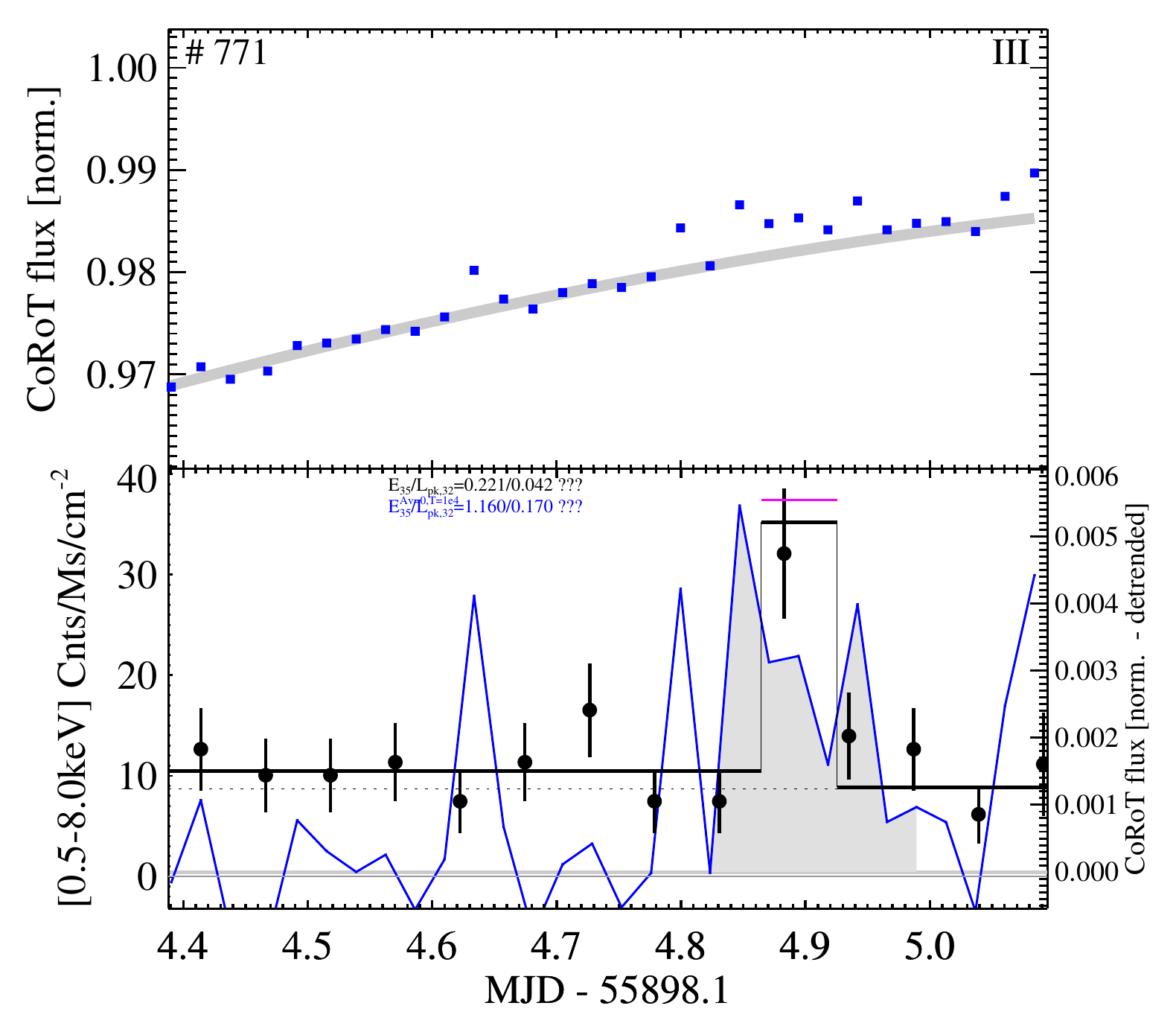}
\includegraphics[width=6.0cm]{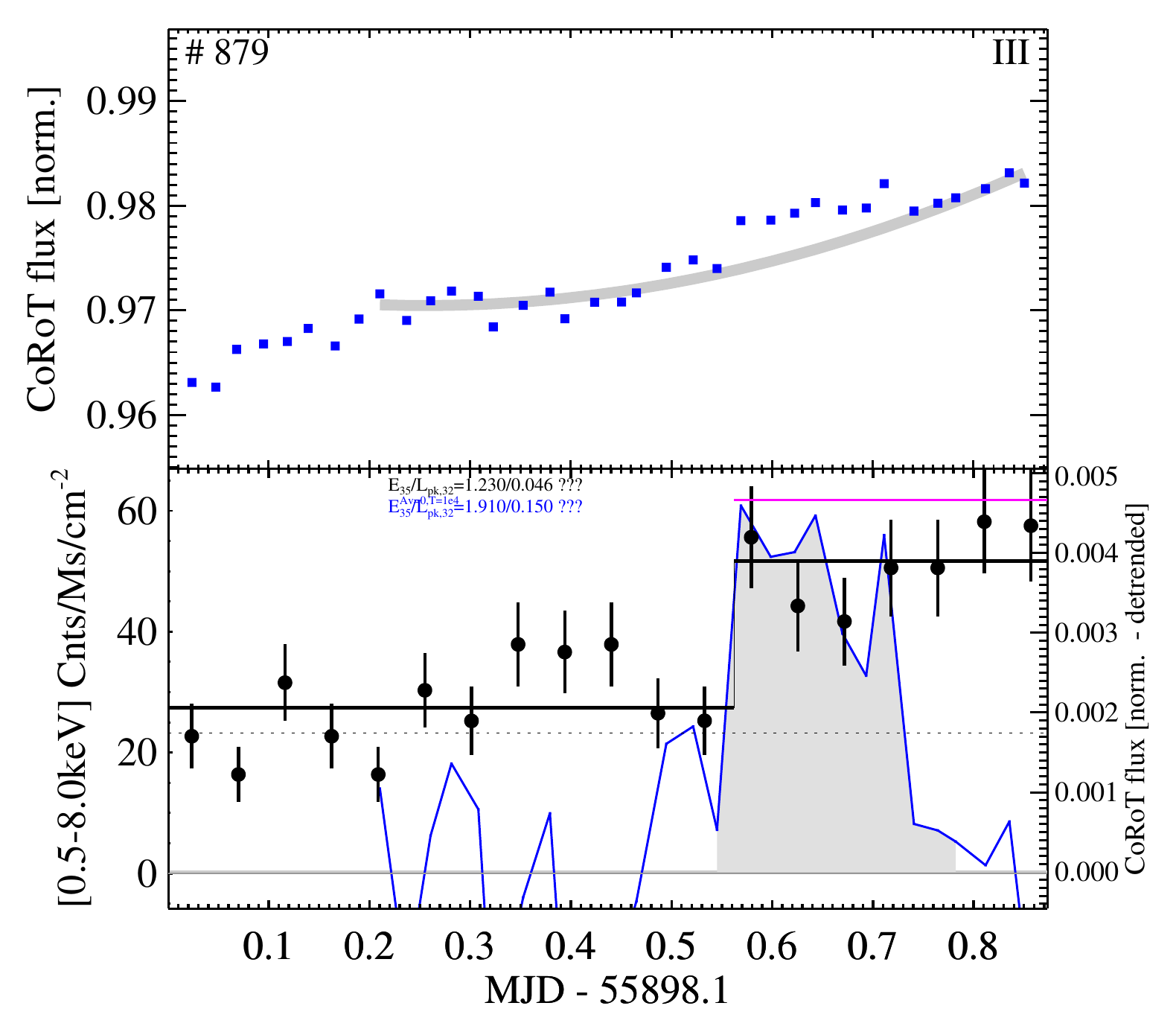}
\includegraphics[width=6.0cm]{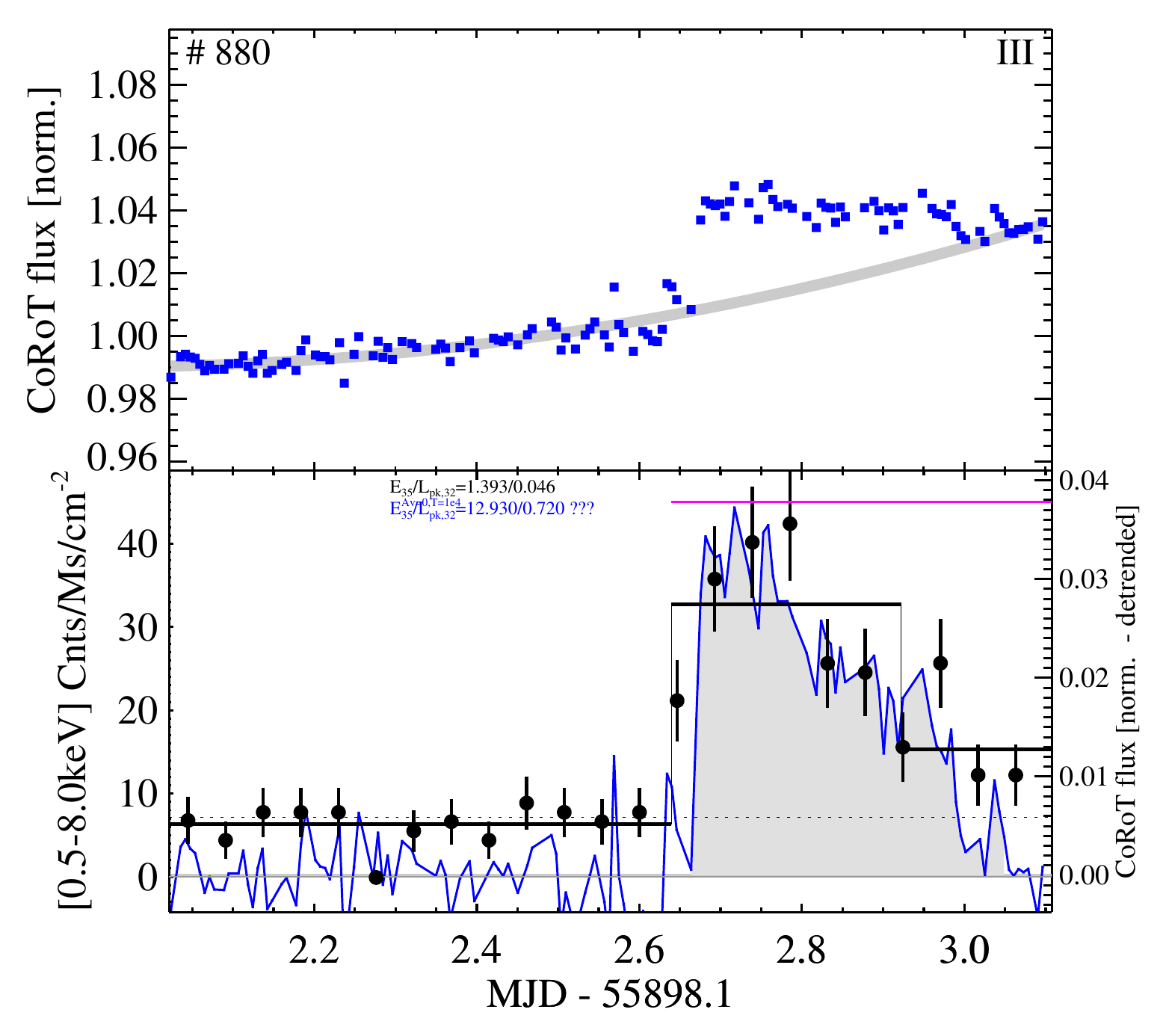}
\includegraphics[width=6.0cm]{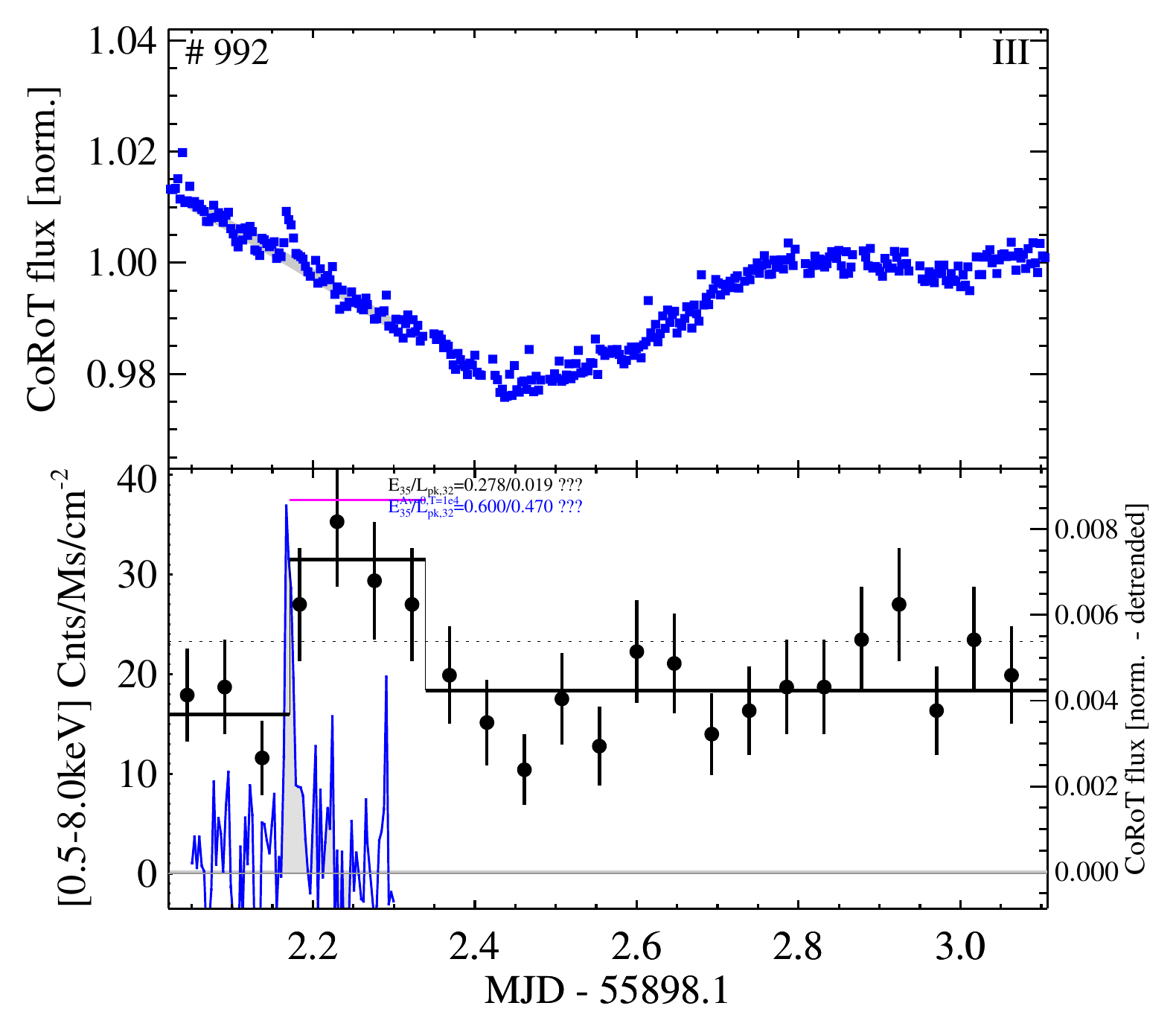}
\includegraphics[width=6.0cm]{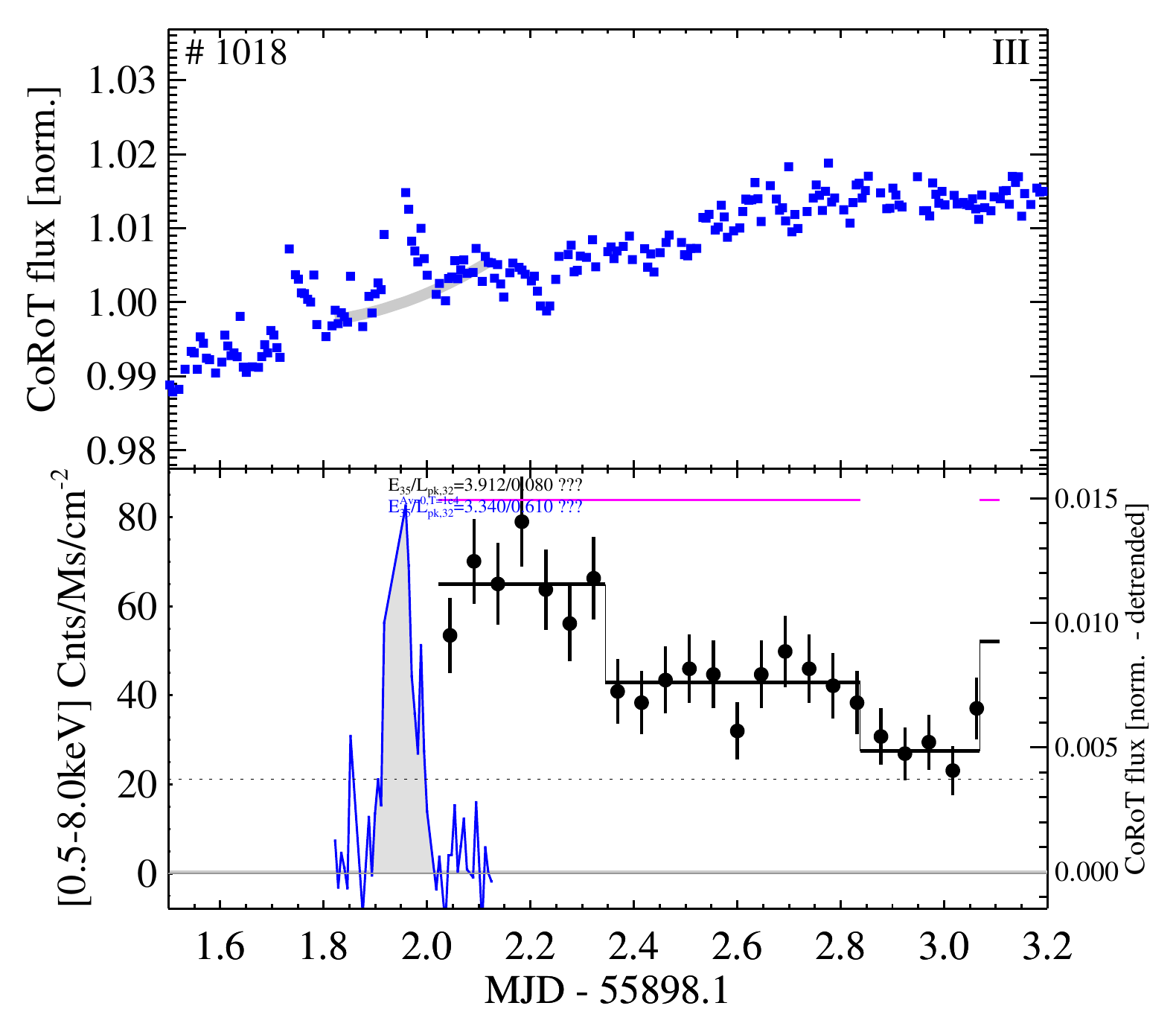}
\includegraphics[width=6.0cm]{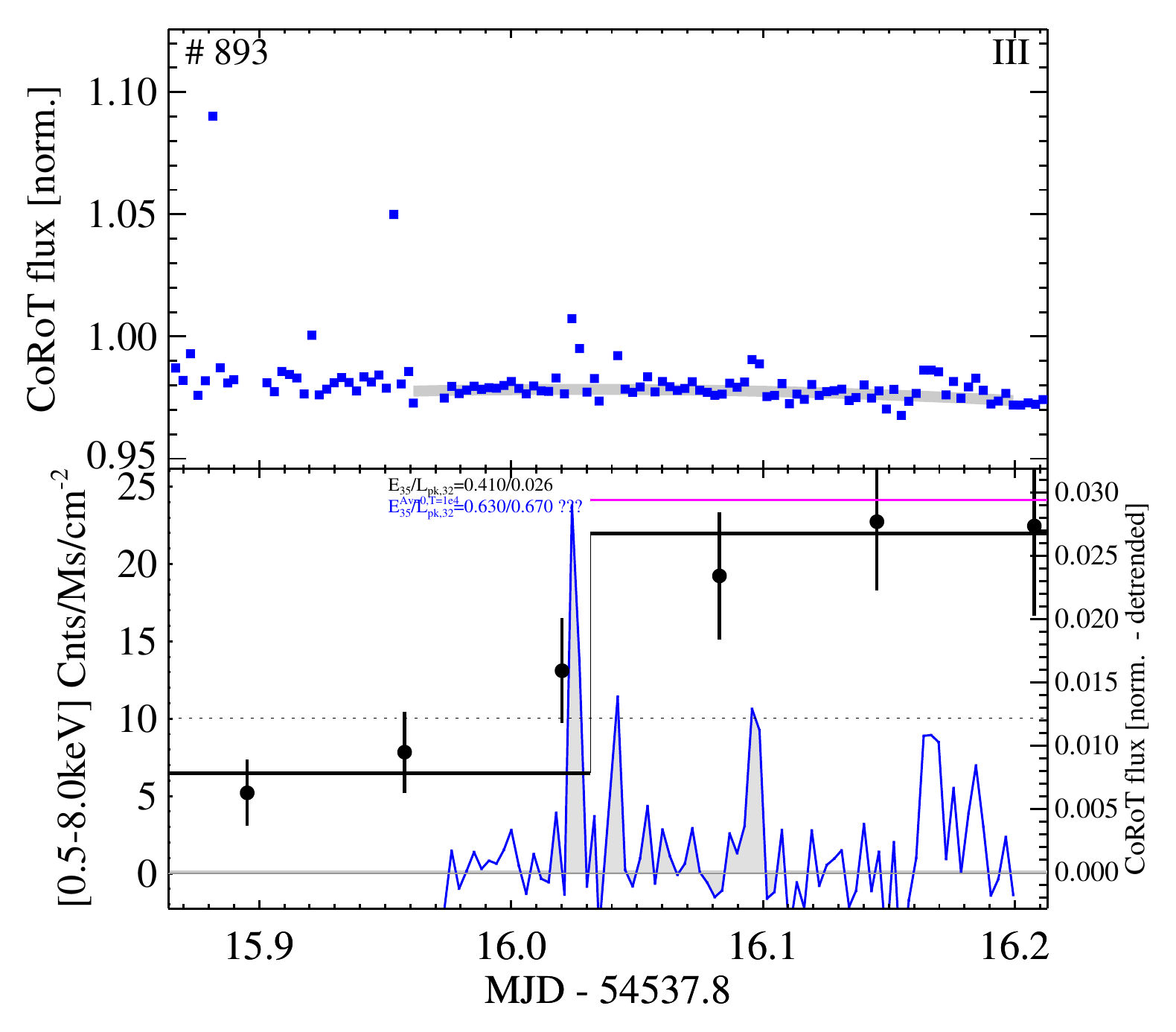}
\caption{(continued)}
\label{fig:}
\end{figure*}

\addtocounter{figure}{-1}

\begin{figure*}[!t!]
\centering
\includegraphics[width=6.0cm]{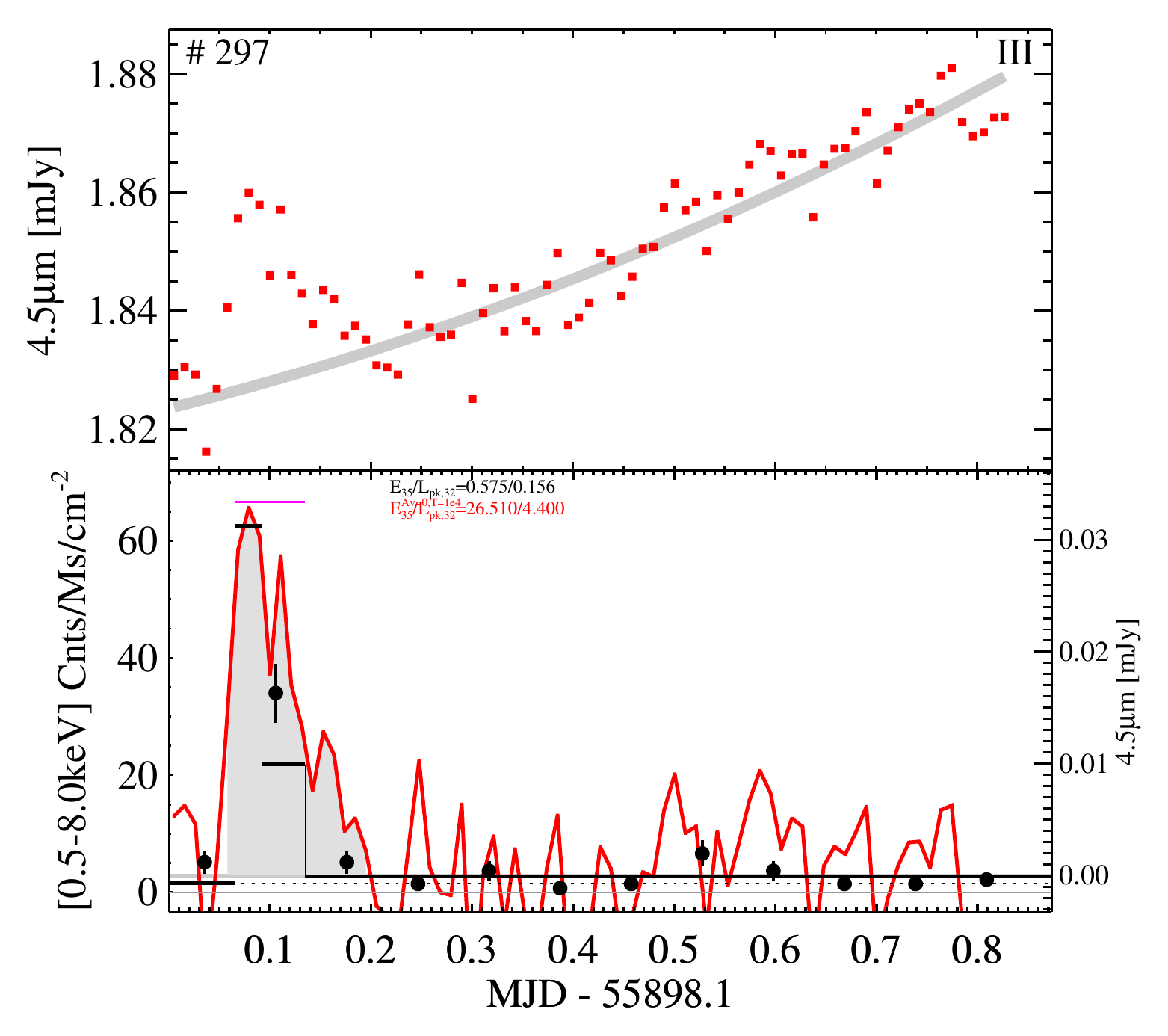}
\includegraphics[width=6.0cm]{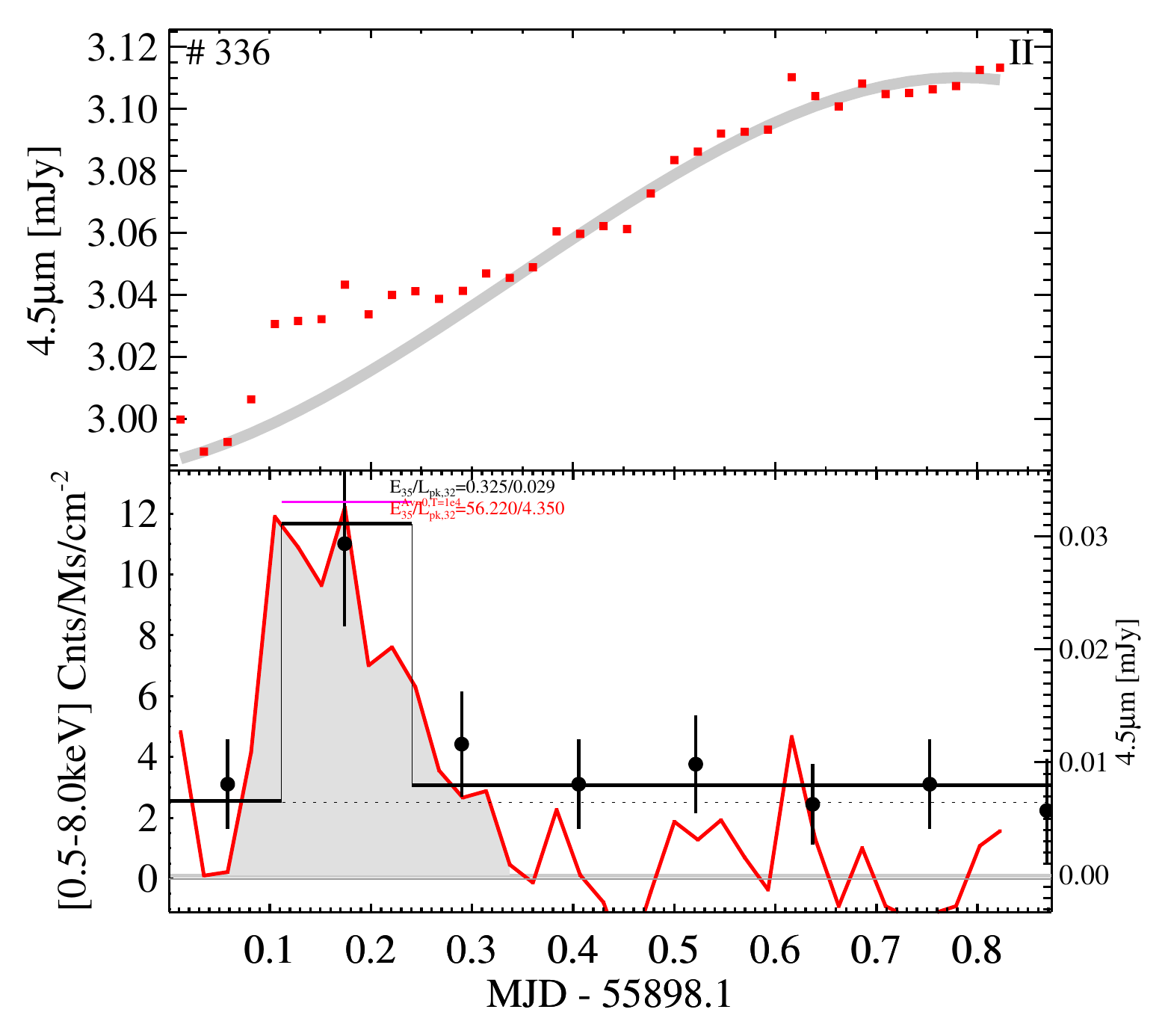}
\includegraphics[width=6.0cm]{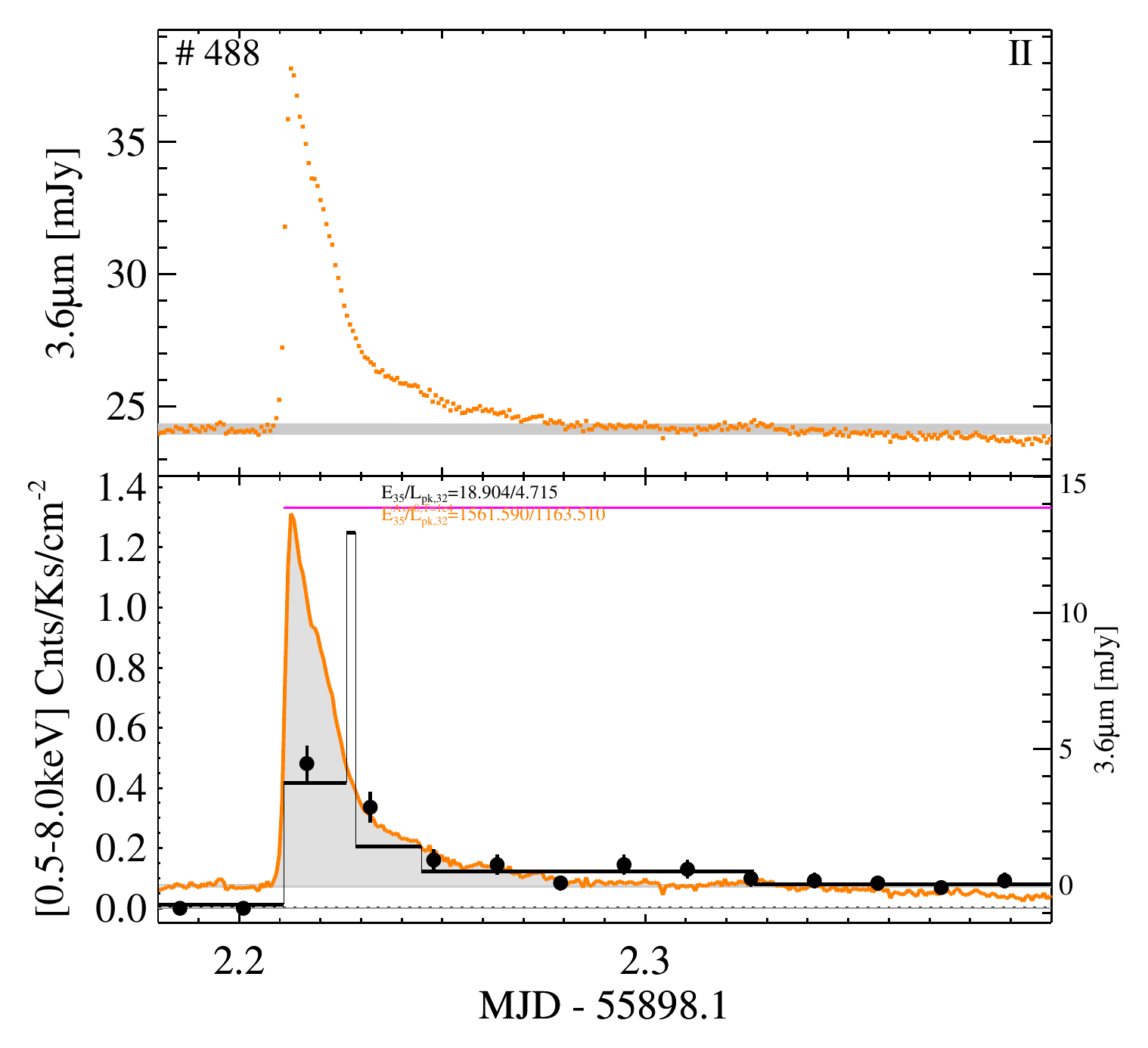}
\includegraphics[width=6.0cm]{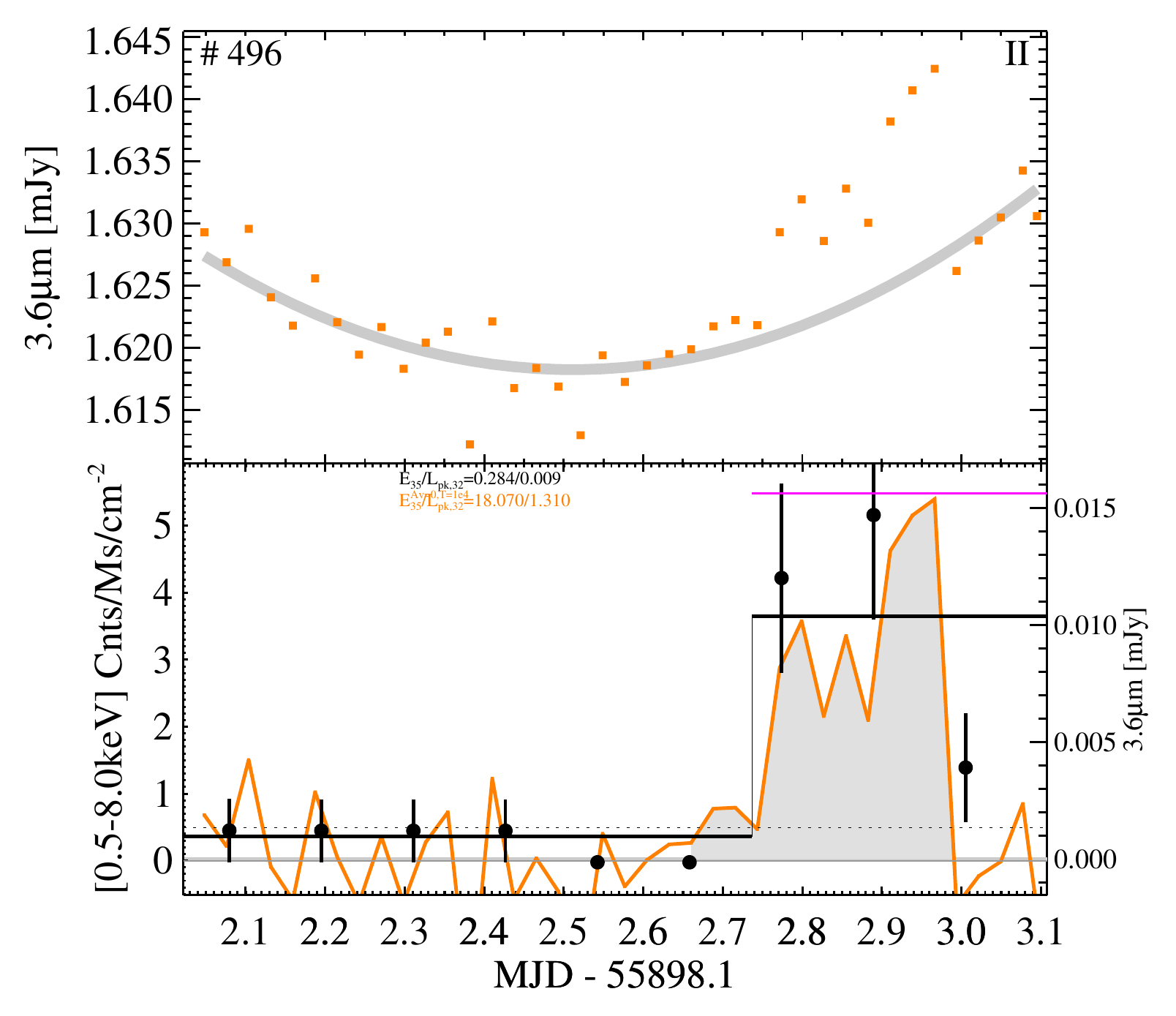}
\includegraphics[width=6.0cm]{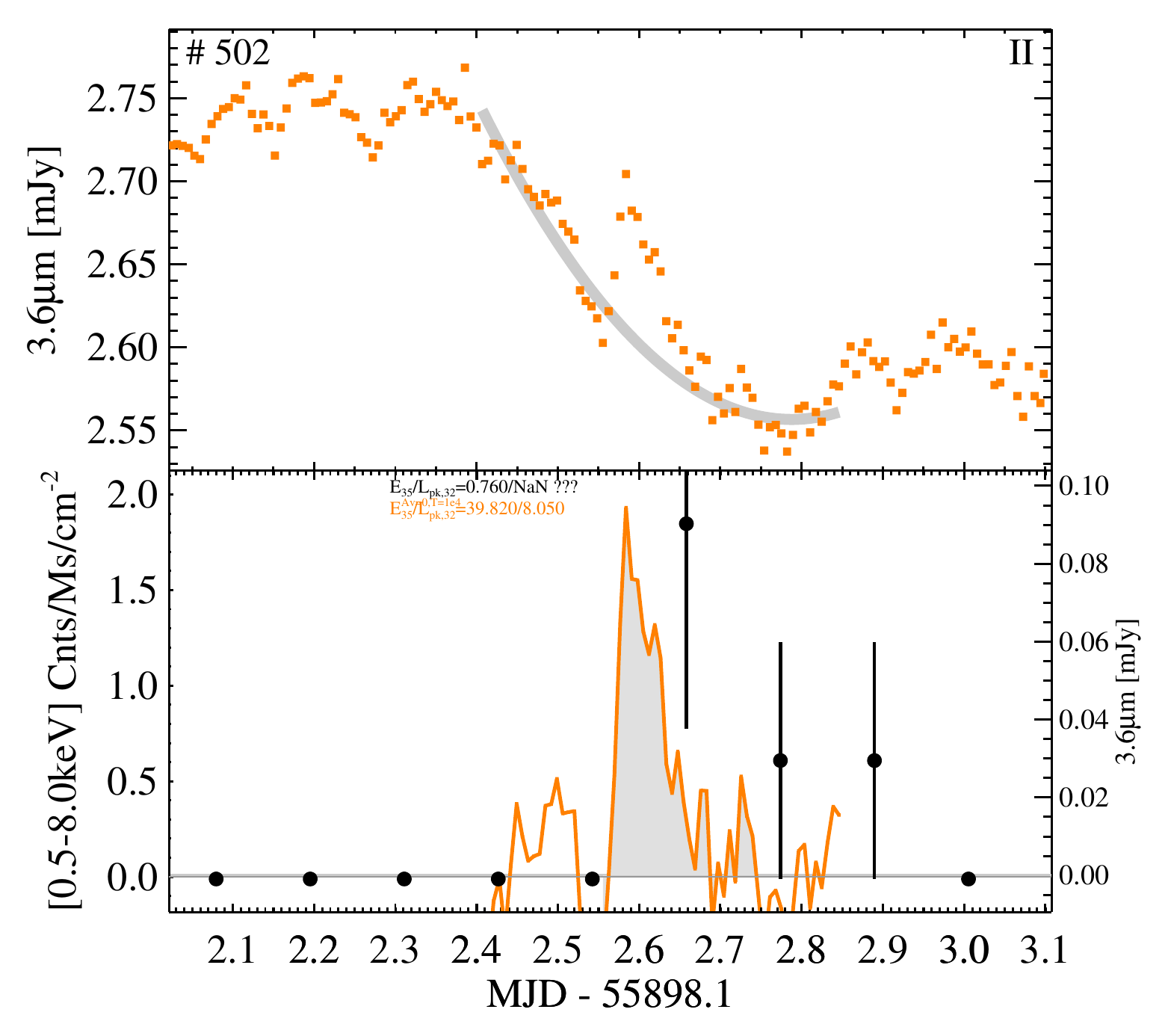}
\includegraphics[width=6.0cm]{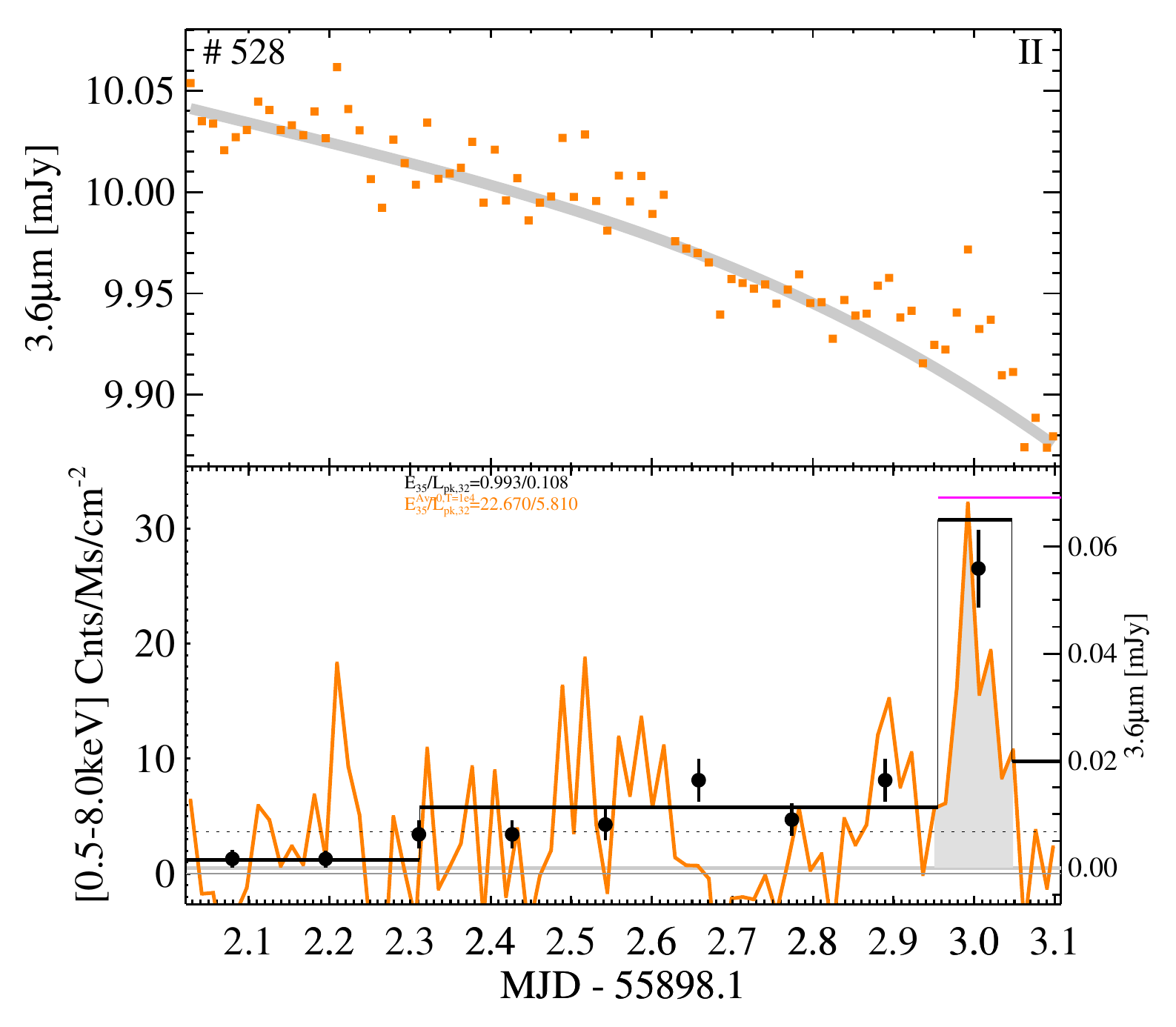}
\includegraphics[width=6.0cm]{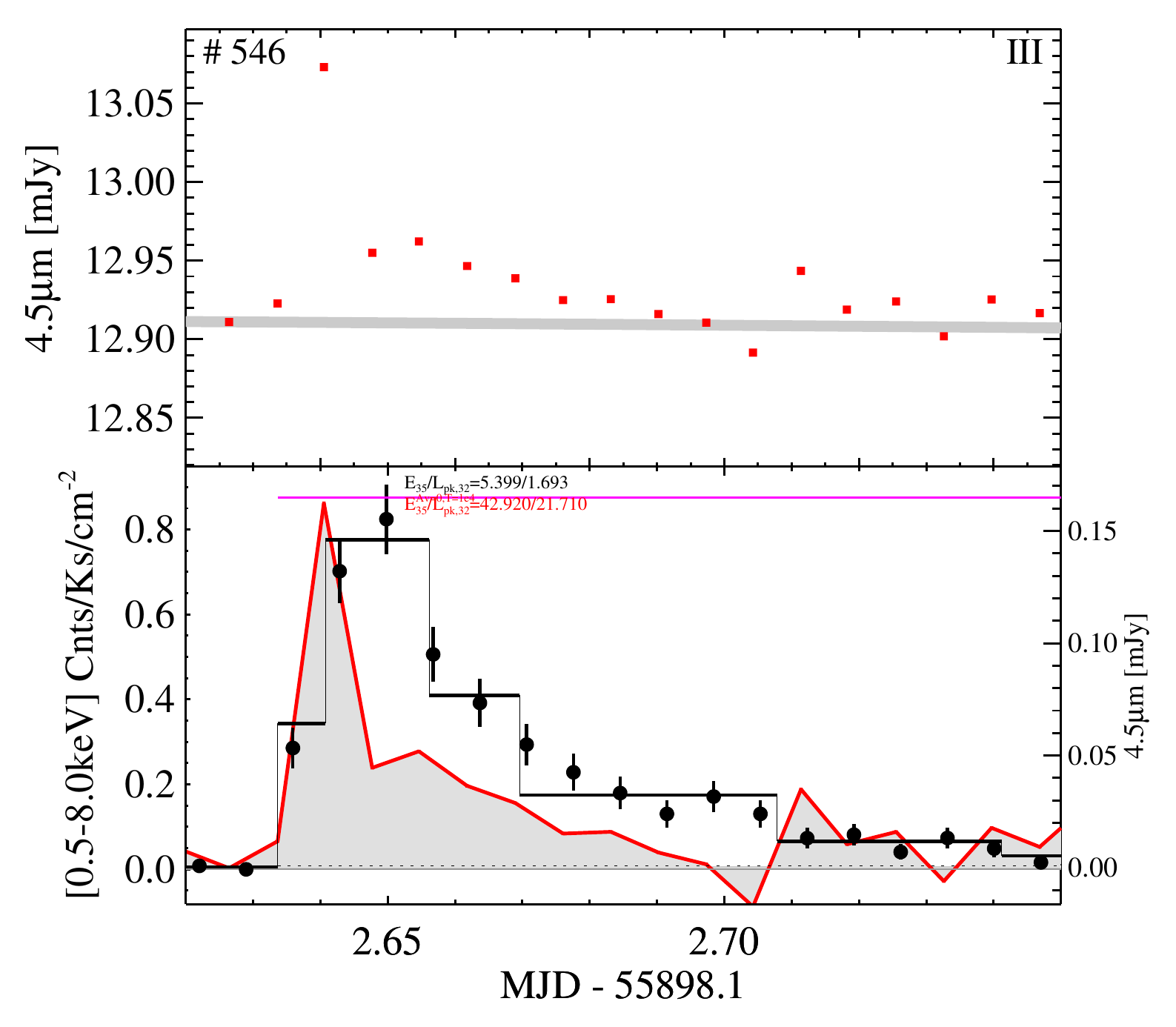}
\includegraphics[width=6.0cm]{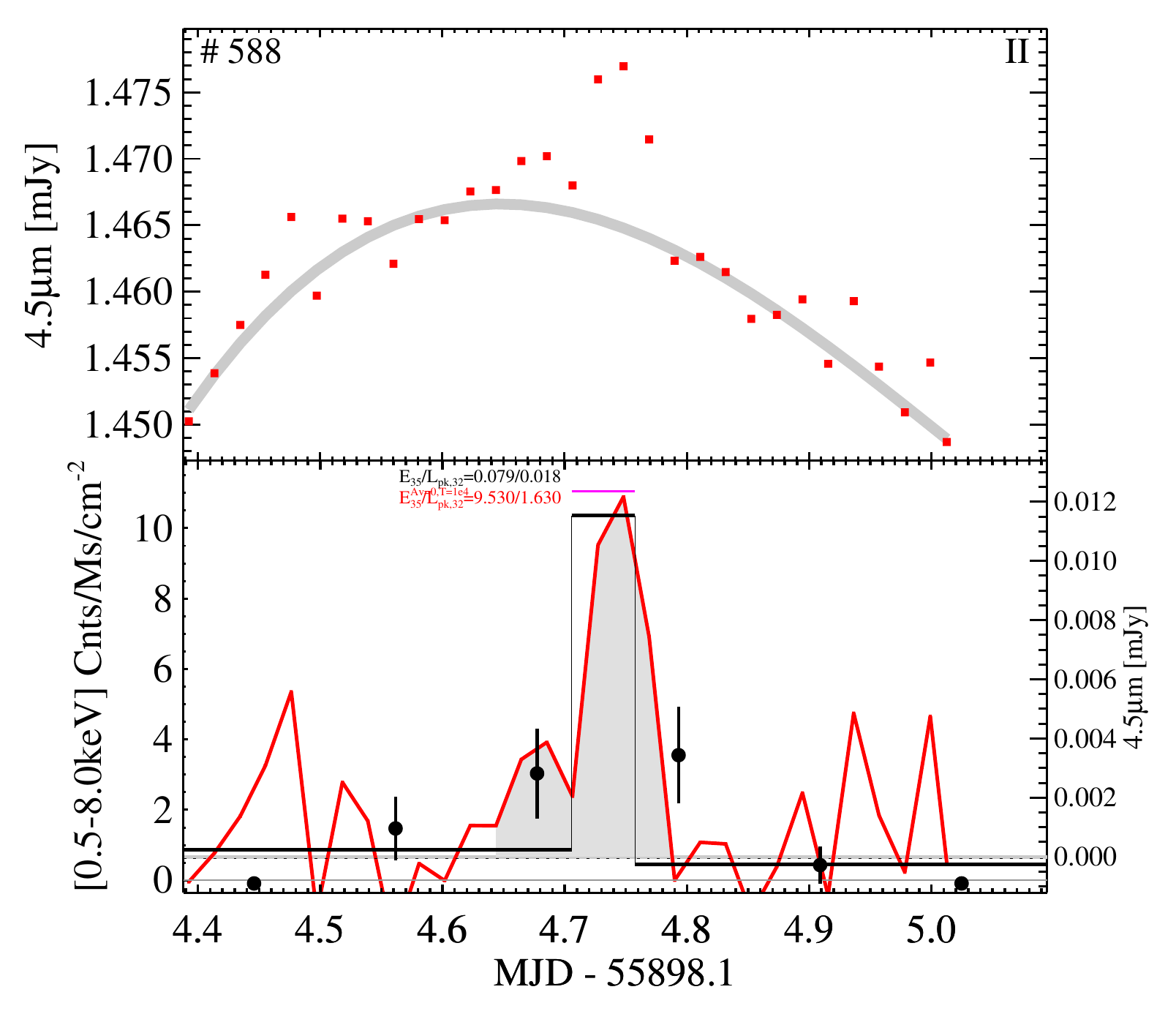}
\includegraphics[width=6.0cm]{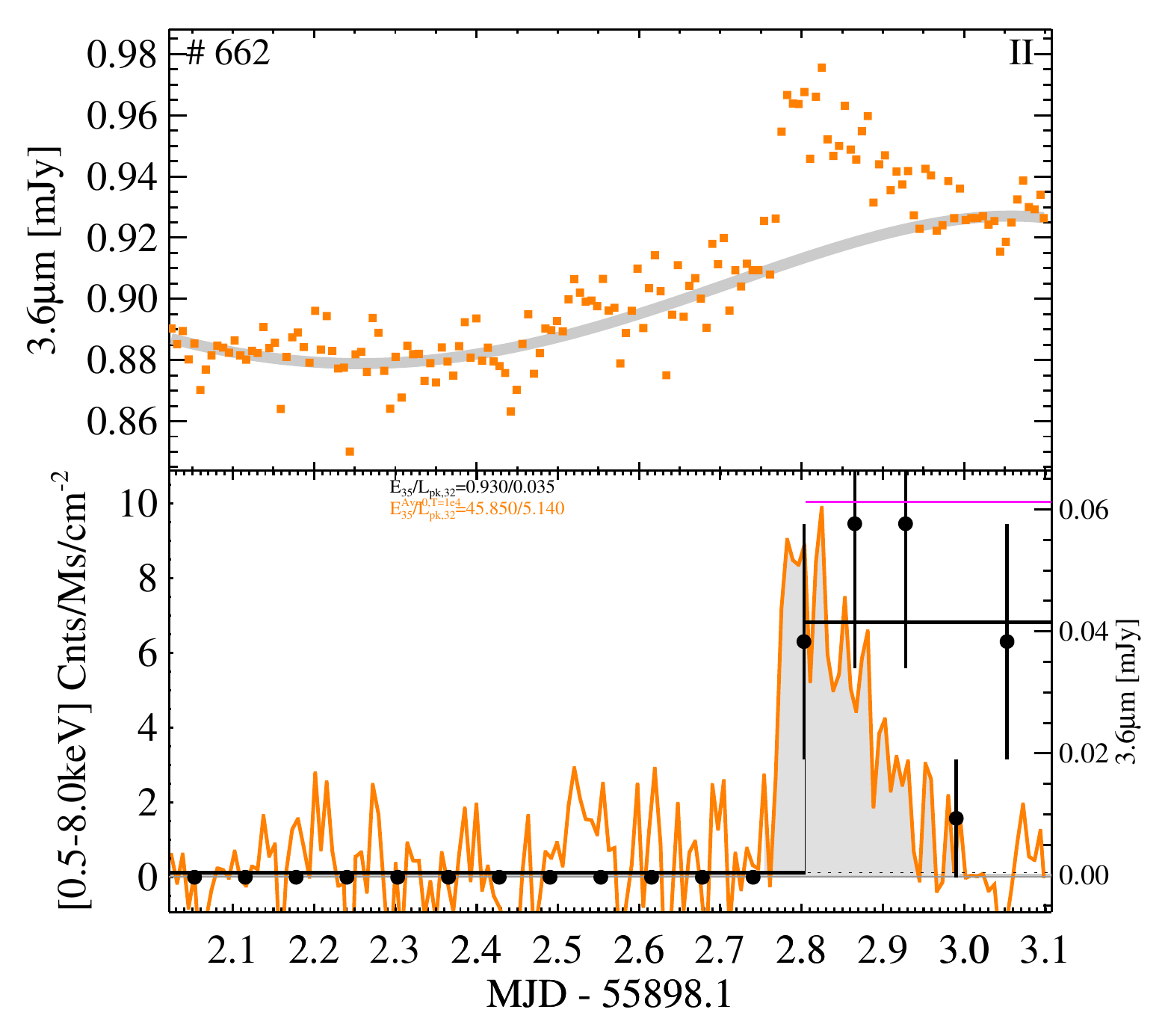}
\includegraphics[width=6.0cm]{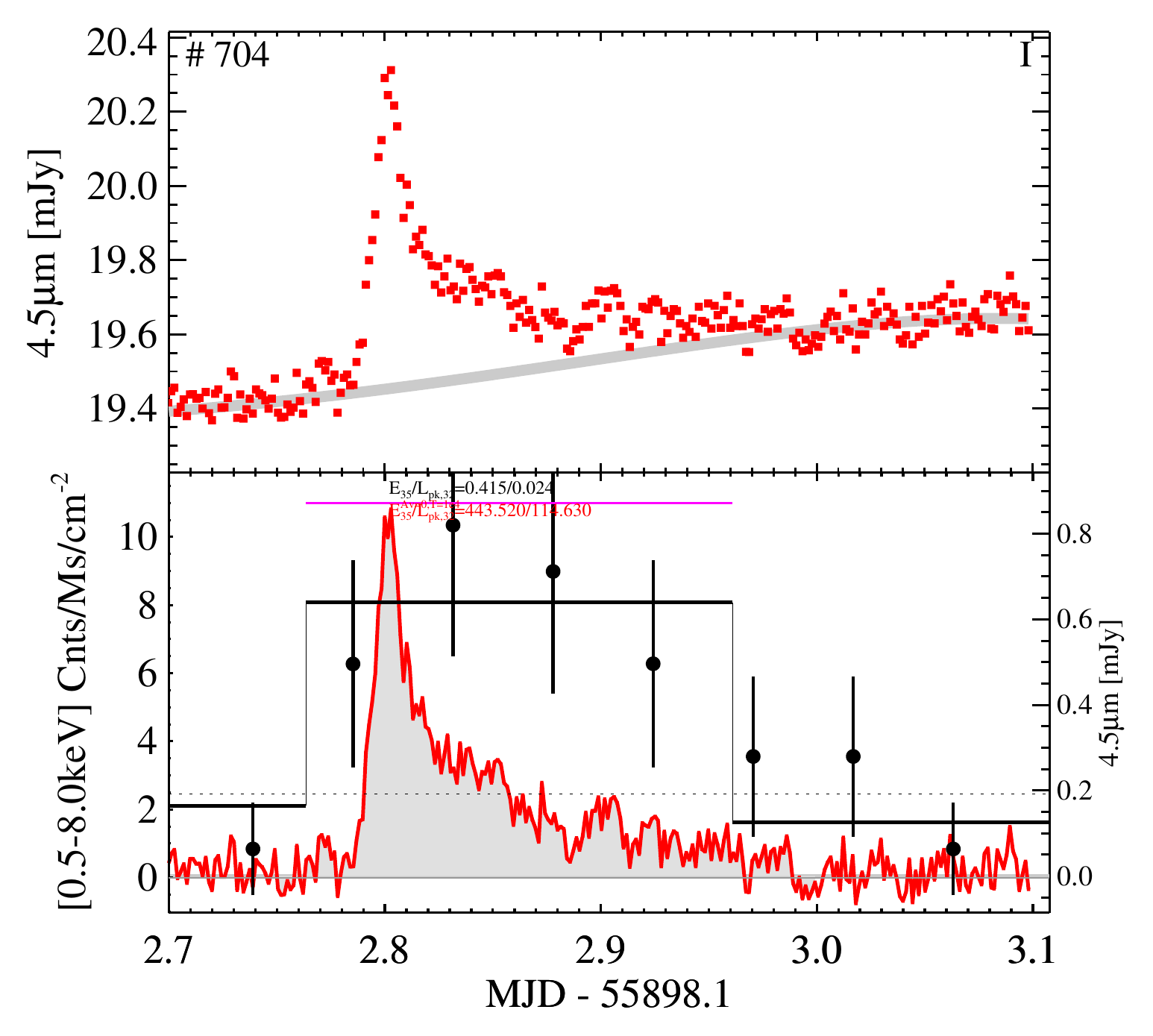}
\includegraphics[width=6.0cm]{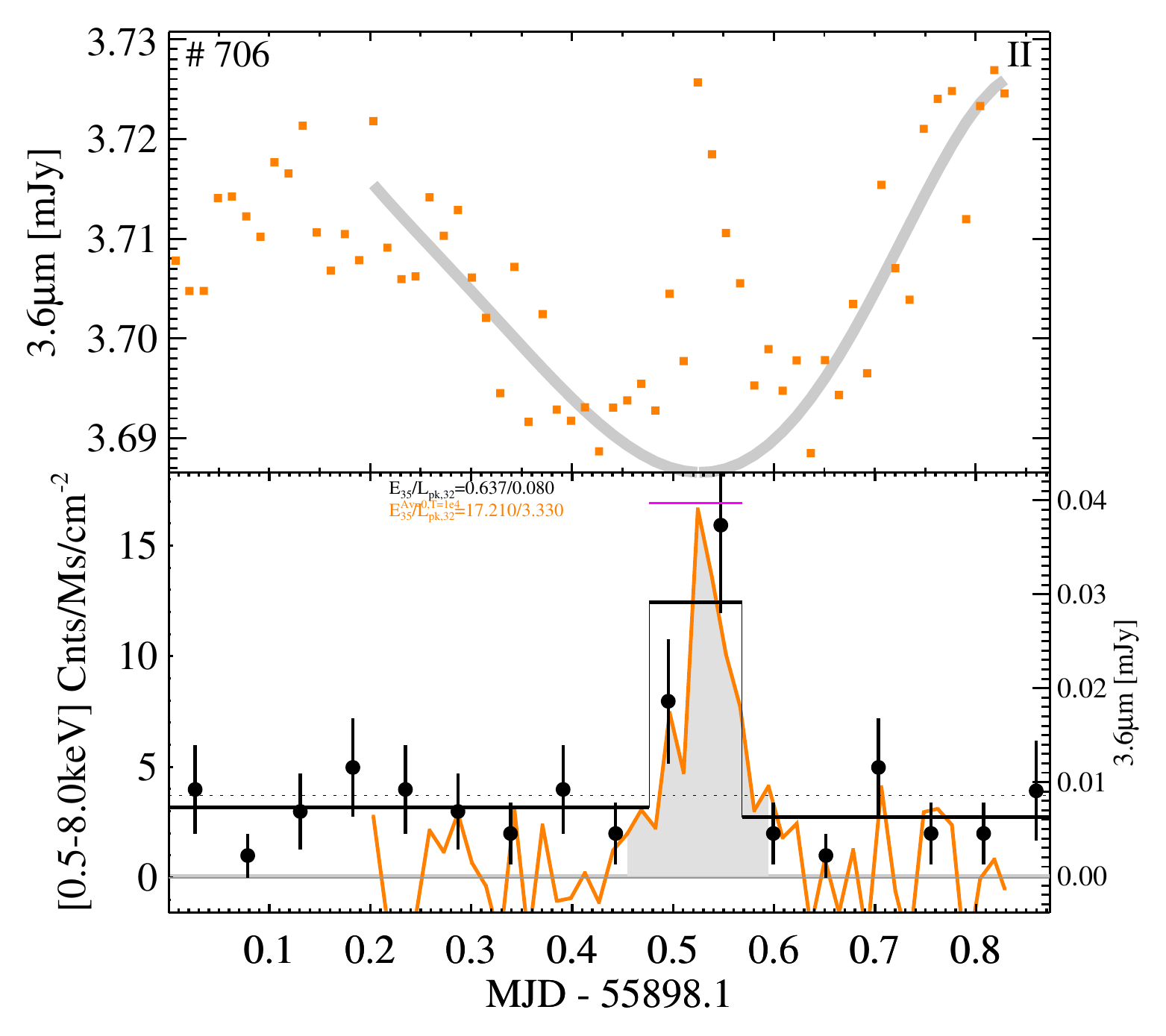}
\caption{(continued)}
\label{fig:}
\end{figure*}

\addtocounter{figure}{-1}

\begin{figure*}[!t!]
\centering
\includegraphics[width=6.0cm]{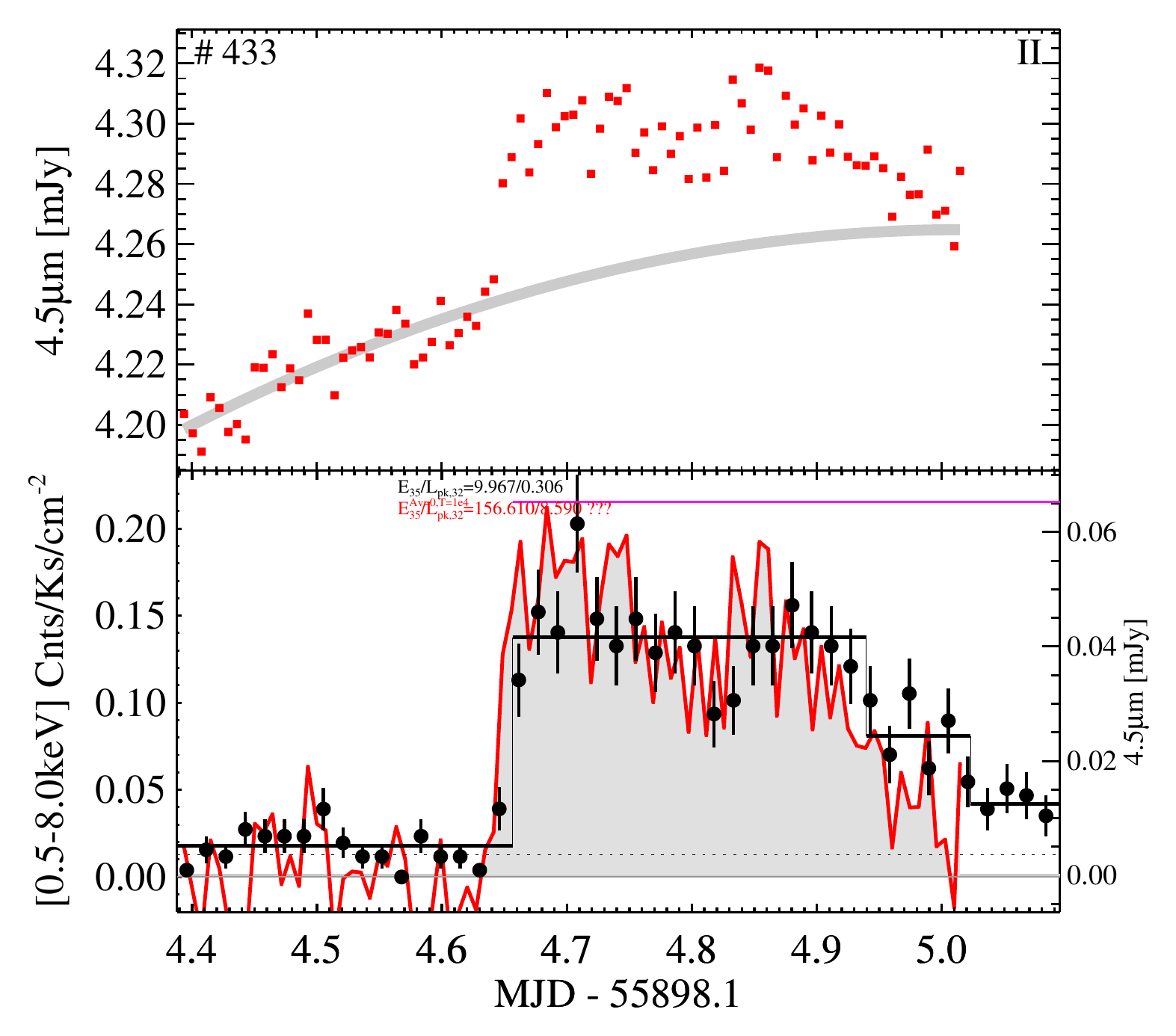}
\includegraphics[width=6.0cm]{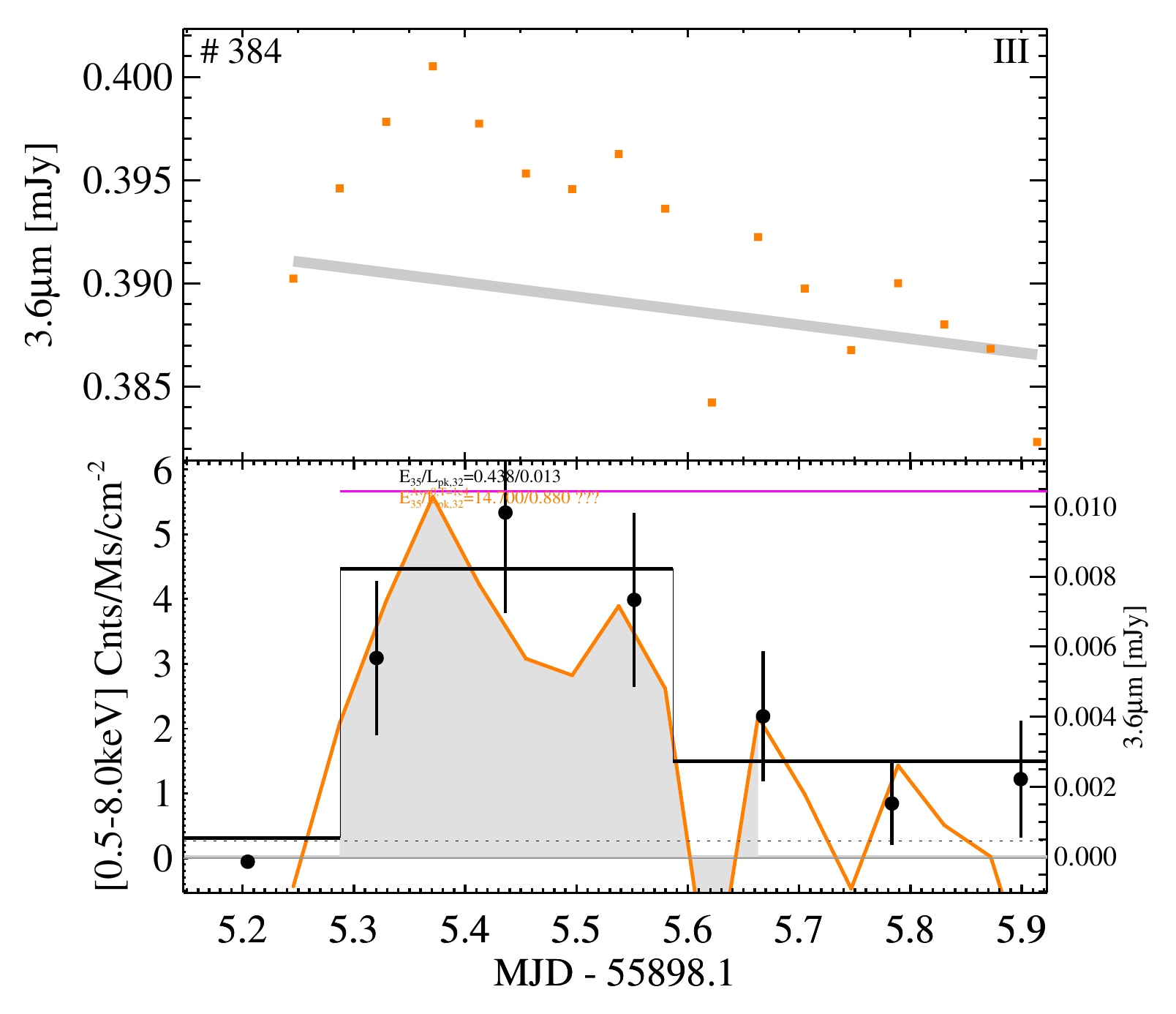}
\includegraphics[width=6.0cm]{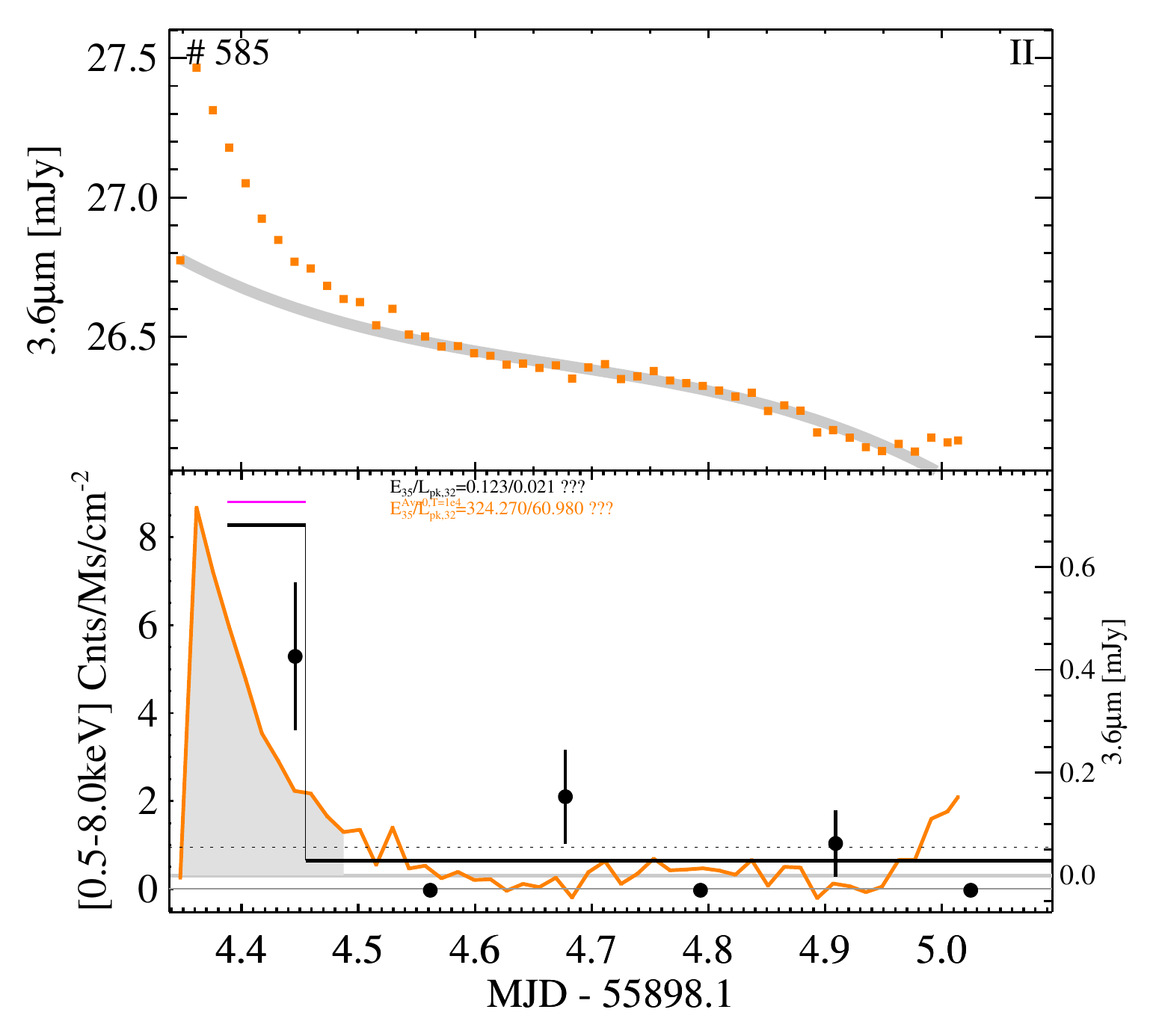}
\includegraphics[width=6.0cm]{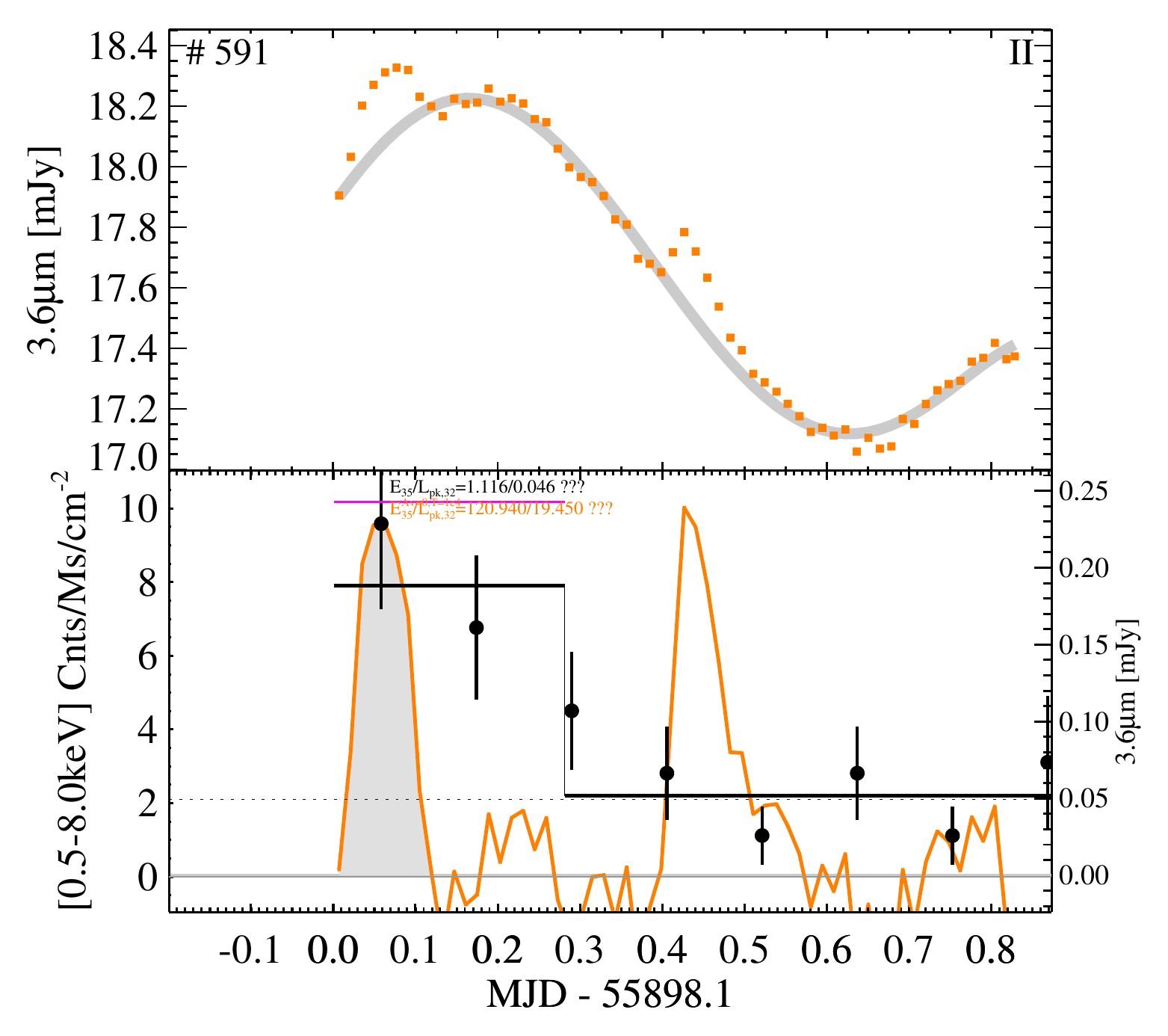}
\includegraphics[width=6.0cm]{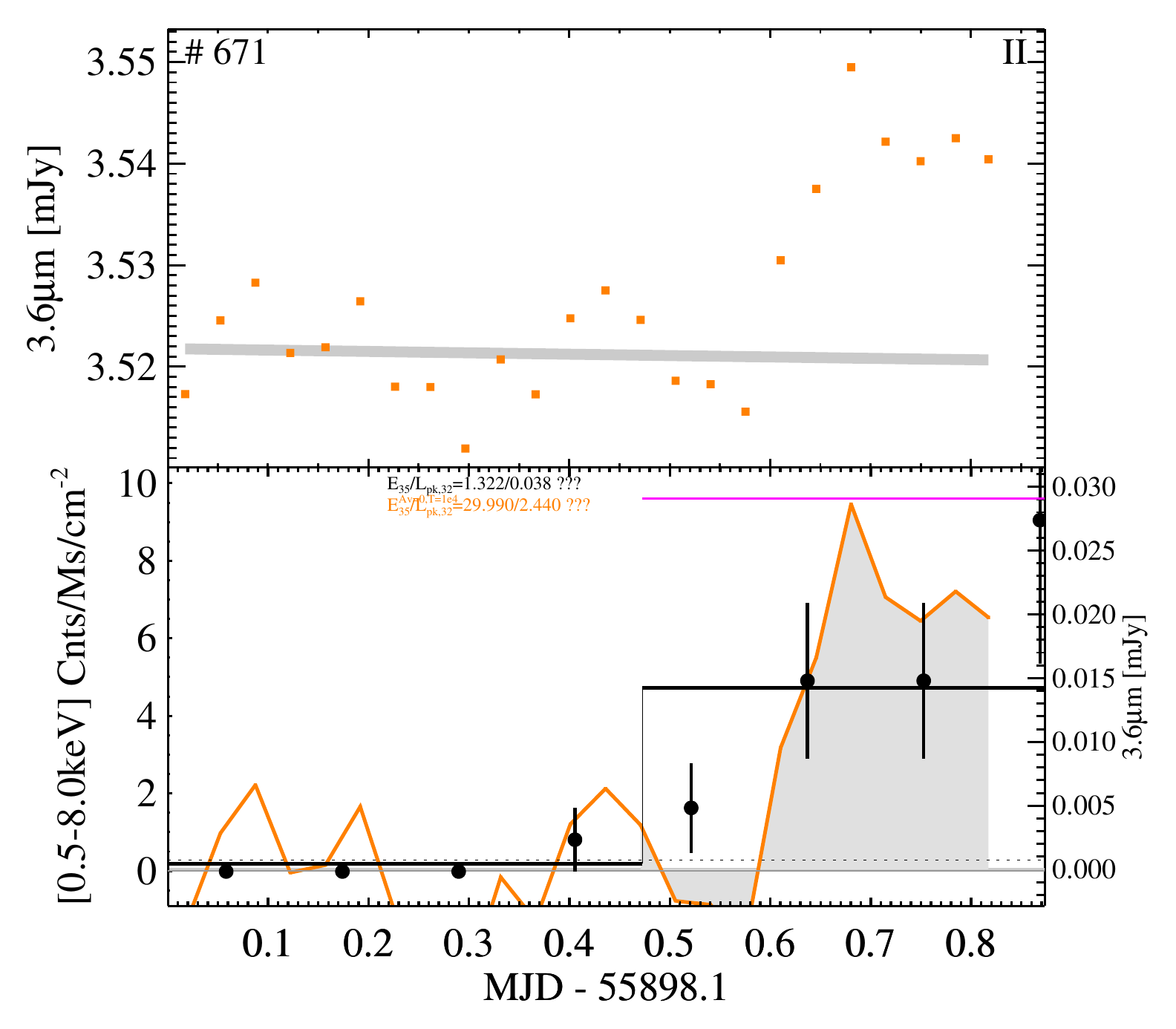}
\includegraphics[width=6.0cm]{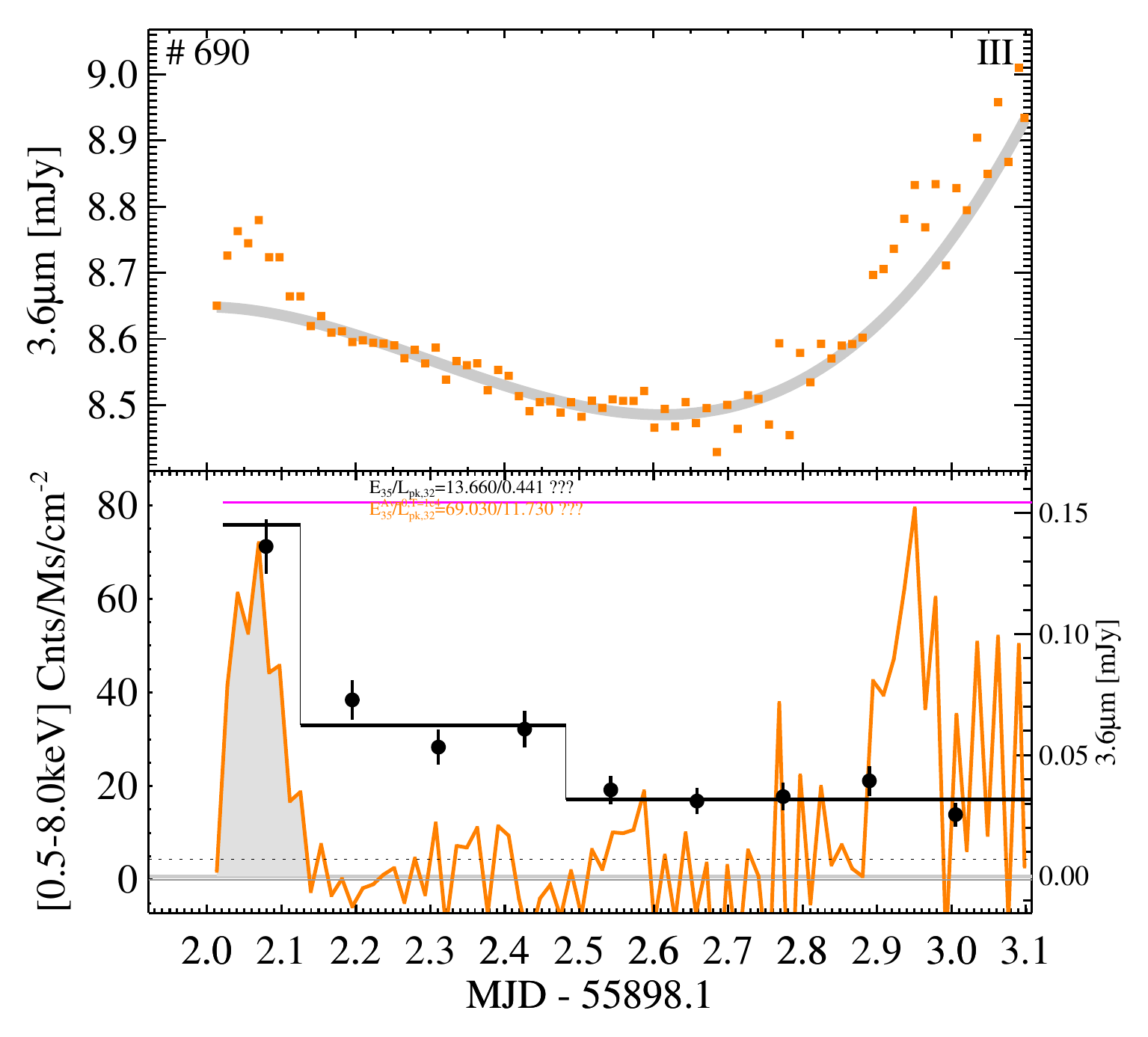}
\includegraphics[width=6.0cm]{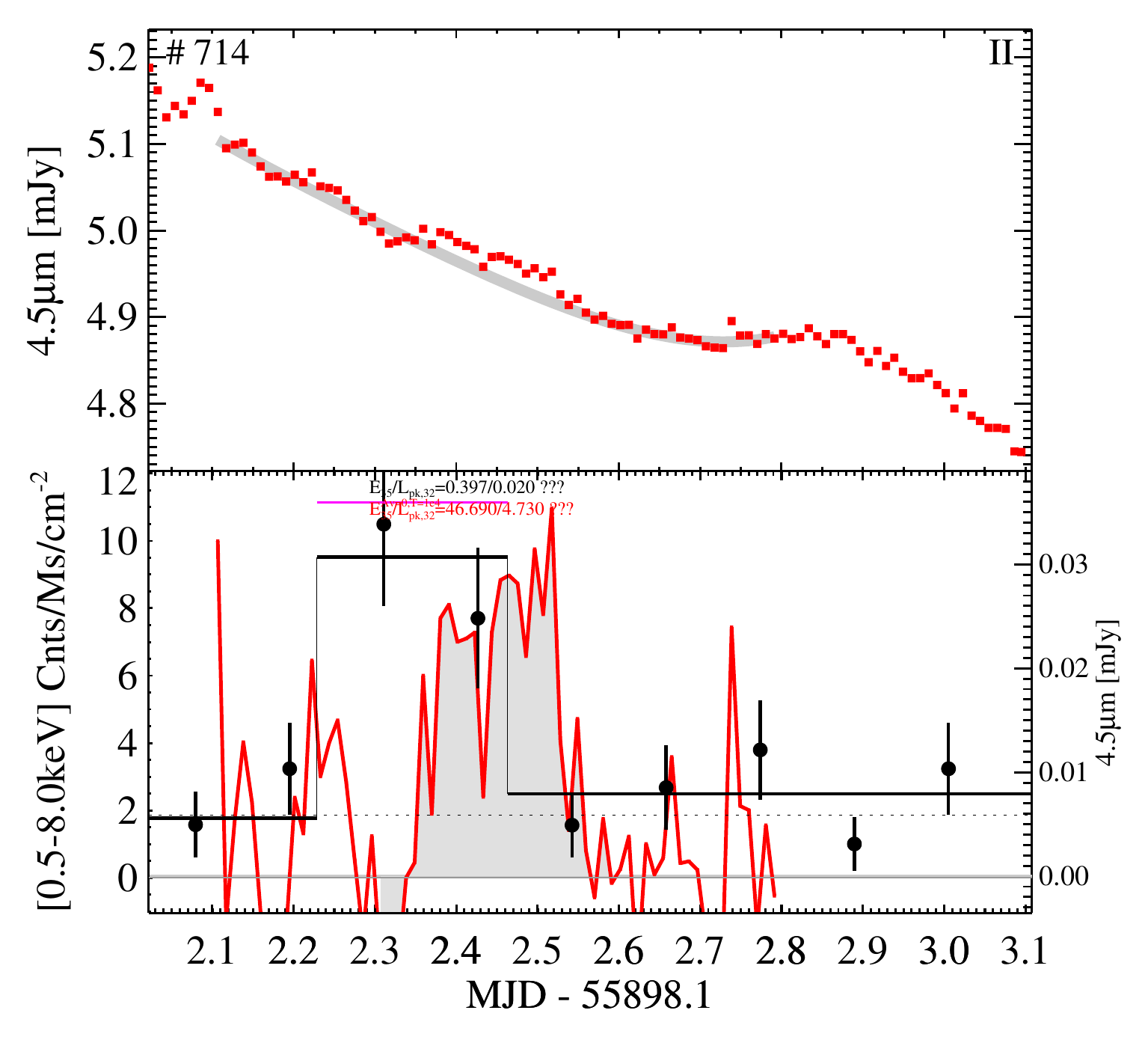}
\includegraphics[width=6.0cm]{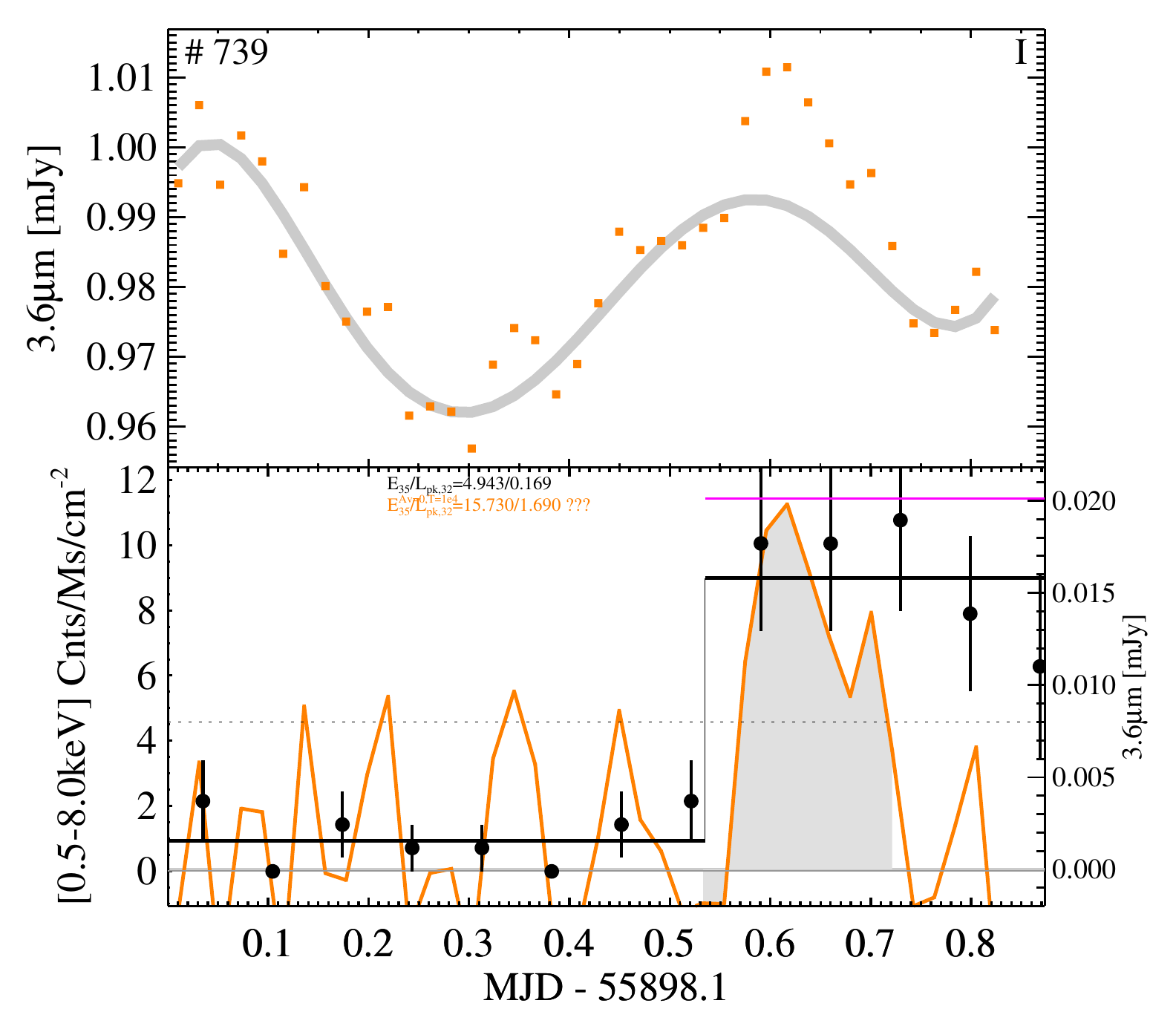}
\caption{(continued)}
\label{fig:}
\end{figure*}

\clearpage

\section{X-ray flares with a possible CoRoT/{\em Spitzer} counterpart}
\label{app:LC_Xrayonly_det}

\begin{figure*}[!t!]
\centering
\includegraphics[width=6.0cm]{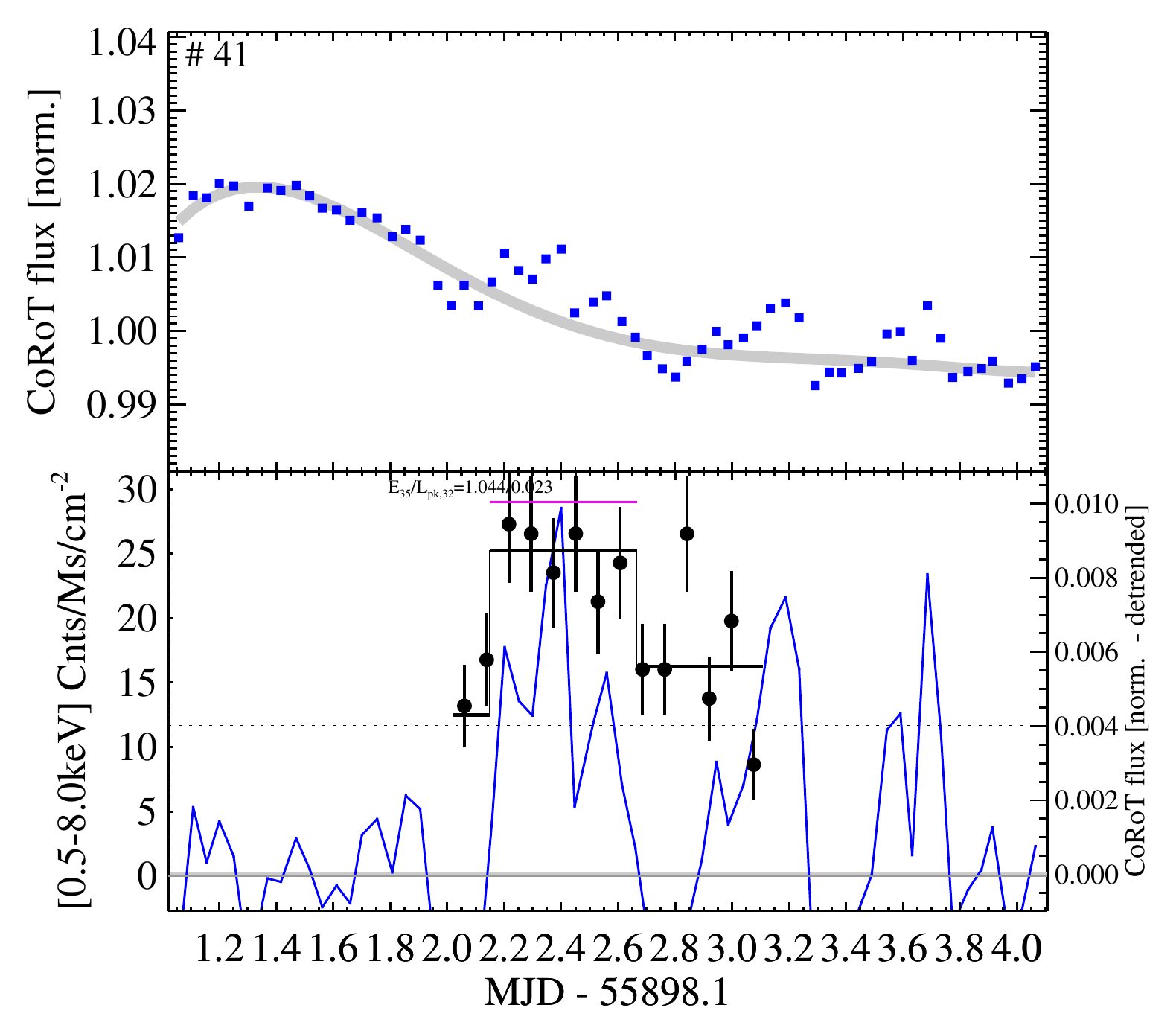}
\includegraphics[width=6.0cm]{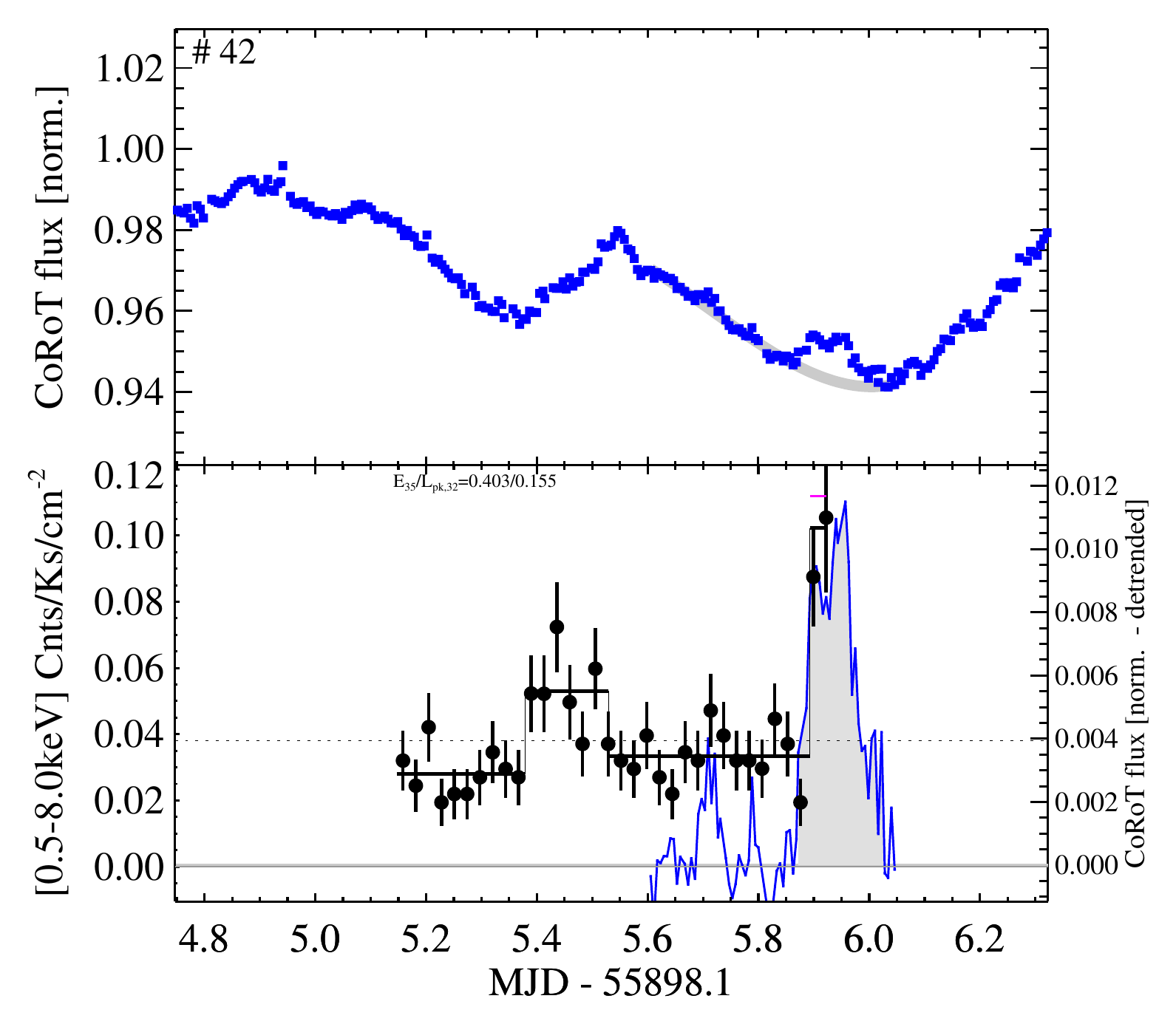}
\includegraphics[width=6.0cm]{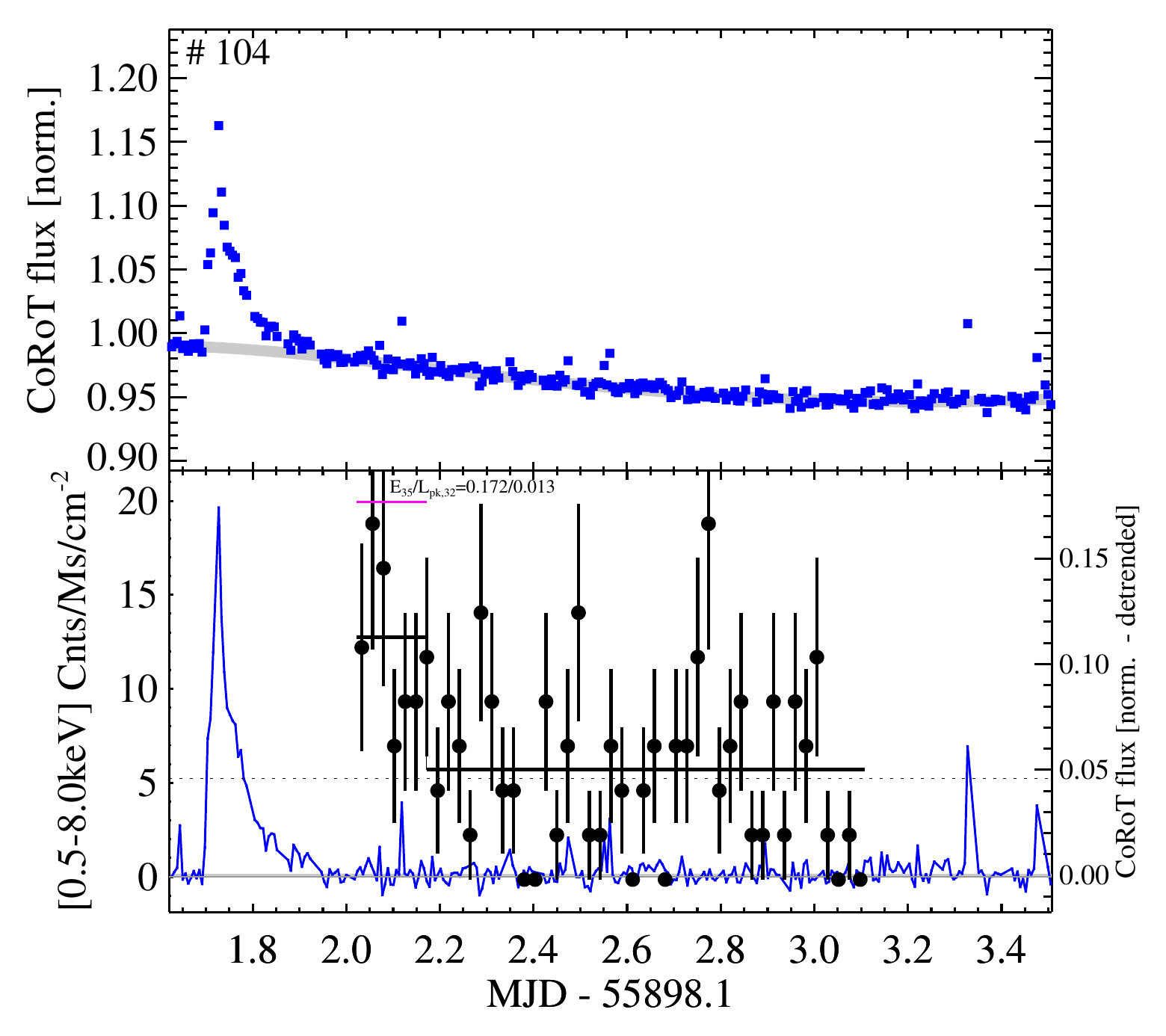}
\includegraphics[width=6.0cm]{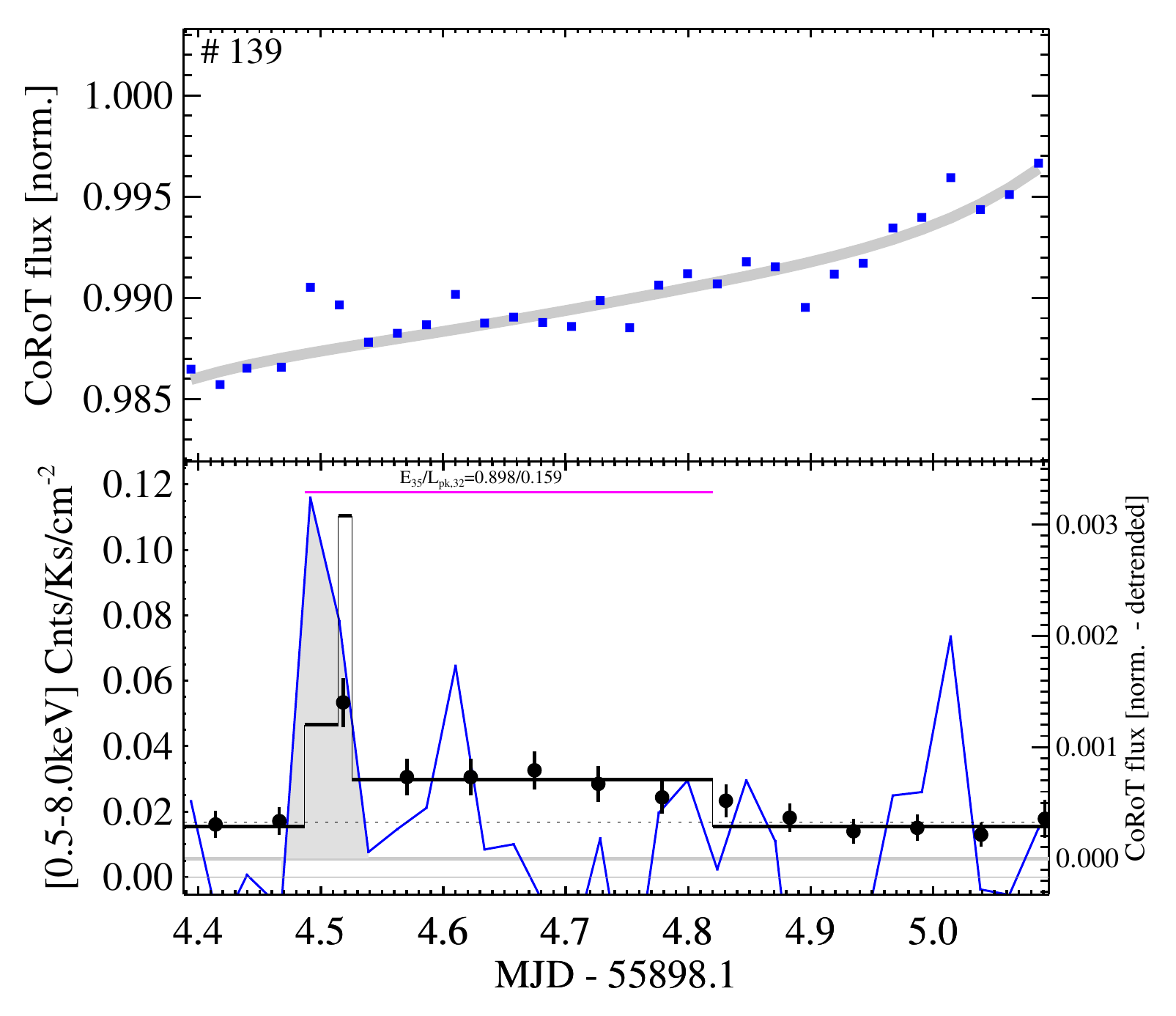}
\includegraphics[width=6.0cm]{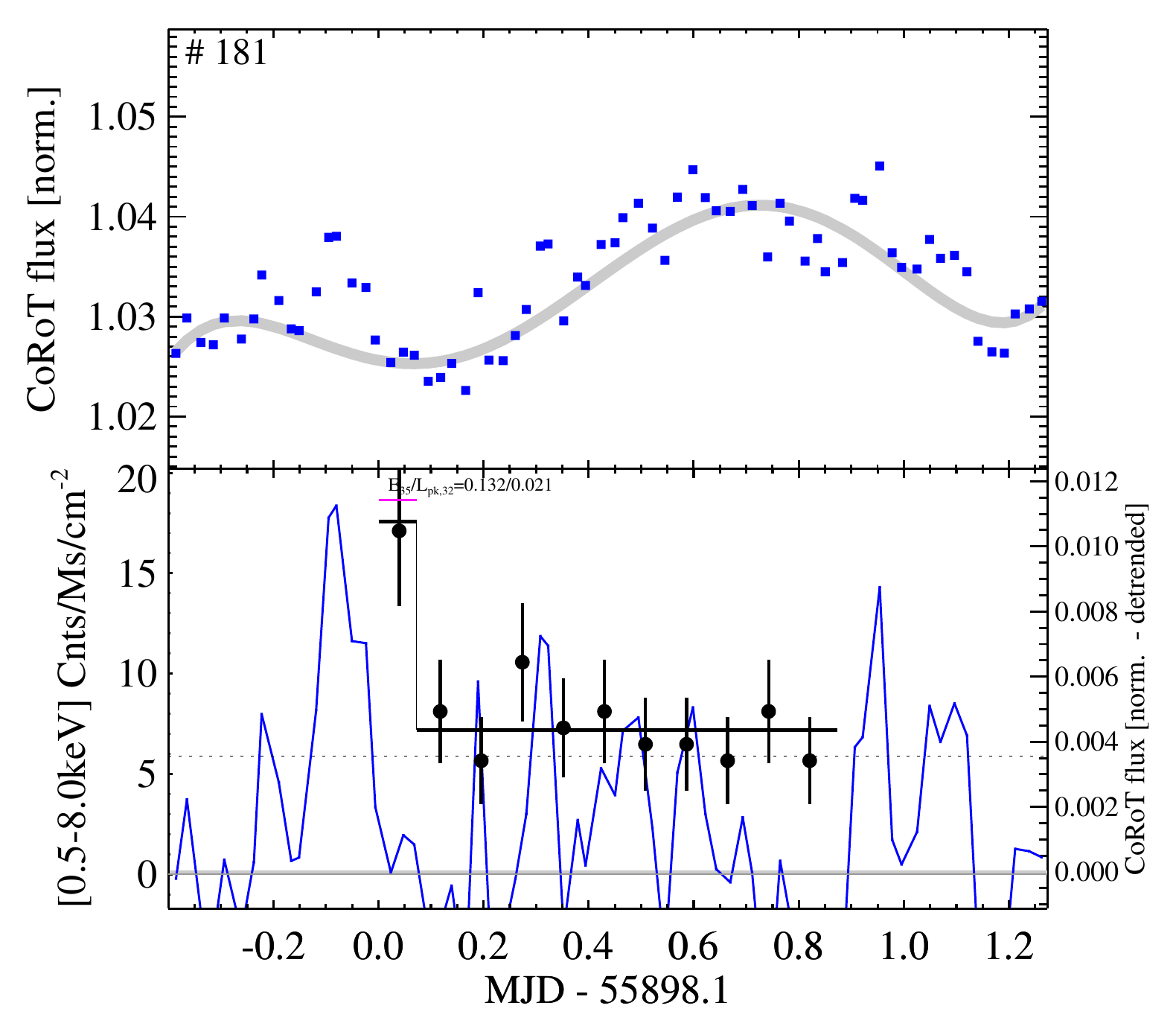}
\includegraphics[width=6.0cm]{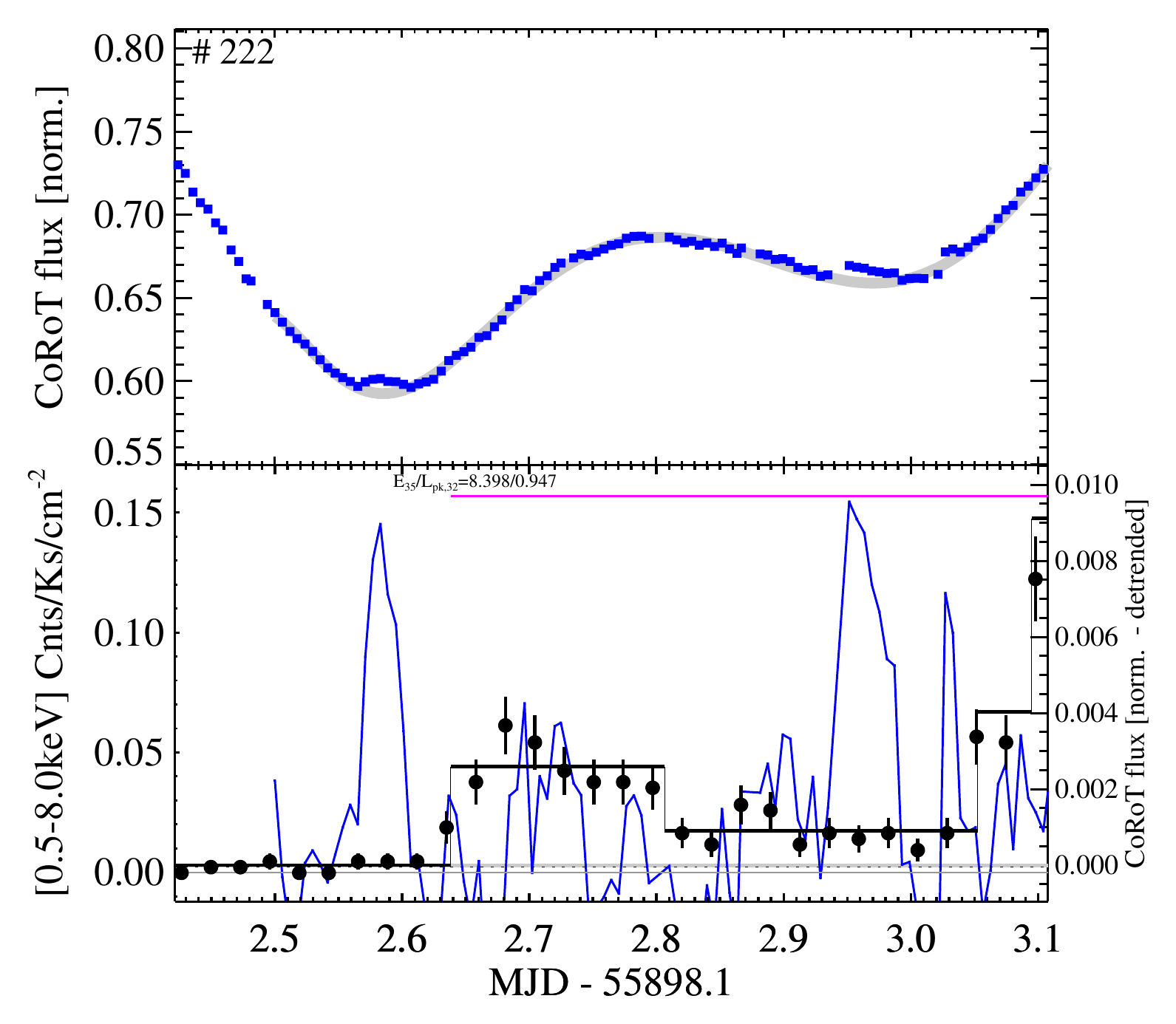}
\includegraphics[width=6.0cm]{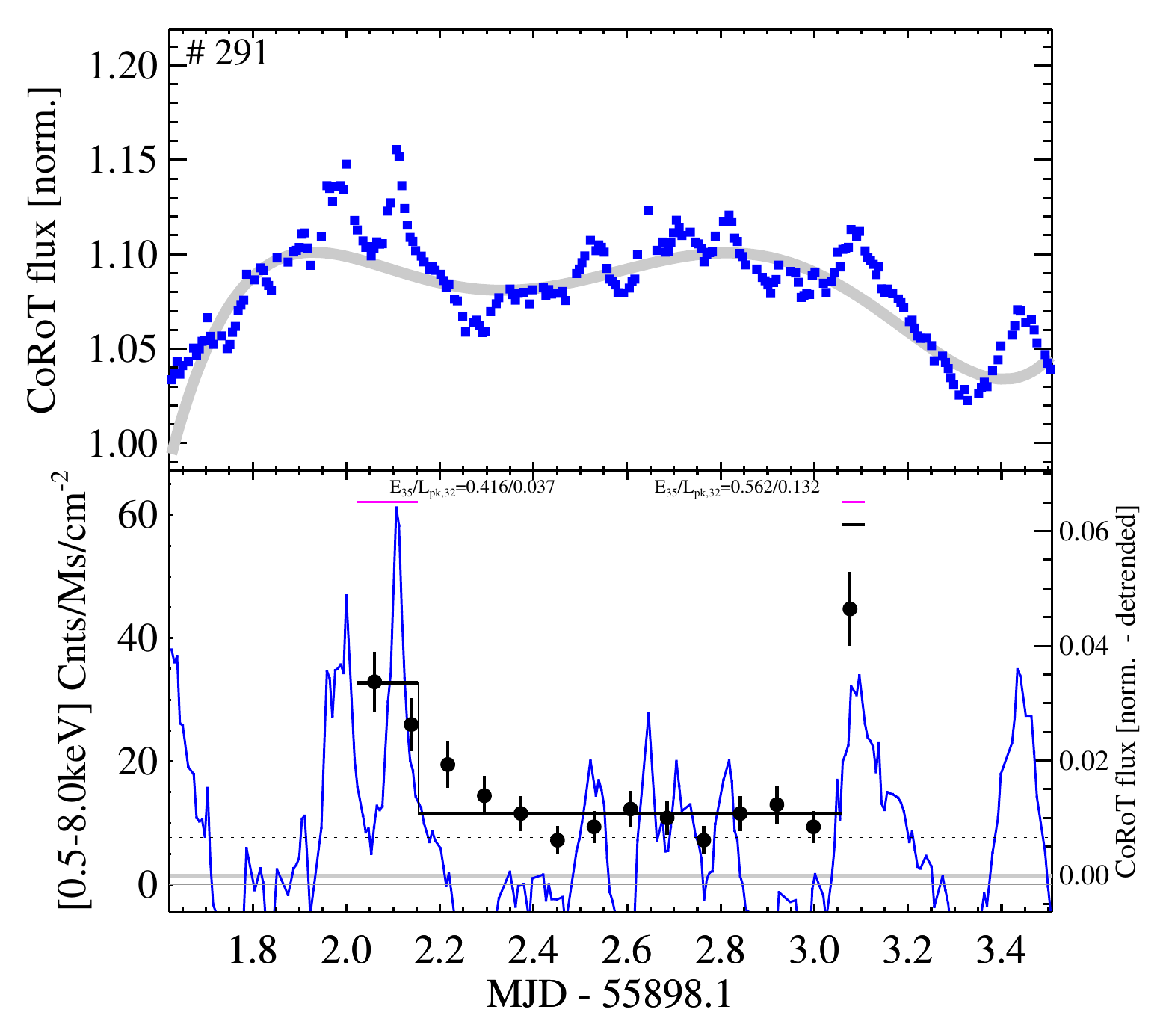}
\includegraphics[width=6.0cm]{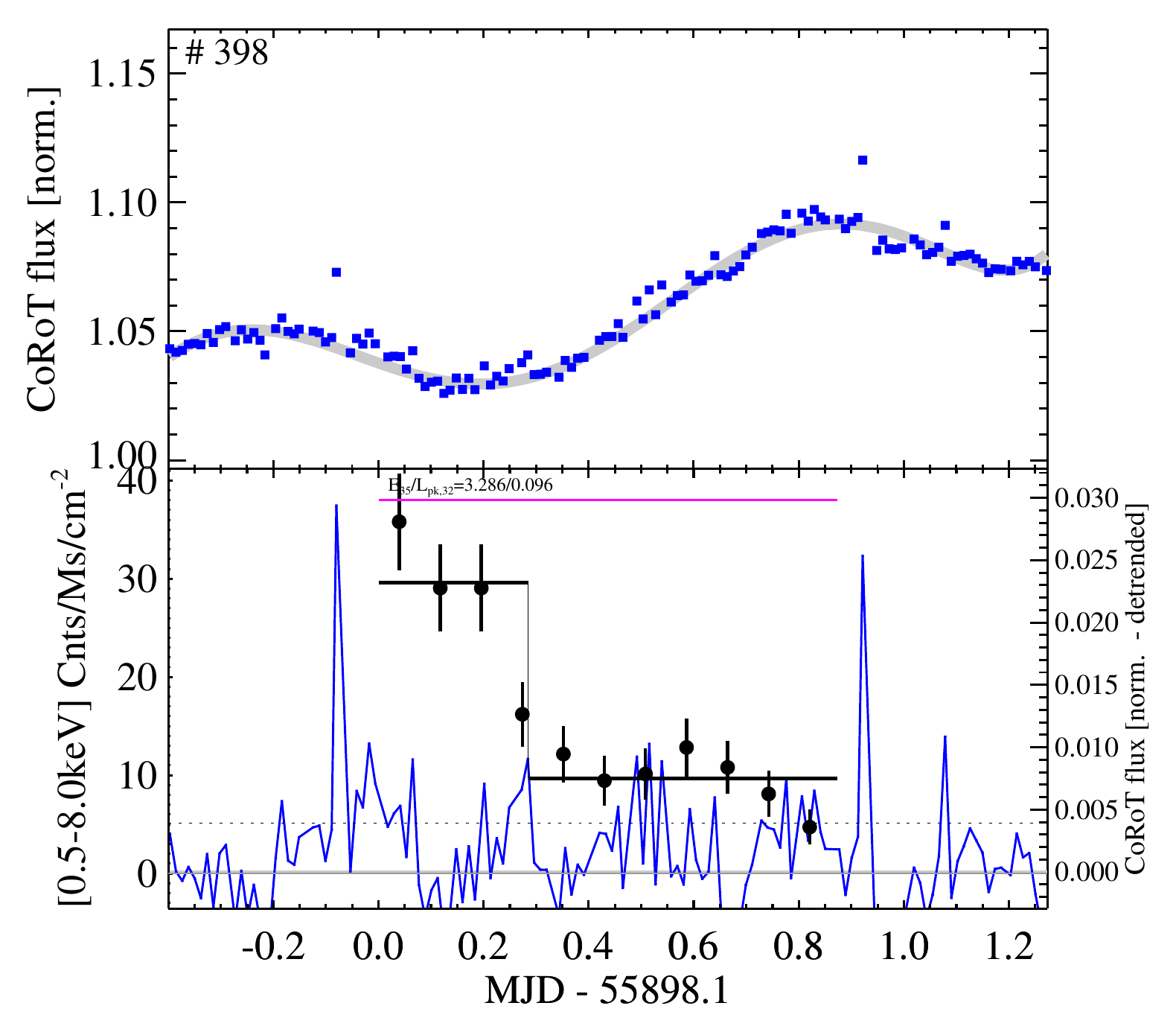}
\includegraphics[width=6.0cm]{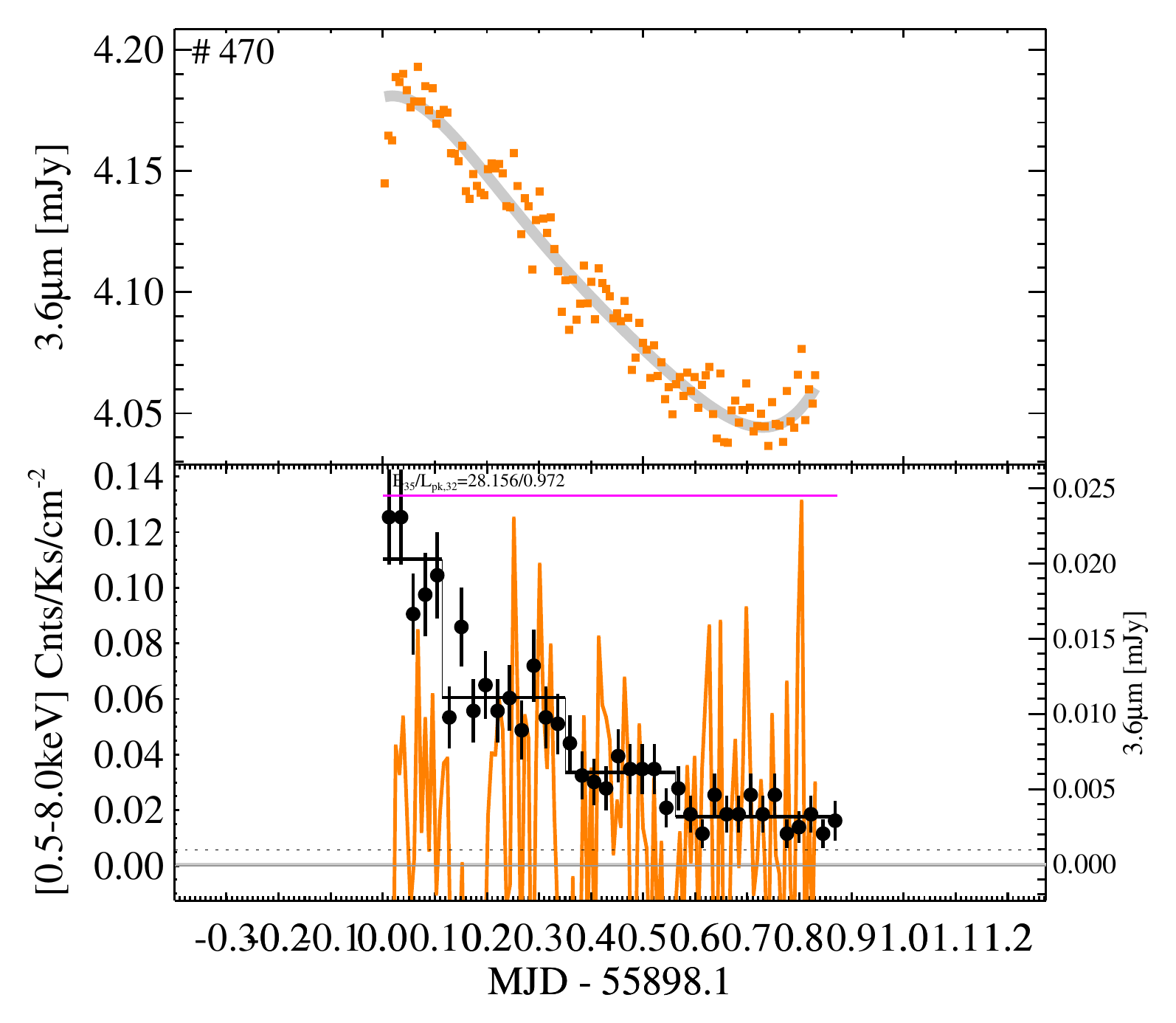}
\includegraphics[width=6.0cm]{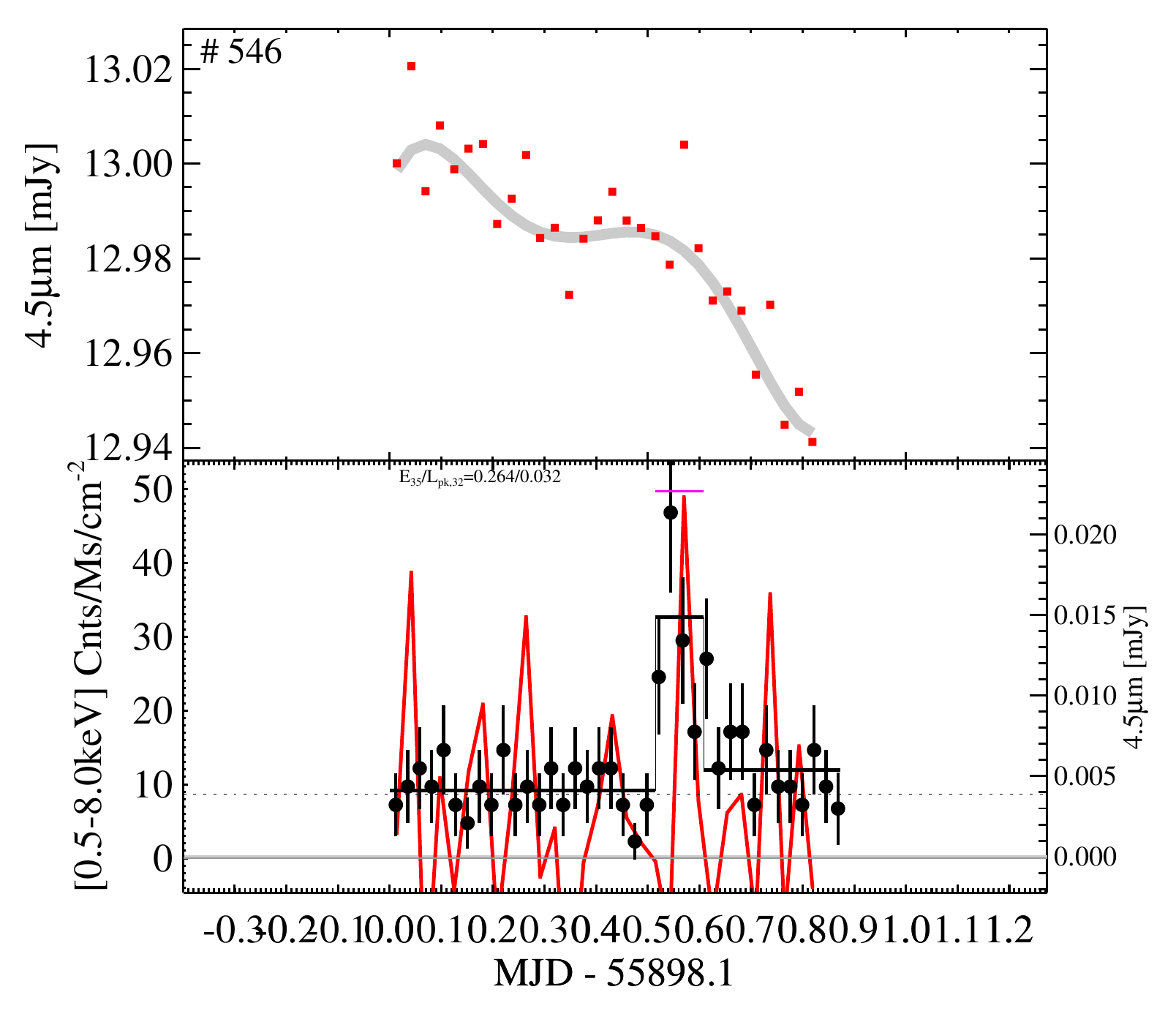}
\includegraphics[width=6.0cm]{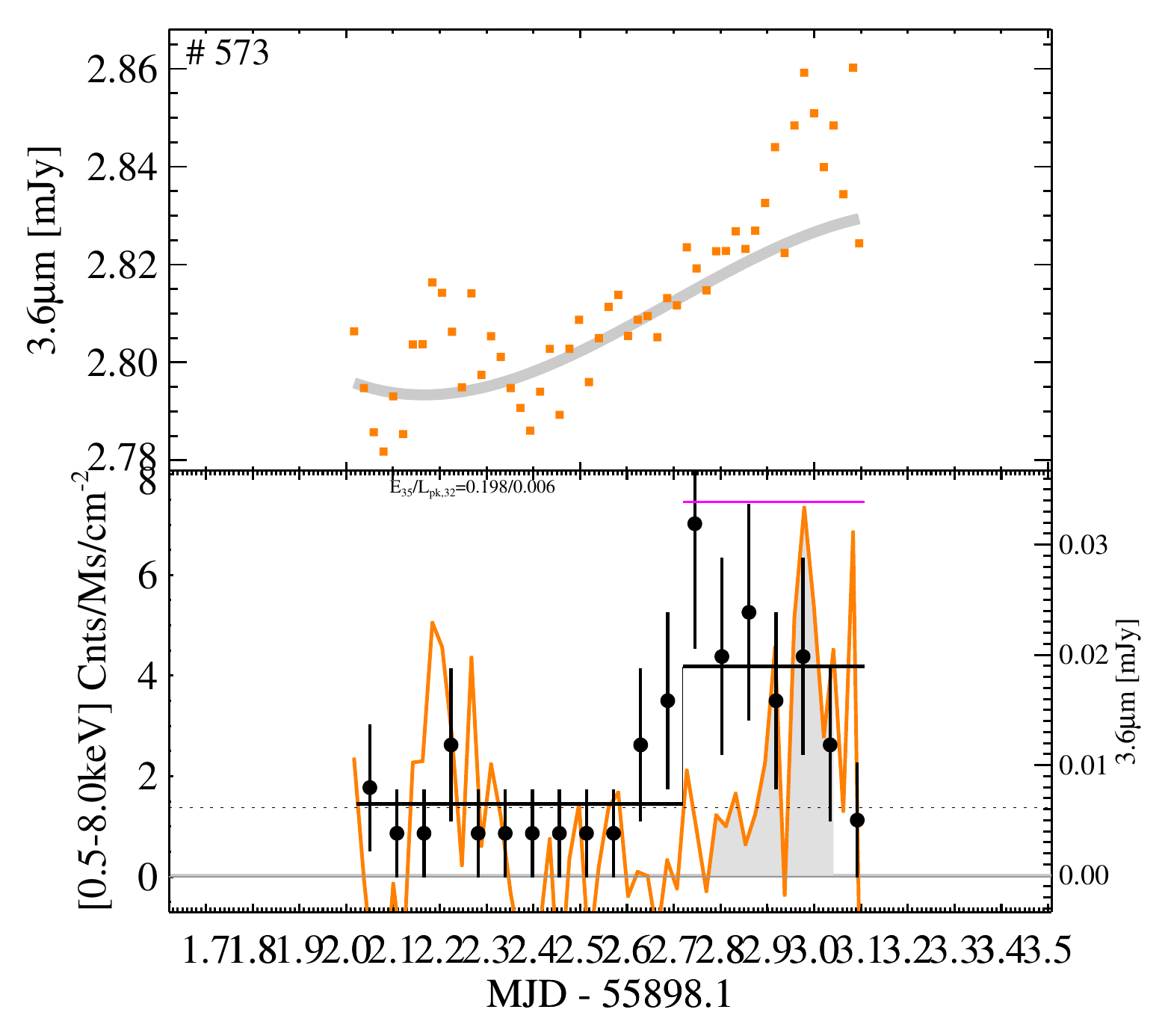}
\includegraphics[width=6.0cm]{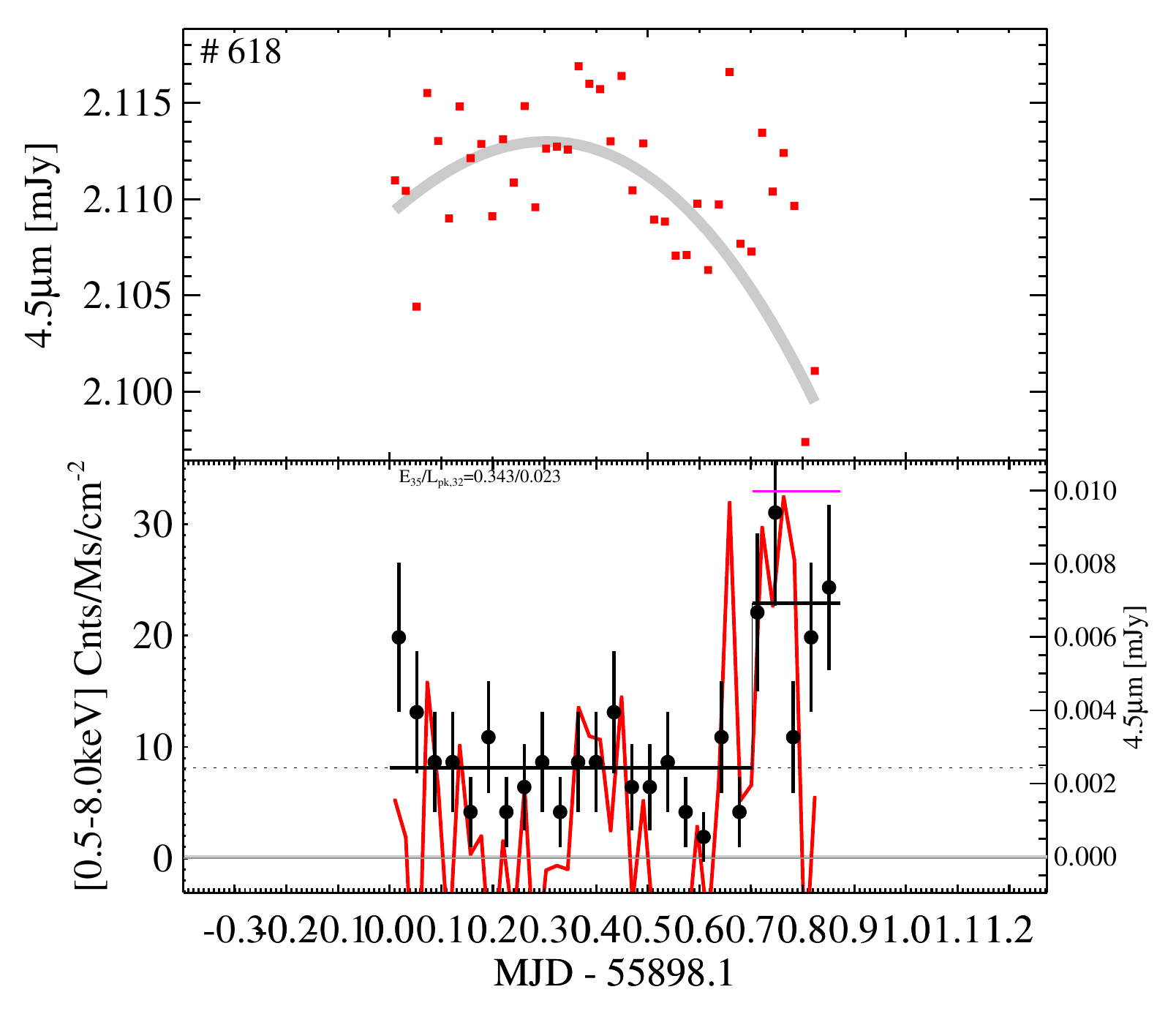}
\label{fig:}
\end{figure*}
\begin{figure*}[!t!]
\centering
\includegraphics[width=6.0cm]{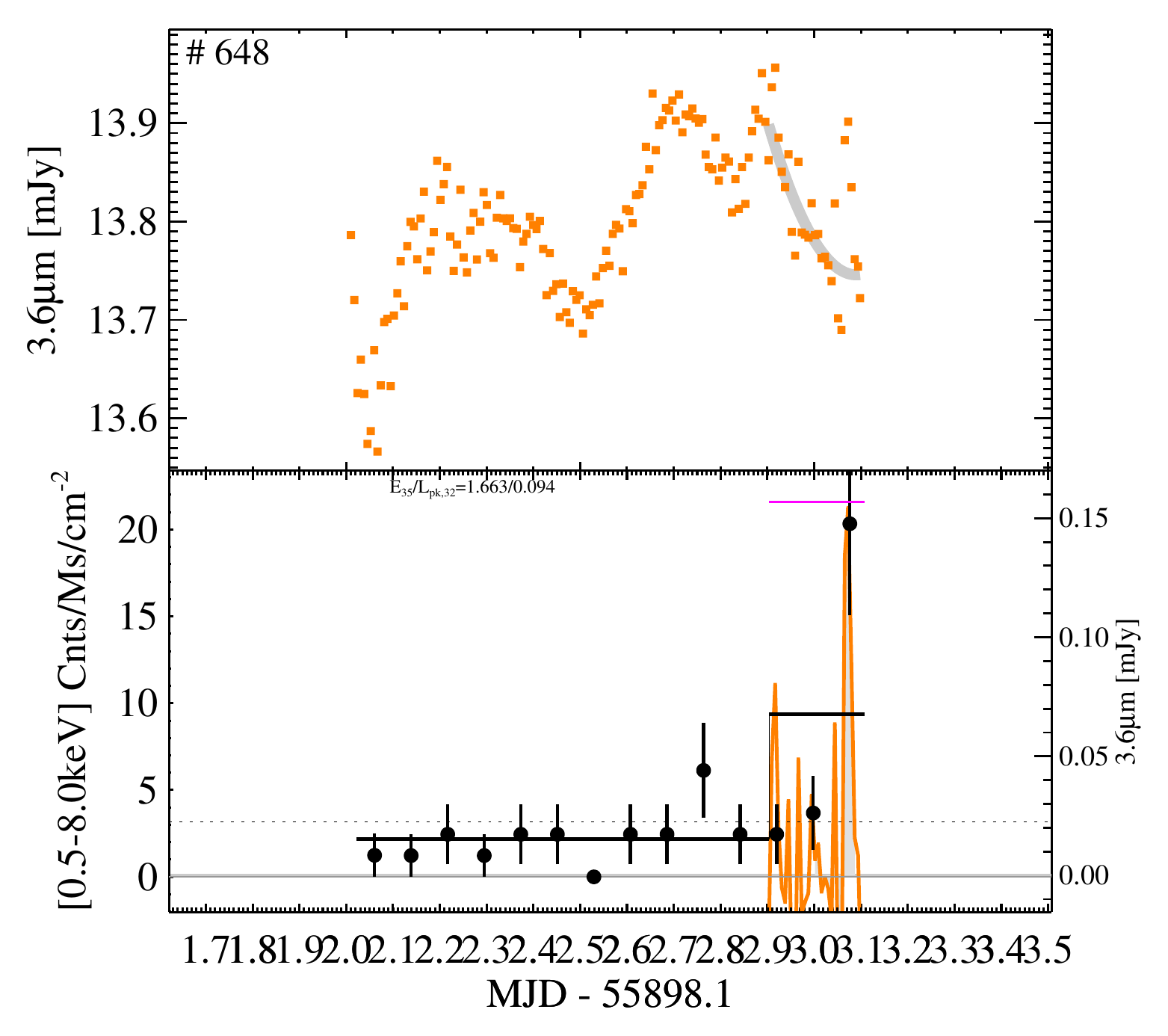}
\includegraphics[width=6.0cm]{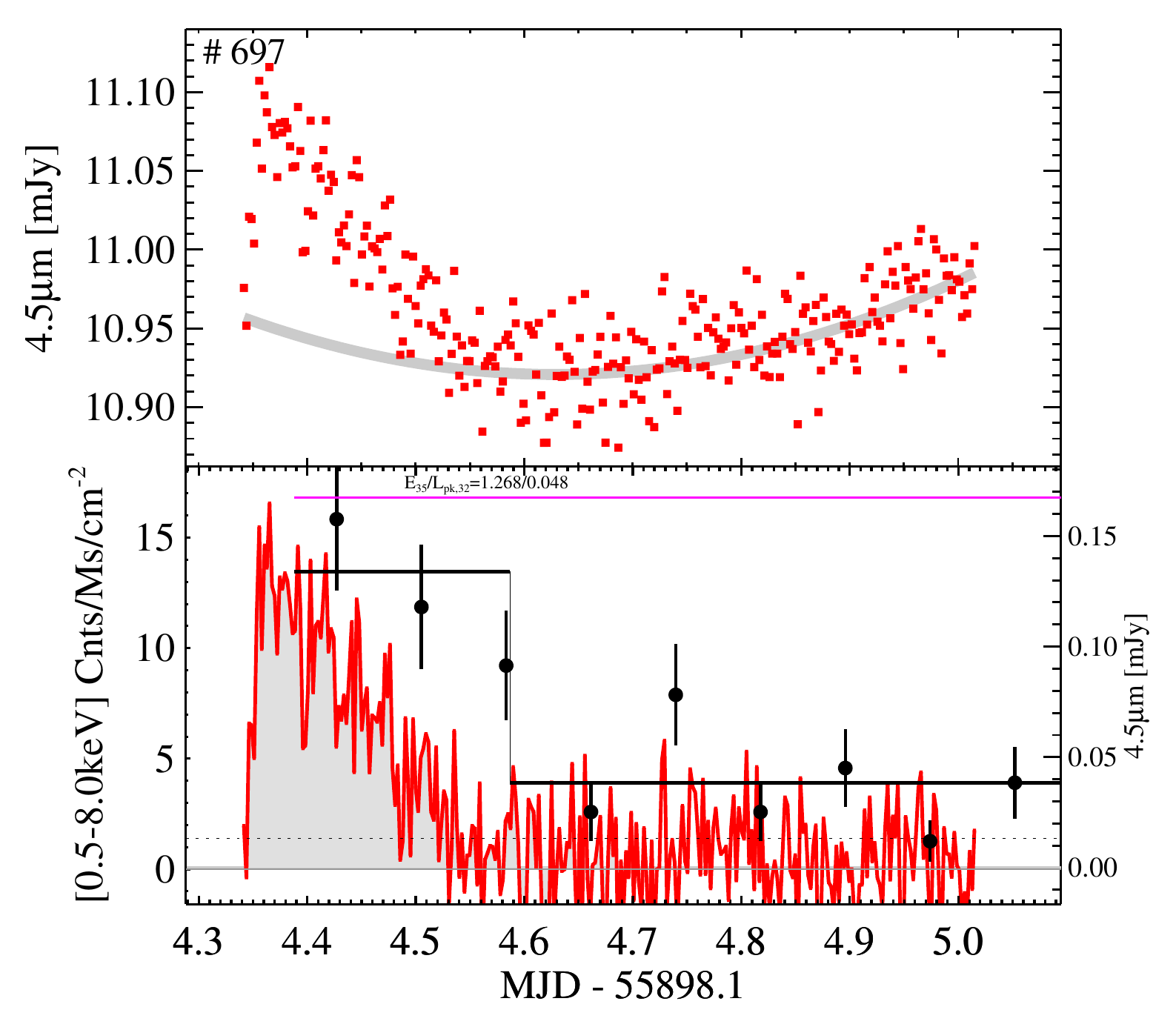}
\includegraphics[width=6.0cm]{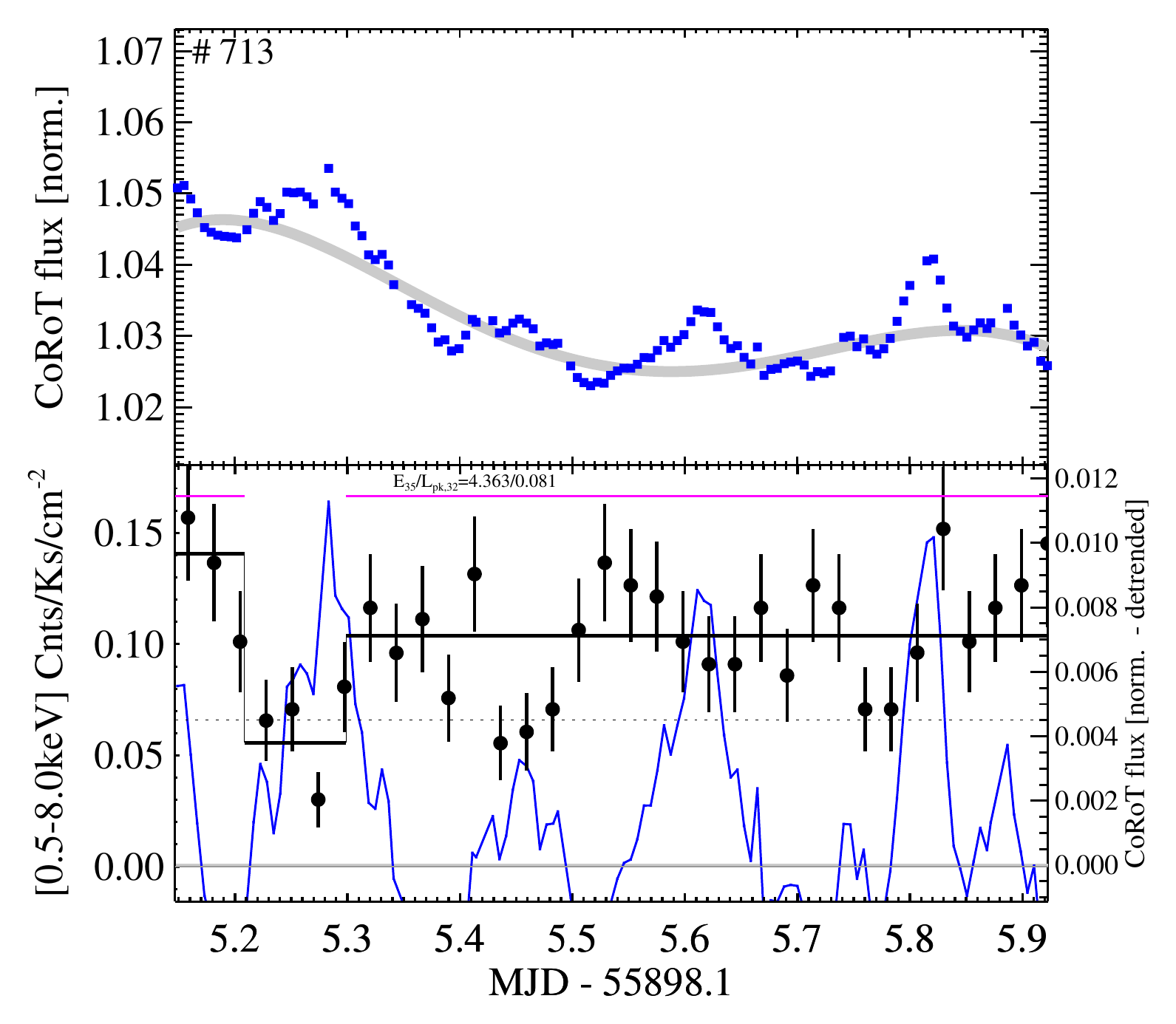}
\includegraphics[width=6.0cm]{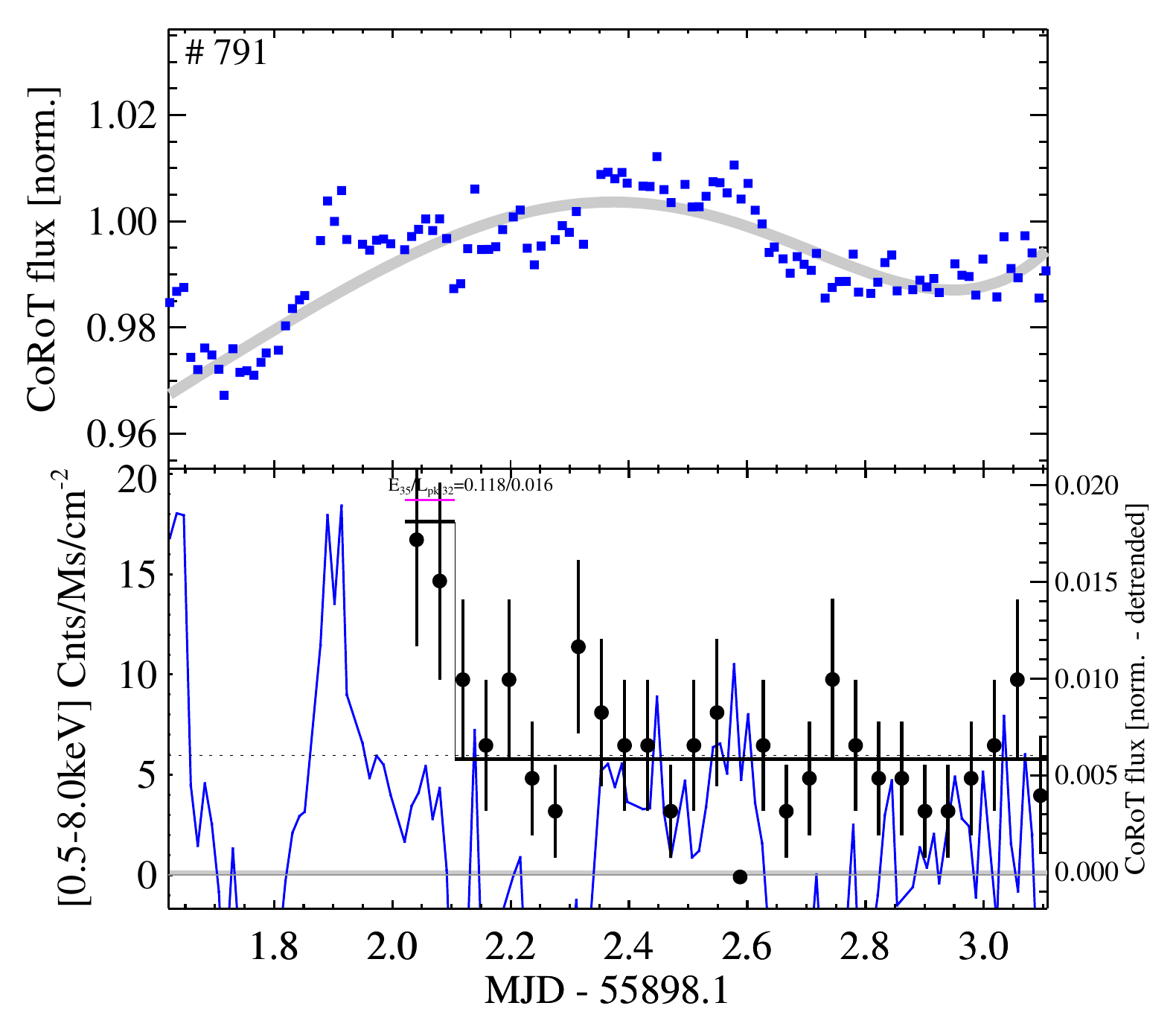}
\includegraphics[width=6.0cm]{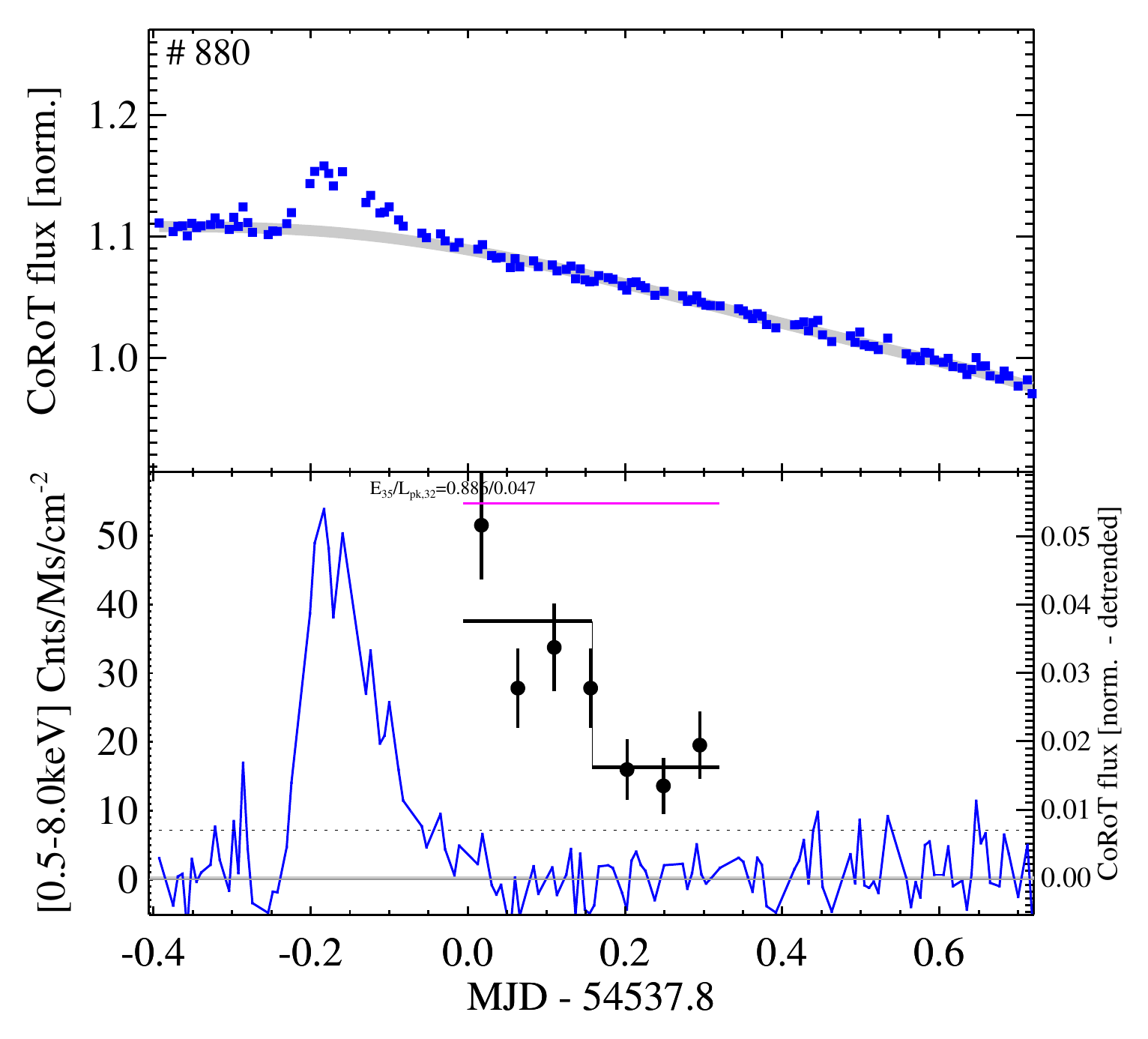}
\includegraphics[width=6.0cm]{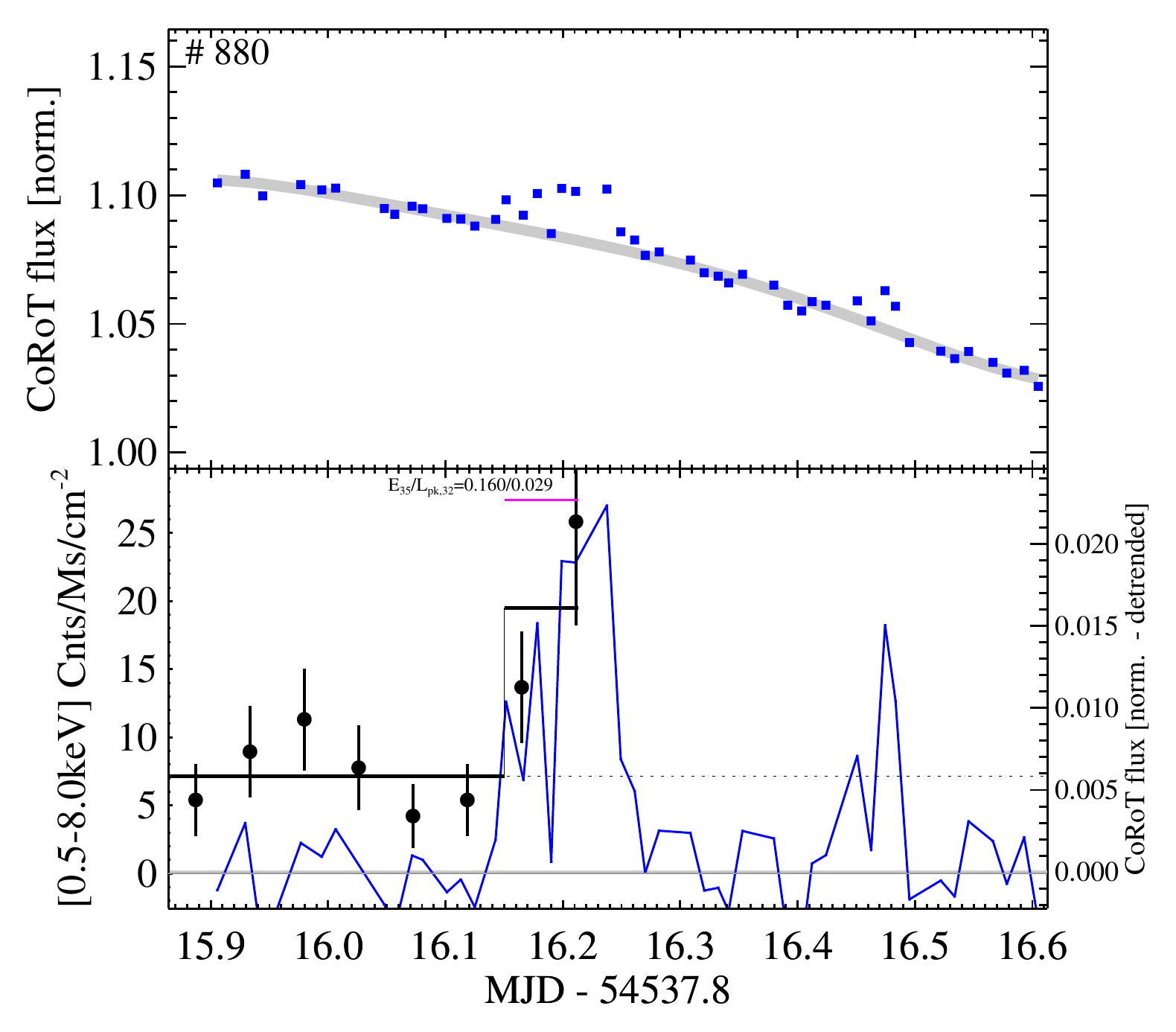}
\includegraphics[width=6.0cm]{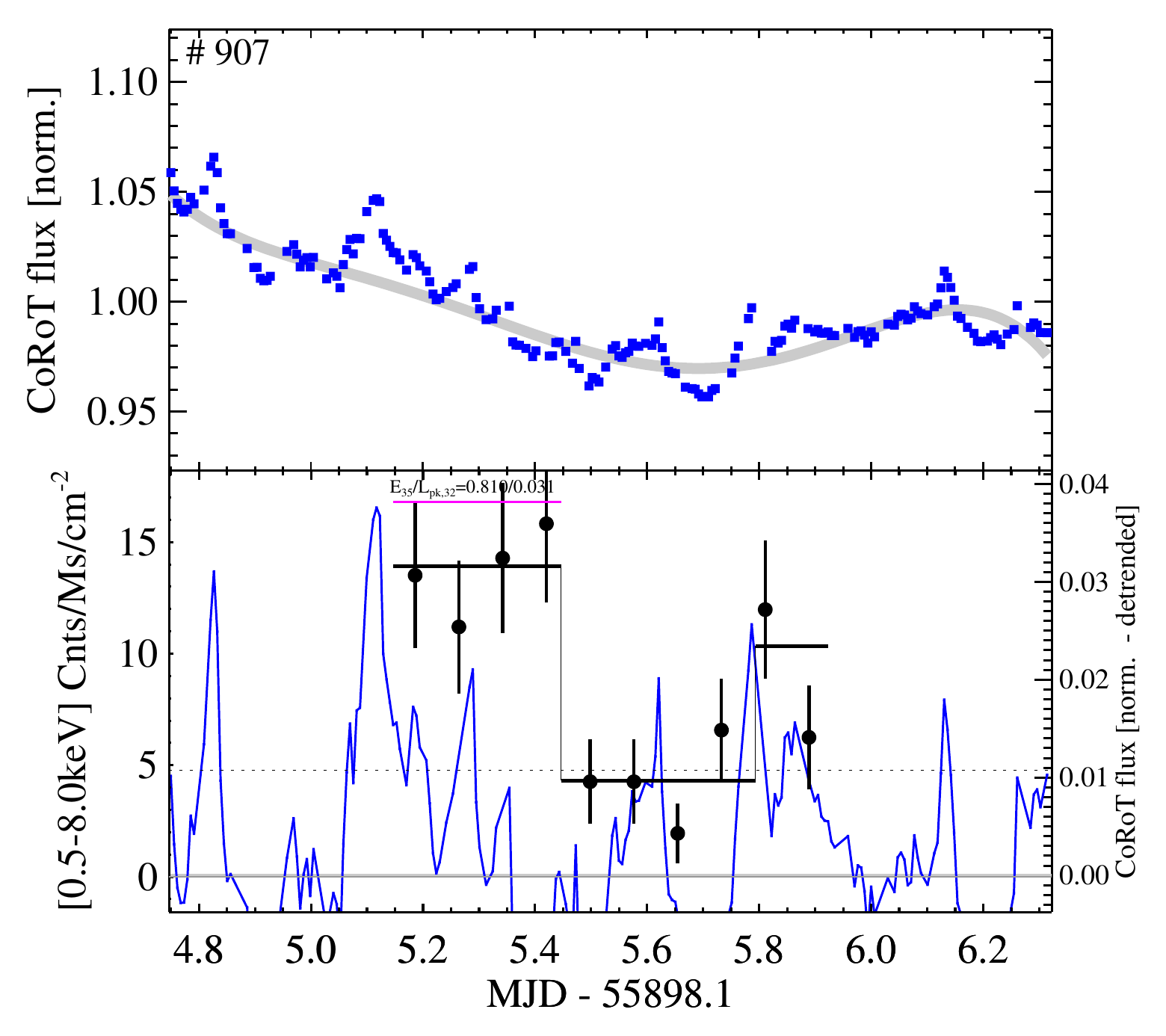}
\includegraphics[width=6.0cm]{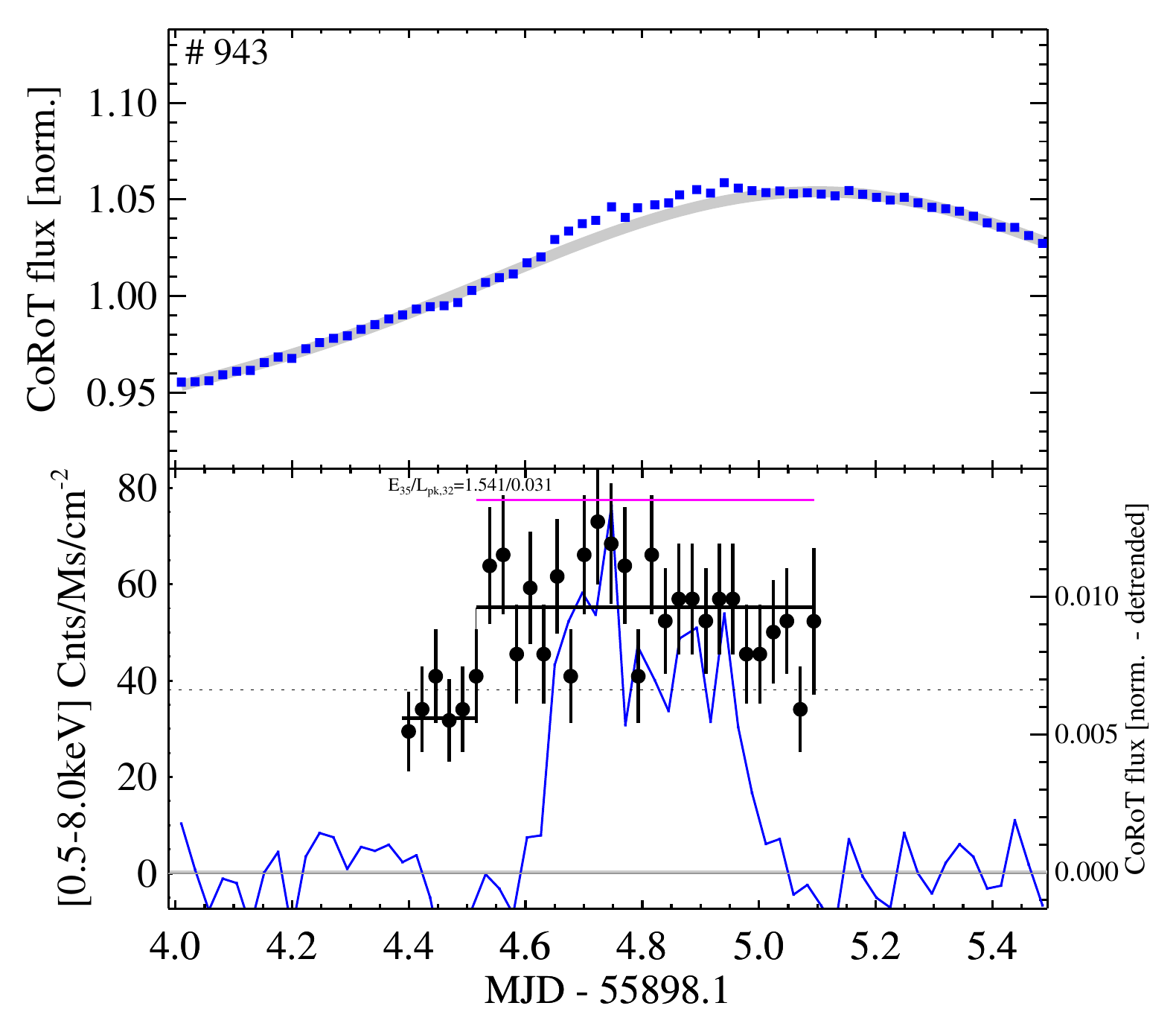}
\label{fig:}
\end{figure*}
\begin{figure*}[!t!]
\centering
\includegraphics[width=6.0cm]{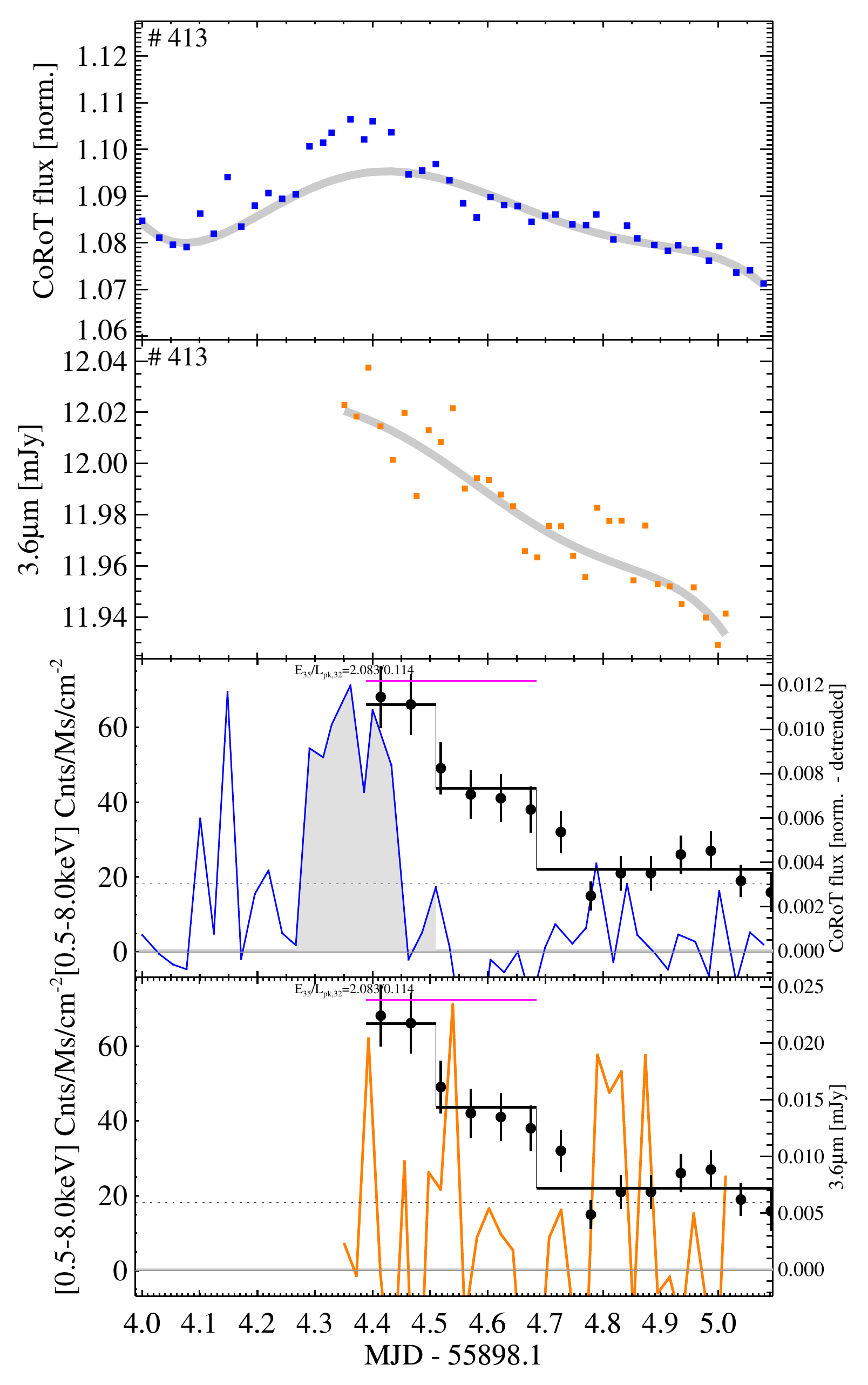}
\includegraphics[width=6.0cm]{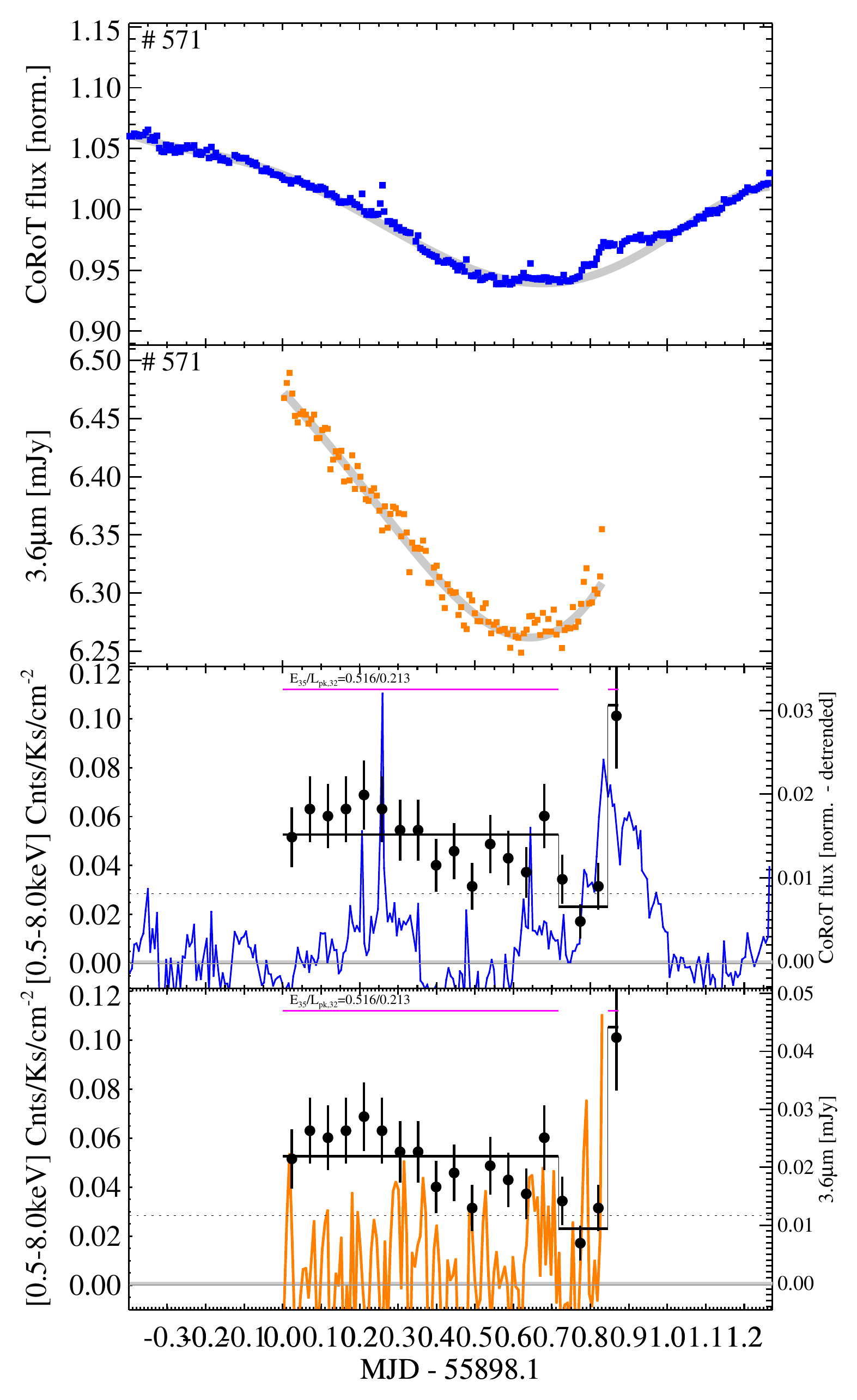}
\includegraphics[width=6.0cm]{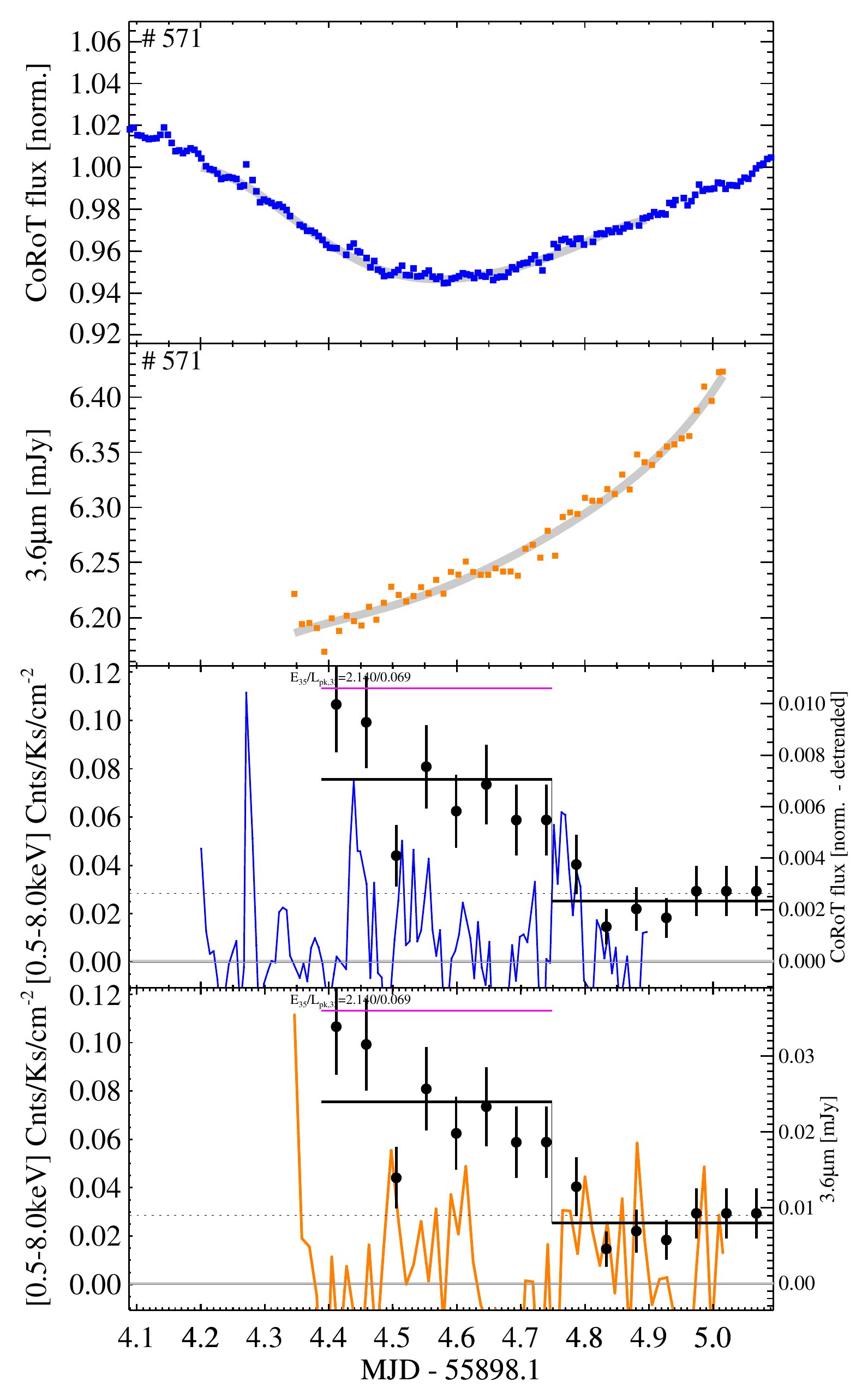}
\label{fig:}
\end{figure*}

\clearpage

\section{X-ray flares within the Chandra observing segments and with no CoRoT  or {\em Spitzer} counterpart}
\label{app:LC_Xrayonly_Oallin}

\begin{figure*}[!t!]
\centering
\includegraphics[width=6.0cm]{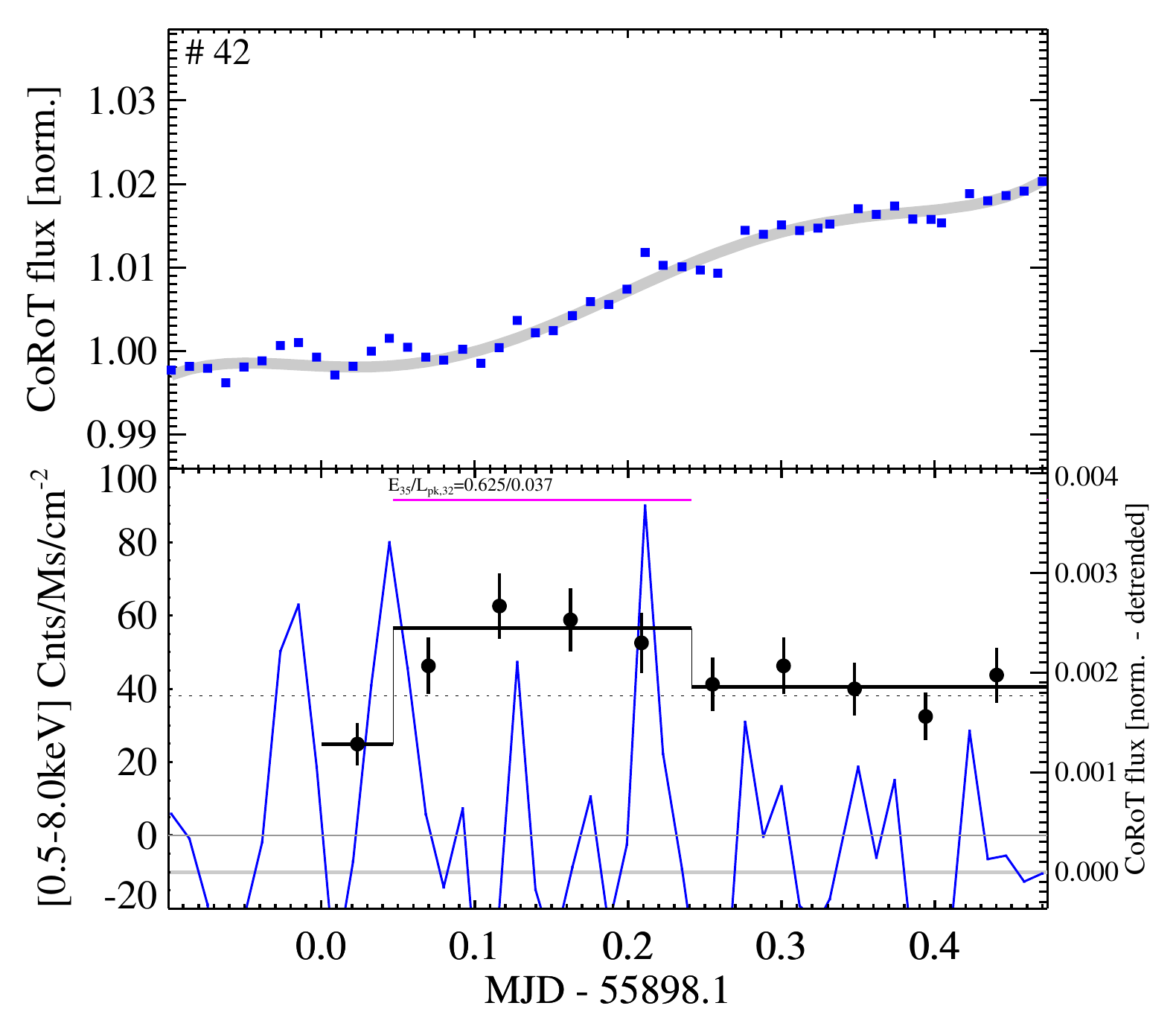}
\includegraphics[width=6.0cm]{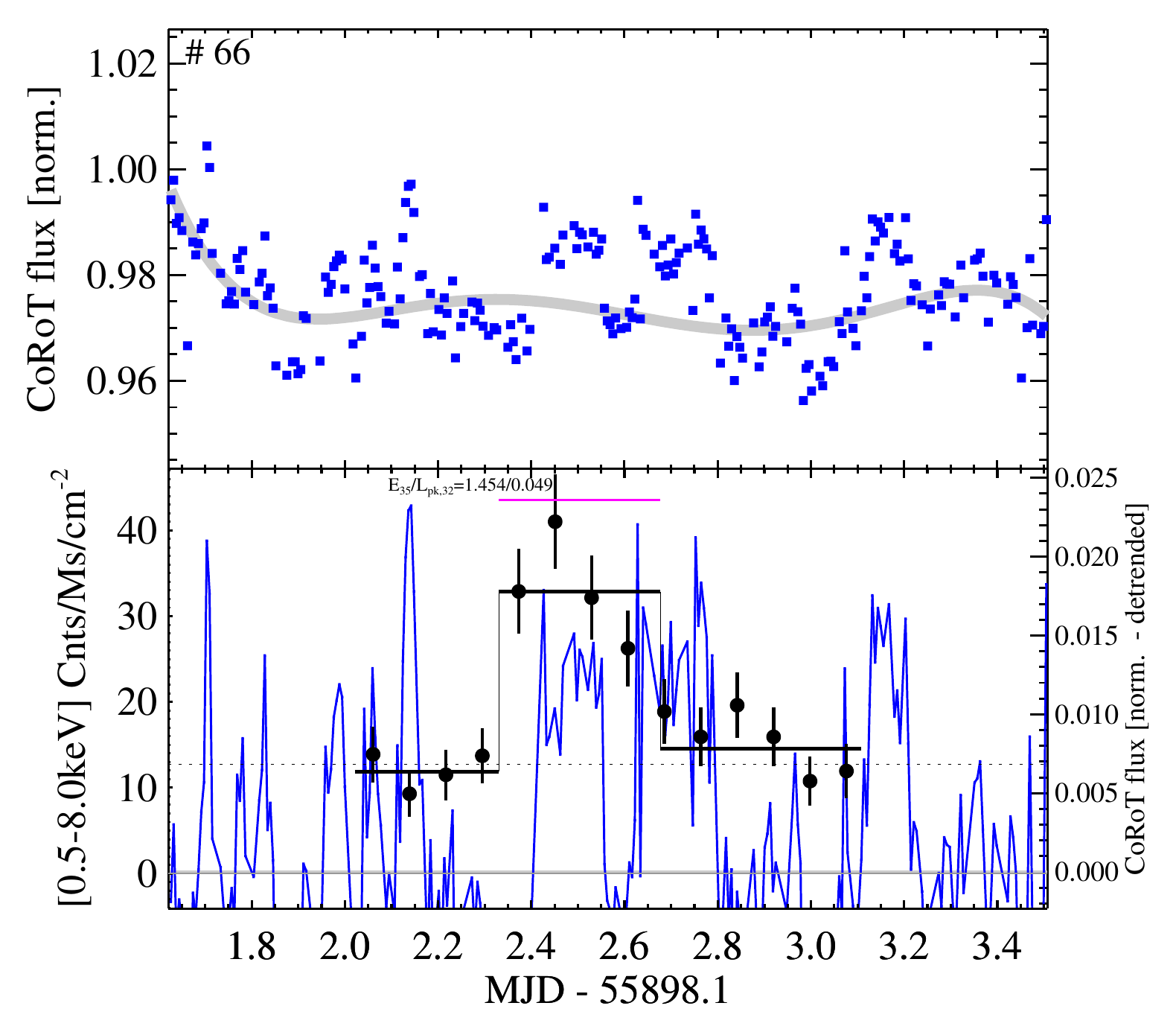}
\includegraphics[width=6.0cm]{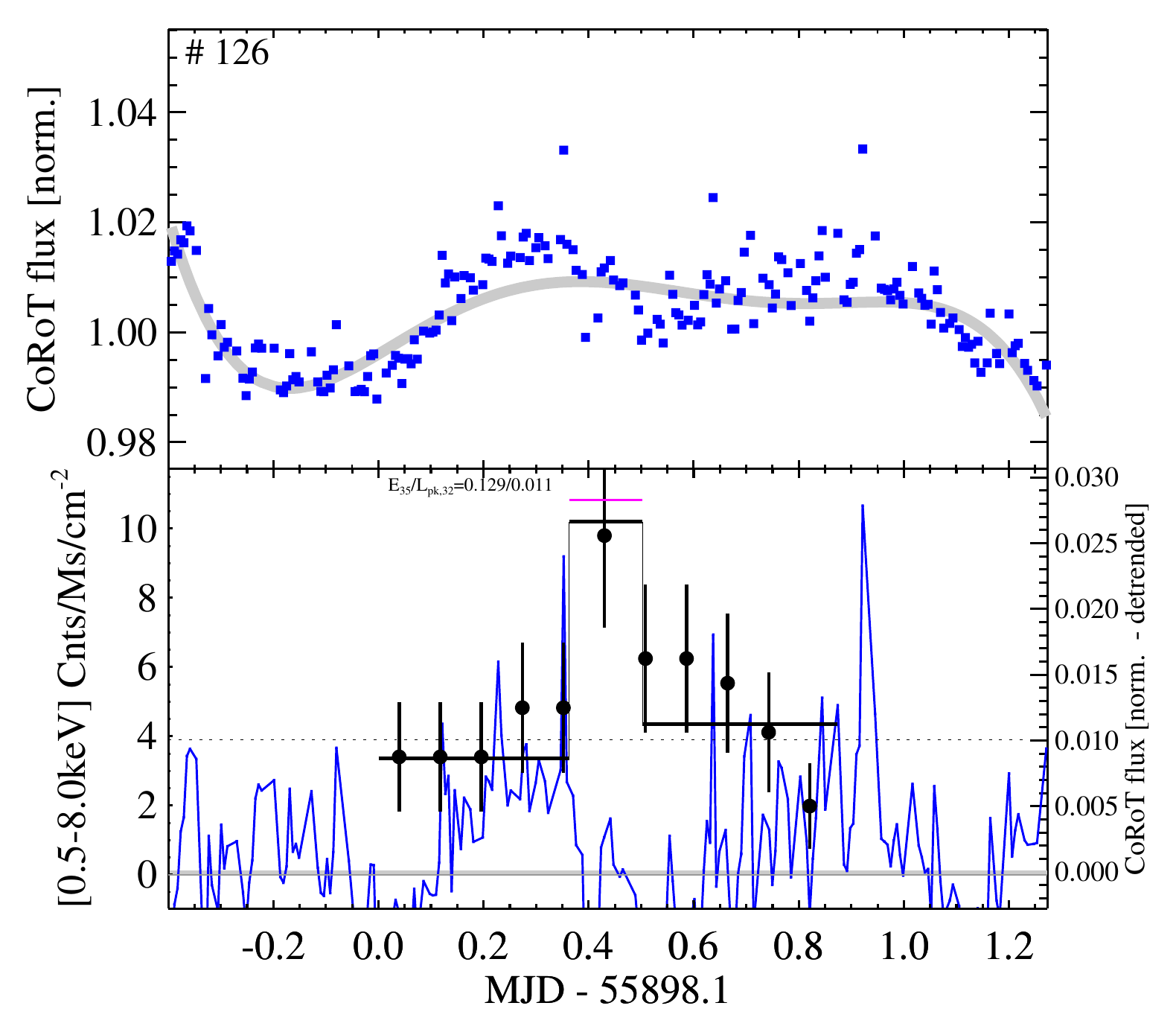}
\includegraphics[width=6.0cm]{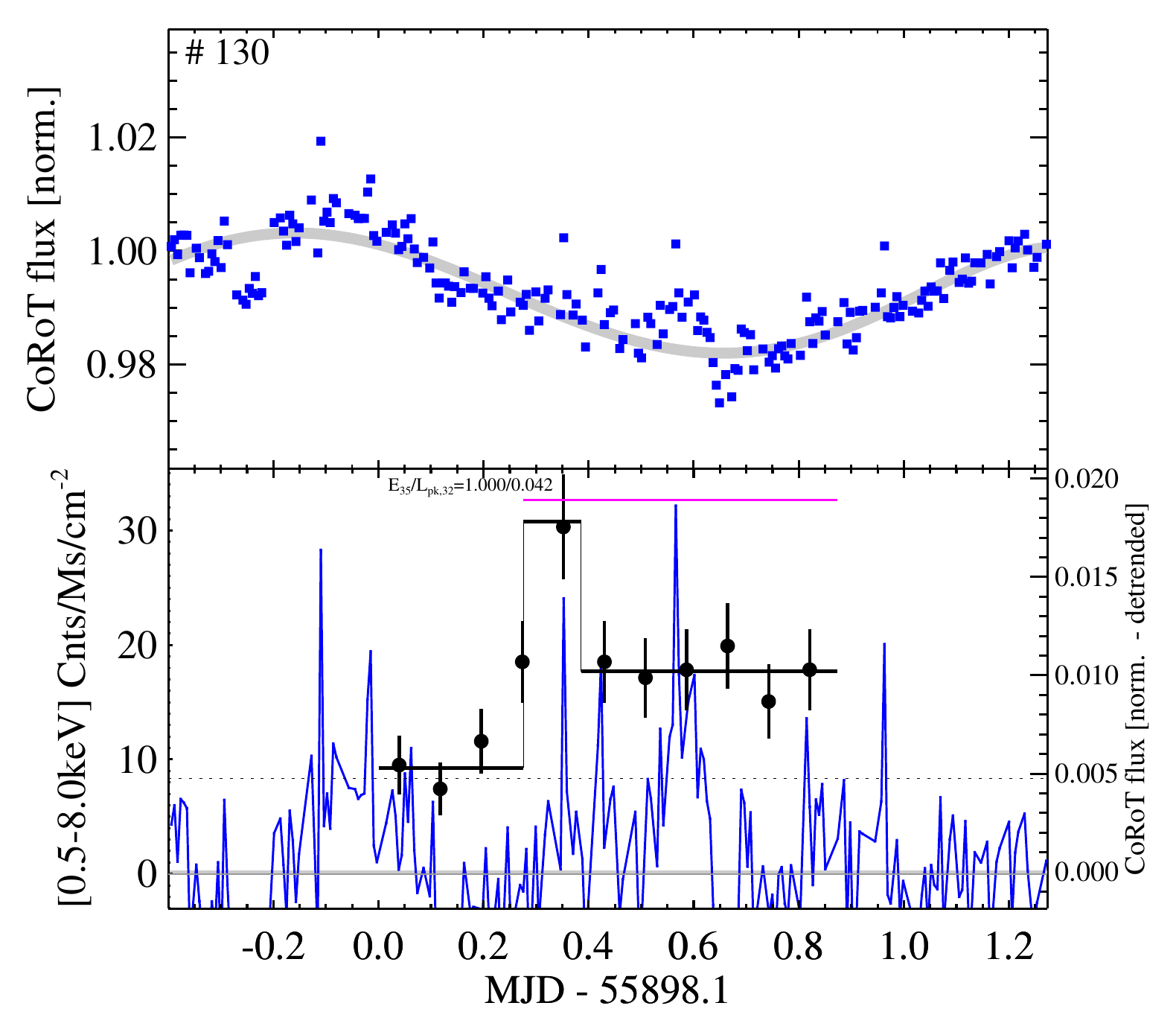}
\includegraphics[width=6.0cm]{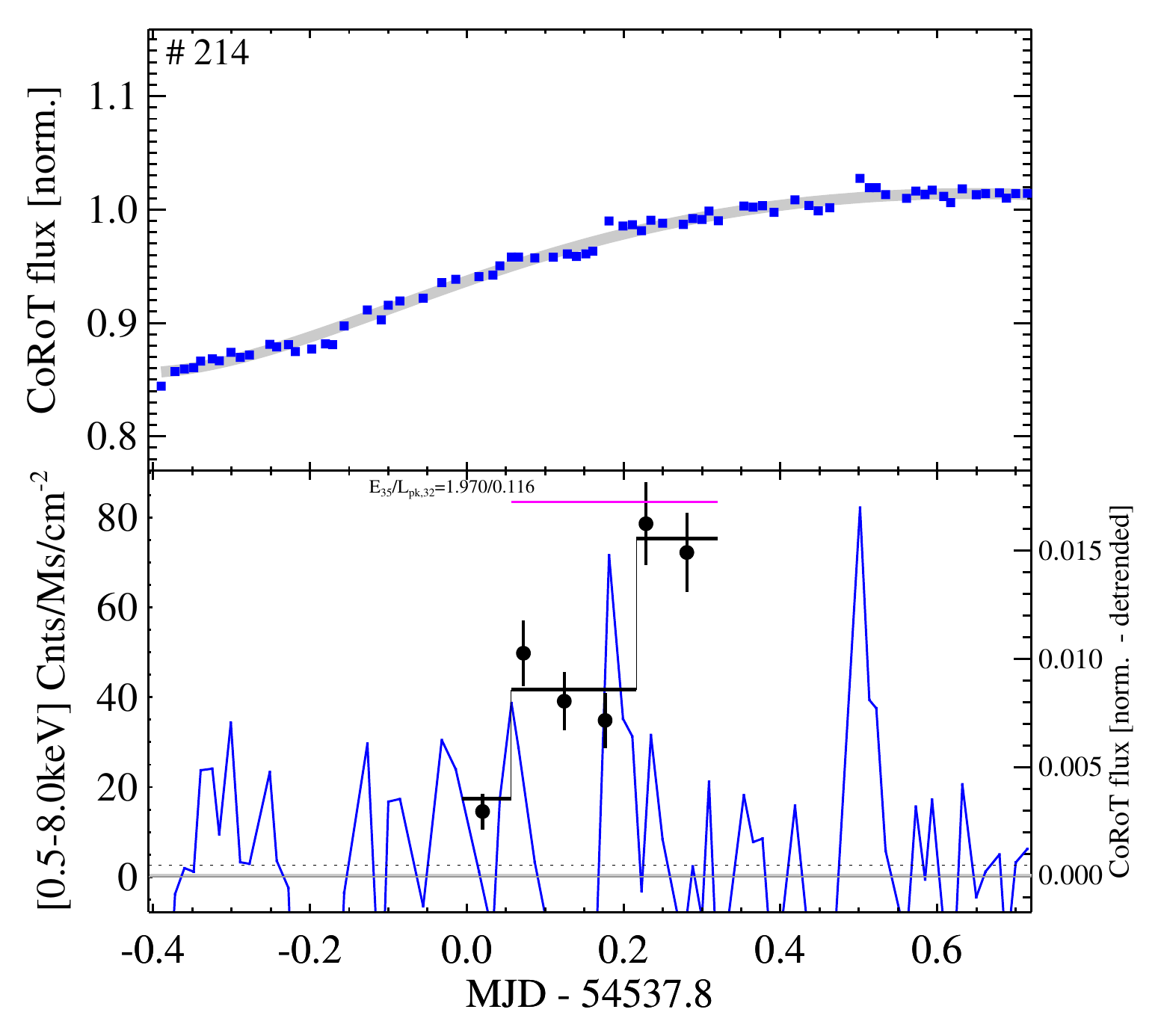}
\includegraphics[width=6.0cm]{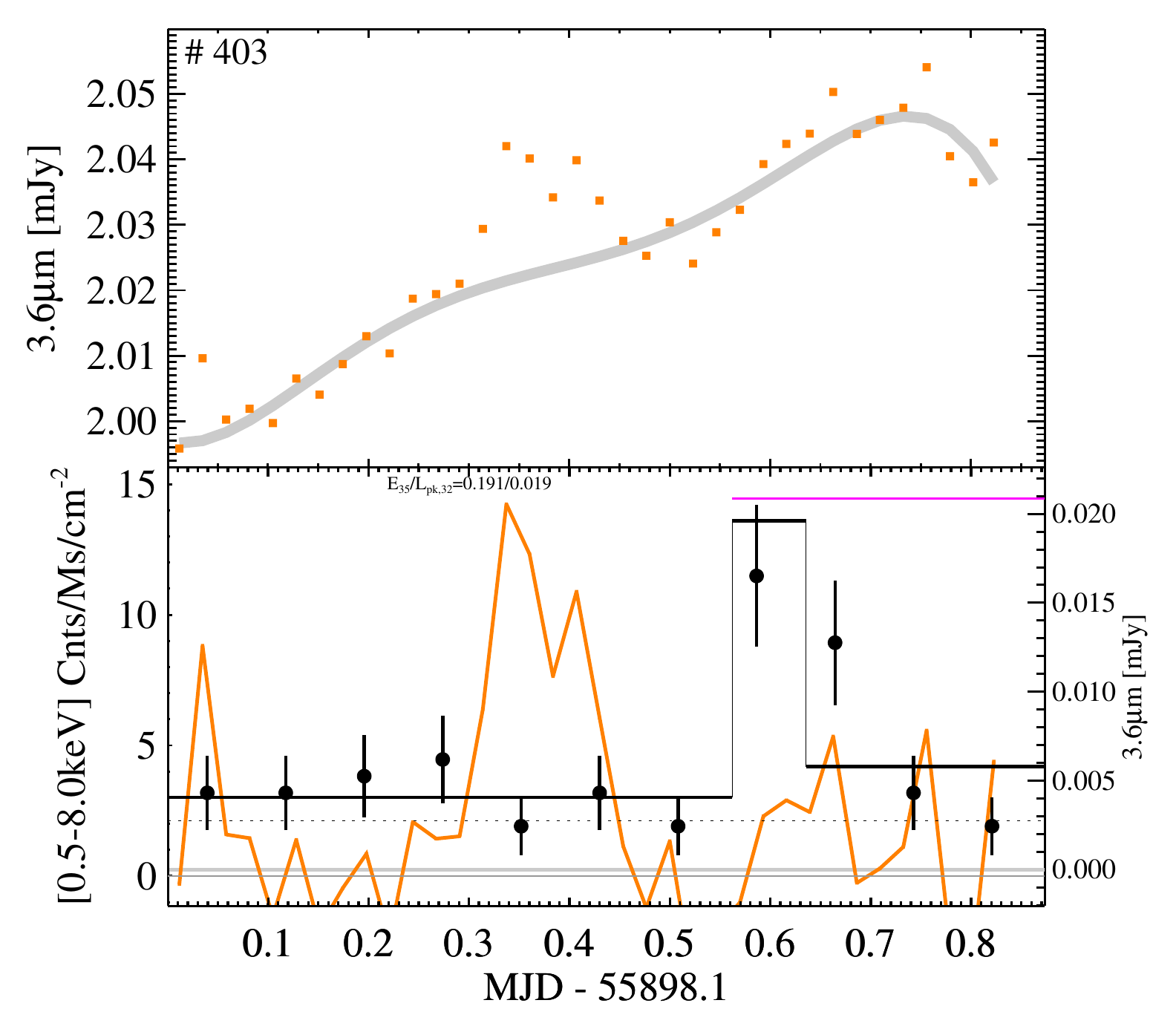}
\includegraphics[width=6.0cm]{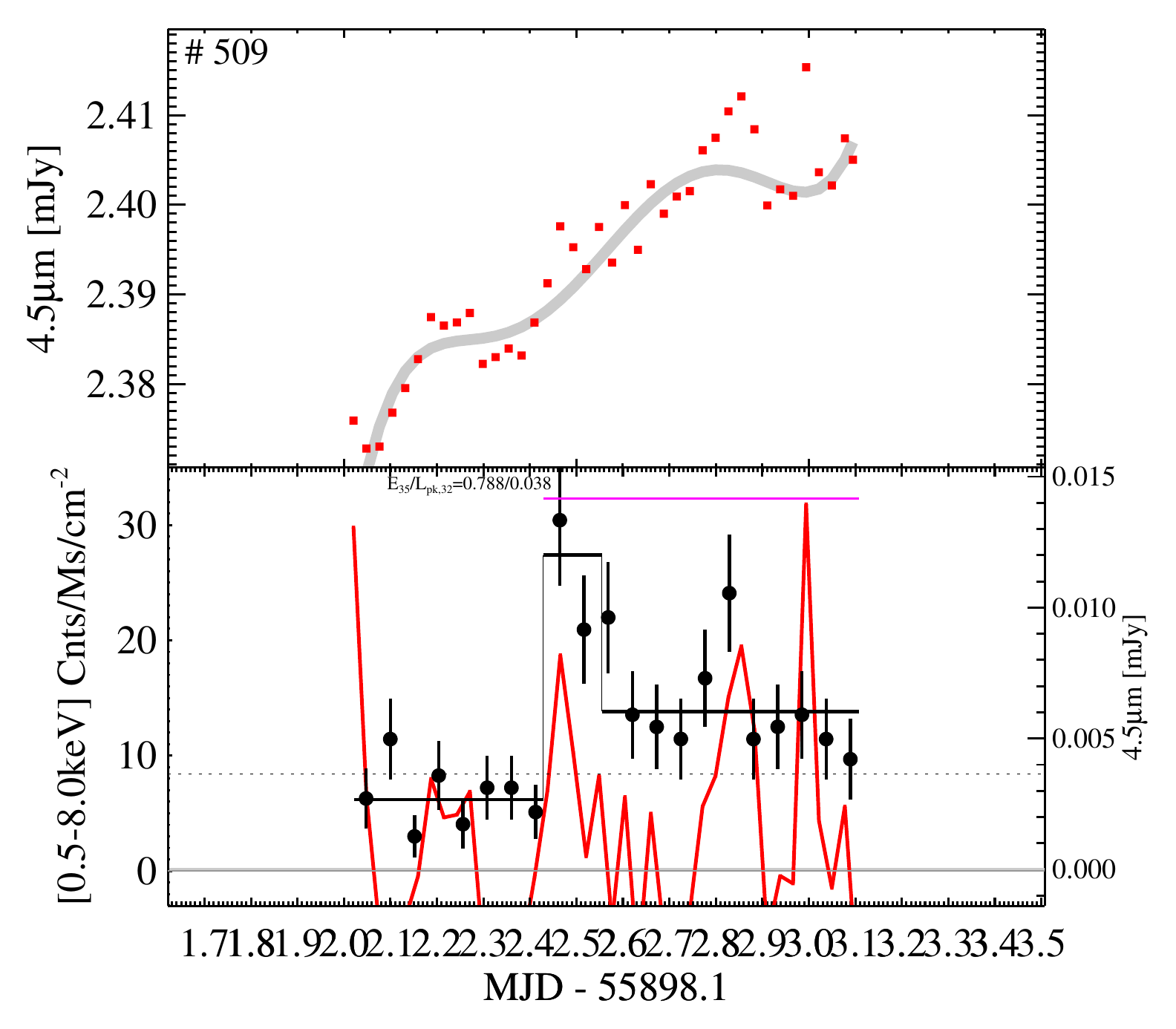}
\includegraphics[width=6.0cm]{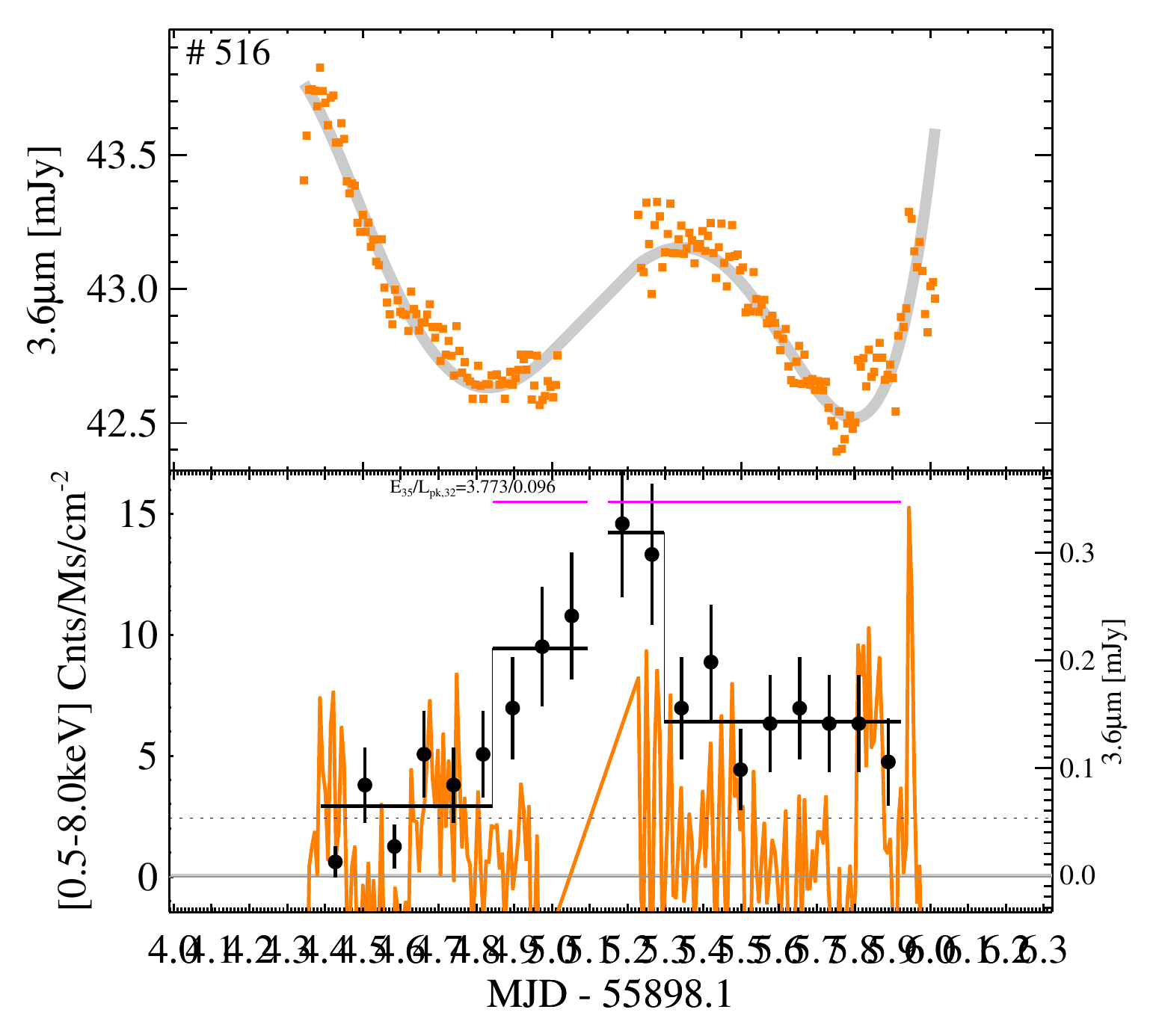}
\includegraphics[width=6.0cm]{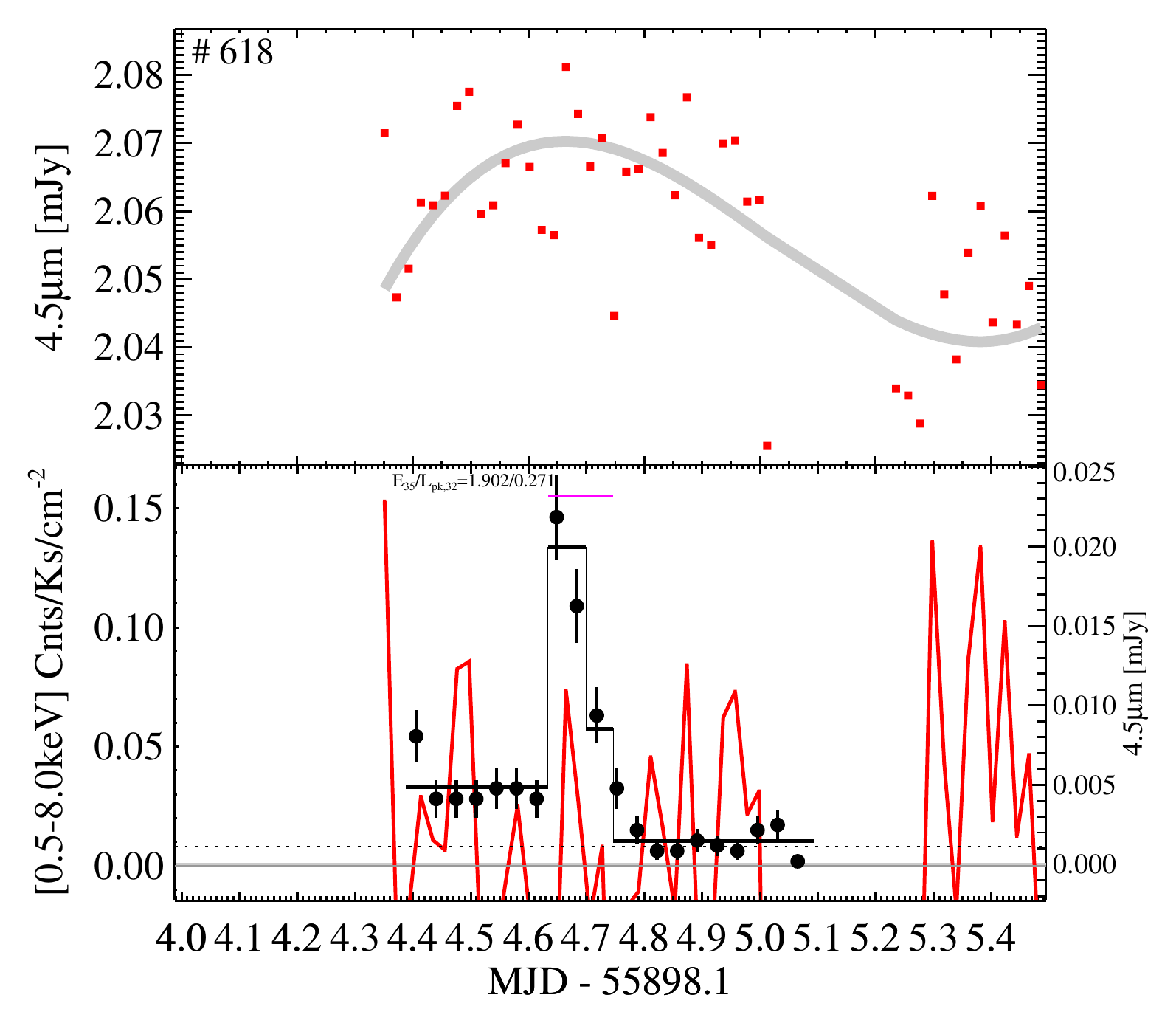}
\includegraphics[width=6.0cm]{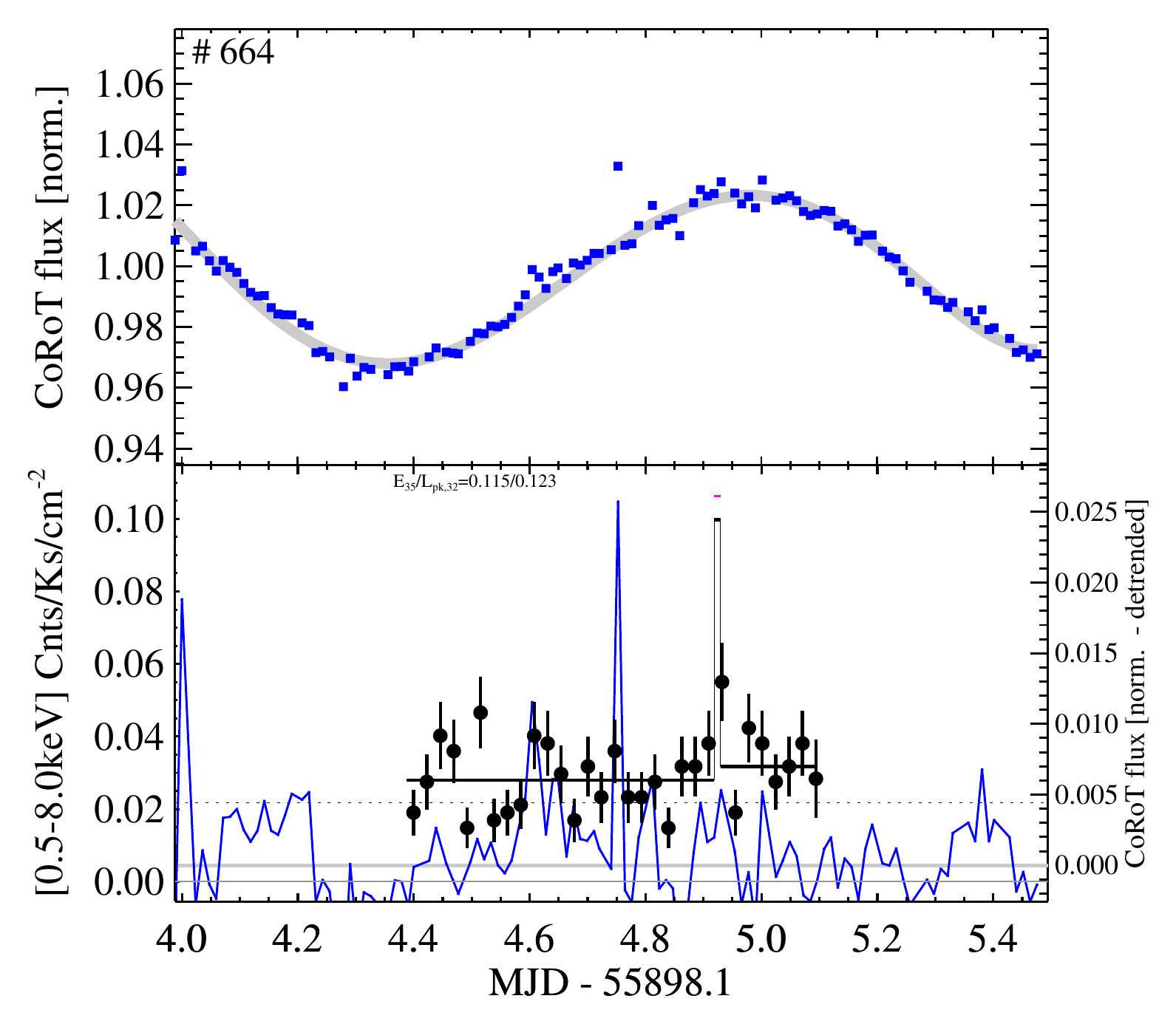}
\includegraphics[width=6.0cm]{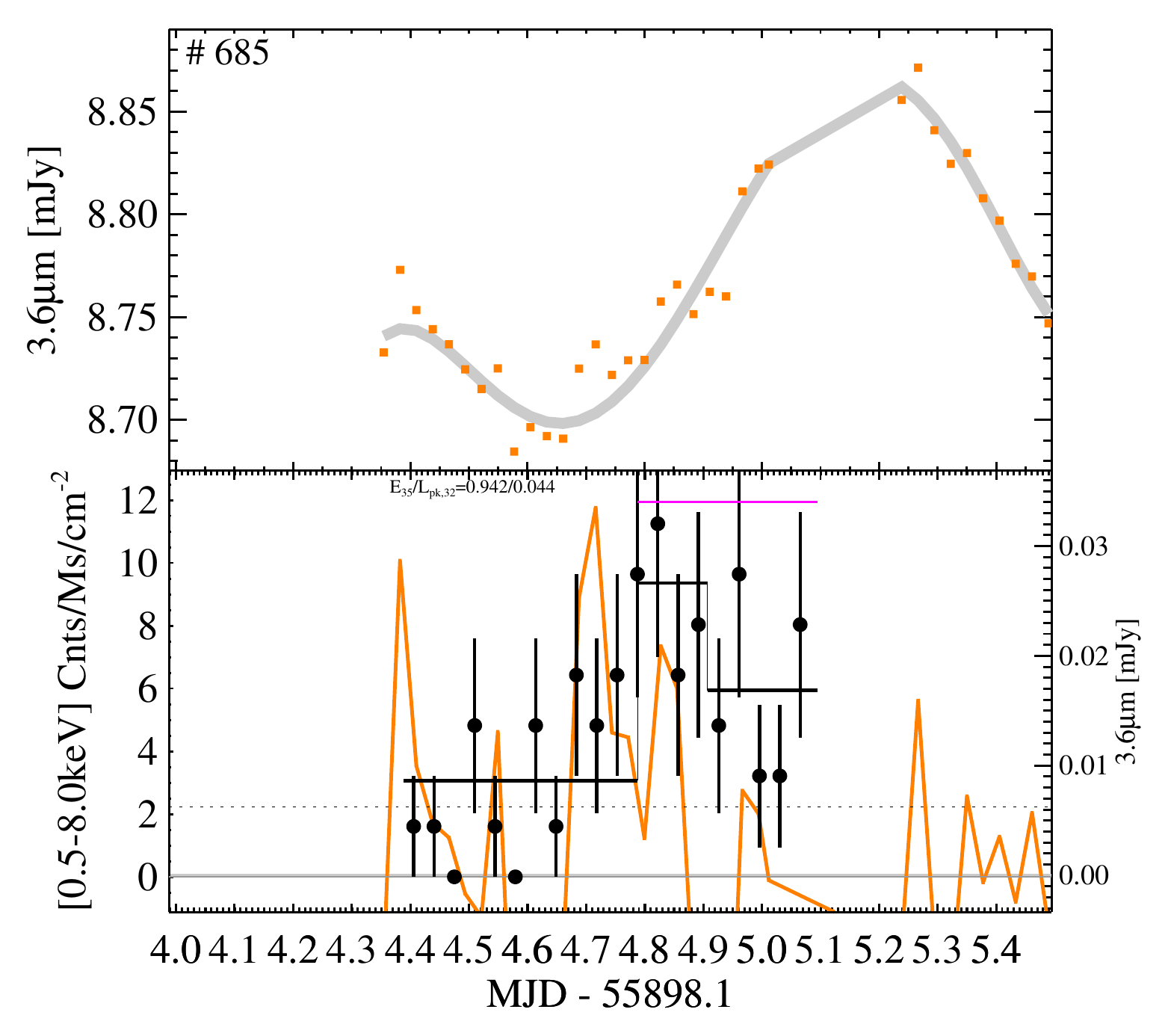}
\includegraphics[width=6.0cm]{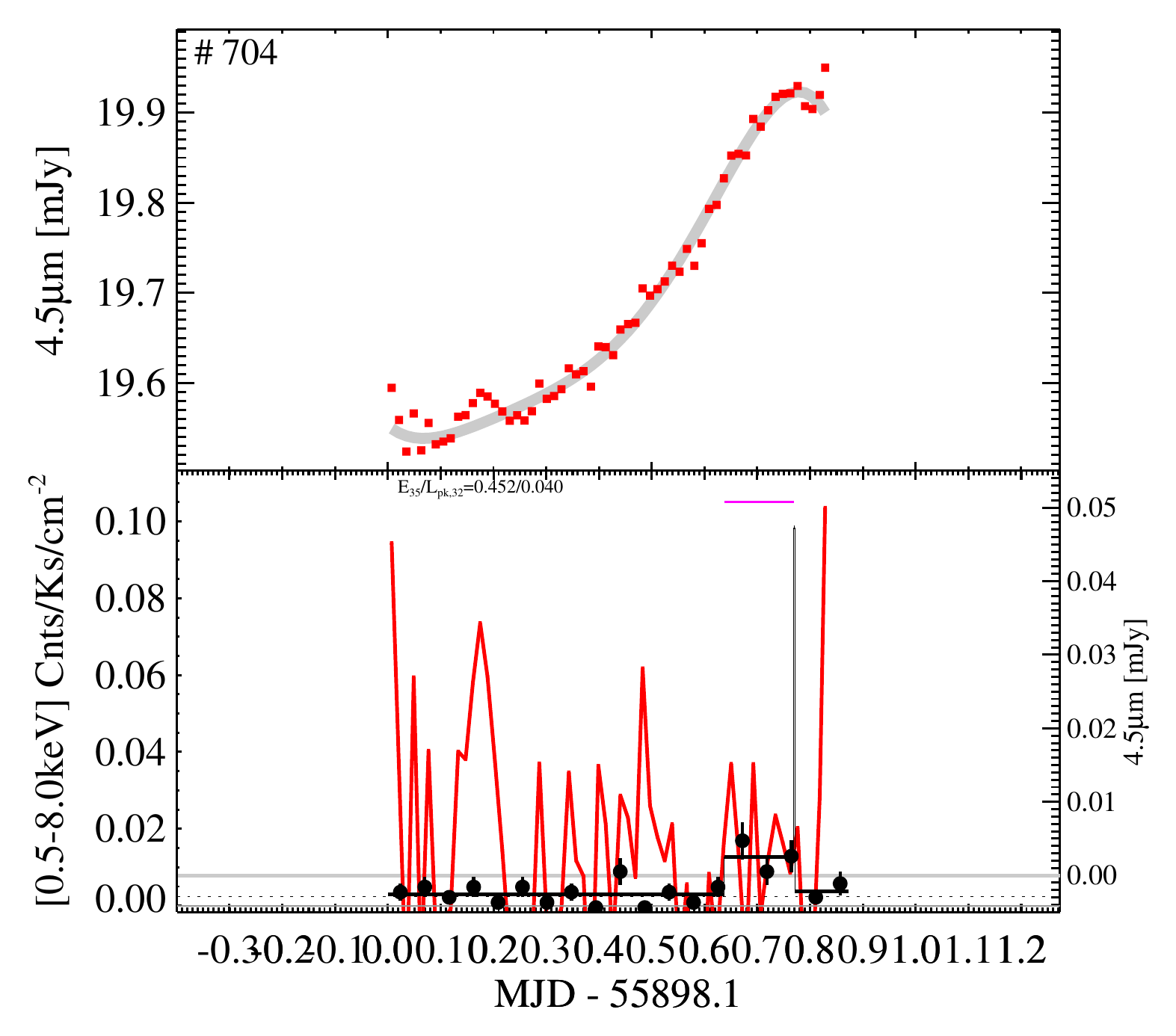}
\label{fig:}
\end{figure*}
\begin{figure*}[!t!]
\centering
\includegraphics[width=6.0cm]{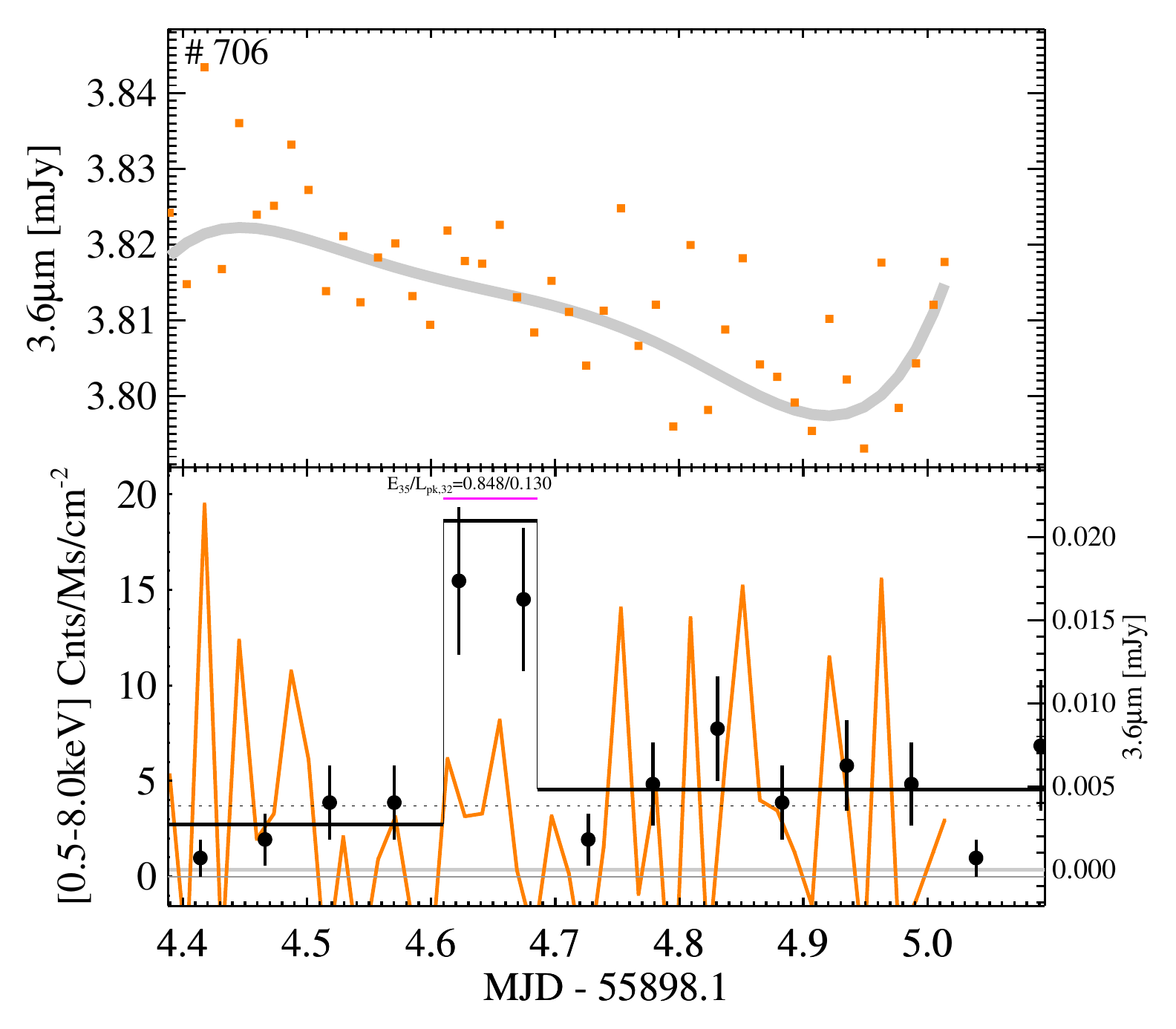}
\includegraphics[width=6.0cm]{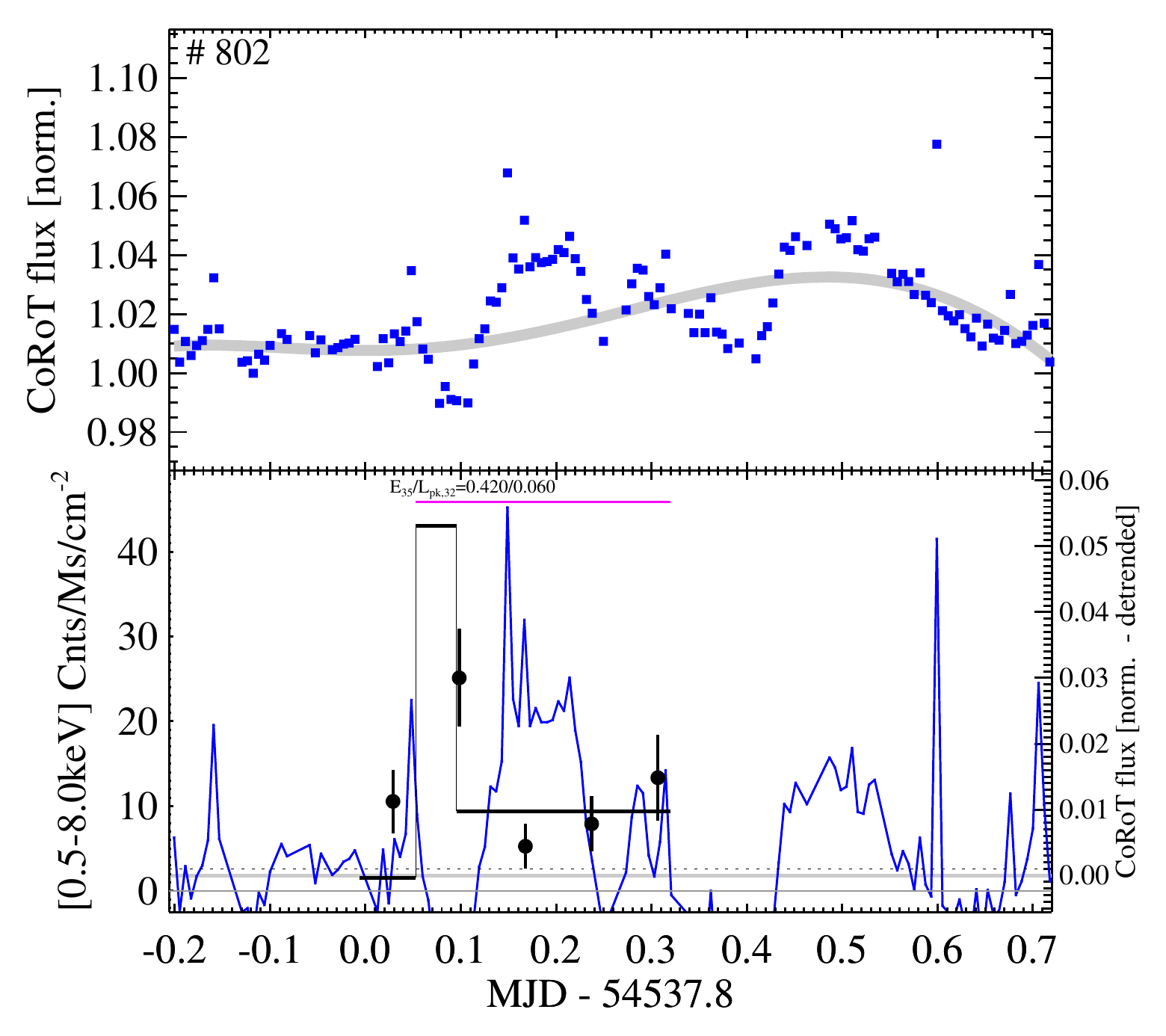}
\includegraphics[width=6.0cm]{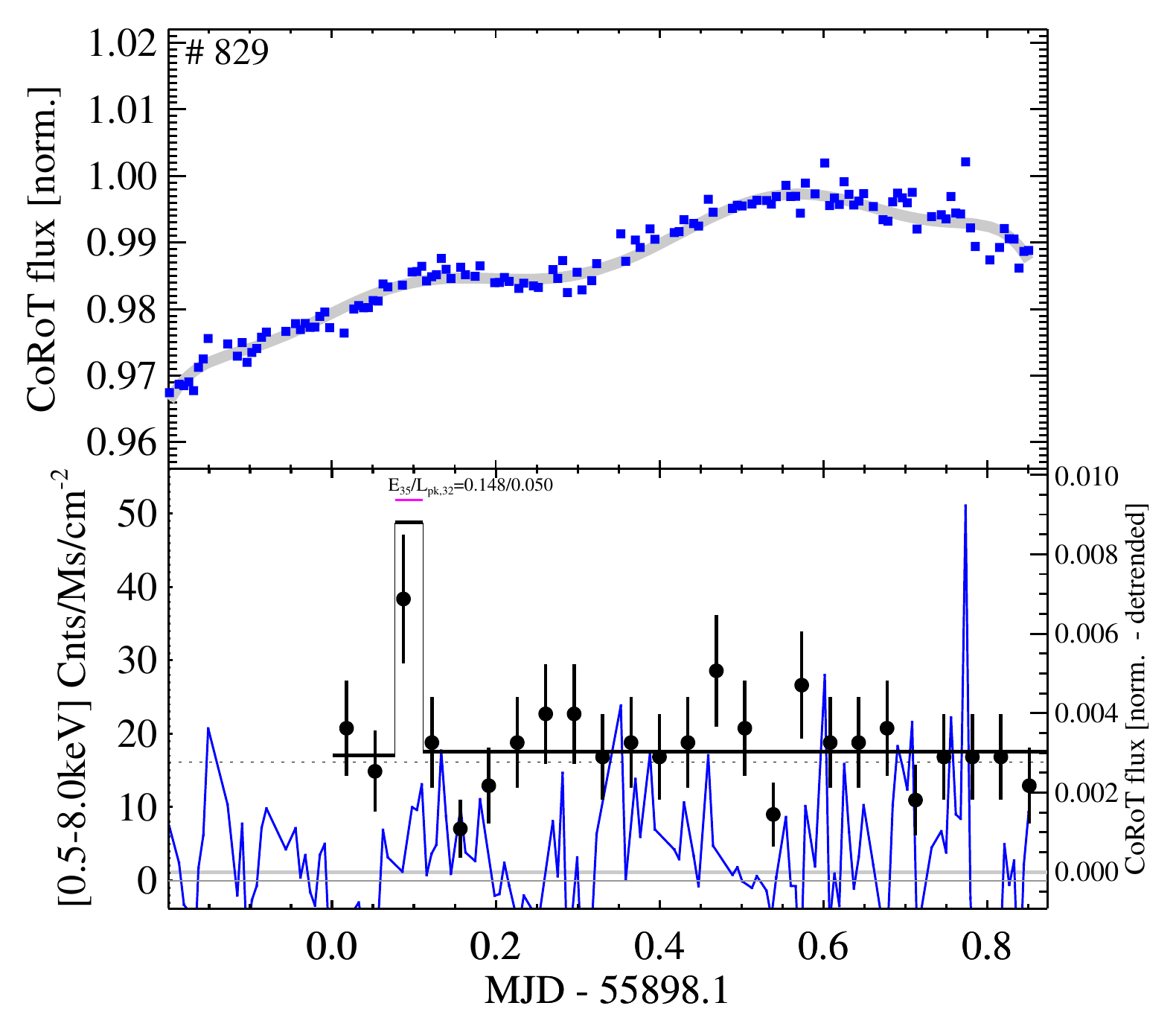}
\includegraphics[width=6.0cm]{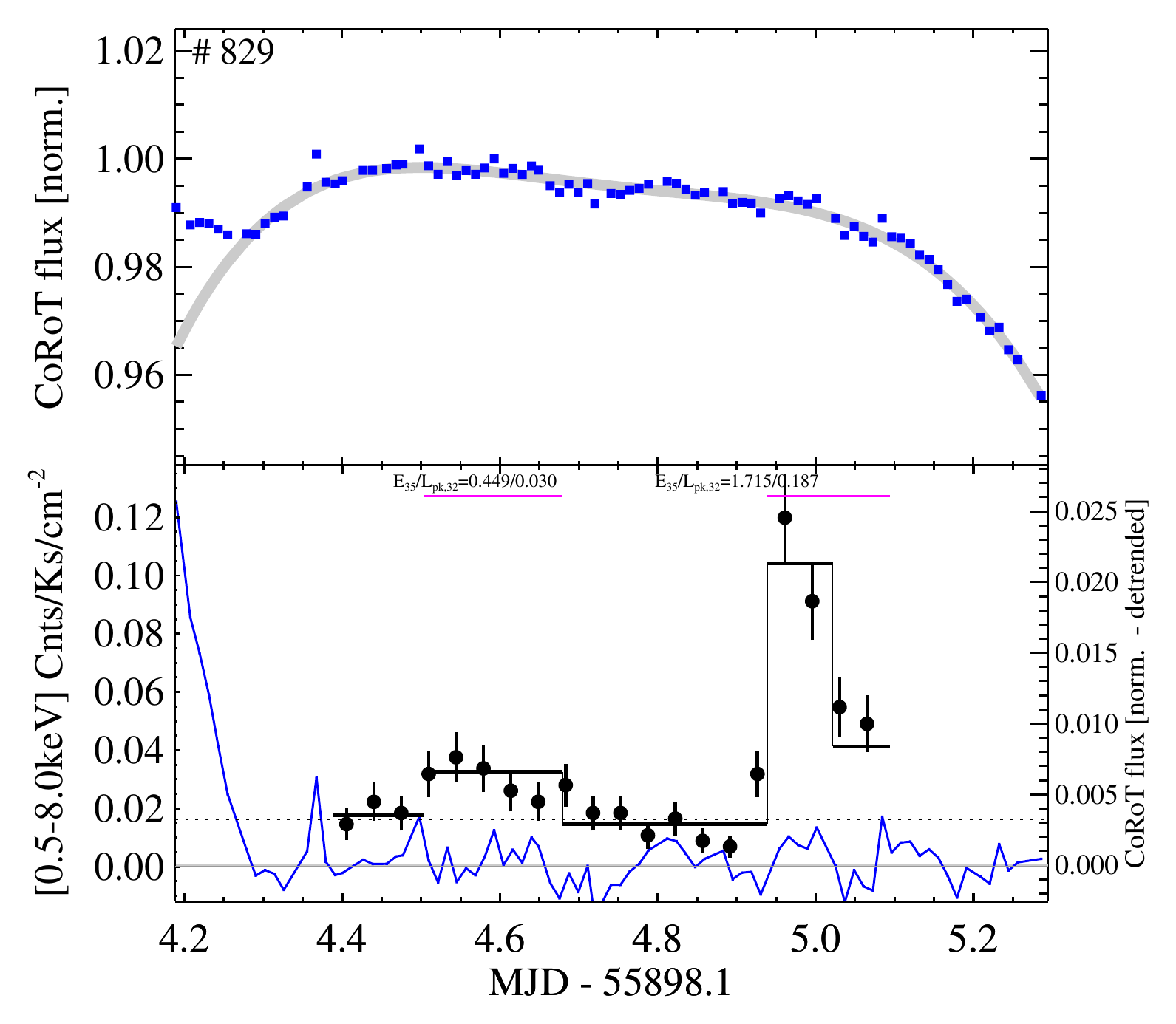}
\includegraphics[width=6.0cm]{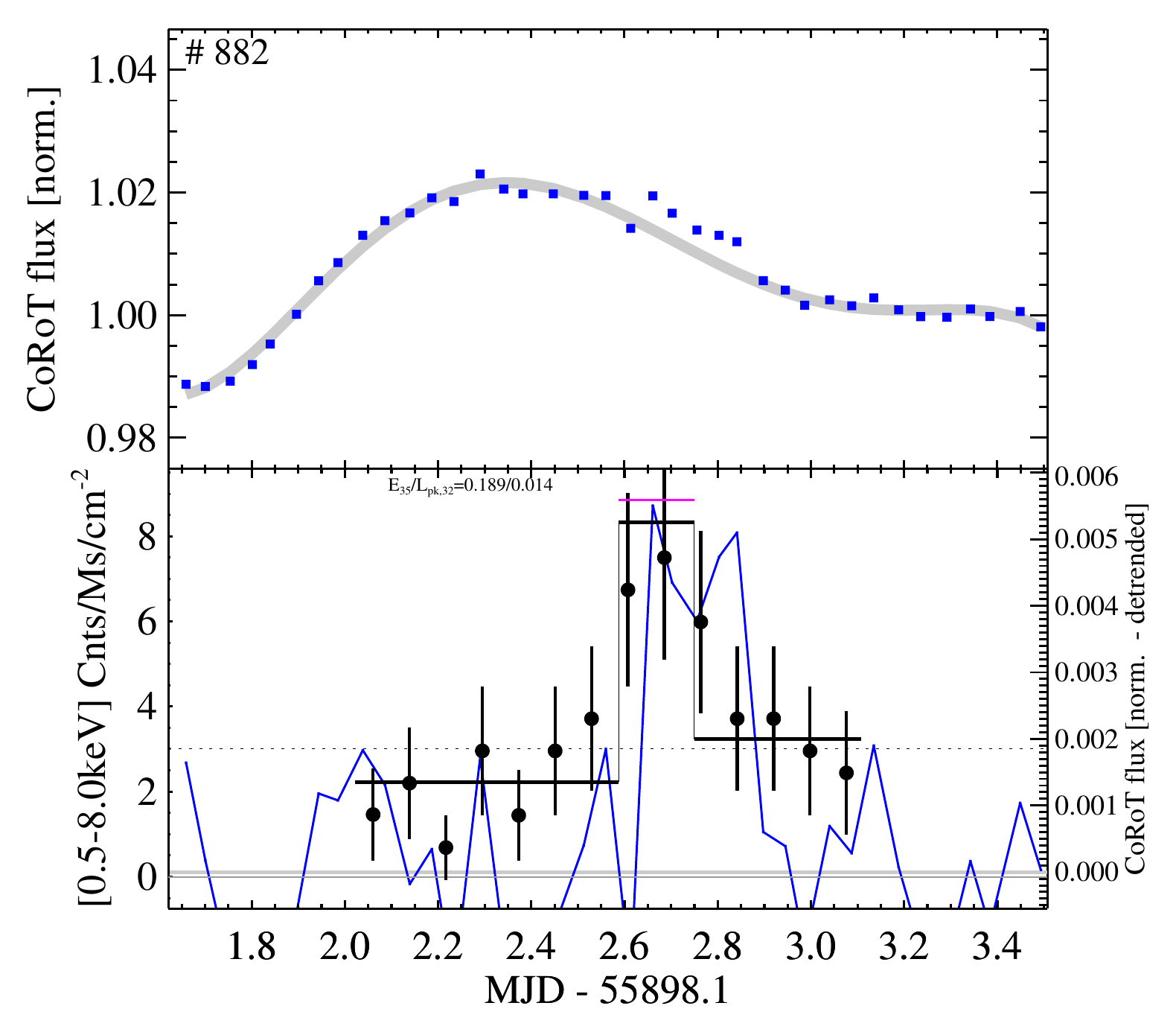}
\includegraphics[width=6.0cm]{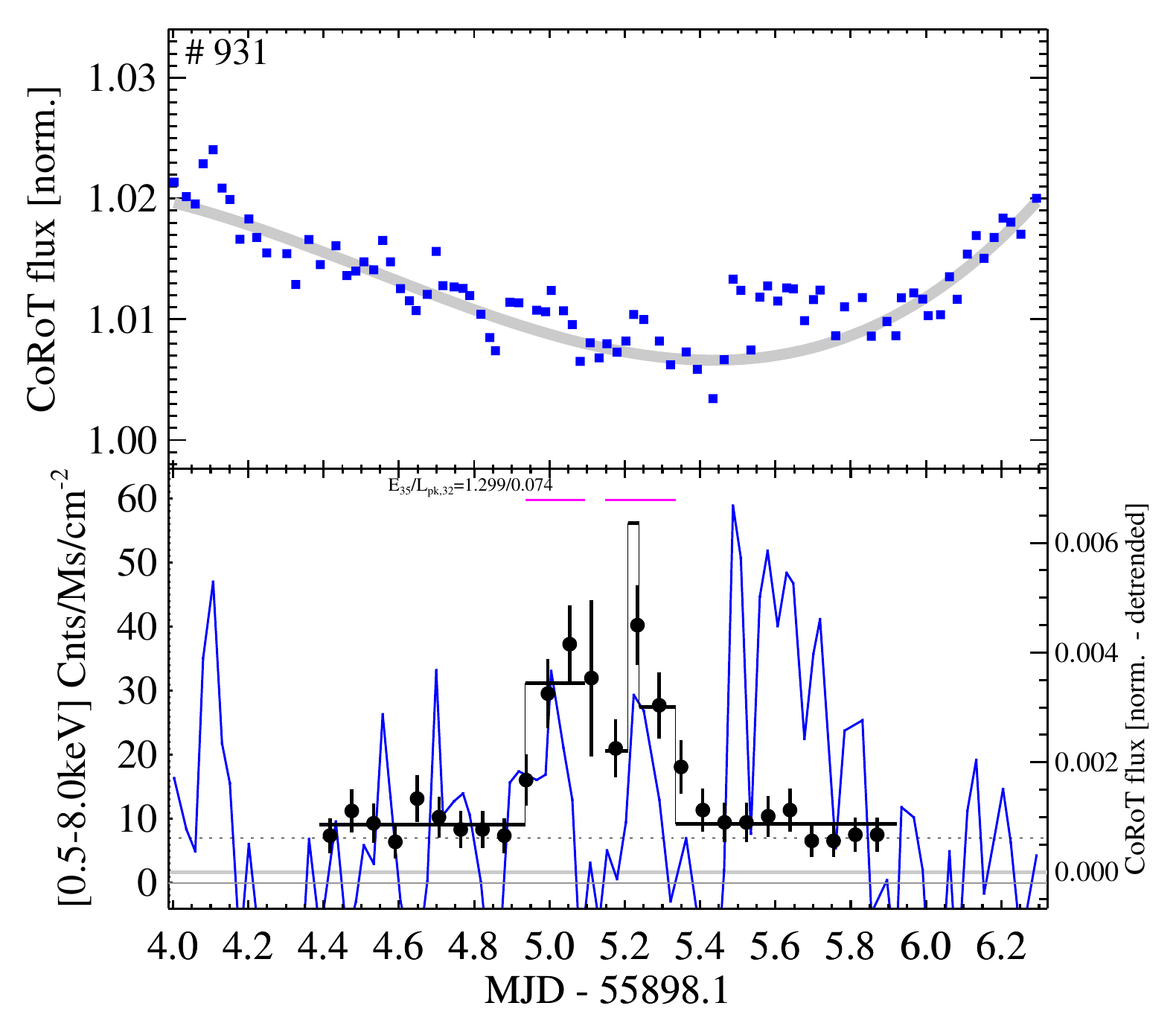}
\includegraphics[width=6.0cm]{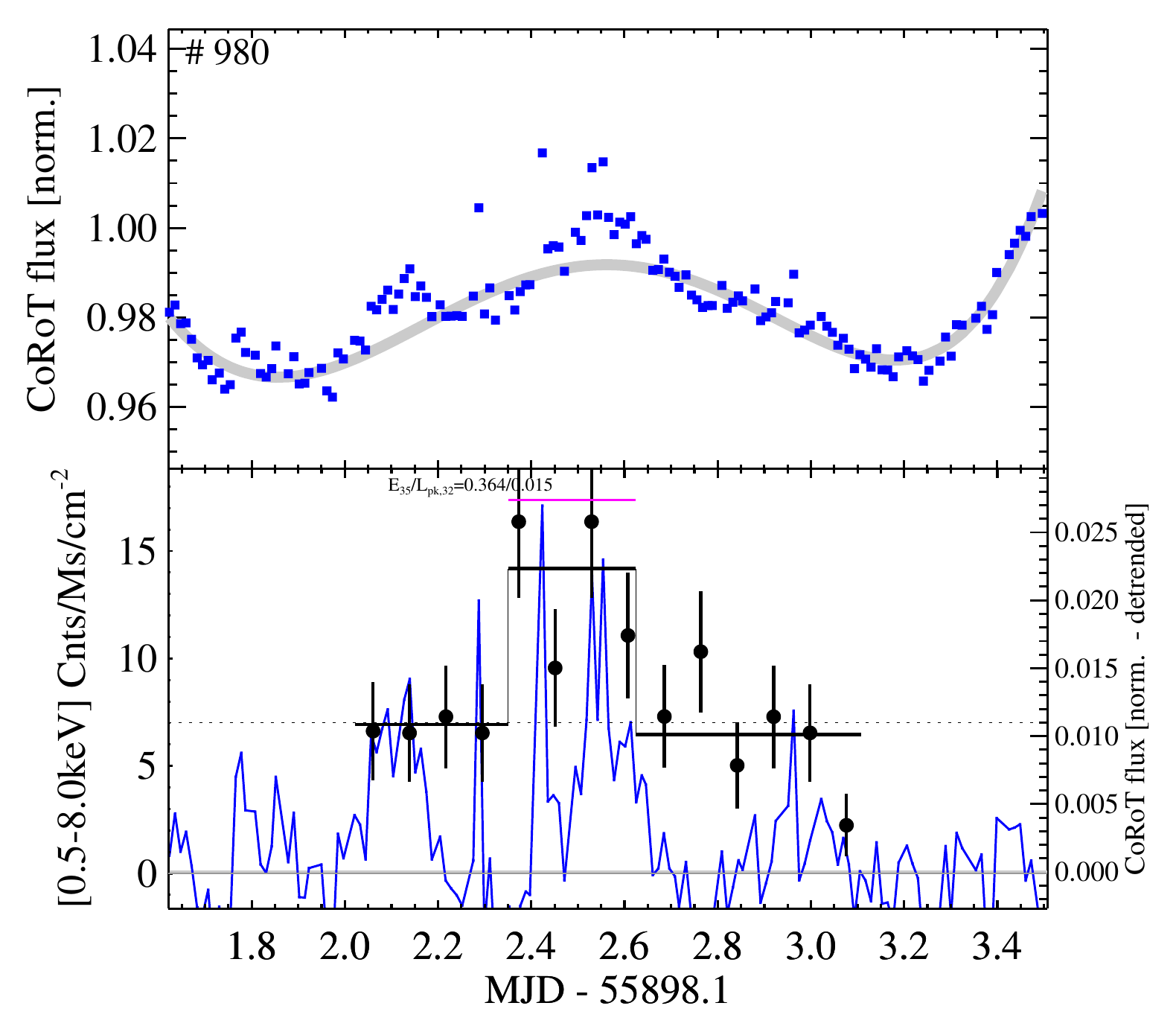}
\includegraphics[width=6.0cm]{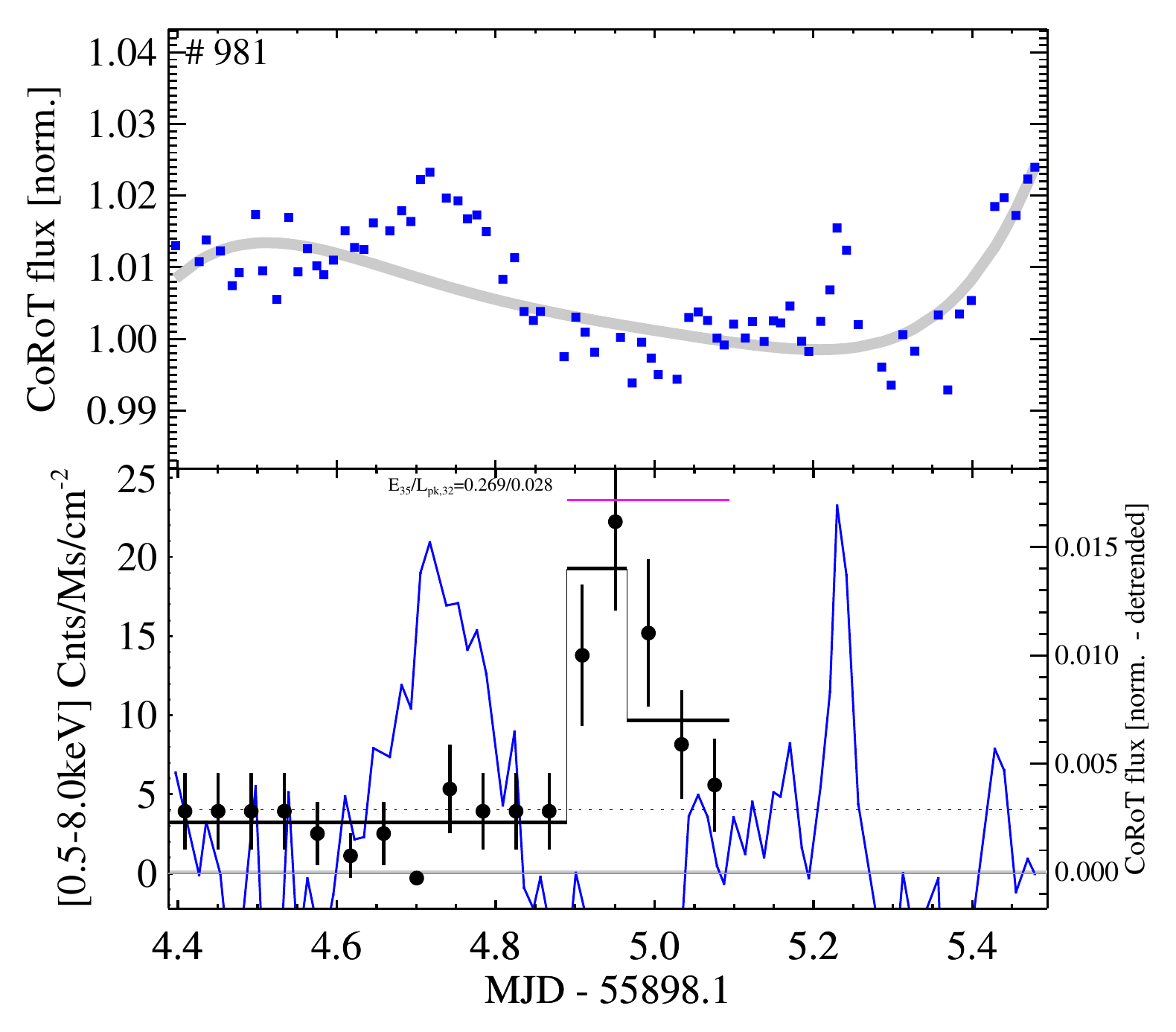}
\includegraphics[width=6.0cm]{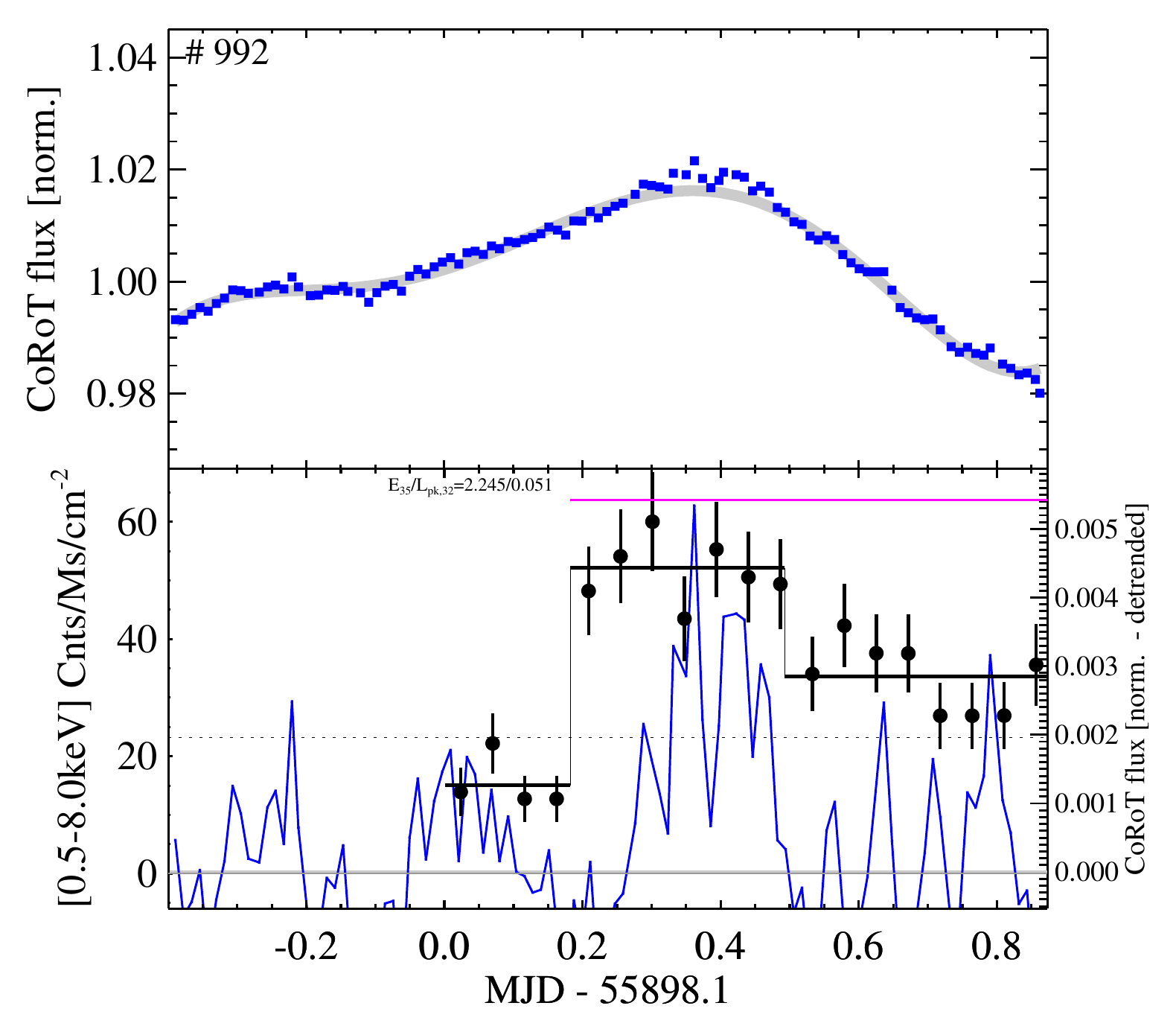}
\label{fig:}
\end{figure*}
\begin{figure*}[!t!]
\centering
\includegraphics[width=6.0cm]{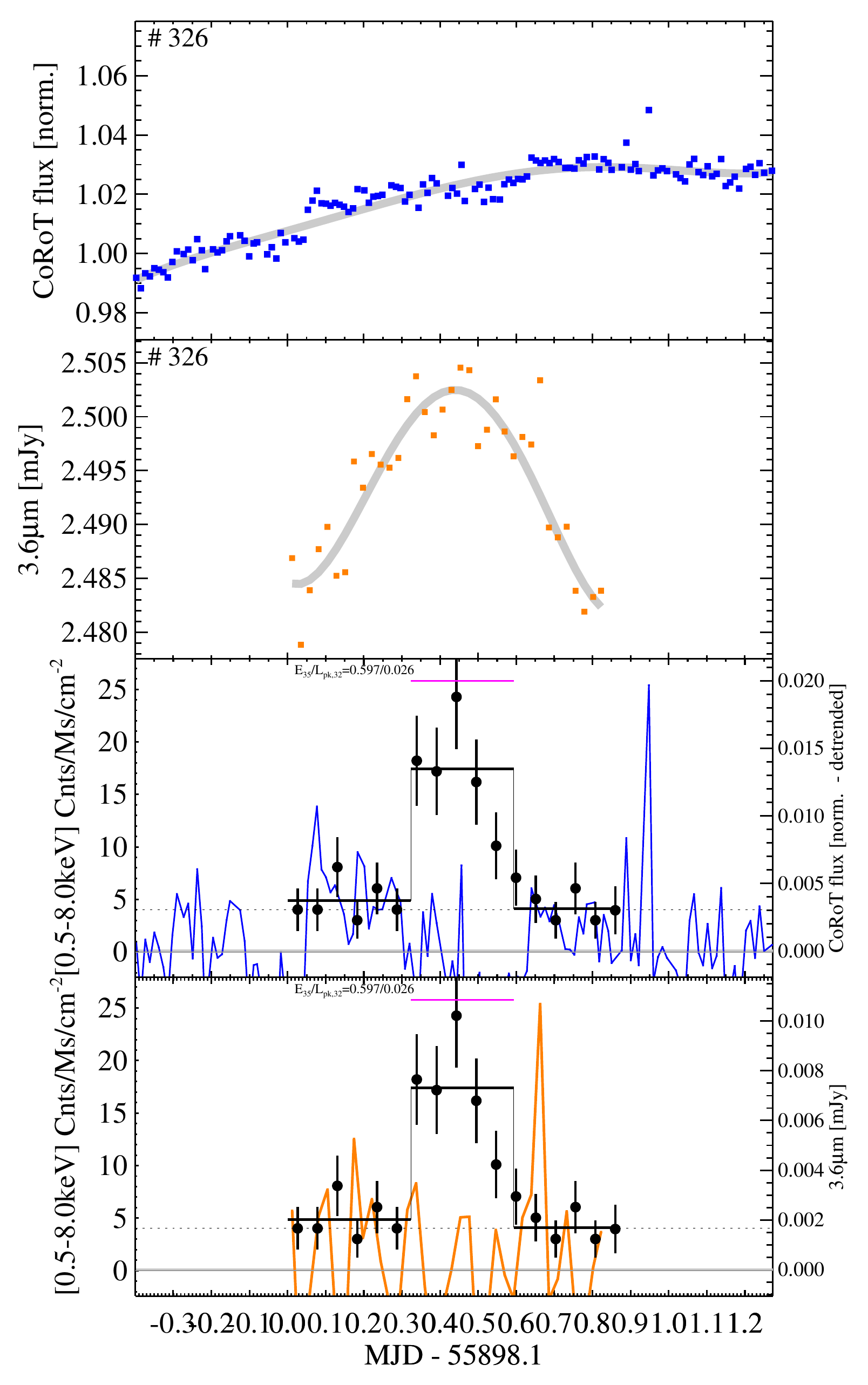}
\includegraphics[width=6.0cm]{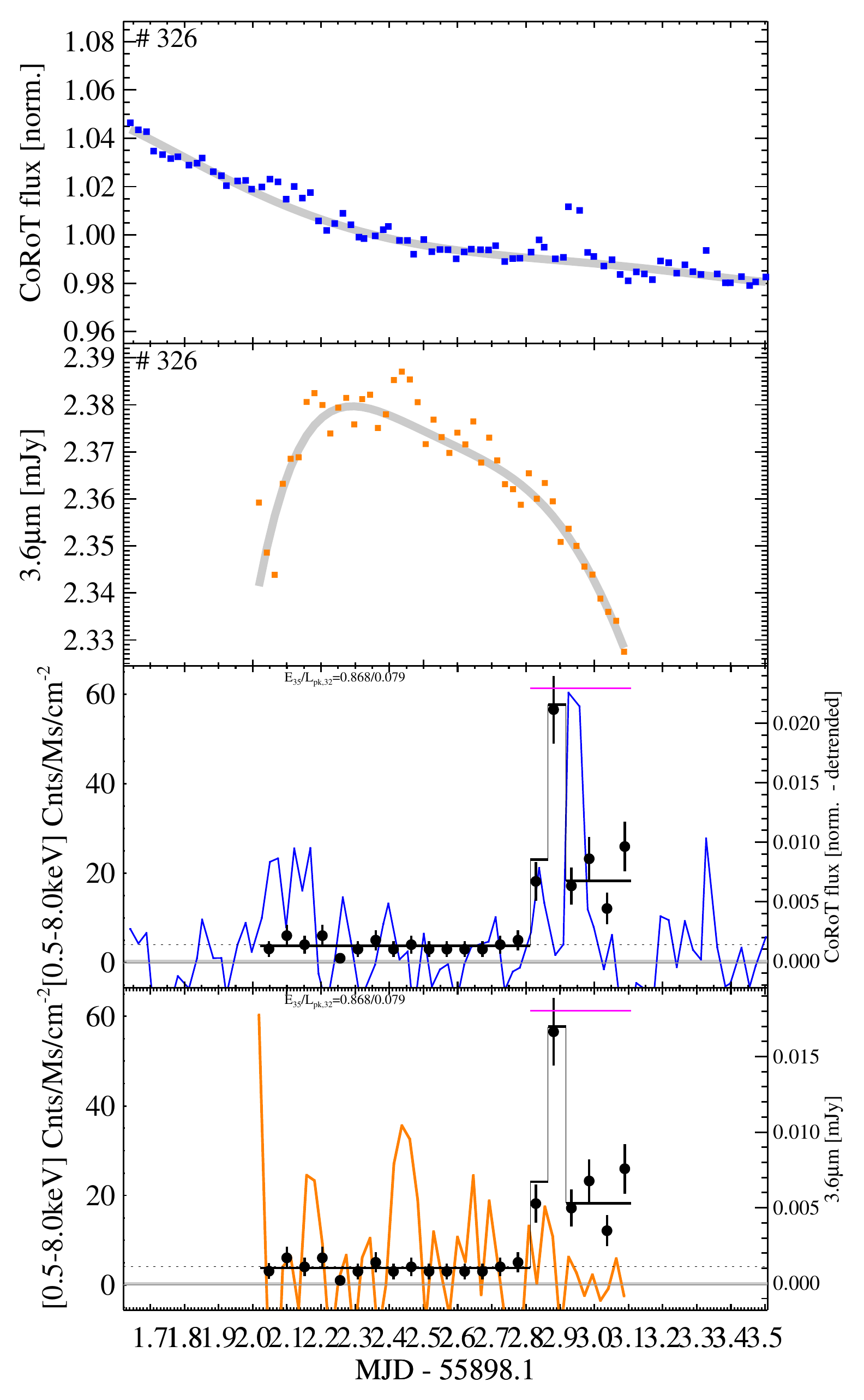}
\includegraphics[width=6.0cm]{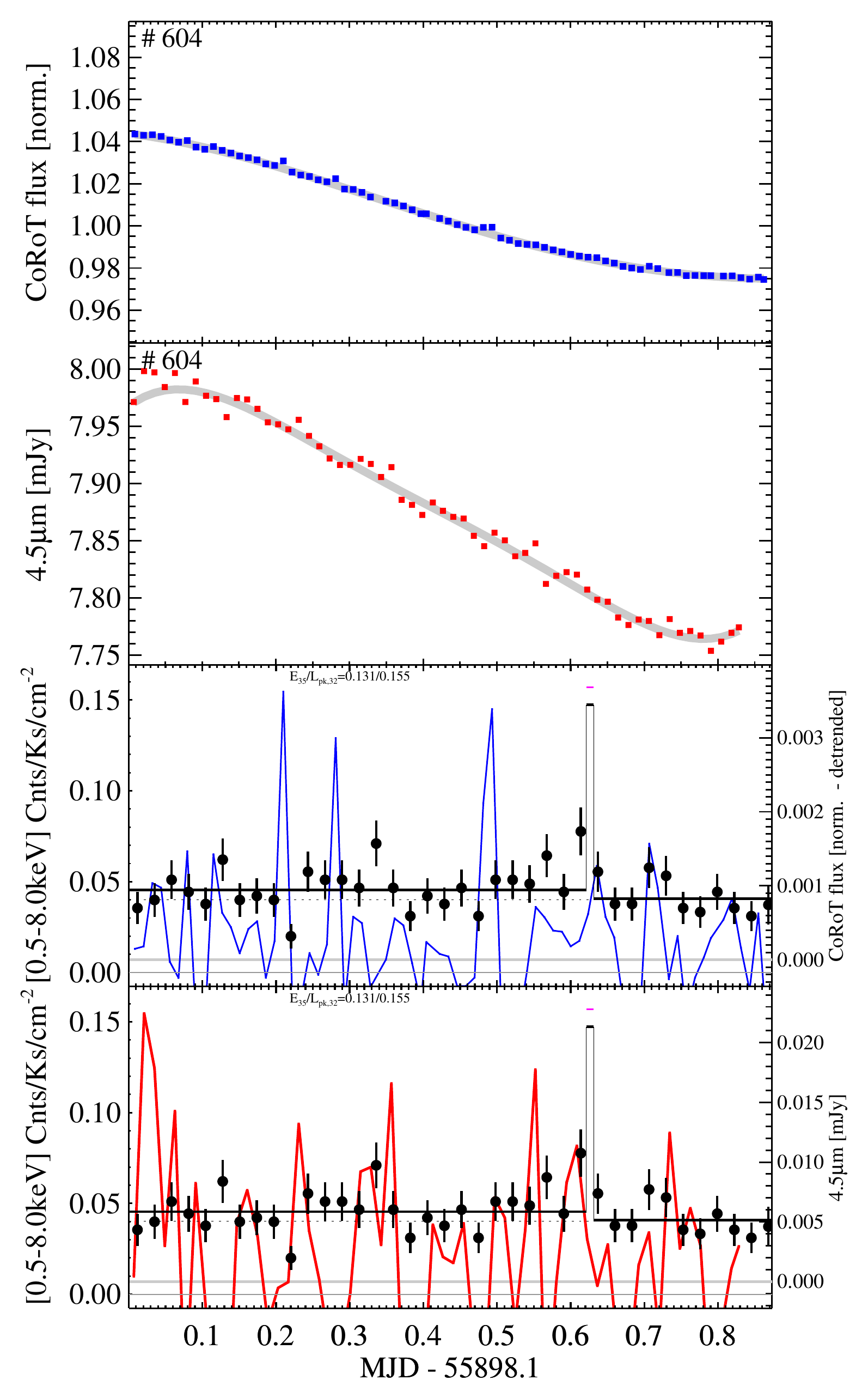}
\includegraphics[width=6.0cm]{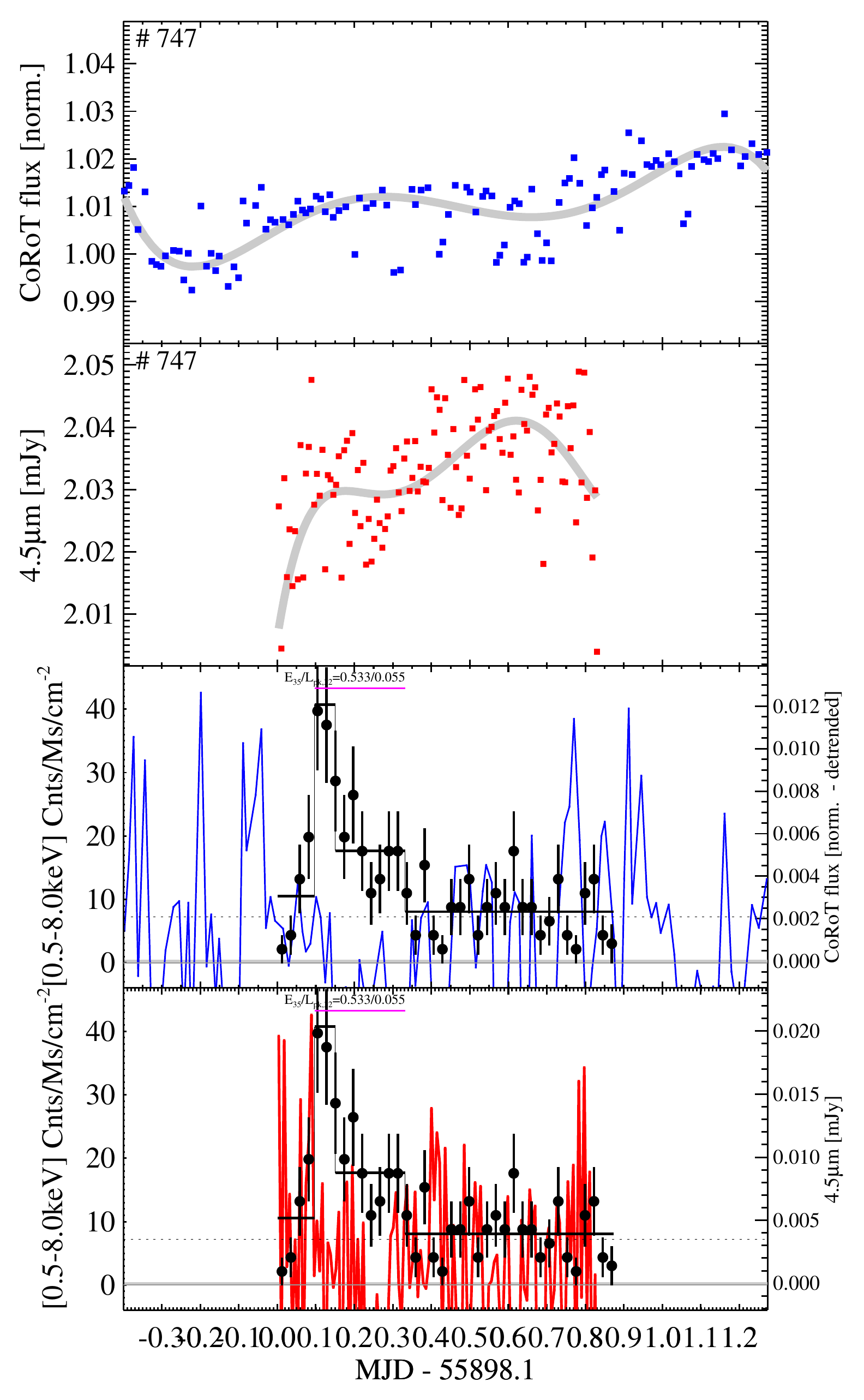}
\includegraphics[width=6.0cm]{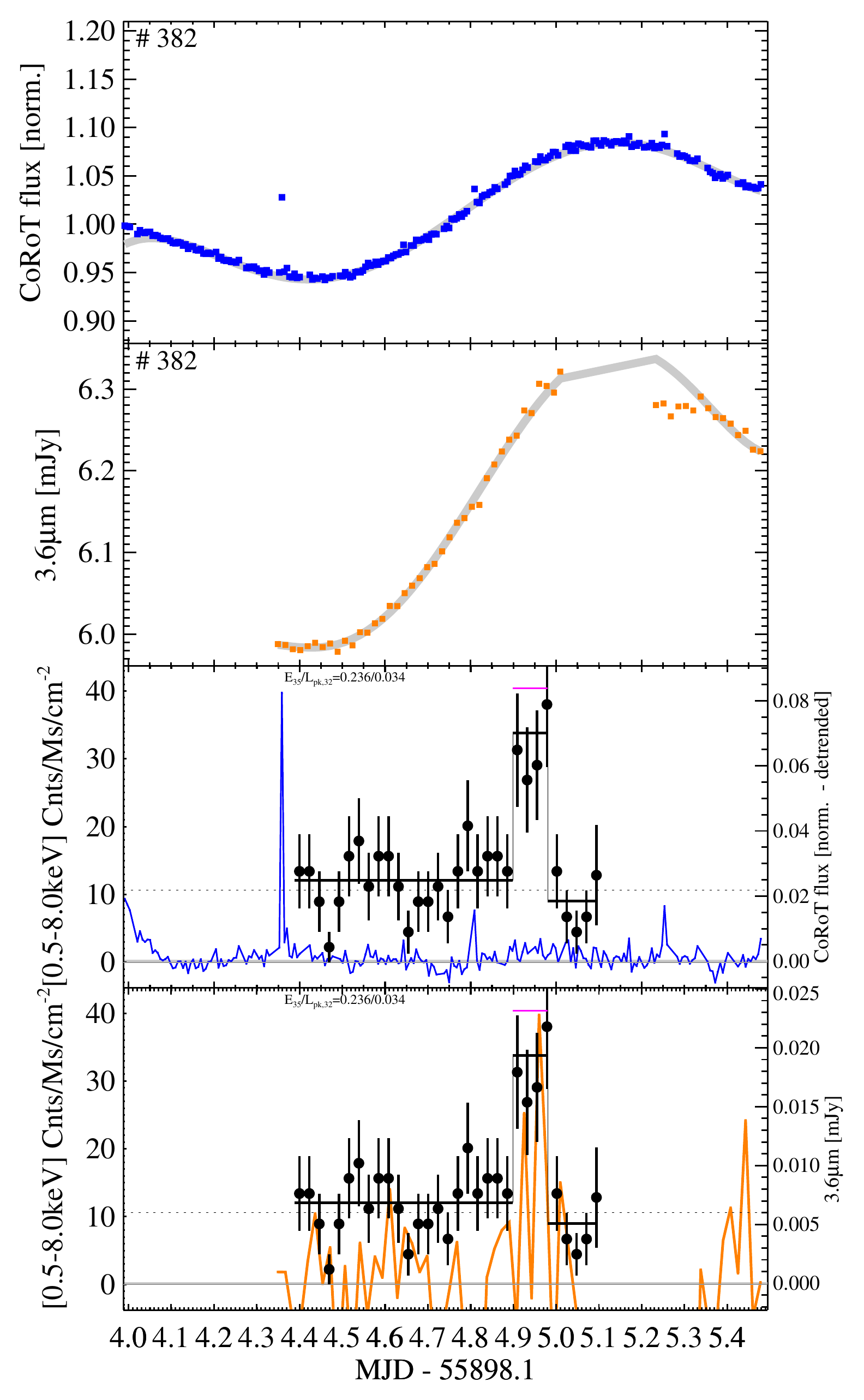}
\includegraphics[width=6.0cm]{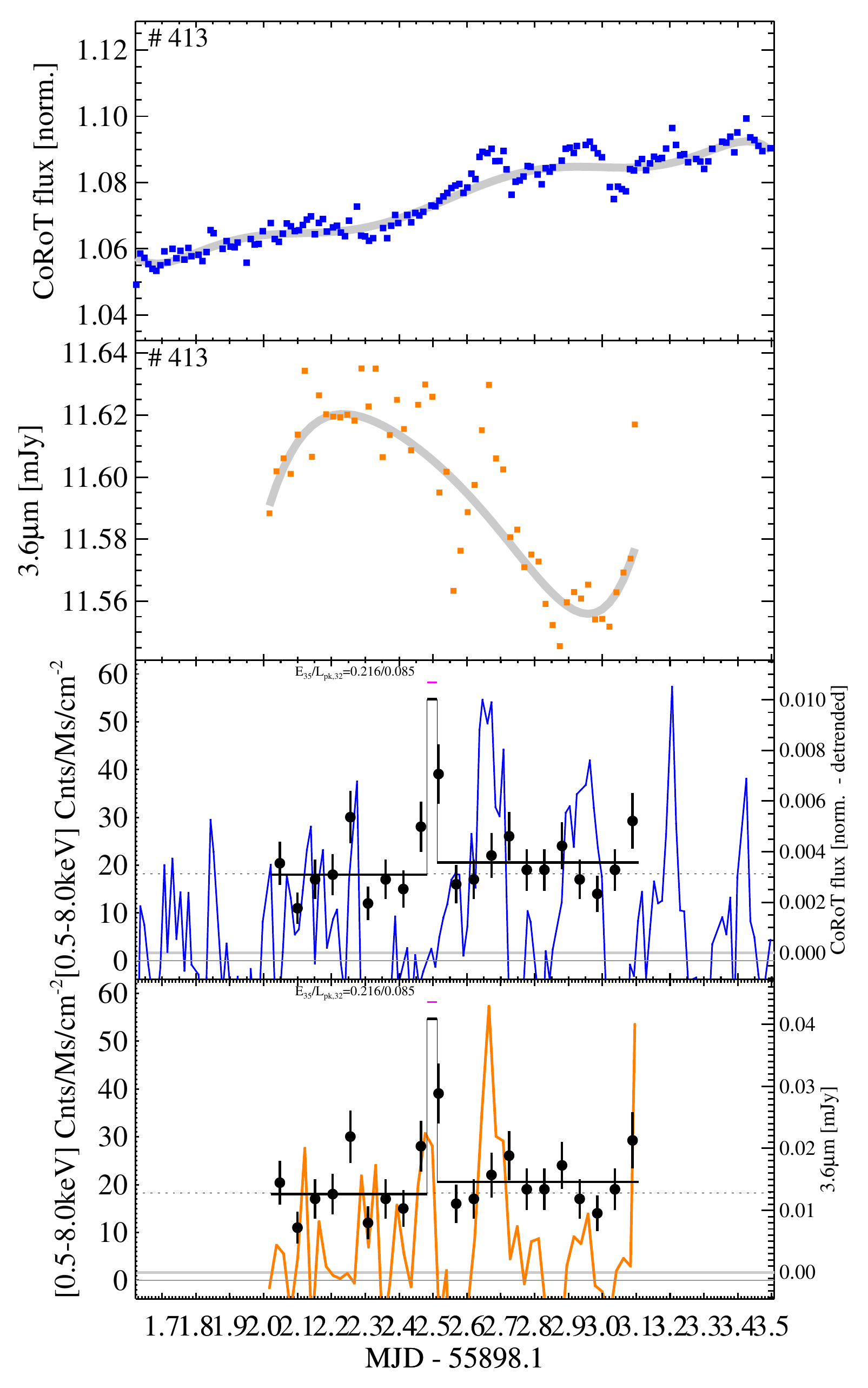}
\label{fig:}
\end{figure*}
\begin{figure*}[!t!]
\centering
\includegraphics[width=6.0cm]{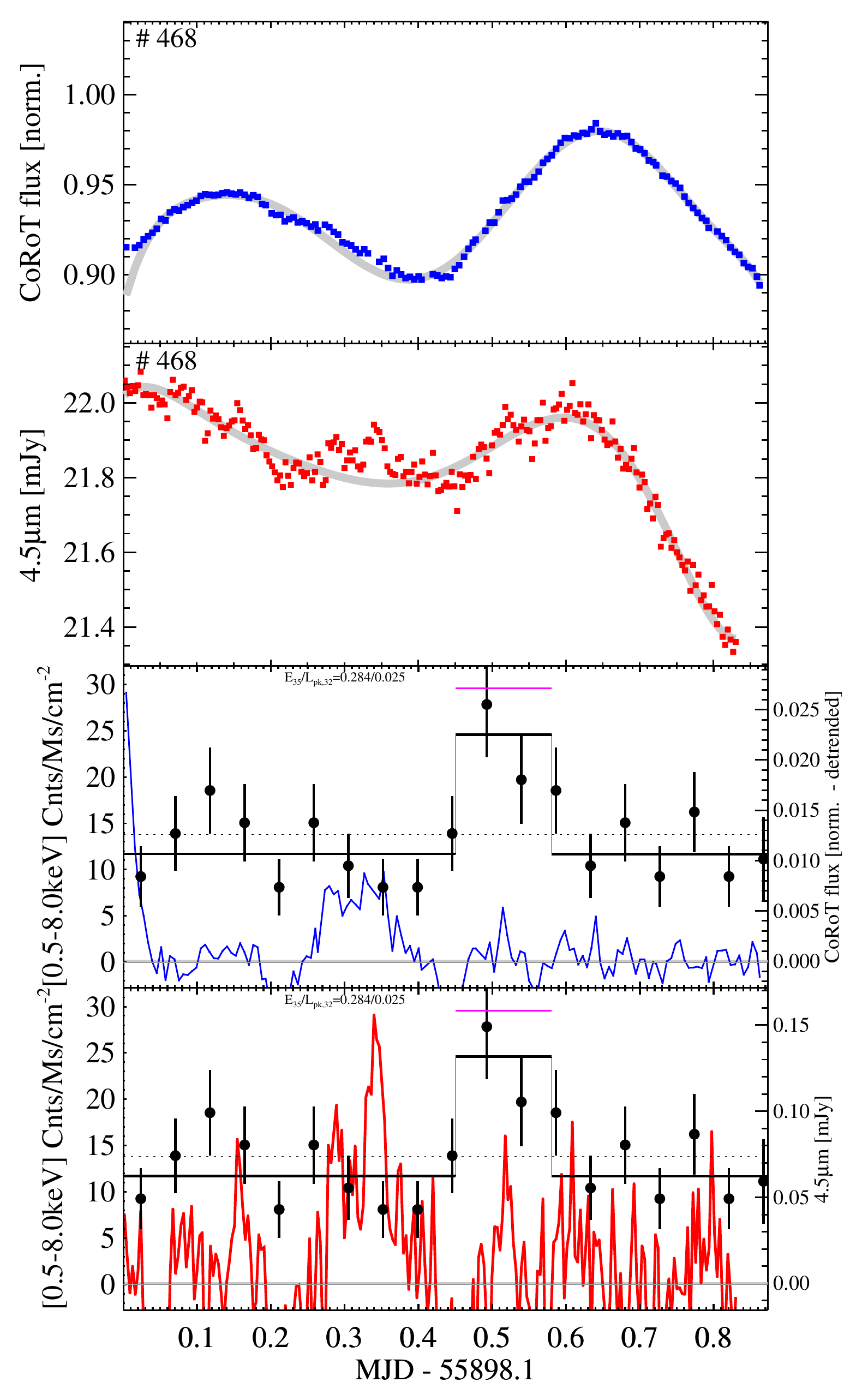}
\includegraphics[width=6.0cm]{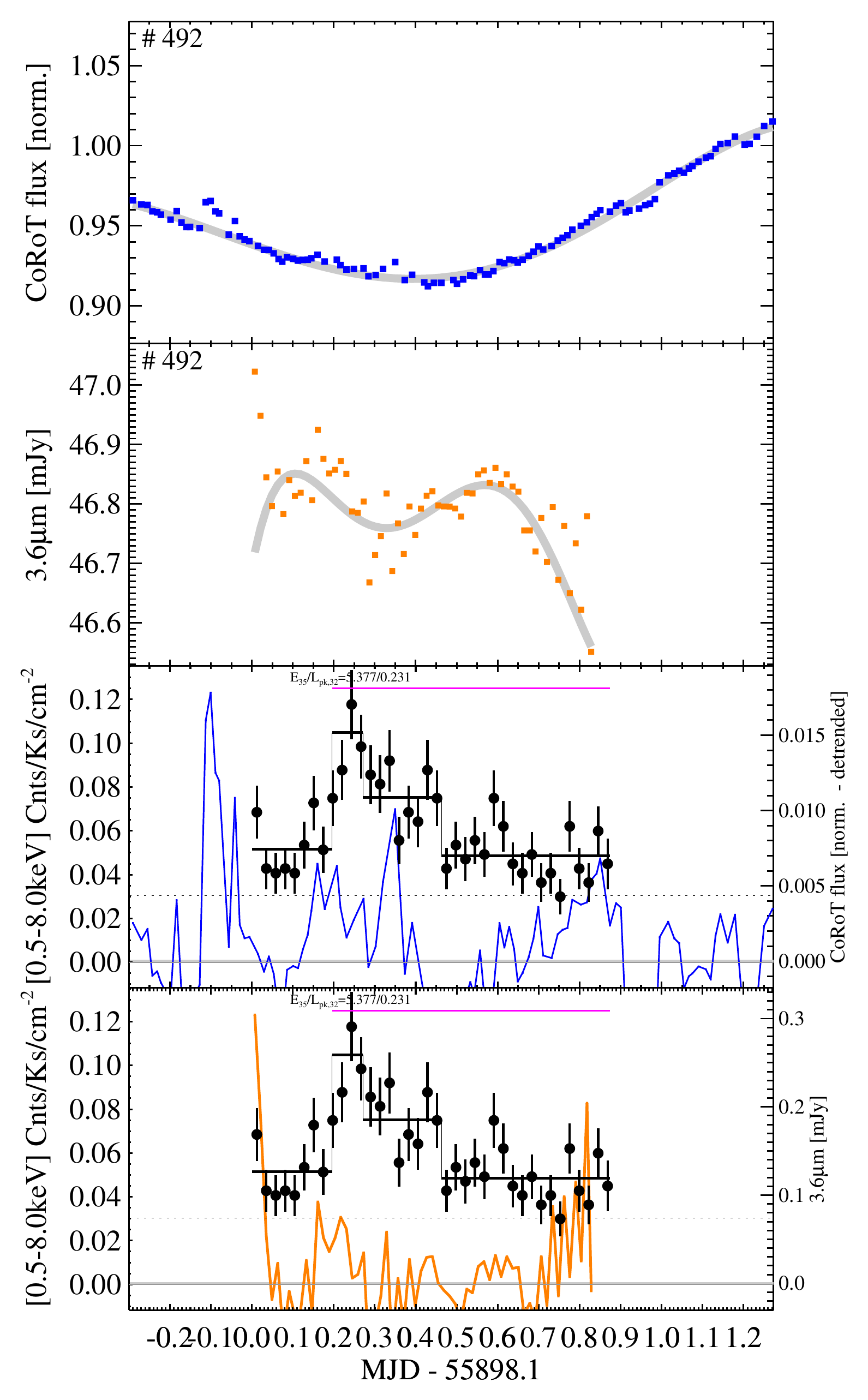}
\includegraphics[width=6.0cm]{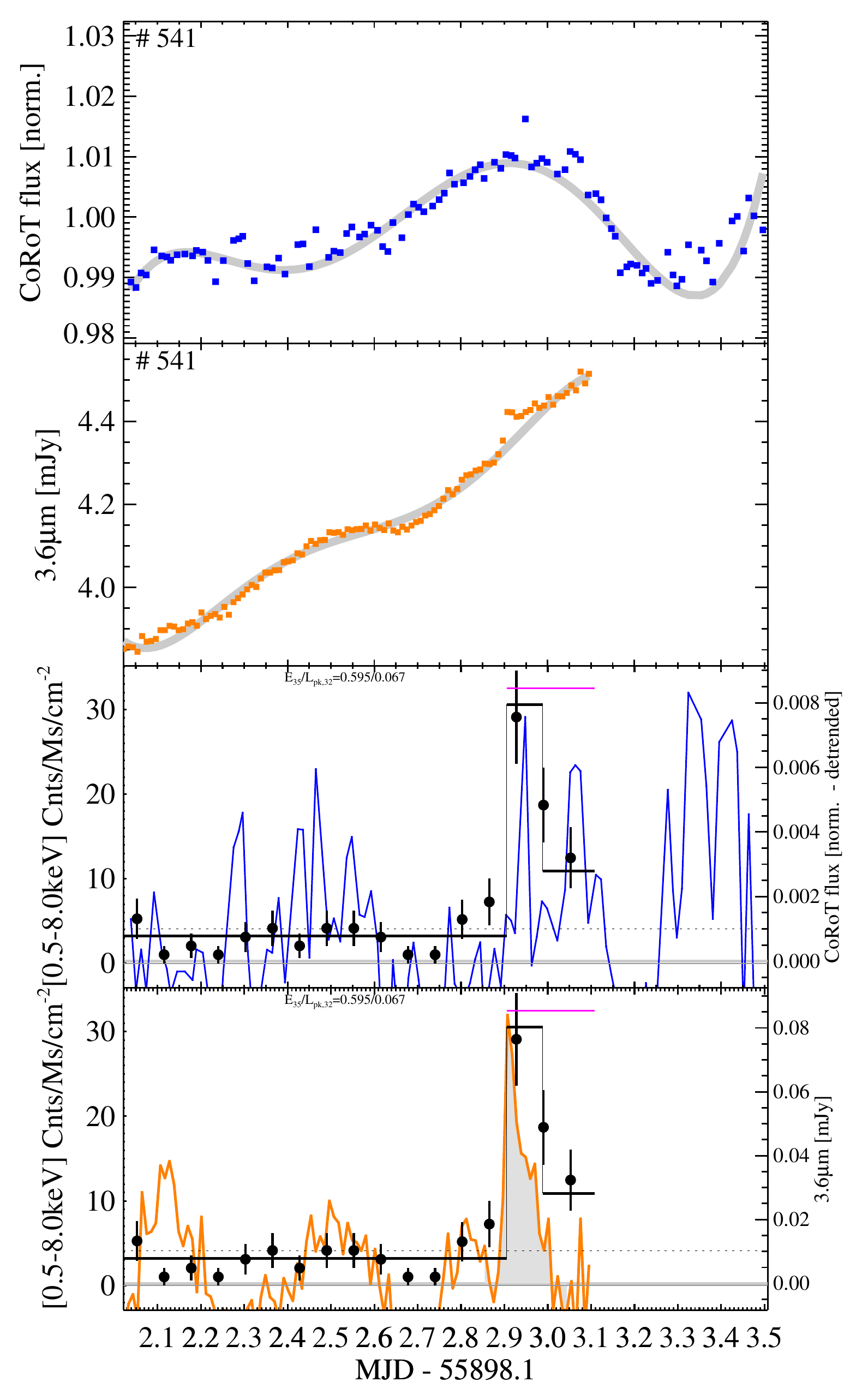}
\includegraphics[width=6.0cm]{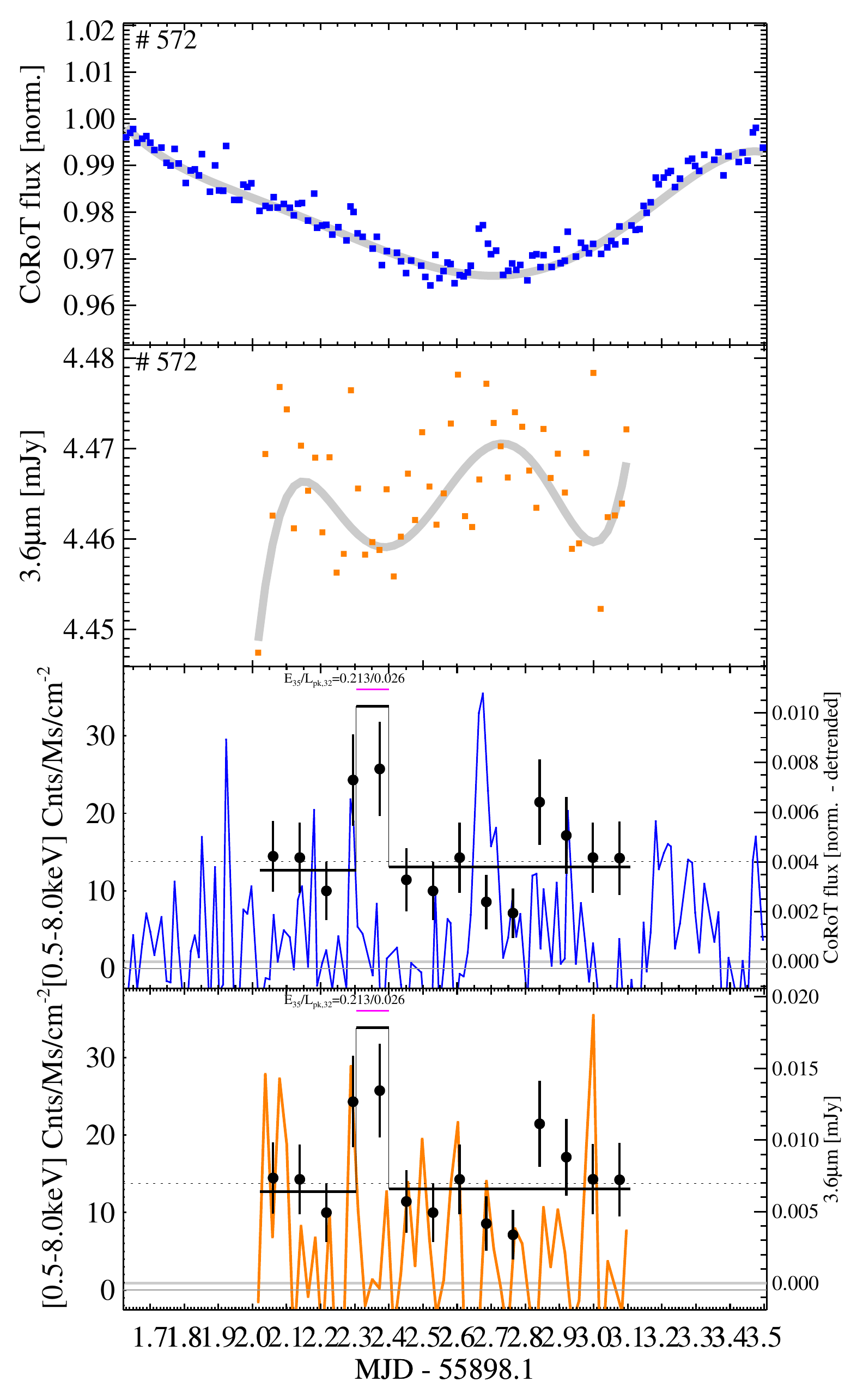}

\label{fig:}
\end{figure*}

\end{document}